\documentclass[a4paper,11pt]{article}
\usepackage{jheppub}
\usepackage[T1]{fontenc}
\usepackage{lmodern}

\usepackage{amstext}
\usepackage{amsfonts}
\usepackage{mathrsfs}
\usepackage{amsmath}
\usepackage{amssymb}
\usepackage{cases}
\usepackage{bm}
\usepackage{extarrows}
\usepackage{tensor}
\usepackage{xcolor}
\usepackage{float}
\usepackage{bbm, dsfont}
\usepackage{stmaryrd}
\usepackage{tocbasic}
\DeclareTOCStyleEntry[
beforeskip=0.3em plus 1pt,
pagenumberformat=\textbf
]{tocline}{section}
\usepackage{subcaption}
\usepackage{graphicx}
\makeatletter
\newsavebox\myboxA
\newsavebox\myboxB
\newlength\mylenA
\newcommand*\xoverline[2][0.75]{%
\sbox{\myboxA}{$\m@th#2$}%
\setbox\myboxB\null
\ht\myboxB=\ht\myboxA%
\dp\myboxB=\dp\myboxA%
\wd\myboxB=#1\wd\myboxA
\sbox\myboxB{$\m@th\overline{\copy\myboxB}$}
\setlength\mylenA{\the\wd\myboxA}
\addtolength\mylenA{-\the\wd\myboxB}%
\ifdim\wd\myboxB<\wd\myboxA%
\rlap{\hskip 0.9\mylenA\usebox\myboxB}{\usebox\myboxA}%
\else
\hskip -0.8\mylenA\rlap{\usebox\myboxA}{\hskip 0.55\mylenA\usebox\myboxB}%
\fi}
\newcommand{\xDownarrow}[1]{%
{\left\Downarrow\vbox to #1{}\right.\kern-\nulldelimiterspace}
}
\newcommand{\xuparrow}[1]{%
{\left\uparrow\vbox to #1{}\right.\kern-\nulldelimiterspace}
}

\definecolor{Green}{RGB}{199,238,206}
\newcommand{\lrb}[1]{\hs\llbracket\hs #1 \hs\rrbracket\hs}
\newcommand{\Fr}[2]{\mbox{$\frac{\,{#1}\,}{#2}$}}
\newcommand{\eqrefe}{Eq.\eqref}
\newcommand{\beq}{\begin{equation}}
\newcommand{\eeq}{\end{equation}}
\newcommand{\ba}{\begin{array}}
\newcommand{\ea}{\end{array}}
\newcommand{\beqa}{\begin{eqnarray}}
\newcommand{\eeqa}{\end{eqnarray}}
\newcommand{\beqs}{\begin{subequations}}
\newcommand{\eeqs}{\end{subequations}}
\def\dis{\displaystyle}

\newcommand{\bla}{\big\langle}
\newcommand{\bra}{\big\rangle}
\newcommand{\fr}[2]{\frac{{#1}}{#2}}
\renewcommand{\rm}{\mathrm}

\def\diag{\text{diag}}

\def\leqq{\leqslant}
\def\geqq{\geqslant}
\def\({\left(}
\def\){\right)}
\def\[{\left[}
\def\]{\right]}
\def\LB{\left\{}
\def\RB{\right\}}
\def\nn{\nonumber}
\def\pd{\partial}
\def\pp{\prime}
\def\to{\rightarrow}
\def\ito{\!\rightarrow\!}

\def\under{\underline}
\def\over{\overline}
\def\ba{\bar{a}}
\def\hA{\hat{A}}

\def\A{\mathcal{A}}

\def\CC{\mathcal{C}}

\def\D{\mathcal{D}}
\def\td{\text{d}}

\def\hF{\hat{F}}
\def\FF{\mathcal{F}}
\def\ff{\mathsf{f}}
\def\fft{\tilde{\mathsf{f}}}
\def\hg{\hat{g}}
\def\hh{\hat{h}}
\def\ii{\text{i}}
\def\KK{\mathcal{K}}
\def\KKt{\widetilde{\mathcal{K}}}

\def\La{\mathcal{L}}
\def\M{\mathcal{M}}
\def\MT{\widetilde{\mathcal{M}}}
\def\NN{\mathcal{N}}
\def\NNt{\widetilde{\mathcal{N}}}
\def\mO{\mathcal{O}}

\def\SS{\mathcal{S}}
\def\TT{\mathcal{T}}
\def\tT{\widetilde{\mathcal{T}}}
\def\uu{\mathsf{u}}

\def\vt{\tilde{v}}
\def\vv{\mathsf{v}}
\def\VV{\mathcal{V}}
\def\ww{\mathsf{w}}

\def\ZZ{\mathbb{Z}_2^{}}
\def\al{\alpha}
\def\be{\beta}

\def\Ga{\Gamma}
\def\ka{\kappa}

\def\ab{\alpha\beta}
\def\mn{\mu\nu}

\def\heta{\hat{\eta}}
\def\ep{\epsilon}
\def\vep{\varepsilon}
\def\lam{\lambda}

\def\hka{\hat{\kappa}}
\def\si{\sigma}
\def\MG{\mathbb{M}}
\def\MGn{\mathbb{M}_n^{}}
\def\MGnn{\mathbb{M}_n^2}
\def\MGm{\mathbb{M}_m^{}}
\def\MGmm{\mathbb{M}_m^2}
\def\MGl{\mathbb{M}_\ell^{}}
\def\MGll{\mathbb{M}_\ell^2}

\def\MGqq{\mathbb{M}_q^2}
\def\MGj{\mathbb{M}_j^{}}
\def\MGjj{\mathbb{M}_j^2}
\def\Mn{M_n^{}}
\def\Mnn{M_n^2}
\def\Mm{M_m^{}}
\def\Mmm{M_m^{2}}
\def\Ml{M_\ell^{}}
\def\Mll{M_\ell^{2}}
\def\Mq{M_q^{}}
\def\Mqq{M_q^{2}}
\def\Mj{M_j^{}}
\def\Mjj{M_j^{2}}
\def\vs{\vspace*{1mm}}

\def\hs{\hspace*{0.3mm}}
\def\hsx{\hspace*{0.5mm}}
\def\hsm{\hspace*{-0.3mm}}
\def\hsmx{\hspace*{-0.5mm}}

\def\ct{c_\theta^{}}
\def\st{s_\theta^{}}

\def\cct{c_\theta^2}
\def\ctt{c_{2\theta}^{}}
\def\cttt{c_{3\theta}^{}}
\def\ctf{c_{4\theta}^{}}

\def\Rxi{R_{\xi}^{}}
\def\sz{s_0^{}}
\def\tz{t_0^{}}
\def\uz{u_0^{}}
\def\rc{r_{\hspace*{-0.3mm}\rm{c}}^{}}
\def\rh{\hat{r}}
\def\LBB{\bar{L}}
\def\Mtt{\tilde{M}}
\def\sss{\widehat{\tt s}}
\def\Ch{\hat{C}}
\def\dnm{\delta_{nm}^{}}
\def\xin{\xi^{}_n}
\def\phin{\phi_n^{}}
\def\phim{\phi_m^{}}
\def\phil{\phi_\ell^{}}

\def\FT{\widetilde{\mathbb{F}}}
\def\bH{\mathbf{H}}
\def\bK{\mathbf{K}}
\def\bX{\mathbf{X}}
\def\bD{\boldsymbol{\mathcal{D}}}

\allowdisplaybreaks[4]
\def\End{\end{document}}

\title{Structure of Massive Gauge/Gravity Scattering Amplitudes, 
Equivalence Theorems, and Extended Double-Copy with Compactified Warped Space}

\author[a]{Yanfeng Hang,}
\emailAdd{yfhang@northwestern.edu}

\author[b]{~Wei-Wei Zhao,}
\emailAdd{weiwei\_zhao@sjtu.edu.cn}

\author[b,c]{~Hong-Jian He,\footnote{Corresponding author.}}
\emailAdd{hjhe@sjtu.edu.cn}

\author[b]{~Yin-Long Qiu,}
\emailAdd{silverdragon@sjtu.edu.cn}

\affiliation[a]{Department of Physics and Astronomy,\\
Northwestern University, Evanston, IL, USA}

\affiliation[b]{T.~D.~Lee Institute \&  School of Physics and Astronomy,\\
Key Laboratory for Particle Astrophysics and Cosmology,\\
Shanghai Key Laboratory for Particle Physics and Cosmology,\\
Shanghai Jiao Tong University, Shanghai, China}

\affiliation[c]{Department of Physics, 
Tsinghua University, Beijing, China; 
\\
Center for High Energy Physics, Peking University, Beijing, China
}

\abstract{\\
We study the structure of scattering amplitudes of massive Kaluza-Klein (KK) states in the compactified 5-dimensional warped gauge and gravity theories.\
We present systematic formulations of the gauge theory equivalence theorem (GAET) and the gravitational equivalence theorem (GRET) for warped KK theories in $R_\xi^{}$ gauge, where the GAET connects the scattering amplitudes of longitudinal KK gauge bosons to that of the corresponding KK Goldstone bosons and the GRET connects the scattering amplitudes of KK gravitons of helicity-zero (helicity-one) to that of the corresponding gravitational scalar (vector) KK Goldstone bosons.\ 
We analyze the structure of 3-point and 4-point scattering amplitudes of massive KK gauge bosons and of massive KK gravitons as well as of the KK Goldstone bosons at tree level.\ 
We first prove the GAET and GRET explicitly for the fundamental 3-point KK gauge/gravity scattering amplitudes.\ 
We then demonstrate that the validity of the GAET and GRET for 4-point gauge/gravity scattering amplitudes can be reduced to the validity of GAET and GRET for 3-point gauge/gravity scattering amplitudes.\ 
With these, we study the double-copy construction of KK scattering amplitudes in the warped gauge/gravity theories.\ We newly realize the double-copy for massive 3-point full gauge/gravity amplitudes under the color-kinematics correspondence and the gauge-gravity coupling correspondence at tree level, whereas we can construct extended double-copy for 4-point KK gauge/gravity scattering amplitudes to the leading order (LO) of high energy expansion.\  
We further demonstrate that this LO double-copy construction can be extended 
to $N$-point KK scattering amplitudes with $N\!\hsm \geqq\!4\hs$.\ 
\\[3mm]
JHEP 02 (2025) 001  [arXiv:2406.12713 [hep-th]].
}

\begin{document}

\maketitle

\setcounter{page}{3}

\vspace*{-6mm}
\section{\hspace*{-2.5mm}Introduction}
\label{sec:1}

Scattering amplitude is an elementary means for studying fundamental forces in nature, and
can bridge theories with experiments.\ There have been substantial efforts\,\cite{Amp-Rev} for studying  
various massless scattering amplitudes over the past four decades.\ 
Studies on massive scattering amplitudes have attracted increasing interests in recent years,
including to consider some models of massive gauge bosons and gravitons.\ 
Among these, the compactified higher dimensional gauge and gravity theories 
\`{a} la Kaluza-Klein (KK)\,\cite{KK}\cite{KK1} stand out as a fundamental approach
for {\it consistent mass generation} of the excited KK states 
of gauge bosons and gravitons.\footnote{%
The KK theories also have important impacts on the experimental side
because the KK compactification predicts an infinite tower of massive KK states 
for each known particle of the Standard Model (SM) and Generality Relativity (GR).\ 
The low-lying KK states in such KK theories have intrigued 
great phenomenological and experimental efforts over
the past two decades\,\cite{Exd}-\cite{Exd-Rev}, 
as they may provide the first signatures for the new physics beyond
the SM, ranging from the KK states of the SM particles to the spin-2 KK gravitons and possible 
dark matter candidate.}  

\vspace*{0.6mm}

Simple models of massive gauge bosons and gravitons include such as the 
massive Yang-Mills (YM) gauge theory\,\cite{mYM} and massive Fierz-Pauli (FP) gravity\,\cite{FP}
and alike\,\cite{FPRev}, which add mass terms by hand and spoil the gauge symmetry or diffeomorphism 
invariance.\ The former is plagued with nonrenormalizability and the latter further suffers 
the van\,Dam-Veltman-Zakharov (vDVZ) discontinuity in the massive graviton propagator\,\cite{vDVZ}\cite{vDVZ2}.\
Moreover, such massive gauge and gravity theories generally exhibit bad high energy behaviors in their
scattering amplitudes involving longitudinally polarized gauge bosons or gravitons.\ 
The reason behind such inconsistencies is because the nonconservation of physical degrees of freedom
prohibits a smooth massless limit.\ 
In contrast, the KK gauge and gravity theories provide a truly consistent geometric ``Higgs'' mechanism
for mass generations of KK gauge bosons and KK gravitons whose additional mass-induced physical 
degrees of freedom originate from absorbing the extra-dimensional components of the 
KK gauge bosons\,\cite{Chivukula:2001esy}\cite{Chivukula:2002ej}\cite{He:2004zr}   
and KK gravitons\,\cite{Duff}\cite{Hang:2021fmp}\cite{Hang:2022rjp}.\ 
Hence, this mass-generation mechanism naturally conserves physical degrees of freedom for KK gauge bosons
and KK gravitons, and thus has smooth massless limit.\
But this geometric mechanism for mass generations of KK gauge bosons and KK gravitons 
is realized through the compactification itself and does not invoke any additional Higgs boson 
of the conventional Higgs mechanism\,\cite{higgsM}.  


The geometric mechanism for mass generations of KK gauge bosons and KK gravitons can be 
formulated at the level of scattering $S$-matrix through the KK Equivalence Theorem (ET), 
as done previously for the 
flat 5d KK gauge theories\,\cite{Chivukula:2001esy}\cite{Chivukula:2002ej}\cite{He:2004zr} 
and the flat 5d KK gravity theories\,\cite{Hang:2021fmp}\cite{Hang:2022rjp}.\   
It is desirable to extend these KK ET formulations to {\it both} the KK gauge theories and
KK gravity theories with warped 5d compactification\,\cite{RS1}\cite{RS-gauge}\cite{RS-gauge2},
as we briefly discussed in Refs.\,\cite{He:2004zr}\cite{Hang:2022rjp} 
(cf.\ also a recent related study on the   warped gravity KK ET\,\cite{Chivukula:2023qrt}).\ 
The warped 5d KK gravity of Randall and Sundrum (RS1)\,\cite{RS1} holds a promise to resolve
the gauge hierarchy problem of the SM and to explain the vast discrepancy between the gravity force 
and other three gauge forces.\ 
It also predicts distinctive KK mass spectrum and KK signatures
which are important for the on-going experimental searches.\  

\vs 

On the other hand, it is conjectured that the scattering amplitudes of gravitons and gauge bosons 
are connected by the striking relation of double-copy, 
GR$\,=\,$(Gauge~Theory)$^2$,
which points to the common root of both the gravity force and gauge forces.\ 
It also provides a powerful means for efficiently computing the highly complex 
scattering amplitudes of spin-2 gravitons.\ 
Such double-copy relation was first found in string theories by Kawai-Lewellen-Tye (KLT)\,\cite{KLT},
which links the scattering amplitudes of massless closed strings to the
products of scattering amplitudes of massless open strings at tree level,
and further leads to the double-copy relation between the scattering amplitudes of massless gravitons 
and of massless gauge bosons in the field theory limit.\ 
With the inspiration of KLT, Bern, Carrasco and Johansson (BCJ) proposed the double-copy method 
in the quantum field theory framework via color-kinematics (CK) duality\,\cite{BCJ}\cite{BCJ-Rev}, 
which constructs the massless graviton amplitudes from the (squared) scattering amplitudes 
of massless gauge bosons.\ 
It is highly nontrivial to extend the massless double-copy formulation to the massive gauge
and gravity theories.\ In a recent series of works, the extended massive KLT-type and BCJ-type double-copies
were constructed for the compactified KK string theories\,\cite{Li:2021yfk} and for the KK
gauge/gravity field theories under toroidal compactification of flat 5d 
(without or with orbifold)\,\cite{Hang:2021fmp}\cite{Hang:2022rjp}\cite{Li:2022rel}.\ 
But, it is even more challenging to extend double-copy to the KK gauge and gravity theories with 
warped extra dimension such as the RS1.\ This is because the double-copy for $N$-point 
($N\!\!\geqq\hsm\!4$) massive KK gauge/gravity scattering amplitudes was proved  
to directly work out only for the toroidal compactification
with flat extra dimensions\,\cite{Li:2022rel}.\ 
In this work, we will demonstrate that extended double-copy can be realized
for the 3-point full scatttering amplitudes of warped KK gauge/gravity theories,  
whereas for the $4$-point warped KK gauge/gravity scattering amplitudes the extended double-copy  
can hold at the leading order (LO) of high energy expansion (which are just what we need to
formulate the ET in both the warped KK gauge and gravity theories).\footnote{%
In passing, the extensions of double-copy  
for the topologically massive Chern-Simons (CS) gauge/gravity theories
were presented in \cite{Hang:2021oso}\cite{Gonzalez:2021bes}.\  
There are other recent works in the literatue tried to extend
the massless double-copy formulation to the massive double-copies,
including the 4d massive YM theory versus Fierz-Pauli-like
gravity\,\cite{dRGT}\cite{DC-4dx1},
the spontaneously broken YM-Einstein supergravity models
with adjoint Higgs fields\,\cite{SUGRAhiggs},
the KK-inspired effective gauge theory with extra global
U(1)\,\cite{Momeni:2020hmc},
the 3d CS gauge/gravity theories
with or without supersymmetry\,\cite{3dCS-susy}-\cite{3dCS1},
and the massless QCD/gravity coupled 
to massive scalars/fermions/vectors\,\cite{Ochirov}-\cite{1911.06785}. 
}

In this work, we study the structure of scattering amplitudes of massive KK states 
in the compactified 5-dimensional warped gauge and gravity theories.\
We present systematic formulations of the gauge theory equivalence theorem (GAET) and the
gravitational equivalence theorem (GRET) for warped KK gauge/gravity theories 
up to loop level and in the $R_\xi^{}$ gauge.\ 
The GAET connects the scattering amplitudes of longitudinal KK gauge bosons to that 
of the corresponding KK Goldstone bosons, whereas the GRET connects the scattering amplitudes 
of KK gravitons of helicity-0 (helicity-1) to that of the corresponding gravitational 
scalar (vector) KK Goldstone bosons.\ 
As a key idea of this work, {\it we take the GAET of KK Yang-Mills gauge theories as the
truly fundamental ET formulation,} from which we can {\it reconstruct the GRET} of 
the corresponding KK gravity theories by using the double-copy 
at leading order (LO) of high energy expansion.\  
Then, we analyze the structure of 3-point and 4-point scattering amplitudes of massive 
KK gauge bosons and of massive KK gravitons, as well as their corresponding KK Goldstone bosons.\  
We first prove explicitly the GAET and GRET 
for the fundamental 3-point massive KK gauge/gravity scattering
amplitudes.\ We then demonstrate that the validity of the GAET and GRET for 4-point KK gauge/gravity 
scattering amplitudes can be {\it reduced to} the validity of GAET and GRET for 3-point KK gauge/gravity 
scattering amplitudes.\ With these, we further study 
the extended double-copy construction of the KK scattering 
amplitudes in the warped gauge/gravity theories.\ 
We newly realize the extended double-copy for the 
3-point full massive gauge/gravity amplitudes at tree level under proper color-kinematics correspondence
and gauge-gravity coupling correspondence without high energy expansion.\ 
Then, we construct extended double-copy for 4-point massive KK gauge/gravity scattering amplitudes 
to the leading order (LO) of high energy expansion.\
We can derive the GRET from GAET by using the LO double-copy construction for
3-point and 4-point KK scattering amplitudes.\ 
Finally, we demonstrate that this LO double-copy construction can be extended 
to the $N$-point KK gauge/gravity scattering amplitudes with $N\!\hsm \geqq\!4\hs$.\ 

\begin{figure}[t]
\centering
\includegraphics[width=13cm]{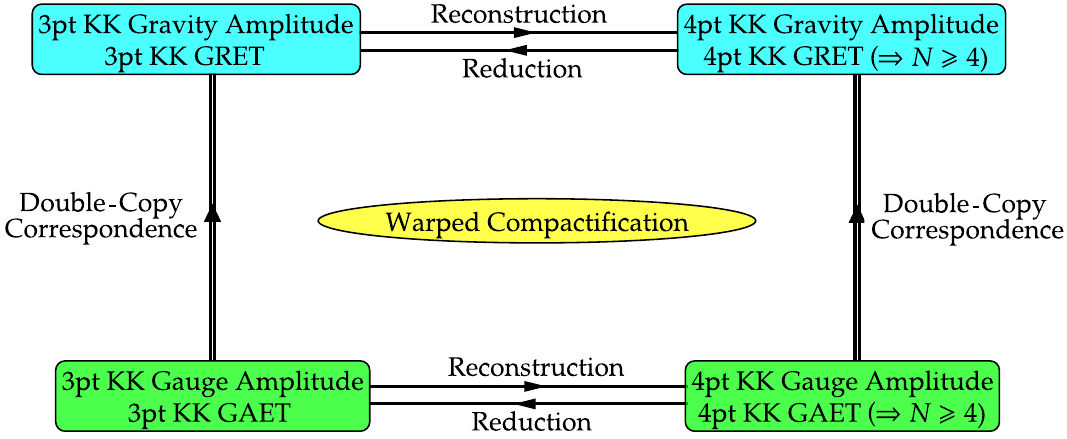}
\caption{\small Equivalence Theorem and Double-Copy Correspondences from 3-point (3pt) scattering amplitudes 
to 4-point (4pt) scattering amplitudes and from massive KK gauge scattering amplitudes to massive KK 
gravitational scattering amplitudes in the warped gauge and gravity theories.}
\label{fig:DC-ET}
\label{fig:1}
\vspace*{-1mm}
\end{figure}

\subsection*{Outline and Summary}

To make the presentation more transpareent for reading, we provide a summary of our main logic and key results 
as shown in the schematic plot of Fig.\,\ref{fig:1}.\
\begin{itemize}

\item 
Fig.\,\ref{fig:1} shows the correspondence of the Equivalence Theorems from the GAET to the GRET, 
the correspondence of Double-Copies from the 3-point KK scattering amplitudes to the 4-point KK scattering amplitudes, and the correspondence from the massive KK gauge scattering amplitudes to the massive KK gravitational scattering amplitudes.\ 

\item 
In the lower-left corner of Fig.\,\ref{fig:1}, we start from analyzing the most fundamental 3-point KK gauge boson scattering amplitudes  and establish the GAET between 3-point scattering amplitudes of longitudinal KK gauge bosons and of the corresponding KK Goldstone bosons to the LO of high energy expansion.\  
This analysis will be presented in Section\,\ref{sec:3.1.1}.\ 

\item 
Then, along the bottom horizontal arrow we reach the lower-right corner where we compute the 
4-point scattering amplitudes of longitudinal KK gauge bosons and of the KK Goldstone bosons, as well as 
establishing the GAET for these 4-point amplitudes at the LO of high energy exoansion.\ Here we use 
the word ``Reconstruction'' or ``Reduction'' to indicate that the validity of the GAET for 
4-point KK gauge/Goldstone boson amplitudes can be reduced to (proved from) the validity 
of the GAET for 3-point KK gauge/Goldstone boson amplitudes.\
This analysis will be presented in Section\,\ref{sec:3.2.1}.\ 

\item 
Next, in Fig.\,\ref{fig:1}, we can go from the lower-left corner to the upper-left corner by the double-copy construction
along the left vertical arrow and from the 3-point KK gauge boson amplitudes to the 3-point 
KK graviton amplitudes, as well as from the LO 3-point KK Goldstone boson amplitudes to the LO
3-point gravitational KK Goldstone amplitudes.\ 
This analysis will be presented in Section\,\ref{sec:4.1}.\

\item 
Then, along the top horizontal arrow we go from the upper-left corner for 3-point KK graviton amplitudes 
to reach the upper-right corner where the 4-point KK graviton amplitudes and 
gravitational KK Goldstone amplitudes are computed;
and thus we establish the GRET for these 4-point gravitational KK amplitudes.\   
Here we use the word ``Reconstruction'' or ``Reduction''  
to indicate that along the top horizontal arrow 
the validity of the GRET for 4-point KK graviton/Goldstone boson amplitudes can be reduced to (proved from) 
the validity of the GRET for 3-point KK graviton/Goldstone boson amplitudes.\ 
These analyses will be presented in Sections\,\ref{sec:3.1.2} and \ref{sec:3.2.2}.\ 

\item 
Finally, we can go from the lower-right corner to the upper-right corner by the double-copy construction
along the right vertical arrow and from the LO 4-point KK gauge (Goldstone) boson amplitudes to the LO 4-point 
KK graviton (Goldstone) amplitudes.\
This analysis will be presented in Section\,\ref{sec:4.2}.\ 
Moreover, we will extend the LO double-copy to the $N$-point 
KK gauge/gravity scattering amplitudes with $N\!\hsm \geqq\!4\hs$
in Section\,\ref{sec:4.3}, and we give the general prescriptions for
the warped doubel-copy at LO in Section\,\ref{sec:4.4}.\ 
  
\end{itemize}

\vspace*{-12mm}

The rest of this paper is organized as follows.\ 
In Section\,\ref{sec:2}, after introducing the warped 5d compactification 
with $S^1\!/\ZZ$ orbifold (Section\,\ref{sec:2.1}),
we present the formulation of GAET in the warped KK gauge theory (Section\,\ref{sec:2.2}),
and then the formulations of the GRET of type-I,II in the warped KK gravity theory (Section\,\ref{sec:2.3}).\ 
In Section\,\ref{sec:3}, we study the structure of massive scattering amplitudes from the warped GAET and GRET.\ 
These include to first compute the fundamental 3-point scattering amplitudes 
of KK gauge bosons (KK Goldstone bosons) 
and of KK gravitons (KK gravitational Goldstone bosons), 
and to establish the GAET and GRET for 3-point
massive KK amplitudes at the leading order of high energy expansion (Section\,\ref{sec:3.1}).\
Then, we present the 4-point scattering amplitudes of KK gauge bosons (KK Goldstone bosons) 
and of KK gravitons (KK gravitational Goldstone bosons), and establish the GAET and GRET for 4-point
massive KK amplitudes at the leading order of high energy expansion (Section\,\ref{sec:3.2}).\
Our key point is to demonstrate that the validity of the GRET 
for 4-point massive KK amplitude can be reduced to 
(proved from) the validity of the GAET for basic 3-point massive KK amplitudes.\ 
In Section\,\ref{sec:4}, we study the double-copy construction of the massive KK scattering 
amplitudes in the warped gauge and gravity theories.\ We realize the extended double-copy for the 
3-point full gauge/gravity amplitudes at tree level under the proper correspondence of color-kinematics 
and the gauge-gravity coupling correspondence without high energy expansion, 
whereas the extended double-copy can be constructed for 
the 4-point KK gauge/gravity amplitudes to the LO of high energy expansion.\
We further study the $N$-point KK gauge/gravity scattering amplitudes (with $\!N\!\! \geqq\!4\hs$) 
at the leading order and their LO double-copy construction.\
Finally, we conclude in Section\,\ref{sec:5}.\ 
The four Appendices present supporting materials for our analyses in the main text.\
Among these, Appendix\,\ref{app:A} provides the kinematics of three and four KK-particle scattering processes;
Appendix\,\ref{app:B} presents the Becchi-Rouet-Stora-Tyutin (BRST) quantization 
for warped 5d KK gravity theory and the detailed derivations of the GRET of type-I and type-II;
Appendix\,\ref{app:C} derives the Feynman Rules for warped KK gauge and gravity theories
as needed for the analyses of Section\,\ref{sec:3};
Appendix\,\ref{app:D} proves the relevant 3-point and 4-point warped identities 
for KK masses and KK couplings.\ 
Appendix\,\ref{app:E} explicitly computes the inelastic longitudinal KK graviton scattering amplitude of $nn\ito mm$
and demonstrates the nontrivial large energy cancellations of $E^{10}\!\ito E^2$.\

\vspace*{1mm}
\section{\hspace*{-2.5mm}Formulation of Gauge and Gravitational Equivalence Theorems\\ 
\hspace*{-2.5mm}with Compactified Warped Space}
\label{sec:2}
\vspace*{1mm}

In this section, we formulate the gauge and gravitational equivalence theorems for the 
compactified 5d warped gauge and gravity theories.\ In section\,\ref{sec:2.1},
we present the warped 5d compactification for RS1.\
In section\,\ref{sec:2.2}, we formulate the gauge theory ET (GAET) for the warped KK gauge theory.\
In section\,\ref{sec:2.3}, we first present the BRST quantization (Sec.\,\ref{sec:2.3.1}), and
then formulate the gravitational ET (GRET) of type-I (Sec.\,\ref{sec:2.3.3}) and the GRET of type-II
(Sec.\,\ref{sec:2.3.2}) for the warped KK gravity theory.\

\subsection{\hspace*{-2.5mm}Warped 5d Compactification}
\label{sec:2.1}
\vspace*{1.5mm}

We consider the warped 5d compactification \`{a} la Randall and Sundrum 
(RS1)\,\cite{RS1},  
where a warped background metric generates
weak scale from Planck scale through an exponential suppression.\ 
The background metric involves a five-dimensional non-factorizable warped geometry, derived from a slice of $\rm{AdS}_5$ spacetime.\  
This framework is realized with two 3-branes situated at fixed points of 
the $S^1\!/\ZZ$ orbifold compactification, where one 3-brane is the 
UV brane (Planck brane) at the fixed point $\hs y\!=\!0\hs$ 
and another 3-brane is IR brane (TeV brane) at the fixed point 
$\hs y\!=\!\LBB\hs$.\ Here the physical coordinate of the fifth dimension
is denoted as $y$, so we have $\,y\!\in\![0,\LBB]$ with  
$\LBB\!=\hsm\pi\rc\hs$ denoting the size of fifth dimension 
and $\rc$ denoting the compactification radius.\

\vs

The 5-dimensional warped background metric in the physical coordinate takes the following nonfactorizable form:
\begin{equation}
\td s^2 = {e^{-2ky}\eta_{\mn}^{}}\td x^\mu \td x^\nu + \td y^2 ,
\end{equation}
where $e^{-2ky}$ is the warp factor and $\hs k\!=\!\!\sqrt{-\Lambda/6\,}\,$ 
is the AdS curvature scale and around the order of Planck mass.\  
(The parameter $\Lambda\,$ denotes a negative bulk cosmological constant,
as will be defined in Sec.\,\ref{sec:2.3.1}.)\
Then, we can transform the physical coordinates $(x^\mu\hsm ,y)$ 
into the conformal coordinate $\hs (x^\mu\hsm , z)$ by 
imposing the relation
$\hs \td y= e^{A(z)}\td z\hs$.\ 
Thus, we have the conformally flat 5-dimensional spacetime metric: 
\begin{equation}
\label{eq:ds2-CFz}
\td s^2\hs =\, e^{2A(z)} 
\big(\eta_{\mn}\td x^\mu \td x^\nu\! + \td z^2\hs\big)\hs ,
\end{equation}
where we have set $A(z)\!=\! -ky\hs$.\ By imposing the relation
$\hs \td y\!=\! e^{A(z)}\td z\hs$, we deduce $\,y\!=\!(1\hsm /k)\ln(1\!+\!kz)$.\  
So we can further resolve the function $A(z)$ in conformal coordinates as follows\,\cite{Exd-Rev}:
\begin{equation}
A(z) = -\ln(1+kz) \hs.
\end{equation} 
In the conformal coordinates, we have 
$z\!\in\![0,{L}]$ with ${L}\equiv [e^{\hs k\LBB}\!-\!1]/k\,$,
which corresponds to the physical coordinate $\,y\!\in\![0,\LBB]$ with  
$\LBB\!=\hsm\pi\rc\hs$.\

\vs

The \eqrefe{eq:ds2-CFz} shows that the 5-dimensional warped metric tensor
$\hs\hg_{MN}^{}$ is conformally flat and can be expressed as
\begin{equation}
\hg_{MN}^{} =\hs e^{2A(z)}\heta_{MN}^{} \,,
\end{equation}
where the flat 5d Minkowski metric $\hs\heta_{MN}^{}$ takes the form:
\begin{equation}
\heta_{MN}^{} = \heta^{MN}= \diag(-1,1,1,1,1) \hs .
\end{equation}
For clarity of notations, we denote the 5d Lorentz indices by 
uppercase Latin letters (such as $M,N\!=\!\mu,5$) and the 
4d Lorentz indices by the lowercase Greek letters 
(such as $\mu,\nu = 0, 1, 2, 3$).\

\vs

In the next subsections, we will consider the RS1 under $S^1\!/\ZZ$
compactification for both warped 5d Yang-Mills gauge theory and 
the warped 5d GR theory.\ With these, we formulate the KK equivalence 
theorems for the warped 5d Yang-Mills gauge theory and 
the warped 5d GR theory.\

\vspace*{1.5mm}
\subsection{\hspace*{-2.5mm}Formulating GAET for Warped KK Gauge Theory}
\label{sec:2.2}
\vspace*{1mm}

The warped bulk gauge theories in RS1 were discussed 
in the literature\,\cite{RS-gauge}\cite{RS-gauge2}\cite{Randall:2001gb}.\ 
We consider the Yang-Mills Lagrangian in warped 5d spacetime:
\begin{equation}
\label{eq:5dYMLa-1}
\hat{\La}_{\rm{YM}}^{} \,=\, \sqrt{-\hg\,}\hs
\Big(\hsm\! -\!\frac{1}{\,4\,}\hg^{MP} \hg^{NQ}\hF_{MN}^{a} \hF_{PQ}^{a} 
\Big),
\end{equation}
where the square root of the determinant of the warped 
5d background metric tensor $\hg_{MN}^{}$ is given by
$\sqrt{-\hg\,}=\exp[5A(z)]\hs$.\ 
The 5d field strength tensor $\hF_{MN}^{a}$ takes the form: 
\begin{equation}
\hF_{MN}^{a} = \pd_M^{} \hA_N^{a} - \pd_N^{} \hA_M^{a} + g_5^{} f^{abc} \hA_M^{b}\hA_N^{c} \,,
\end{equation}
where $(a,b,c)$ denote the gauge-group indices, $f^{abc}$ denotes 
the gauge-group structure constant, and $g_5^{}$ is the 5d gauge coupling constant having mass-dimension $-\frac{1}{\,2\,}\hs$.  

\vs

We expand the 5d Lagrangian \eqref{eq:5dYMLa-1} to the quadratic order 
as presented in Eq.\eqref{Aeq:5dYMLa-2} of Appendix\,\ref{appx:B.1}.\ 
By inspecting Eq.\eqref{Aeq:5dYMLa-2}, 
we can remove the mixing term between the 5d components $\hA_\mu^a$ and $\hA_5^a$ 
by imposing the following $R_\xi^{}$ gauge-fixing term:
\begin{equation}
\label{eq:LGF-gauge-5d}
\hat{\La}_{\rm{GF}}^{} = 
-\frac{~e^{A(z)}\hs}{2\hs\xi}\big(\hat\FF^a\big)^{\!2},
~~~~~
\hat\FF^a=\pd^\mu\hsm \hA_\mu^a \!+\hsm \xi ( A'\hsm\!+\hsm\pd_z)\hA_5^a \,,
\end{equation}
where $\hs\xi\hs$ is the gauge-fixing parameter and 
$A'\!\hsm =\hsm\pd_z A(z)\hs$.\ 
We note that the gauge-fixing term \eqref{eq:LGF-gauge-5d} 
contains an $\hA_\mu^a$-$\hA_5^a$ mixing term which precisely cancels
the corresponding mixing term in the original Lagrangian 
\eqref{Aeq:5dYMLa-2} upon integration by part.\
The fifth component $\hA_5^a$ will be identified as the would-be Goldstone boson 
via a geometric ``Higgs'' mechanism of the KK 
compactification\,\cite{Chivukula:2001esy}\cite{He:2004zr}.
Using Eqs.\eqref{Aeq:5dYMLa-2} and \eqref{eq:LGF-gauge-5d}, we derive the 5d 
equations of motion (EOM) for the free fields $(\hA_\mu^a,\hA_5^a$)
as in Eqs.\eqref{Aeq:5dEOM-Amu}-\eqref{Aeq:5dEOM-A5}.\  

\vs 

Next, we compactify the warped 5d space under $S^1\!/\ZZ$
orbifold, which corresponds to compactifying the fifth-dimension to 
a line segment $\hs y\!\in\![0,\LBB]\hs$ or
$\hs z\!\in\![0,L]\hs$,  
with $\LBB\!=\!\pi\hs \rc$ and ${L}\equiv (e^{\hs k\LBB}\hsm\!-\!1)\hsm /k\,$,
where $\rc$ denotes the compactification radius.\ 
We impose the gauge-covariant boundary conditions
on the 5d gauge field strength, 
$\hat{F}_{M5}^a\!=\!\hat{F}_{5M}^a\!=\hsm 0\,$ at $z=0,L\hs$.\ 
This leads to the Neumann and Dirichlet boundary conditions on
the 5d gauge field components 
$\hA_{\mu}^a$ and $\hA_5^a$ as follows:
\begin{equation}
\label{eq:BC-Amu-A5}
\partial_z^{}\hA^a_{\mu}({x}, z)\Big|_{z=0,L}^{}\! =\hs 0\,,
\hspace*{10mm}
\hA^a_{5}({x}, z)\Big|_{z=0,L}^{}\! =\hs 0\,.
\end{equation}
Then, we can make KK expansions for the 5d gauge fields 
$\hA_{\mu}^a$ and $\hA_5^a$ 
in terms of the eigenfunctions of warped 5d space:
\beqs
\label{eq:AA5Exp}
\begin{align}
\hA^{a\mu}(x, z) &\,=\,
\frac{1}{\sqrt{L\,}\,}
\sum_{n=0}^{\infty} A^{a\mu}_{n}({x})\,\ff_n^{}(z) \,,
\\[1mm]
\hA^{a5} (x, z) &\,= \,\frac{1}{\sqrt{L\,}\,}
\sum_{n=1}^{\infty} A^{a5}_{n}({x}) \,\fft_n^{}(z)\,.
\end{align}
\eeqs
Substituting the KK expansions \eqref{eq:AA5Exp} into Eq.\eqref{eq:BC-Amu-A5}, 
we obtain the boundary conditions imposed on  $(\ff_n^{},\fft_n^{})$:
\begin{equation}
\label{eq:BC-YM}
\pd_z^{} \ff_n^{}(z)\big|_{z=0,L}=0\,,  \qquad
\fft_n^{}(z)\big|_{z=0,L}=0 \,. 
\end{equation}
The 5d wavefunctions $(\ff_n^{},\fft_n^{})$ 
also obey the orthonormal conditions shown in Eq.\eqref{Aeq:Normalize-YM}.

\vs 

Under the KK expansions for the 5d gauge fields \eqref{eq:AA5Exp}, 
the 4d effective KK Lagrangian is derived by integrating over 
the fifth coordinate $z$ on the interval $[\hs0,L\hs]$.\
The quadratic KK Yang-Mills Lagrangian is given by 
\begin{equation}
\La_{\rm{YM}}^{(2)} = \!\int_0^{L}\!\!\hsm\td z \, 
\hat{\La}_{\rm{YM}}^{(2)}
=-\frac{1}{\,4\,}\!\sum_{n=0}^{\infty}\!
\(\!\pd_{\mu}A_{\nu}^{an}\!-\!\pd_{\nu} A_{\mu}^{an}\!\)^{\!2}
\!-\!\frac{1}{\,2\,}\!\sum_{n=1}^{\infty}\!
\(\!\pd_\mu A_5^{an}\!+\!\Mn A_\mu^{an}\!\)^{\!2}.
\end{equation}
Then, using Eq.\eqref{eq:LGF-gauge-5d} and making the KK expansions \eqref{eq:AA5Exp}, 
we can derive the following $R_{\xi}^{}$ gauge-fixing term for the KK Lagrangian:
\\[-4mm]
\beqs 
\label{eq:LGF-gauge-4d}
\begin{align}
\label{eq:LGF1-gauge-4d}
\La_{\rm{GF}}^{} & \,=\, \int_0^{L}\!\!\td z \, \hat{\La}_{\rm{GF}}^{} 
\,=\sum_{n=0}^{\infty} \frac{-1}{\,2\hs\xi_n^{}} (\FF^{a}_n)^2 \hs,
\\[1mm]
\FF^{a}_n & \,=\,  \pd_\mu^{} A^{a\mu}_{n}\!+\hsm\xin \Mn A_n^{a5}
= \mathbf{K}_{n}^T\mathbf{A}_n^a \hs,\label{eq:LGF2-gauge-4d}
\end{align}
\eeqs 
in agreement with Refs.\cite{Chivukula:2001esy}\cite{He:2004zr}, where
$\,\mathbf{K}_{n}^{}\!=\!(\pd_\mu^{},\hs\xi_n^{}\hsm \Mn)^T$ and 
$\mathbf{A}_n^a \!=\!(A_n^{a\mu},\hs A_n^{a5})^T$.\ 
In the above gauge-fixing term, 
we introduce a gauge-fixing parameter $\xin$ for each KK level-$n\hs$
and use $\Mn$ to denote the KK gauge boson mass of level-$n$.\ 
By substituting the KK expansions \eqref{eq:AA5Exp} into the 5d EOM
\eqref{Aeq:5dEOM-A} and using the 4d EOM \eqref{Aeq:4d-EOM-AmuA5} for free KK gauge fields,
we can derive the EOMs for the 5d eigenfunctions 
as in Eqs.\eqref{Aeq:EOM-fn-Amu}-\eqref{Aeq:EOM-fnT-A5}.\ 
The KK gauge boson mass $\Mn$ is determined as in Eq.\eqref{Aeq:YM-Mn} by solving the roots 
of the eigenvalue equations \eqref{Aeq:EOM-fn-Amu}-\eqref{Aeq:EOM-fnT-A5}.\ 

\vs

Then, corresponding to the 5d gauge-fixing term \eqref{eq:LGF-gauge-5d}, we have the
following 5d Faddeev-Popov ghost term:
\begin{equation}
\label{eq:FP-gauge-5d}
\hat{\La}_{\rm{FP}}^{} = e^{A(z)}\hat{\bar{c}}^a \widehat{\tt s}\hat{\FF}^a \,,
\end{equation}
where $\sss$ is the Becchi-Rouet-Stora-Tyutin (BRST) operator\,\cite{BRST} 
and the 5d BRST transformations take the form:
\begin{equation}
\label{eq:BRST-5d}
\widehat{\tt s}\hat{A}^{aN}\!= D_b^{aN}\!(\hat{A})\hat{c}^b, 
~~~~~
\sss\hat{c}^a \!= \Fr{1}{2}\hat{g}_5^{}f^{abc}\hat{c}^b\hat{c}^c,
~~~~~
\sss\hat{\bar{c}}^a \!= -\xi^{-1}\hsm\hat\FF^a .
\end{equation}
Accordingly, we make KK expansions for 5d ghost and anti-ghost fields as follows: 
%
\beqs
\label{eq:ghost-exp}
\begin{align}
\hat{c}^a(x, z) &\,=\,
\frac{1}{\sqrt{L\,}\,}
\sum_{n=0}^{\infty} c^{a}_{n}(x)\hs\ff_n^{}(z) \,,
\\
\hat{\bar{c}}^a (x,z) &\,= \,\frac{1}{\sqrt{L\,}\,}
\sum_{n=1}^{\infty}\bar{c}^a_{n}(x) \hs\fft_n^{}(z)\,.
\end{align}
\eeqs
With these, we derive the BRST transformations for the KK ghost fields as in Eqs.\eqref{Aeq:BRST-KK}-\eqref{eq:DDD-KKghosts} of Appendix\,\ref{appx:B.1}.

\vs

Using these, we can further derive a BRST identity for the propagator of the
gauge fields $\mathbf{A}_n^a \!=\!(A_n^{a\mu},\hs A_n^{a5})^T$ as follows:
\begin{equation}
\label{eq:KD=Omega}
\mathbf{K}_n^T \bD_{nm}^{ab}(k) =
-{\bX}_{nm}^{ab}(k)^T \,,
\end{equation}
where we introduce the notations, 
\begin{align}
\bD_{nm}^{ab}(k) &=
\bla 0|T\mathbf{A}_n^a{\mathbf{A}_m^b}^T|0\bra \hs,~~~~~
\SS_{nm}^{}(k)\delta^{ab} =
\bla 0|Tc_n^a{\bar{c}_m^b}|0\bra \hs,
\nn\\[1mm]
{\bX}_{nm}^{ab}(k) &=
\under{\bX}_{mj}^{ab}(k)\SS_{jn}^{}(k)  \,,
\nn\\[-1mm]
\under{\bX}_{mj}^{ab}(k) &=
\bla 0|T\sss\mathbf{A}_m^b |\bar{c}_j^a\bra
=\!\(\!\begin{matrix} 
\bla 0|T\sss A_m^{b\mu} |\bar{c}_j^a\bra
\\[1mm]
\bla 0|T\sss A_m^{b5} |\bar{c}_j^a\bra
\end{matrix}\!\) 
\equiv\!\(\!\begin{matrix} 	
-\ii k^\mu\delta^{ab}
[\delta_{jm}^{}\!+\!\Delta^a_{jm}(k^2)]
\\[1mm]
-\Mm\delta^{ab}
[\delta_{jm}^{}\!+\!\tilde\Delta^a_{jm}(k^2)]
\end{matrix}\) \!,
\end{align}
with $\Delta^a_{jm}(k^2)$ and $\tilde{\Delta}^a_{jm}(k^2)$
being loop-level quantities. 

\vs

In the following, we derive the 
KK gauge boson equivalence theorem (KK GAET)
for the compactified warped 5d Yang-Mills gauge theory.\  
The KK GAET was discussed\,\cite{He:2004zr} 
for the compactified 5d Yang-Mills gauge theories with a general metric 
in connection to its deconstruction formulation,
as an extension of the formulations in the flat 
5d KK gauge theories\,\cite{Chivukula:2001esy}\cite{Chivukula:2002ej}.\ 
Our current derivation differs from \cite{He:2004zr}
since we focus on the warped RS1 and choose the conformal coordinate frame of $(x^\mu,z)$
instead of the physical coordinate frame of $(x^\mu, y)$.\ 
Besides, we will further study explicitly the three-point and four-point KK gauge (Goldstone) boson 
scattering amplitudes in Section\,\ref{sec:3} and their double-copy in Section\,\ref{sec:4}.\  

Due to the 5d gauge invariance, we can derive a Slavnov-Taylor (ST) type of identity:
\beq
\label{eq:5D-ET1}
\bla 0|T
\hat\FF^{\hs a_1^{~}}(\hat{x}_1^{})
\hat\FF^{\hs a_2^{}}(\hat{x}_2^{})
\cdots
\hat\FF^{\hs a_{\!N}^{}}(\hat{x}_{\!N}^{}){\hat{\tilde{\Phi}}} |0\bra \hs =\, 0\,,
\eeq
where $\hat{\tilde{\Phi}}$ denotes any possible external physical states in 5d spacetime.\ 
This is analogous to the ST identity (with external lines being gauge-fixing functions) 
in the 4d SM having spontaneous electroweak gauge symmetry breaking\,\cite{He:1993yd}.\ 
Then, after making the KK expansion, we obtain the following ST identities for
the external KK gauge-fixing functions:
\beq
\label{eq:5D-ETI-KK}
\bla 0|T
\FF^{\hs a_1^{}}_{n_1^{}}(x_1^{}) 
\FF^{\hs a_2^{}}_{n_2^{}}(x_2^{}) \cdots
\FF^{\hs a_{\!N}^{}}_{n_{\!N}^{}}(x_{\!N}^{}){\tilde{\Phi}} |0\bra \,=\, 0\,.
\eeq
Using the BRST identity \eqref{eq:KD=Omega}, 
we can amputate each external $\FF^{an}$ function
in the momentum space and obtain 
a corresponding identity for the $S$-matrix amplitude:
\beq
\label{eq:5D-ETI2-KK}
\TT [\under{\FF}^{\hs a_1^{}}_{n_1^{}}\hsm (k_1^{}),\hs 
\under{\FF}^{\hs a_2^{}}_{n_2^{}}\hsm (k_2^{}),\hs \cdots\!,\hs
\under{\FF}^{\hs a_{\!N}^{}}_{n_{\!N}^{}}\hsm (k_{N}^{}),{\Phi} ] \,=\, 0\,,
\eeq
where each external momentum $k_j^{}$ ($j\!=\!1,2,\cdots$)
is set on-shell with $k_j^2\!=\!\!-M_{n_j}^2$,
each $\under{\FF}^{\hs a_{j}^{}}_{n_j}\!(k_j^{})$ represents 
the amputated external gauge-fixing function, and ${\Phi}$
denotes any possible physical states 
after the LSZ (Lehmann-Symanzik-Zimmermann) amputation.\ 
We can express the amputated gauge-fixing function as follows:
\beqs 
\begin{align}
\under{\FF}^{\hs a}_n\hsm &= 
k_\mu^{} A^{a\mu}_{n}
\!-\hsm \Mn \Ch^a_{nm} A_m^{a5}
=  \Mn (A^{an}_S\!-\hsm\Ch^a_{nm} A_5^{am}) 
= \Mn (A_L^{an} \!-\hsm{\mathbb{Q}}^a_n )\hs,
\\[1mm]
{\mathbb{Q}}^{\hs a}_n &= \Ch^a_{nm}A_5^{am}\!+v_n^a\hs,~~~
A_S^{an}\hsm =\ep_S^{\mu}A_{\mu}^{an} ,~~~
A_L^{an}\hsm =\ep_L^{\mu}A_{\mu}^{an} ,~~~
v^{a}_n\hsm =v_n^{\mu}A_{\mu}^{an} , 
\end{align}
\eeqs
with notations defined as 
\beqs 
\begin{align} 
\Ch^a_{nm} &= \ii (\Mm/\Mn)
\big[(\mathbf{1}\!+\!\boldsymbol{\tilde\Delta}^{\!a}(k^2))
(\mathbf{1}\!+\!\boldsymbol{\Delta}^{\!a}(k^2))^{-1}\big]_{mn} 
= \ii\hs\delta_{nm}^{}\!+\hsm O(\rm{loop}) \hs,
\\[1mm]
\ep_S^{\mu} &= k^{\mu}\!/\Mn
\!=\ep_L^{\mu} \!- v_n^{\mu} \hs,~~~
v_n^{\mu}\!=O(\Mn/E_n)\hs,
\label{eq:epL-epS-v}
\end{align}
\eeqs 
where the matrix form is used such that 
$\big(\boldsymbol{\tilde\Delta}^{\!a}\big)_{\!jj'}\!\equiv\hsm\tilde\Delta^{\!a}_{jj'}$
and
$\big(\boldsymbol{\Delta}^{\!a}\big)_{\!jj'}\!\equiv\hsm\Delta^{\!a}_{jj'}\hs$.\ 
In the above, $\ep_L^{\mu}$ and $\ep_S^{\mu}$ denote the KK gauge boson's 
longitudinal polarization vector and scalar polarization vector
respectively.\ 
Thus, from the identity \eqref{eq:5D-ETI2-KK} we deduce
the KK gauge equivalence theorem (GAET) identity: 
\beqs 
\label{eq:5D-ETI3-ALQ}
\begin{align}
\label{eq:5D-ETI3-ALQ1}
&\TT[\hs A_{L\hs n_1^{}}^{a_1^{}},\hs\cdots\!,\hs 
A_{L\hs n_{\!N}^{}}^{a_{N}^{}}\!,{\Phi}\hs ]
= 
\TT[\hs{\mathbb{Q}}^{\hs a_1^{}}_{n_1^{}},\hs\cdots\!,\hs 
{\mathbb{Q}}_{n_{\!N}^{}}^{\hs a_{N}^{}}\!,{\Phi}\hs]
\\[1mm]
\label{eq:5D-ETI3-ALA5Tv}
& \hspace*{3.6cm}
= \Ch_{\rm{mod}}^{n_j^{}m_j^{}}
\TT[\,A_5^{a_1^{}m_1^{}}\!,\hs\cdots\!,\hs A_5^{a_N^{}m_N^{}}\!,{\Phi}\,] 
+ \TT_v^{}\,,
\\
\label{eq:KK-ET1-Tv}
& \TT_v^{} = \sum_{\ell=1}^N \tilde{C}_{\!\rm{mod},\ell}^{n_j^{}m_j^{}}\hs
\TT[\,v^{a_1^{}n_1^{}},\cdots\!,v^{a_\ell^{}n_\ell^{}},
A_5^{a_{\ell+1}^{}m_{\ell+1}^{}},\cdots\!,A_5^{a_N^{}m_N^{}},{\Phi}\,]\,,
\end{align}
\eeqs 
where the multiplicative modification factors are given by
\beqs 
\label{eq:5D-GAET-Cmod}
\begin{align}
\Ch_{\rm{mod}}^{n_j^{}m_j^{}} &=
\Ch^{a_1^{}}_{n_1^{}m_1^{}}\!\cdots\Ch^{a_N^{}}_{n_{\!N}^{}m_{\!N}^{}} 
=\,\ii^N\delta_{n_1^{}m_1^{}}^{}\!\!\cdots\hs\delta_{n_{\!N}^{}m_{\!N}^{}}^{} 
\!+O(\rm{loop})\hs, 
\\
\tilde{C}_{\!\rm{mod},\ell}^{n_j^{}m_j^{}} &=	
\Ch^{a_{\ell+1}^{}}_{n_{\ell+1}^{}m_{\ell+1}^{}}\!
\cdots\Ch^{a_N^{}}_{n_{\!N}^{}m_{\!N}^{}} 
=\,\ii^{N-\ell}
\delta_{n_{\ell+1}^{}m_{\ell+1}^{}}^{}\!\!\cdots\hs\delta_{n_{\!N}^{}m_{\!N}^{}}^{} 
\!+O(\rm{loop}) \hs.
\end{align}
\eeqs
Such radiative modification factors also exist in the ET formulation 
for the 4d electroweak gauge theory of the SM\,\cite{He:1993yd}-\cite{He:1997zm}.\
Note that the residual term $\TT_v^{}$ in the KK-ET identity is suppressed
under high energy expansion due to 
$v_n^{\mu}\!=O(\Mn/E_n^{})$.\  
With these, we derive the KK Equivalence Theorem for warped gauge theory
(KK GAET) for the high energy scattering: 
\begin{equation}
\label{eq:KK-ET1-N}
\TT[\,A_L^{a_1^{}n_1^{}},\hs\cdots\!,\hs A_L^{a_N^{}n_N^{}}\!,{\Phi}\,] 
\hs =\, \Ch_{\rm{mod}}^{n_j^{}m_j^{}}
\TT[\,A_5^{a_1^{}m_1^{}}\!,\hs\cdots\!,\hs A_5^{a_N^{}m_N^{}}\!,{\Phi}\,] 
+ O(M_n^{}/E_n^{})\hs,
\end{equation}
where $\Phi$ denotes any other external physical state(s) 
{after LSZ amputation}, and at tree-level
the multiplicative factor $\Ch_{\rm{mod}}^{n_j^{}m_j^{}}$
just reduces to a simple factor
$(\ii^N\delta_{n_1^{}m_1^{}}^{}\!\!\cdots\hs
\delta_{n_{\!N}^{}m_{\!N}^{}}^{})\hs$.\

\vspace*{3mm}
\subsection{\hspace*{-2.5mm}Formulating GRET for Warped KK Gravity Theory}
\vspace*{2mm}
\label{sec:2.3}

In this subsection, we first present the BRST quantization 
for the warped KK gravity theory in Section\,\ref{sec:2.3.1}.\ 
Then, we formulate the warped Gravitational Equivalence Theorem (GRET)
for the KK gravitons with helicity\,$\pm1$ (called GRET of type-II) in Section\,\ref{sec:2.3.2} 
and for the longitudinal KK gravitons with helicity-0 (called GRET of type-I)   
in Section\,\ref{sec:2.3.3}, respectively.\  
Using the GRET and our generalized energy-power counting method\,\cite{Hang:2021fmp}, 
we further prove the different large energy cancellations for the
scattering amplitudes of KK gravitons with helicity-0 and with
helicity\,$\pm1$ around the end of this subsection.\   


\subsubsection{\hspace*{-2.5mm}Warped KK Gravity Theory with BRST Quantization}
\vspace*{1.5mm}
\label{sec:2.3.1}

In this subsection, we consider the 5d RS1 with the fifth spatial dimension 
compactified on an interval $[0,\hs{L}]$.\ 
The setup of RS1 contains warped 5d Einstein-Hilbert action and is 
compactified on a 5d interval $[0,\hs{L}]$ with the two 3-branes 
located at the boundaries.\ 
The relevant part of the warped 5d Lagrangian can be written as 
\begin{equation}
\label{eq:LEH-5d}
\hat\La_{\rm{RS}}^{} =\frac{1}{2}\sqrt{-\hg\,}\hat{M}_{\rm{P}}^3
\big(\hat{R}-2\Lambda\big) \hs,
\end{equation}
where $\hat{R}$ and $\Lambda\,(<0)$ are the 5d Ricci scalar curvature and cosmological constant respectively.\  The 5d Planck mass $\hat{M}_{\rm{P}}^{}$ is given by 
$\hat{M}_{\rm{P}}^{}\!=\!M_{\rm{P}}^{\frac{2}{3}}\bar{L}^{-\frac{1}{3}}$, 
where  $M_{\rm{P}}^{}$ denotes the 4d reduced Planck mass
and is connected to the 4d Newton constant $G$ via
$M_{\rm{P}}^{}\!=\!(8\hs\pi\hs G)^{-1/2}$.\ 
The possible brane terms\,\cite{RS1} are irrelevant for the current analysis.

\vs

To study the linearized theory, we apply the following weak field expansion 
for the 5d Lagrangian \eqref{eq:LEH-5d}:
\begin{equation}
\hg_{MN}^{} \,=\, 
e^{2A(z)}\big(\heta_{MN}^{} + \hat{\ka}\hs \hh_{MN}^{}\big) \,,
\end{equation}
where the 5d gravitational coupling $\hat{\ka}$ has mass-dimension $-3/2$ 
and is connected to the 4d gravitational coupling ${\ka}$ via
$\,\hat{\ka} = \ka\sqrt{L\,}\!=2M_{\rm{P}}^{-1}\hsmx\sqrt{L\,}\,$.\   
(Here $M_{\rm{P}}^{}$ is the 4d reduced Planck mass.)
The 5d graviton field $\hh_{MN}^{}$ can be further parametrized 
as follows\,\cite{Hang:2021fmp}\cite{Hang:2022rjp}:
\begin{equation}
\label{eq:hDecom}
\hh_{MN}^{}=
\begin{pmatrix}
\hh_{\mn}^{} \!-\!\frac{1}{2} \eta_{\mn} \hat{\phi} &~~ \hat{\VV}_{\mu}
\\[2mm]
\hat{\VV}_{\nu} &~~ \hat{\phi}
\end{pmatrix} \!,
\end{equation}
where we denote the (off-diagonal) vector component as 
$\,\hat{\VV}_{\mu} \hsm\equiv\hsm \hh_{\mu5}\hs$ and 
the (diagonal) scalar component as $\hs\hat{\phi}\equiv\hh_{55}^{}\hs$.\

\vs

Then, we derive the 5d Lagrangian at quadratic order, which is presented 
in Eq.\eqref{Aeq:LagLOExp} of Appendix\,\ref{appx:B.2}.\ 
To eliminate the mixing terms appearing in Eq.\eqref{Aeq:LagLOExp}, 
we choose the 5d gauge-fixing terms as follows:
\begin{equation}
\label{eq:L-GF-KKgravity} 
\hat{\La}_{\rm{GF}}^{} = 
-\frac{\,e^{3A(z)}\,}{\xi} \big(\hat{\FF}_\mu^2 + \hat{\FF}_5^2\,\big)\hs,
\end{equation}
where $\,\xi\,$ is a gauge-fixing parameter and the gauge-fixing functions 
$\hat{\FF}_\mu^{}$ and $\hat{\FF}_5^{}$ take the following form:
\\[-7mm]
\beqs
\label{eq:FmuF5-5d}
\begin{align}
\hat{\FF}_\mu^{} &=\pd^\nu \hh_{\mn}
\!-\!\(\!1\!-\!\frac{1}{\,2\xi\,}\)\!\pd_\mu \hh
+\xi (3A'\!+\hsm\pd_z^{}) \hat{\VV}_{\mu} \,,
\\[1mm]
\hat{\FF}_5^{} &=\frac{1}{\,2\,}\! 
\[\pd_z^{} \hh-3\hs\xi (2A'\!+\pd_z) \hat{\phi}-\hsm 2\pd_\mu^{}\hsm \hat{\VV}^{\mu}\]\!.
\end{align}
\eeqs
In the above, we denote the gravitational gauge-fixing parameter 
by $\,\xi\,$ for notational convenience, 
but it is independent of the gauge-fixing parameter $\,\xi\,$ 
for gauge theories as defined in Eq.\eqref{eq:LGF-gauge-5d}.\  

\vs

For the BRST quantization, the 5d gravitational ghost fields 
$(\hat{c}^M\!,\hs\hat{\bar{c}}^M)$ are introduced to compensate 
the gravitational gauge-fixing term \eqref{eq:L-GF-KKgravity}  
in the path integral formulation.\ 
The 5d Faddeev-Popov Lagrangian takes the following form:
\begin{equation}
\hat{\La}_{\rm{FP}}^{} \,=\, 
e^{3A(z)}\, 
\hat{\bar{c}}^M \sss \hat{\FF}_M^{} \,,
\end{equation}
where the gauge-fixing functions,
$\hat{\FF}_M^{}\hsm\!=\!(\hat{\FF}_\mu^{},\,\hat{\FF}_5^{})$, are defined as in \eqrefe{eq:FmuF5-5d}, 
with the 5d Lorentz index $M\!=\!\mu,5\hs$.\

\vs 

Under the $S^1\!/\ZZ$ orbifold compactification, 
the tensor component of graviton field $\hh_{\mn}^{}$ 
and the scalar component $\hat{\phi}$ are assigned to be even under the $\ZZ$ parity, 
whereas the vector component $\hat{\VV}_{\mu}^{}$ is $\ZZ$ odd:
%
\begin{align}
\label{eq:5d-BC-hVphi}
\hh_{\mn}(x, z) = \hh_{\mn}(x,-z)\hs,~~~~
\hat{\VV}_{\mu}(x, z) = -\hat{\VV}_{\mu}(x, -z) \hs,~~~~
\hat{\phi}(x, z) = \hat{\phi}(x, -z)  \hs.
\end{align}
%
We find that the ghost fields $(\hat{c}^{\hs\mu},\hat{c}^{\hs 5})$ 
obey the same equations of motion as that of the gravitational fields 
$(\hat{h}^{\mn},\hs \hat{\VV}^{\mu})$ respectively.\ 
Thus, we assign the following $\ZZ$ parities for the ghost fields:
\beq 
\label{eq:5d-BC-cmu-c5}
\hat{c}^{\mu}(x, z) = \hat{c}^{\mu}(x,-z)\hs,~~~~
\hat{c}^{5}(x, z) = -\hat{c}^{5}(x,-z)\hs,
\eeq 
and their corresponding anti-ghost fields share the same $\ZZ$ parities.\
With these, we make the following KK expansions for 
the 5d graviton fields in terms of Fourier series containing the zero-modes and KK-modes:
\beqs
\label{eq:KK-Exp-GR}
\begin{align}
\hh^{\mn}(x, z) & = \frac{1}{\sqrt{L\,}\,}
\sum_{n=0}^{\infty} h^{\mn}_n(x)\, \uu_n^{}(z)\,, 
\\
\hat{\VV}^\mu(x, z) & =\frac{1}{\sqrt{L\,}\,}\sum_{n=1}^{\infty} \VV^\mu_n(x)\, \vv_n^{}(z)\,,
\\
\hat{\phi}(x, z) & =\frac{1}{\sqrt{L\,}\,}\sum_{n=0}^{\infty} \phi_n^{}(x)\, \ww_n^{}(z)\,,
\end{align}
\eeqs
where the 5d wavefunctions 
$\big\{\uu_n^{}(z),\hs\vv_n^{}(z),\hs\ww_n^{}(z)\big\}$
obey the orthonormal conditions \eqref{Aeq:Normalization-GR}.\
These 5d wavefunctions also satisfy the following boundary conditions,
\beq 
\left.\pd_z \uu_n^{}(z)\right|_{z=0,L}=0\hs,  ~~~~
\left.\vv_n^{}(z)\right|_{z=0,L}=0 \hs,  ~~~~
\left.(2A'\!+\pd_z) \ww_n^{}(z)\right|_{z=0,L}=0\hs,
\eeq 
which can be derived by imposing the orbifold symmetry and integrating around the fixed points 
to obtain the junction conditions\,\cite{Charmousis:1999rg}, 
or by introducing the Gibbons-Hawking extrinsic curvature term\,\cite{Gibbons:1976ue} and 
applying the variational principle\,\cite{Lalak:2001fd}\cite{Carena:2005gq}.\ 

\vs

Making KK expansions for the 5d Lagrangian \eqref{Aeq:LagLOExp}  
and integrating $z$ over the interval $[\hs 0,\hs L\hs]$, 
we derive the 4d KK Lagrangian to the quadratic order 
as given in Eq.\eqref{Aeq:LagLOKK}.\ 
Thus, we can derive the following 4d $R_{\xi}^{}$-type gauge-fixing terms under the KK expansions:
\begin{equation}
\label{eq:GF-4d}
\La_{\rm{GF}}^{} = \int_0^L \!\!\td z\, \hat{\La}_{\rm{GF}}^{} 
\hs =\hs -\sum_{n=0}^{\infty}\!\frac{1}{\,\xin \hs}\!
\[(\FF_n^\mu)^2\!+\!(\FF_n^5)^2\] \!,
\end{equation}
where $\hs\xin$ is the gauge-fixing parameter for KK level-$n\hs$, and 
the gauge-fixing functions $\FF_n^\mu$ and $\FF_n^5$ take the following forms:
\beqs
\label{eq:Fmun-F5n2}
\begin{align}
\label{eq:Fmun2}
\FF_n^\mu &\,=\,\pd_{\nu} h_n^{\mn}\! -\!\(\!1\!-\!
{\frac{1}{\,2\hs\xin}\,}\!\)\!\pd^{\mu} h_n^{}
\!+\hsm\frac{1}{\sqrt{2\,}\,} \xin \MGn \VV^{\mu}_n  \,,
\\[0mm]
\label{eq:F5n2}
\FF_n^5 &\,=\hs {\frac{1}{\hs 2\hs}}\,
\!\MGn h_n^{}\!-\! {\sqrt{\frac{3}{2}}}\,\xin\MGn\phin 
\!+\!{\frac{1}{\sqrt{2\,}\,}}\, \pd_{\mu}^{}\VV^\mu_n\,,
\end{align}
\eeqs
where $\MGn$ denotes the mass eigenvalue for the KK gravitons of level-$n$.\ 
In the above, to ensure proper normalization of the kinetic terms of $\VV_n^\mu$ and $\phin$,
we have made the field rescaling,
$\VV^{\mu}_n \!\ito\! \frac{1}{\sqrt{2\,}\,} \VV^{\mu}_n$ and 
$\phin  \!\ito\! \sqrt{\frac{2}{3}\,}\phin\hs$
as defined in Eq.\eqref{Aeq:rescaling-Vn-phin}.\ 
With these, we can derive the propagators of KK fields $(h_n^{\mn},\hs \VV_n^\mu,\hs\phi_n^{})$
at tree level, which take simple diagonal forms 
as shown in Eqs.\eqref{Aeq:Prophh-xi=1}-\eqref{Aeq:Prop55-xi=1}.\ 
Using the KK decompositions \eqref{eq:KK-Exp-GR}, 
we derive the equations of motion (EOM) 
for the 5d gravitational KK wavefunctions as in Eq.\eqref{Aeq:EOM-GR}.\ 
The KK mass eigenvalue $\MGn$ can be determined by solving Eq.\eqref{Aeq:GR-Mn},
and differs from the KK gauge boson mass $\Mn$ solved from Eq.\eqref{Aeq:YM-Mn}.\ 

\vs 

As shown in Eqs.\eqref{eq:5d-BC-hVphi} and \eqref{eq:5d-BC-cmu-c5},
the 5d ghost fields $(\hat{c}^{\hs\mu},\hat{c}^{\hs 5})$ 
share the same $\ZZ$ parities as that of the gravitational fields 
$(\hat{h}^{\mn},\hs \hat{\VV}^{\mu})$ respectively.\ 
Thus, we make the following KK expansions for the gravitational 
ghost and anti-ghost fields:
\beqs
\begin{align}
\hat{c}^\mu (x,z) &= \frac{1}{\sqrt{L\,}\,}\! 
\sum_{n=0}^{\infty}\! c^\mu_n(x)\uu_n(z) \hs,\qquad
\hat{\bar{c}}^\mu (x,z) =
\frac{1}{\sqrt{L\,}\,}\!\sum_{n=0}^{\infty}\! \bar{c}_n^\mu(x)\uu_n(z)\hs,
\\
\hat{c}^5 (x, z) &=
\frac{1}{\sqrt{L\,}\,}\!\sum_{n=1}^{\infty}\! c_n^5(x)\vv_n(z) \hs,\qquad
\hat{\bar{c}}^5 (x, z)=
\frac{1}{\sqrt{L\,}\,}\!\sum_{n=1}^{\infty}\! \bar{c}_n^5 (x)\vv_n(z) \hs. 
\end{align}
\eeqs
We have further derived the BRST transformations for the KK fields, 
which are presented in Eq.\eqref{Aeq:BRST-KK-Fields} of Appendix\,\ref{app:B}.\

Then, we derive a general (Slavnov-Taylor-type) identity in the momentum space
for the gravitational gauge-fixing functions \eqref{eq:Fmun2}-\eqref{eq:F5n2}
as external lines of the Green function\,\cite{Hang:2021fmp}\cite{Hang:2022rjp}:
\begin{equation}
\label{eq:F-identityP}
\bla 0\hs |\hs\FF_{n_1}^{\hs\mu_1^{}}\hsm (k_1^{})\hs\FF_{n_2^{}}^{\hs\mu_2}\hsm (k_2^{})\cdots
\FF_{m_1^{}}^{\hs 5}\!(p_1^{})\hs\FF_{m_2^{}}^{\hs 5}\!(p_2^{})\cdots \Phi
\hs |\hs 0\bra =\, 0\,,
\end{equation}
where each external momentum ($k_j^{}$ or $p_j^{}$) is chosen as incoming and 
$\Phi$ denotes any possible external physical states 
after the LSZ reduction.\

\vspace*{2mm}
\subsubsection{\hspace*{-2.5mm}Formulating GRET for Warped KK Gravity Theory\,(I)}
\vspace*{1.5mm}
\label{sec:2.3.2}

In this subection, we first formulate a KK Gravitational Equivalence Theorem (GRET of type-II)\footnote{%
For convenience, we call this GRET (for KK gravitons with helicity\,$\pm1$) the GRET of type-II, and
call the other GRET (for KK gravitons with helicity-0) the GRET of type-I 
which will be given in Section\,\ref{sec:2.3.3}.}  
to quantitatively connect the scattering amplitude of KK gravitons $h^{\mn}_n$ with helicity\,$\pm1$ 
to that of the corresponding gravitational KK vector Goldstone bosons $\VV_n^\mu$.\
For this purpose, we have derived the general identity \eqref{eq:F-identityP} 
for the gravitational gauge-fixing functions
$\FF_{n}^{\hs\mu}$ and $\FF_{n}^{\hs 5}$ of Eq.\eqref{eq:Fmun-F5n2}.\ 

\vs

We first reexpress the gauge-fixing function \eqref{eq:Fmun2}  into the following matrix form:
\beqs
\label{eq:Fmu=KH}
\begin{align}
\FF_n^\mu &= -\frac{\,\ii\MGn\,}{\sqrt{2\,}\,} \mathbb{F}_n^{\mu} \,, \quad
\mathbb{F}_n^{\mu} =  \bK_n^T \bH_n^{} \,,
\\
\bK_n^{}\hsm &=\hsm \Big(\!
\frac{1}{\,\sqrt{2\,}\MGn}\!\big[2k_\be^{}\tensor{\eta}{_\al^\mu}\!- k^\mu(2\!-\hsm\xi_n^{-1})\eta_{\ab}^{}\big],\, \ii\xin \Big)^{\!T},
\quad
\bH_n^{}\!=\hsm \big(h_n^{\ab},\,\VV_{n}^\mu \big)^{\!T}.
\end{align}
\eeqs
Then, we derive the following BRST identity including the gauge-fixing function 
$\mathbb{F}_n^{\mu}\hs$,
\begin{equation}
\label{eq:FnHm1}
\bla0\big| T\, \mathbb{F}_n^{\mu} \bH_m^{T}\big|0\bra = 
-\frac{\ii\xin}{\sqrt{2}\hs\MGn}\bla 0\big|T\, \sss \bH_m^{T}\bar{c}^\mu_n\big|0\bra \,.
\end{equation}
Using Eq.\eqref{eq:Fmu=KH}, we can rewrite \eqrefe{eq:FnHm1} as follows:
\begin{equation}
\label{eq:KD-X-1}
\bK_n^T \bD_{nm}^{}(k)=-\bX^T_{nm}(k)\,,
\end{equation}
where we have defined full propagator in the matrix form  
$\bD_{nm}^{}(k) \!=\! \bla0\big|T\hs\bH_n^{} \bH_m^{T}\big|0\bra\hsm (k)$,
and the matrix quantity $\bX_{nm}^{}$ is given in 
\eqref{Aeq:app-DX-KD=X-2} of Appendix\,\ref{appx:B.4}.

\vs

Then, we analyze the identity \eqref{eq:F-identityP} and construct an external state 
as a combination of gauge-fixing functions
($\FF_n^\mu$ or $\mathbb{F}_n^{\mu}$) as given in Eq.\eqref{eq:Fmu=KH}.\ 
Thus, we amputate this external state by using the BRST identity \eqref{eq:KD-X-1}
and obtain: 
\begin{align}
\label{eq:Fn-mu-LSZ}
& 0 \,=\, \bla 0 |\hs\mathbb{F}_n^{\hs\mu}(k)\cdots {{\Phi}}\hs | 0\bra 
= -\bX_{nm}^{T}(k)\hs
\M\hsm\big[\under{\mathbf{H}}_m^{}(k),\cdots\!,{{\Phi}}\hs\big] .
\end{align} 
From the above, we can derive the following identity: 
\begin{align}
\M\hsm\big[\under{\mathbb{F}}_n^{\hs\mu}\hsm (k),\cdots\!,{\Phi}\hs\big]
= 0 \,,
\label{eq:Fn-mu-ID1}
\end{align}
where $\M[\cdots]$ denotes the amputated scattering amplitude.\ 
In \eqrefe{eq:Fn-mu-ID1}, the amputated external state $\under{\mathbb{F}}_n^{\hs\mu}$
takes the following form (after removing an irrelevant overall coefficient): 
\begin{equation}
\label{eq:Fnmu-Amp1}
\under{\mathbb{F}}_n^{\hs\mu}(k) =
\frac{\sqrt{2\,}\,}{\,\MGn}\hs k_{\nu}^{} 
h_{n}^{\mu\nu}\!
-\hat{C}^{\hs nm}\hs{\eta}^{\hs\mn}\VV^{}_{m,\nu} \,,
\end{equation}
where the loop-induced modification factor is derived as follows:
\begin{equation}
\label{eq:Cmod-GRET-2}
\hat{C}^{}_{\hs nm}(k^2) 
= -\frac{\,\ii\MGm\,}{\MGn} 
\!\!\[\!\frac{\,{\mathbf{1}}\!+\!\tilde{\boldsymbol{\Delta}}^{\!(2)}(k^2)\,}
{\,{\mathbf{1}}\!+\!\boldsymbol{\Delta}^{\!(1)}(k^2)\,}\!\]_{mn}
\!= -\ii\hs\delta_{nm}^{}+O(\rm{loop}) \hs.
\end{equation}
In the above, the matrix form is used such that the loop-level quantities 
$(\boldsymbol{\Delta}^{\!(1)}_{})_{jj'}^{}\!=\hsm {\Delta}^{\!(1)}_{jj'}$ and
$(\boldsymbol{\tilde{\Delta}}^{\!(2)})_{jj'}^{}\!=\!\tilde{\Delta}^{\!(2)}_{jj'}\hs$,
where the loop quantities $({\Delta}^{\!(1)}_{jj'},\,\tilde{\Delta}^{\!(2)}_{jj'})$
are defined in Eq.\eqref{Aeq:Delta-12}.\ 
The derivations of Eqs.\eqref{eq:Fn-mu-LSZ}-\eqref{eq:Cmod-GRET-2} are given 
in detail around 
Eqs.\eqref{Aeq:app-Fn-mu-LSZ}-\eqref{Aeq:app-Fn-mu-ID1} of Appendix\,\ref{app:B}.\ 

\vs

Then, we utilize a transverse polarization vector $\ep^\mu_{\pm}$ to contract with 
$\under{\mathbb{F}}_n^{\hs\mu}(k)$ in \eqrefe{eq:Fnmu-Amp1} and obtain the formula: 
\begin{equation}
\label{eq:Fn-Amp1}
\under{\mathbb{F}}_n^{}(k) \equiv 
\ep_\mu^{\pm}\hs\under{\mathbb{F}}_n^{\hs\mu}(k) =
h_n^{S,\pm1} - \hat{C}^{}_{\hs nm}\hs\VV^{\pm1}_m \,,
\end{equation}
where we have used the fact $\hs\ep^\mu_{\pm}\hs k_\mu^{}\!=\hsm 0\hs$, 
and defined the external states $h^{S,\pm 1}_n$ and $\VV^{\pm1}_m$ as follows: 
\begin{equation}
h_n^{S,\pm1} \!= \vep^{S,\pm 1}_{\mn} h^{\mn}_n \hs,~~~~
\VV^{\pm1}_m \!= \ep_\mu^{\pm}\hs\VV_{m}^{\mu}\hs,~~~~
\vep^{S,\pm1}_{\mn} 
\!=\! {\frac{1}{\sqrt{2\,}\,}}\hsm\big(\ep^S_\mu\ep_\nu^{\pm} \!+\hsm \ep^S_\nu\ep_\mu^{\pm}\big) ,
\label{eq:hS-polS+-1}
\end{equation}
with $\ep^S_\mu\!\!=\!\!k_\mu^{}/\MGn$ being the scalar-polarization vector.\ 
A massive spin-2 KK graviton has 5 physical helicity states $(\lam \!=\!\pm 2,\pm 1, 0)$.\ 
The polarization tensors of KK graviton with helicities $\pm 1$ are given by
\beq
\label{eq:hpol+-1}
\vep_{\pm 1}^{\mn} =
{\frac{1}{\sqrt{2\,}\,}}\!\(\ep_{\pm}^{\mu}\ep_{L}^{\nu}\!+\hsm\ep_{L}^{\mu}\ep_{\pm}^{\nu}\)  \!,
\eeq 
where $(\ep_{\pm}^\mu,\,\ep_L^\mu)$ denote the 3 polarization vectors of a massive spin-1 KK gauge boson.\ 
Under high energy expansion, we have $\ep_{L}^{\mu}\!=\hsm\ep_{S}^{\mu}\hsm +\hsm v^\mu$
with $\ep_{S}^{\mu}\!=\!k^\mu\!/\MGn$ and $v^\mu\!=\hsm O(\MGn/E_n^{})\hs$.\ 
Thus, we can decompose the polarization tensor $\vep_{\pm 1}^{\mn}$ as follows:
\beqs 
\begin{align}
\vep_{\pm 1}^{\mn} &= \vep_{S,\pm 1}^{\mn} + v_{\pm1}^{\mn}\,,~~~~
\vep_{S,\pm 1}^{\mn} =\! \Fr{1}{\sqrt{2\,}\,}\hsm
\big(\ep_S^\mu\ep^\nu_{\pm} \!+\hsm \ep_S^\nu\ep^\mu_{\pm}\big) ,
\\[1mm]
v_{\pm1}^{\mn} &=\! \Fr{1}{\sqrt{2\,}\,}\hsm
\big(v^\mu\ep^\nu_{\pm} \!+\hsm v^\nu\ep^\mu_{\pm}\big)\hsm =\hsm O(\MGn/\hsm E)\hs. 
\end{align}
\eeqs 
With these, we can reexpress the function $\under{\mathbb{F}}_n^{}\hsm (k)$ 
of \eqrefe{eq:Fn-Amp1} in the form:
\begin{align} 
\label{eq:Fn=hn1-Omega}
\under{\mathbb{F}}_n^{}\hsm (k) =
h_n^{\pm1}\! -\Theta_n^{}\hs,~~~~
\Theta_n^{} \!= \hat{C}^{\hs nm}\VV_m^{\pm1}	\!+ v_n^{\pm1}\hs, 
\end{align} 
where we have used the notations,
\begin{equation}
h_n^{\pm1} \!= \vep^{\pm1}_{\mn} h^{\mn}_n , \quad~
\VV_m^{\pm1} \!= \ep^{\pm}_{\mu} \VV^{\mu}_m
\hs, \quad~
v_n^{\pm1} \!=\hsm v_{\mn}^{\pm1} h^{\mn}_n \hs,
\end{equation}
with the tensor $v_{\pm1}^{\mn}\!=\hsm O(\MGn/\hsm E_n^{})$ under high energy expansion and 
thus $v_n^{\pm1}$ is suppressed by an energy factor of $O(\MGn/\hsm E_n^{})\hs$.\ 
For later analysis, we can also multiply $\,v_{}^\mu \!=\! \ep_L^\mu\!-\hsm\ep_S^\mu\,$ 
on both sides of Eq.\eqref{eq:Fn-mu-ID1} and use the expression \eqref{eq:Fnmu-Amp1} to
obtain the following identity:
\beq 
\ep_\mu^Sv_\nu^{}\hs\M\hsm\big[h_n^{\mn}\hsm (k),\cdots\!,{\Phi}\hs\big] 
= \hat{C}^{}_{nm}\hs v_\mu^{}\hs\M \big[\VV_{m}^\mu\hsm (k),\cdots\!,{\Phi}\hs  \big],
\label{eq:Fn-v-ID}
\eeq
where we see that the right-hand side is suppressed by the external factor
$v_\mu^{}\!=\hsm O(\MGn/\hsm E_n^{})\hs$.\  
Hence the above identity \eqref{eq:Fn-v-ID} ensures that its left-hand side must receive a suppression
such that its overall energy-power dependence is the same as the right-hand side
[despite that the external factor of the left-hand side is  
$\ep_\mu^Sv_\nu^{}\!=\hsm O(E^0_n)$].\   

\vs

Then, we reformulate the identity \eqref{eq:Fn-mu-ID1} and extend it to the general case of $N$ external amputated
gauge-fixing functions: 
\beq 
\label{eq:GRET2-Fn1}
\M\hsm\big[\under{\mathbb{F}}_{n_1^{}}^{}\hsm\!(k_1^{}),\cdots\!,
\under{\mathbb{F}}_{n_{\!N}^{}}^{}\hsm\!(k_{\!N}^{}),
{\Phi}\hs\big]
= 0 \hs. 
\eeq 
Using this identity, we can further derive the gravitational equivalence theorem (GRET) identity 
which connects the $h_n^{\pm1}$ amplitude to the amplitude of $\Theta_{n}^{}$:
\begin{align}
\label{eq:GRET2-hn1=Theta}
\M\hsm\big[h_{n_1^{}}^{\pm1}\hsm (k_1^{}),\cdots\!,h_{n_{\!N}^{}}^{\pm1}\hsm (k_{N}^{}), 
{\Phi}\hs\big]\hs =\hs 
\M\hsm\big[\Theta_{n_1^{}}^{}\hsm\!(k_1^{}),\cdots\!, \Theta_{n_{\!N}^{}}^{}\hsm\!(k_{N}^{}),
{\Phi}\hs\big] \hs.
\end{align}
The above identity can be proven by computing the amplitude on its right-hand side and 
using the relation  $\Theta_n^{}\!\!=\!h_n^{\pm1}\!-\!\under{\mathbb{F}}_n^{}$ in Eq.\eqref{eq:Fn=hn1-Omega}.\
Expanding $\Theta_n^{} (=\! \hat{C}^{}_{\hs nm}\VV_m^{\pm1} \hsm\!+\! v_n^{\pm1})$
for each external state on the right-hand side of the identity \eqref{eq:GRET2-hn1=Theta},  
we derive the following gravitational equivalence theorem (GRET) identity:
\beqs
\label{eq:KK-GRET-hV}
\begin{align}
& \M [\,h^{\pm1}_{n_1^{}},
\cdots\!, h^{\pm1}_{n_N^{}},\Phi\,]
\,=\,\Ch_{\hsm\rm{mod}}^{n_j^{}m_j^{}}
\M [\,{\VV^{\pm1}_{m_1^{}},\cdots\!, \VV_{m_N^{}}^{\pm1}},\Phi\,] + \M_v^{}\hs,
\label{eq:KK-GRET-hV1}
\\[1mm]
& \M_v^{} = \sum_{\ell=1}^{N}
\tilde{C}_{\hsm\rm{mod},\ell}^{n_j^{}m_j^{}}\,
\M[\,v_{n_1^{}}^{\pm1},\cdots\!, v_{n_{\ell}^{}}^{\pm1},
\VV_{m_{\ell+1}^{}}^{\pm1},\cdots\!,\VV_{m_N^{}}^{\pm1},\Phi\,] \hs,
\label{eq:KK-GRET-Mv}
\end{align}
\eeqs
where the multiplicative modification factors are given by
\beqs
\label{eq:Cmod-GB+Res}
\begin{align}
\label{eq:Cmod-GB}
\Ch_{\hsm\rm{mod}}^{n_j^{}m_j^{}} &=\,
\Ch^{}_{n_1^{}m_1^{}}\!\cdots\Ch^{}_{n_{\!N}^{}m_{\!N}^{}}
=\,(-\ii)^N\delta_{n_1^{}m_1^{}}^{}\!\!\cdots\hs\delta_{n_{\!N}^{}m_{\!N}^{}}^{}
\!+O(\rm{loop})\hs,
\\[1mm]
\label{eq:Cmod-Res}
\tilde{C}_{\hsm\rm{mod},\ell}^{n_j^{}m_j^{}} &=\,	
\Ch^{}_{n_{\ell+1}^{}m_{\ell+1}^{}}\!
\cdots\Ch^{}_{n_{\!N}^{}m_{\!N}^{}}
=\,(-\ii)^{N-\ell}
\delta_{n_{\ell+1}^{}m_{\ell+1}^{}}^{}\!\!\cdots\hs\delta_{n_{\!N}^{}m_{\!N}^{}}^{}
\!+O(\rm{loop}) \hs.
\end{align}
\eeqs
Under the high energy expansion, the residual term is suppressed by the energy factor of 
$O(\MGn/\hsm E)\hs$.\ Hence, from the identity \eqref{eq:KK-GRET-hV1} 
we derive the following warped GRET of type-II:
\beq
\M [\,h^{\pm1}_{n_1^{}},
\cdots\!, h^{\pm1}_{n_N^{}},\Phi\,]
\,=\,\Ch_{\hsm\rm{mod}}^{n_j^{}m_j^{}}\M [\,{\VV^{\pm1}_{m_1^{}},\cdots\!, \VV_{m_N^{}}^{\pm1}},\Phi\,] 
\hs +\hs 
O({\MGn}/\hsm{E_n^{}}\hsm\rm{-suppressed})\hs .
\label{eq:KK-GRET-hV1f}
\eeq 
This connects the scattering amplitude of the KK gravitons ($h_n^{\pm1}$) with helicity\,$\pm1$ to the
corresponding gravitational KK vector-Goldstone bosons ($\VV_m^{\pm1}$).

\vspace*{2mm}
\subsubsection{\hspace*{-2.5mm}Formulating GRET for Warped KK Gravity Theory\,(II)}
\vspace*{1.5mm}
\label{sec:2.3.3}

In this subsection, we further derive the KK GRET to connect the scattering amplitude of 
helicity-0 longitudinal KK gravitons ($h_L^n$) to that of the corresponding gravitational KK 
scalar Goldstone bosons ($\phin$) under high energy expansion.\ 
This GRET will be called GRET of type-I.

\vs

We use the combination of gauge-fixing functions 
$\pd_\mu^{}\FF_n^\mu \!-\xin\MGn\FF_n^5\,$ 
to eliminate the KK vector-Goldstone field $\,\VV_n^{\mu}\,$ and 
derive an expression in momentum space:
\\[-3mm]
\begin{equation}
\label{eq:Fmu-F5-0}
-\ii k_\mu^{}\FF_n^\mu \!-\xin\MGn\FF_n^5 \,=\,
-k_\mu^{}k_\nu^{} h_n^{\mn}
\!+\Fr{1}{2}\!\[\!(2\!-{\xi_n^{-1}})k^2\!-\xin\MGnn\]\!h_n^{}
\!+\!\sqrt{\!\Fr{3}{2}\,}{\xi_n^2}\MGnn\phin \,,
\end{equation}
where we have made the rescaling of $\hs\phin$ 
as in Eq.\eqref{Aeq:rescaling-Vn-phin}.\ 
Imposing the on-shell condition $k^2\!=\!-\MGnn\hs$, 
we can rewrite the above formula as follows:
\beqs 
\label{eq:Fmu-F5-1}
\begin{align}
\label{eq:Fmu-F5-1a}
&\ii k_\mu^{}\FF_n^\mu \!+ \xin\MGn\FF_n^5 
\,=\sqrt{\!\Fr{3}{2}\,}\MGnn\hs\FT_n^{}\,,
\\
\label{eq:Fmu-F5-1b}
& \FT_n^{} = \sqrt{\!\Fr{2}{3}\,}h_n^{S}
+\Fr{1}{\sqrt{6\,}\,}\big(2\hsm +\hsm\xi_n^{}\!-\xi_n^{-1}\big)h_n^{}-\xi_n^2\phin
=\mathbf{K}^T_n\mathbf{H}_n^{}
\\
\label{eq:Fmu-F5-1c}
&\hspace*{4.1mm}
= \sqrt{\!\Fr{2}{3}\,}\big(h_n^{S}
+h_n^{}\big)\!-\phi_n^{}\hs, \hspace*{7mm}
(\text{for}~\xin\!=\! 1)\hs,
\\
& \mathbf{K}^{}_n\!= \Big(\!\sqrt{\!\Fr{2}{3}\,}\vep^S_{\mn}
\!+\!\Fr{1}{\sqrt{6\,}\,}\big(2\hsm +\hsm\xin\!-\xi_n^{-1}\big)\eta_{\mn}^{},\,-\xi_n^2\Big)^{\!\!T}\hsm\!,~~~~
\mathbf{H}_n^{} = (h_n^{\mn},\,\phin)^{T}\hs, 
\end{align}
\eeqs
where we have defined the scalar KK graviton 
$\,h_n^S\!=\!\varepsilon_{\mn}^S{h_n^{\mn}}\,$ with its scalar-polarization tensor
$\vep_{\mn}^S\!=\! k_\mu^{}k_\nu^{}/\MGnn\hsx$.\
Then, we inspect the KK graviton's longitudinal polarization tensor  $\vep_L^{\mn}$
and make high energy expansion of it as follows:
\beqa
\label{eq:epLmunu}
\vep_L^{\mn} =\frac{1}{\sqrt{6\,}\,}\!
\(\ep_+^\mu\ep_-^\nu \!+\ep_+^\nu\ep_-^\mu \!+ 2\ep_L^\mu\ep_L^\nu\)
= \sqrt{\frac{2}{3}\,}\vep_S^{\mn} \!+ \vt^{\mn},
\eeqa
where the longitudinal polarization vector
$\ep_L^\mu \!=\! (k^0\!/\MGn)(|\vec{k}|/k^0\hsm,\,\vec{k}/|\vec{k}|)
\!=\!\ep_S^{\mu}\!+\!v^\mu$\, with $\,\ep_S^\mu =k^\mu\!/\MGn\,$  
and $\,v^\mu \!=\! \mO(\MGn/E_n)\hs$.\
In the above, the residual term $\vt^{\mn}$ is given by
\beq 
\label{eq:v-munu}
\vt^{\mn}\equiv\, \vep_L^{\mn}\!\!-\!\sqrt{\frac{2}{3}\,}\vep_S^{\mn}
= \frac{1}{\sqrt{6\,}\,}\!\[\hsm\big(\ep_+^\mu\ep_-^\nu\!+ \ep_+^\nu\ep_-^\mu\big) 
\!+2\big(\ep_S^\mu v^\nu \!+\!\ep_S^\nu v^\mu \!+\! v^\mu v^\nu\big) \hsm\]\!,
\eeq 
where $\ep_S^\mu\!=\! k^\mu/\MGn$ and $\,\vt^{\mn}\!\!=\!\mO (E_n^0)$
under high energy expansion. 

\vs

According to Appendix\,\ref{appx:B.3}, we derive a BRST identity for the full propagator
of the matrix field $\mathbf{H}_n^{}\!=\!(h_n^{\mn},\,\phin)^{T}$,
\begin{equation}
\label{eq:KD-X-2} 
\bK_n^T \bD_{nm}^{}(k)=-\bX_{nm}^T(k)\,,
\end{equation}
where we have defined full propagator in the matrix form 
$\bD_{nm}^{}(k) \!=\! \bla0\big|T\hs\bH_n^{} \bH_m^{T}\big|0\bra\hsm (k)$,
and the other matrix quantity $\bX_{nm}^{}$ is given in 
\eqref{Aeq:app-X-Delta34} of Appendix\,\ref{appx:B.3}.\  
We also note that the content of Eq.\eqref{eq:KD-X-2} actually differs from that of
Eq.\eqref{eq:KD-X-1} although they are expressed in the same notations.\

\vs 

Using the identity \eqref{eq:F-identityP} and Eq.\eqref{eq:Fmu-F5-1} together with
Eq.\eqref{eq:KD-X-2},  we can derive a new identity with
external state $\FT_n^{}(k)\hs$,
%
\begin{align}
\label{eq:Fn-ampLSZ-X}
& 0 \,=\, \bla 0\hs |\,\FT_n^{}(k)\cdots \Phi\,|\hs 0\bra 
= -\bX_{nm}^{T}(k)\hs
\M\hsm\big[\under{\mathbf{H}}_m^{}(k),\cdots\!,\Phi\hs\big] ,
\end{align}
from which we derive the following identity,
\begin{align}
\M\hsm\big[\under{\FT}_n^{}\hsm (k),\cdots\!,\Phi\hs\big]
= 0 \,,
\label{eq:Fn-AmpLSZ}
\end{align}
where the amputated external state 
$\under{\FT}_n^{}\hsm (k)$ is given as follows:
\beqs 
\begin{align}
\label{eq:Fn-Gfin}
\under{\FT}_n^{} &= 
\sqrt{\!\Fr{2}{3}\,}(h_n^S\!-\!\Fr{1}{2}h_n^{})\hsm -\hsm C_{nm}^{}\phim
= h_n^L- \Omega_n^{} \hs,
\\[1mm]
\label{eq:Omega-Gfin}
\Omega_n^{} & = C_{nm}^{}\hs\phi_m^{}\!+\hsm\tilde{\Delta}_n^{}\hs,~~~~
\tilde{\Delta}_n^{} = \Fr{1}{\sqrt{6\,}\,}h_n^{}\!+\hsm\vt_n^{} \hs,~~~~
\vt_n^{}\!=v_{\mn}^{}h_n^{\mn} \hs.
\end{align}
\label{eq:Fn-2}
\eeqs 
In the above, $C_{nm}^{}$ is a multiplicative modification factor
induced at loop level:
\begin{equation}
\label{eq:Cmod-GRET-1}
C_{nm}^{}(k^2) = 
\[\!\frac{~\mathbf{1}\!+\!\boldsymbol{\tilde{\Delta}}^{\!(4)}(k^2)~}
{\,\mathbf{1}\!+\!\boldsymbol{\Delta}^{\!(3)}(k^2)}\!\]_{mn}
\!\equiv\, \delta_{nm}^{}+O(\rm{loop}) \hs,
\end{equation}
where the matrix form is used such that 
$(\boldsymbol{\Delta}^{\!(3)}_{})_{jj'}^{}\!\!=\hsm\!{\Delta}^{\!(3)}_{jj'}$ and
$(\boldsymbol{\tilde{\Delta}}^{\!(4)})_{jj'}^{}\!\!=\!\!\tilde{\Delta}^{\!(4)}_{jj'}$
with the matrix elements $({\Delta}^{\!(3)}_{jj'},\hs\widetilde{\Delta}^{\!(4)}_{jj'})$
given by Eq.\eqref{Aeq:app-X-Delta34}.\ 
More details of the derivations are given in Appendix\,\ref{app:B.2}.\

Then, using Eq.\eqref{eq:Fn-AmpLSZ}  
we can derive the following general identity for 
the gravitational equivalence theorem (GRET):
\beq
\label{eq:GRET-hL-Omega}
\M\hsm\big[\under{\FT}_{n_1^{}}^{}\!(k_1^{}),\cdots\!,\under{\FT}_{n_{\!N}^{}}^{}\!(k_N^{}), 
{\Phi}\hs\big] = 0 \,.
\eeq 
From this, we further derive the following GRET identity:
\begin{align}
\label{eq:GRET-Fn}
\M\hsm\big[h_{n_1^{}}^L\hsm\!(k_1^{}),\cdots\!,h_{n_{\!N}^{}}^L\hsm\!(k_{N}^{}),{\Phi}\hs\big]
\hs =\hs 
\M\hsm\big[\Omega_{n_1^{}}^{}\hsm\!(k_1^{}),\cdots\!, 
\Omega_{n_{\!N}^{}}^{}\hsm\!(k_{N}^{}),{\Phi}\hs\big] \hs,
\end{align}
which can be proven by computing the amplitude on its right-hand side and using 
the relation  $\Omega_n^{}\!=\!h_n^L\!-\under{\FT}_n^{}$
in Eq.\eqref{eq:Fn-Gfin}.\ 
Expanding the right-hand side of the GRET identity \eqref{eq:GRET-Fn},
we connect the scattering amplitude of helicity-zero longitudinal KK gravitons ($h_n^L$) 
to that of the corresponding KK Goldstone bosons $(\phim)\hs$:
\beqs
\label{eq:KK-GRET}
\begin{align}
\label{eq:GET-ID0} 
& \M [\hs h^L_{n_1^{}},\cdots\!, h^L_{n_{\!N}^{}},\Phi\hs ] 
\,=\, {C}_{\hsm\rm{mod}}^{n_j^{}m_j^{}} 
\M\hsm [\hs\phi_{m_1^{}}^{},\cdots\!,\phi_{m_{\!N}^{}}^{}\hsm ,\Phi\hs ]+\M_{\!\Delta}^{}\,,
\\
\label{eq:RT-G}
& \M_{\!\Delta}^{}=\,\sum_{\ell=1}^N \!\widetilde{C}_{\hsm\rm{mod},\ell}^{n_j^{}m_j^{}}\,
\M\hsm\big[\tilde{\Delta}_{n_1^{}}\hsm, \cdots\!,\tilde{\Delta}_{n_\ell^{}},
\phi_{m_{\ell+1}^{}}^{},\cdots\!,\phi_{m_N^{}}^{}\hsm,\Phi\hs\big] ,
\\[-9mm]
\nn 
\end{align}
\eeqs
where $\tilde{\Delta}_n^{} \hsm\!=\! \frac{1}{\sqrt{6\,}\,}h_n^{}\hsm\hsm +\vt_n^{}$  
with $\vt_n^{}\!=\!v_{\mn}^{}h_n^{\mn}$, and 
$\Phi$ denotes any other external on-shell physical state(s) after the LSZ reduction.\
In the above, the multiplicative modification factors are defined as follows:
\beqs 
\label{eq:Cmod-GRET1}
\begin{align}
C_{\hsm\rm{mod}}^{n_{\!j}^{}m_{\!j}^{}} &=\, 
C_{n_1^{}m_1^{}}^{}\!\!\cdots C^{}_{n_{\!N}^{}m_{\!N}^{}}
=\, \delta_{n_1^{}m_1^{}}^{}\!\!\cdots\hs\delta_{n_{\!N}^{}m_{\!N}^{}}^{} 
\!+O(\rm{loop})\hs, 
\\
\widetilde{C}_{\hsm\rm{mod},\ell}^{n_{\!j}^{}m_{\!j}^{}} &=\, 	
C^{}_{n_{\ell+1}^{}m_{\ell+1}^{}}\!\!
\cdots C^{}_{n_{\!N}^{}m_{\!N}^{}} 
=\,
\delta_{n_{\ell+1}^{}m_{\ell+1}^{}}^{}\!\!\cdots\hs\delta_{n_{\!N}^{}m_{\!N}^{}}^{} 
\!+O(\rm{loop})\hs,
\end{align}
\eeqs
which are renormalization-scheme-dependent and can be normalized to one 
under proper setup of the renormalization scheme.\ 
They reduce to one at tree level.\
We can further justify that in the residual term \eqref{eq:RT-G} each external state
$\tilde{\Delta}_n^{} \hs(=\! \frac{1}{\sqrt{6\,}\,}h_n^{}\hsm\hsm +\hsm\vt_n^{})$  
is suppressed by an energy factor $\MGn/E_n^{}$ under high energy expansion.\
For this, using the expression \eqref{eq:v-munu} and the normalization condition 
$\sum_{\si}\ep_\si^\mu (\ep_\si^\nu)^*\!=\hsm\eta^{\mn}\!+\hsm k^\mu k^\nu/\MGnn\hs$ 
(with helicity $\si\!=\!\pm,L$), we derive the following:
\beq 
\tilde{\Delta}_n^{} = \Fr{1}{\sqrt{6\,}\,}h_n^{}\!+\hsm\vt_n^{}
= \sqrt{\frac{3}{2}\,}\big(2\ep^S_\mu v^{}_\nu\!+\! v^{}_\mu v^{}_\nu\big)
h_n^{\mn} \,,
\eeq 
where $v^{}_\mu v^{}_\nu\hsm\!=\hsm\!O(\MGnn/\hsm E_n^2)$ and $\ep^S_\mu v^{}_\nu\!\!=\!\!O(E_n^0)$.\ 
The external state of $(v^{}_\mu v^{}_\nu h_n^{\mn})$ is clearly suppressed 
by the extra energy factor
$v^{}_\mu v^{}_\nu\!\sim\!\MGnn/E_n^2$ in the high energy limit.\
As proven by Eq.\eqref{eq:Fn-v-ID}, despite the external factor $\ep_\mu^Sv_\nu^{}\!=\hsm O(E^0_n)$, 
the amplitude with external state 
$\ep^S_\mu v^{}_\nu h_n^{\mn}$ is actually suppressed by an overall energy factor 
$v^\mu\!=\!O(\MGn/E_n^{})\hs$.\  
Thus, as the external state, we can reexpress $\tilde{\Delta}_n^{}$ as follows:
\beq 
\label{eq:Delta-vV+vvh}
\tilde{\Delta}_n^{} =\, \sqrt{\frac{3}{2}\,}\big(\hat{C}_{nm}^{}2\hs v_{\!\mu}^{}\VV^{\mu}_m
\!+\! v_{\!\mu}^{}v_{\nu}^{}h_n^{\mn}\big)\hs, 
\eeq 
where the two terms on right-hand side is suppressed by external factors of 
$v_{\!\mu}^{}\!\!=\!O(E^{-1})$ and $v_{\!\mu}^{}v_{\nu}^{}\!\!=\!O(E^{-2})$, respectively.\ 
Hence, the residual term $\M_{\!\Delta}^{}$ in Eq.\eqref{eq:RT-G} must be suppressed 
(relative to the leading KK Goldstone amplitude) at least by an overall energy factor 
$v^\mu\!=\!O(\MGn/E_n^{})\hs$ under the high energy expansion.\
Thus, we can further formulate the following warped GRET of type-I:
\beq 
\label{eq:KK-GRET1b}
\M [\hs h^L_{n_1^{}},\cdots\!, h^L_{n_{\!N}^{}},\Phi\hs ] 
\,=\, {C}_{\hsm\rm{mod}}^{n_j^{}m_j^{}}\hsm  
\M\hsm [\hs\phi_{m_1^{}}^{},\cdots\!,\phi_{m_{\!N}^{}}^{}\hsm ,
\Phi\hs ]
+ O({\MGn}/\hsm{E_n^{}}\hsm\rm{-suppressed})\hs .
\eeq 
This ensures that the GRET identity \eqref{eq:GET-ID0} can establish  
the equivalence between the helicity-zero longitudinal KK graviton amplitude
$\M[h_{n_1^{}}^L,\cdots]$ and the corresponding gravitational KK Goldstone amplitude 
${C}_{\hsm\rm{mod}}^{n_j^{}m_j^{}}\M[\phi_{m_1^{}}^{},\cdots]$ in the high energy limit.\
Thus,  the identity \eqref{eq:KK-GRET} gives in Eq.\eqref{eq:KK-GRET1b}
the KK gravitional equivalence theorem (GRET) 
for the warped 5d GR theory under high energy expansion, 
which extends our previous formulation of the KK GRET for the 
compactified flat 5d GR \cite{Hang:2021fmp}\cite{Hang:2022rjp}.\

\vs 

We note that the generalized power-counting method 
for the compactified KK gauge theories and KK GR theories  
as established in Ref.\,\cite{Hang:2021fmp} does hold for both flat and warped extra dimensions.\
Thus, we can deduce the leading energy-power dependence of 
$N$-point helicity-zero longitudinal KK graviton amplitudes 
and of the corresponding $N$-point KK Goldstone amplitudes:
\beqs 
\label{eq:E-counting-hnL}
\begin{align} 
\label{eq:E-counting-hnL-or-hn55}
& D_E^{}(Nh^L_n)= 2(N\!+\!L\!+\!1), \hspace*{7mm} 
D_E^{}(N\phin)=2(L\!+\!1),
\\[0.7mm]
& \Longrightarrow~~
D_E^{}(Nh^L_n)\hsm -\hsm D_E^{}(N\phin) = 2N ,
\label{eq:E-counting-hnL-hn55}
\end{align} 
\eeqs 
where $L\!\geqq\! 0\hs$ denotes the number of loops involved.\  
For the scattering amplitudes of $N$ longitudinal KK gravitons 
(with helicity-zero) at tree level ($L\!=\!0$), 
we use the right-hand side of the GRET identity \eqref{eq:GET-ID0} 
and prove the large energy cancellations of $E^{2N+2} \!\ito E^2$
by an energy factor $E^{2N}$
in the $N$ longitudinal KK graviton amplitude.\

The case of 3-point KK graviton amplitudes needs a further discussion.\
According to the above power counting rule \eqref{eq:E-counting-hnL-or-hn55}, the tree-level
3-point longitudinal KK graviton amplitude ($3h_n^L$) have a possible leading energy dependence of 
$E^{2(N+1)}\!\!=\!E^8$  for $N\!\!=\!3\,$.\ 
The tree-level amplitude of 3 gravitational
KK Goldstone bosons ($3\phin$) arises from trilinear KK Goldstone vertex
whose coupling factor contains products of two external momenta 
$\,p_i^{}\cdot p_j^{}$.\  So the naive power counting shows that this KK Goldstone amplitude
would have leading energy dependence of $E^2$ in agreement with Eq.\eqref{eq:E-counting-hnL-or-hn55}.\ 
Thus, according to the GRET \eqref{eq:KK-GRET}, the 3-point tree-level longitudinal KK graviton amplitude 
should have a total energy-power deduction of $E^{8-2}\!=\!E^6$, which agrees to Eq.\eqref{eq:E-counting-hnL-hn55}.\ 
But, as we will further explain at the end of the following Section\,\ref{sec:3.1.2},
the $N\!=\!3$ is a special case where the momentum conservation and on-shell condition of each external state
plays additional roles to reduce the leading energy dependence.\ 
Namely, after imposing the momentum conservation at the trilinear vertex and the on-shell condition 
for each momentum, the individual leading-energy dependence of 3-point longitudinal KK graviton amplitude
reduces to $E^4$ and that of the corresponding KK Goldstone amplitude becomes $E^0$.\ 
Hence the GRET \eqref{eq:KK-GRET} further ensures a nontrivial energy cancellation of
$E^4\ito E^0$.\ 

\vs 

Then, we can further consider the scattering amplitudes of  
$N$ KK gravitons of $h_n^{\pm1}$ (with helicity $\pm1$) and of the 
corresponding gravitational KK vector-Goldstone bosons ($\VV_n^{\pm1}$).\ 
Using the generalized power-counting method\,\cite{Hang:2021fmp}, 
we derive their leading energy-power dependence as follows:
%
\beqs 
\label{eq:E-counting-hn+-1}
\begin{align} 
& D_E^{}(Nh_n^{\pm1})= N\!+\!2(L\!+\!1), \hspace*{6mm} 
D_E^{}(N\VV_n^{\pm1})=2(L\!+\!1),
\\[0.7mm]
& \Longrightarrow~~
D_E^{}(Nh_n^{\pm1})\hsm -\hsm D_E^{}(N\VV_n^{\pm1}) = N . 
\label{eq:E-counting-hn1-Vn1}
\end{align} 
\eeqs 
For the scattering amplitudes of $N$ KK gravitons ($h_n^{\pm1}$) at tree level, 
we use the right-hand side of the GRET identity \eqref{eq:KK-GRET-hV} 
[or \eqref{eq:KK-GRET-hV1f}]  
and prove the large energy cancellations of $E^{N+2} \!\ito E^2$
by an energy factor $E^{N}$
in the scattering amplitude of $N$ KK gravitons (with helicity $\pm1$).

\vs

For the warped KK gauge theories as studied in Section\,\ref{sec:2.2}, 
using the generalized power-counting method\,\cite{Hang:2021fmp} and
the GAET identity \eqref{eq:5D-ETI3-ALQ},  
we derive their leading energy-power dependence for $N$ longitudinal 
as follows:
\beqs 
\label{eq:E-counting-An}
\begin{align} 
& D_E^{}(N\hsm A_L^{a\hs n})= 4\hs, \hspace*{5.8mm} 
D_E^{}(N\hsm A_5^{a\hs n})
= 4\hsm -\hsmx N\hsmx -\hsmx N_v\hsmx -\hsmx \bar{V}_3^{\min}\hs,
\label{eq:E-counting-AL+A5}
\\[0.7mm]
& \Longrightarrow~~
D_E^{}(N\hsm A_L^{a\hs n})\hsm -\hsm D_E^{}(N\hsm A_5^{a\hs n}) 
= N\! +\hsm N_v\! +\hsm \bar{V}_3^{\min} \hs,
\label{eq:E-counting-AL-A5}
\end{align} 
\eeqs 
where $\bar{V}_3^{\min}$ denotes 
the involved minimal number of non-derivative cubic vertices 
in the KK Goldstone boson amplitude, 
with $\bar{V}_3^{\min}=0\,(1)$ for $N\!=\hs$even\,(odd).\ 
Hence, this proves that the GAET identity \eqref{eq:5D-ETI3-ALQ} 
guarantees nontrivial energy cancellations by an energy power of
$E^N$ (or, $E^{N+1}$) for $N\!\!=\hs$even (or, $N\!\!=\hs$odd)
in the scattering amplitude of $N$ longitudinal KK gauge bosons.\ 
In the above, $N_v$ denotes the number of possible
external $v_\mu^{}A_n^{a\mu}$ lines in the KK Goldstone boson amplitude,
where $v_\mu^{}\!\!=\!O(M_n/E)$ in the high energy limit.\   
We find that the above large energy cancellations [as shown in 
Eqs.\eqref{eq:E-counting-hnL-hn55}, \eqref{eq:E-counting-hn1-Vn1} and
\eqref{eq:E-counting-AL-A5}] hold for both the flat and warped 
5d KK gravity (gauge) theories.\  
Before ending this discussion, we further comment on 
the $N\!=\!3$ case for KK gauge boson (KK Goldstone boson) scattering amplitudes.\
The 3-point longitudinal KK gauge boson amplitude ($3A_L^{an}$) has a leading energy dependence
of $E^4$ by the power counting rule \eqref{eq:E-counting-AL+A5},
which arises from the trilinear KK gauge boson vertex
(containing one external momentum) and the three external longitudinal polarization vectors
(each of which scales like $\ep_L^\mu\!\sim\! p_j^\mu/M_{n_j^{}}$ in the high energy limit).\ 
On the other hand, the corresponding 3-point KK Goldstone amplitude vanishes at tree level 
due to the absence of cubic vertex $3A_5^{an}$.\ 
According to the GAET identity \eqref{eq:5D-ETI3-ALQ}, the leading nonzero contribution
on its right-hand side arises from the residual terms with the trilinear vertex having 
two KK Goldstone bosons and one KK gauge boson (contracting with the external vector $v^\mu$).\
So we have $(N,N_v)\!=\!(3,1)$ and $\bar{V}_3^{\min}\!=\!0$, and thus this residual term
has energy power
$D_E^{}\!=\!4\!-\!3\!-\!1\!-\!0\!=\!0$, 
showing that the right-hand side of the GAET identity \eqref{eq:5D-ETI3-ALQ}
has leading energy dependence of $E^0$.\ 
Hence, the total energy reduction of the 3-point longitudinal KK gauge boson amplitude is
$E^4\ito E^0$.\

We also note that for the Goldberger-Wise (GW) model\,\cite{GW}, it further introduces a bulk scalar field 
$\hat{\mathcal S}$ to stabilize the size of the 5d extra dimension of the RS1.\ Its zero-mode $\mathcal{S}_0$ 
can mix with the radion field ${h}_0^{55}$, so that the physical radion after stabilization becomes a mixture
of $\SS_0$ and ${h}_0^{55}$, whereas the scalar KK mode $\SS_n$ mixes with the Goldstone KK mode $h_n^{55}$ 
(denoted as $\phin$ in our notation)
at each KK level, leading to a mass-eigenstate $\SS_n'$ and the true Goldstone KK state $h_n^{55\prime}$.\ 
Thus, the gauge-fixing term \eqref{eq:L-GF-KKgravity}-\eqref{eq:FmuF5-5d} should be extended to
contain the true KK Goldstone boson field ($h_n^{55\prime}$).\ 
Consequently, the GRET \eqref{eq:KK-GRET1b} can be extended to hold the equivalence between 
the scattering amplitudes of the longitudinal KK graviton states ($h_n^L$) 
and of the true KK Goldstone boson states ($h_n^{55\prime}$) 
in the high energy limit.\footnote{%
In passing, Ref.\,\cite{chivukula-GW} studied spin-2 KK scattering amplitudes with 
massive radion in the GW model, but it did not concern the GRET issue.}

In passing, during the finalization of this paper, we received a recent 
paper\,\cite{Chivukula:2023qrt} which studied the tree-level formulation of the
gravitational ET for warped 5d RS1 model in the\,'t\,Hooft-Feynman gauge
and the $O(s)$ high-energy behavior of the 4-point amplitudes therein.\footnote{%
We thank colleagues Sekhar Chivukula and Elizabeth Simmons for discussions on their paper\,\cite{Chivukula:2023qrt}.}\  
Our general formulations in this section significantly differ from \cite{Chivukula:2023qrt} 
in several aspects.\ 
We first derived the new GAET for the warped KK gauge theory in Sec.\,\ref{sec:2.2},
in addition to deriving the general GRET for warped KK gravity theory in Sec.\,\ref{sec:2.3}.\
The GAET of the warped KK gauge theory is the focus of our present work and serves as 
the most fundamental KK ET formulation, because we will prove in Sec.\,\ref{sec:4} 
that the GRET of the warped KK gravity can be derived (reconstructed) from the warped GAET 
by our leading-order double-copy method.\ 
We derived in Sec.\,\ref{sec:2.3} two new BRST identities 
\eqref{eq:KD-X-1} and \eqref{eq:KD-X-2} to generally amputate the {\it mixed} 
external KK graviton states ($h_n^{\mn}$) and KK Goldstone states 
($\phin$ or $\VV_n^\mu$), which are needed because such mixings generally reappear 
at loop levels.\  We presented GRET type-II in general $R_\xi^{}$ gauge up to loop level (Sec.\,\ref{sec:2.3.3})
and GRET type-I up to loop level in 't\,Hooft-Feynman gauge for simplicity (Sec.\,\ref{sec:2.3.2}),
whereas in Appendix\,\ref{sec:B.1} we also formulated the GRET type-I 
(including the LSZ reduction for external states) in the general $R_\xi^{}$ gauge 
at tree level.\ Using the generalized power-counting method\,\cite{Hang:2021fmp}  
for the compactified KK GR theories [with the GRET \eqref{eq:KK-GRET} and \eqref{eq:KK-GRET-hV}] 
and the KK gauge theories [with the GAET \eqref{eq:5D-ETI3-ALQ}], 
we proved the different energy cancellations
for the scattering amplitudes of longitudinal KK gravitions (with helicity-0)
in Eq.\eqref{eq:E-counting-hnL}, of the KK gravitons (with helicity\,$\pm1$) 
in Eq.\eqref{eq:E-counting-hn+-1}, and of the KK gauge bosons in Eq.\eqref{eq:E-counting-An}.\

\vspace*{1mm}
\section{\hspace*{-2.5mm}Structure of Massive Scattering Amplitudes from Warped
\\ 
\hspace*{-2.5mm}Gauge and Gravitational Equivalence Theorems}
\label{sec:3}
\vspace*{1mm}

In this section, we systematically compute the 3-point and 4-point scattering amplitudes
of KK gauge bosons (KK gravitons) and of their corresponding KK Goldstone bosons.\
Then, we explicitly demonstrate the validity of KK GAET \eqref{eq:KK-ET1-N}
and GRET \eqref{eq:KK-GRET-hV1f}\eqref{eq:KK-GRET1b}
by comparing the scattering amplitudes of KK gauge bosons (KK gravitons)
with their corresponding KK Goldstone boson amplitudes
at the leading order of high energy expansion.\
The major motivation of this section is not only to explicitly verify the GAET and GRET
for the 3-point and 4-point massive KK scattering amplitudes,
but also to explicitly analyze and uncover the {\it structures}
of the basic massive KK gauge/gravity scattering amplitudes under the warped compactification,
which are highly intricate due to various nontrivial large energy-cancellations
in these massive amplitudes 
and due to the warped nonlinear KK mass-spectra plus KK couplings.\
Moreover, the analyses of this section provide all the 3-point and 4-point KK 
gauge/gravity amplitudes that are needed for the extended double-copy study 
in Section\,\ref{sec:4}.

\vspace*{1mm}
\subsection{\hspace*{-2.5mm}Warped Gauge and Gravitational ETs for
3-Point KK Amplitudes}
\label{sec:3.1}
\vspace*{1mm}

In this subsection, we systematically compute the basic 
3-point scattering amplitudes of massive KK gauge bosons at tree level 
in Section\,\ref{sec:3.1.1}, and the basic 3-point scattering amplitudes
of KK gravitons in Section\,\ref{sec:3.1.2}.

\subsubsection{\hspace*{-2.5mm}Warped GAET for 3-Point KK Gauge Boson Amplitudes}
\label{sec:3.1.1}
\vspace*{1mm}

To compute the on-shell 3-point KK gauge and Goldstone scattering amplitudes, 
we derive the Feynman rules for cubic KK gauge/Goldstone vertices 
in Appendix\,\ref{app:C1}.\ 
Then, we derive the following general 3-point on-shell 
scattering amplitude of KK gauge bosons:
\begin{equation}
\label{eq:AAA-n1n2n3}
\hspace*{-4mm}
\TT[A^{a n_1}_{\lam_1^{}}\!\hsm A^{b n_2}_{\lam_2^{}}\!\hsm A^{c n_3}_{\lam_3^{}}] 
= -\ii 2\hs g\hs a_{n_1^{}n_2^{}n_3^{}}^{} f^{abc}\hsm 
\[ (\ep_1^{}\!\cdot\hsm\ep_2^{})
(\ep_3^{}\!\cdot\hsm p_1^{}) \hsm +\hsm 
(\ep_2^{}\!\cdot\hsm\ep_3^{})(\ep_1^{}\!\cdot\hsm p_2^{})\hsm +\hsm (\ep_3^{}\!\cdot\hsm\ep_1^{})(\ep_2^{}\!\cdot\hsm p_3^{})\]\!, 
\end{equation}
where the subscript $\lam_j^{}\!\!=\!\hsm\pm,0$ ($j=1,2,3$) denotes the helicity
of each external state and $\ep_j^{\mu}(=\!\ep_{\lam_j^{}}^{\mu})$ is
the shorthand notation for the polarization vector of each external 
KK auge boson.\  All the three external momenta $\{p_j^{}\}$ 
are outgoing and obey the momentum conservation 
$p_1^{}\!+p_2^{}\!+p_3^{}\!=\hsm 0\hs$.\
In the above tree-level scattering amplitude, $g$ is the Yang-Mills gauge coupling and
$a_{n_1^{}n_2^{}n_3^{}}^{}$ denotes coupling coefficient induced by
integrating over the compactified 5th-dimension coordinate
as defined in \eqrefe{app-eq:KKYM-RS-couplings}.\ 

\vs

With these, we compute the 3-point scattering amplitude of 
KK gauge bosons of $LLT$ type at tree level, which contains 
two longitudinal KK gauge bosons [chosen as particles 1 and 2 and
denoted by $A_L^{a\hs n_1^{}}\hsm (p_1^{})$ and $A_L^{b\hs n_2^{}}\hsm(p_2^{})$], 
and one transverse KK gauge boson [chosen as the particle 3 and    
denoted by $A_{\pm}^{c\hs n_3^{}}\hsm (p_3^{})$].\ 
Thus, we derive the following 3-point KK gauge boson amplitude:
\begin{align}
\label{eq:LLT}
\hspace*{-8mm}
\TT[A_L^{an_1^{}}\!A_L^{bn_2^{}}\!A_{\pm}^{cn_3^{}}]
\,&=\, \mp \ii\hs g f^{abc} 
\frac{\,[E_1^2\!+\!E_2^2\!+\!E_3^2\!-\!2E_1^{}E_2^{}(1\!+\!\ctt)\!-\!2k^2]k\hs\st \,}
{\sqrt{2\,} M_{n_1^{}}^{}\! M_{n_2^{}}^{}} 
\hs a_{n_1^{}n_2^{}n_3^{}}^{} 
\nn\\
\hspace*{-8mm}
&=\, \mp \ii\hs g\hs f^{abc} 
\frac{~(M_{n_3^{}}^2 \!\!-\!M_{n_1^{}}^2\!\!-\!M_{n_2^{}}^2)k\hs\st~}
{\sqrt{2\,}M_{n_1^{}}^{}\! M_{n_2^{}}^{}} \hs {a_{n_1^{}n_2^{}n_3^{}}^{}} + O(E^0)
\\
\hspace*{-8mm}
&=\, -\ii\hs g \hs f^{abc} (\ep_3^{}\! \cdot\hsm {p_1^{}}) 
\frac{~(M_{n_3^{}}^2 \!\!-\!M_{n_1^{}}^2\!\!-\!M_{n_2^{}}^2)~}
{M_{n_1^{}}M_{n_2^{}}} \hs {a_{n_1^{}n_2^{}n_3^{}}^{}} + O(E^0) \hs,
\nn 
\end{align}
where each external momentum is on-shell ($\hs p_{j\mu}^{}\hs p_j^\mu\!=\hsm -M_{n_j^{}}^2$) and 
$\st\!=\!\sin\hsm\theta\,$ with $\theta$ denoting the angle 
between the moving directions of particles\,1 and 3\hs.\ 
The cubic coupling coefficient of KK gauge bosons  $a_{n_1^{}n_2^{}n_3^{}}^{}\!$ 
is given by \eqrefe{app-eq:KKYM-RS-couplings}.\ 
We further compute the corresponding 3-point scattering amplitude with 
particles\,1 and 2 replaced by the KK Goldstone bosons, 
$A_5^{a\hs n_1^{}}\hsm (p_1^{})A_5^{a\hs n_2^{}}\hsm (p_2^{})$.\ 
Thus, we derive the following 3-point scattering amplitudes of KK Goldstone 
and gauge bosons:\footnote{From this section and hereafter, 
we denote scattering amplitudes containing external Goldstone states with a tilde symbol, 
such as $\tT$ and $\MT$, to distinguish them from the amplitudes with all external states 
being physical particles.}
\begin{equation}
\label{eq:55T}
\tT[A_5^{an_1^{}}\!A_5^{bn_2^{}}\!A_{\pm}^{cn_3^{}}] 
\,=\, -\ii\hs 2\hs gf^{abc}(\ep_3^{}\!\cdot\hsm p_1^{})\hs  \tilde{a}_{n_1^{}n_2^{}n_3^{}}^{}
= \mp \ii\hs g\hs f^{abc}\hsm\sqrt{2\,}k\hs\st\hs\tilde{a}_{n_1^{}n_2^{}n_3^{}}^{} \hs, 
\end{equation}
where the cubic KK Goldstone-gauge coupling coefficient $\hs\tilde{a}^{}_{n_1^{}n_2^{}n_3^{}}\!$ 
is given by \eqrefe{app-eq:KKYM-RS-couplings}.\

\vs 

According to the KK GAET \eqref{eq:KK-ET1-N}, the equivalence between the two KK scattering amplitudes \eqref{eq:LLT} and \eqref{eq:55T} in the high energy limit, 
\beq
\label{eq:GAET-LLT-3pt}
\TT_0^{}[A_L^{an_1^{}}\!A_L^{bn_2^{}}\!A_{\pm}^{cn_3^{}}]
\hs =\hs -\tT_0^{}[A_5^{an_1^{}}\!A_5^{bn_2^{}}\!A_{\pm}^{cn_3^{}}] \hs , 
\eeq 
where $\TT_0^{}$ and $\tT_0^{}$ denote the leading-order KK amplitudes
under high energy expansion.
The 3-point GAET \eqref{eq:GAET-LLT-3pt} is equivalent to the following 
condition:
\begin{equation}
\label{eq:a-at-3pt}
\(M_{n_1^{}}^2\!\!+\!M_{n_2^{}}^2\!-\!M_{n_3^{}}^2\)
a_{n_1^{}n_2^{}n_3^{}}^{} =\, 
2 {M_{n_1^{}}^{}\! M_{n_2^{}}^{}} \tilde{a}_{n_1^{}n_2^{}n_3^{}}^{}\hs,
\end{equation}
which connects the trilinear KK gauge boson coupling $\,a_{n_1^{}n_2^{}n_3^{}}^{}\hsm$
to the corresponding trilinear KK Goldstone-gauge boson coupling
$\,\tilde{a}_{n_1^{}n_2^{}n_3^{}}^{}$.\ 
The condition \eqref{eq:a-at-3pt} 
does hold for the warped 5d KK Yang-Mills gauge theory 
as we directly prove it in Eq.\eqref{app-eq:a-ta-general} of Appendix\,\ref{app:D}.
For the special case of $(n_1^{},\hs n_2^{},\hs n_3^{})=(n,\hs n,\hs j)$, 
the condition \eqref{eq:a-at-3pt} takes the following simple form:
\begin{equation}
\label{eq:a-at-nnj}
\big(1\!-\!\Fr{1}{2}\hs r_j^2\big)a_{nnj}^{} =\, \tilde{a}_{nnj}^{}\hs,
\end{equation}
where we denote $\,r_j^{}\!=\!\Mj\hsm/\Mn\hs$.\ 
We can also examine the condition \eqref{eq:a-at-3pt} by taking the flat 5d limit.\
Under this limit, we have the KK indices 
$\{n_1^{},n_2^{},n_3^{}\}\!=\!\{n,n,2n\},\{n,n,0\}$, 
the KK mass $M_n\!=\!n/R$, and the KK couplings 
$a_{nn2n}^{}\!=\!-\tilde{a}_{nn2n}^{}\!=\!1/\sqrt{2\,}$ and 
$a_{nn0}^{}\hsm =\hsm\tilde{a}_{nn0}^{}\!=\!1\hs$,
as shown in Table\,\ref{tab-app:1} of Appendix\,\ref{app:C}.\ 
For the case $\{n,n,2n\}$, we can use the coupling relation
$a_{nn2n}^{}\!=\!-\tilde{a}_{nn2n}^{}$ to simplify Eq.\eqref{eq:a-at-3pt} as 
$M_{n_1^{}}^{}\!\!\!+\!M_{n_2^{}}^{}\!\!=\!M_{n_3^{}}^{}$,
which corresponds to the KK number conservation 
$n_1^{}\!+\!n_2^{}\!=\!n_3^{}$.\  
But, under the warped 5d compactification, the 5d momentum is no longer conserved
due to the absence of translation invariance,  
so that both the mass formula 
and coupling relation of the flat 5d limit do not hold,
and the condition \eqref{eq:a-at-3pt} works in a highly nontrivial way
[as proved in Eq.\eqref{app-eq:a-ta-general}].\ 

\vs

Next, we compute the 3-point KK gauge boson scattering amplitude including 
one longitudinal and two transverse external states and the corresponding KK Goldstone-gauge boson 
scattering amplitude:
\beqs
\begin{align}
\TT [A^{an_1^{}}_\pm \! A^{bn_2^{}}_\pm \! A^{cn_3^{}}_L] 
&= -\ii\hs g f^{abc} {a_{n_1^{} n_2^{} n_3^{}}^{}} 
\Big\{\! (\ep_1^{}\hsm\!\cdot\hsm\ep_2^{}) \!\!\[\!
(M_{n_1^{}}^2\!\!-\!M_{n_2^{}}^2)M_{n_3}^{-1} \hsm\!+\!(p_1^{}\hsm\!-\hsm p_2^{}) \!\cdot\!v_3^{}\hsm\]  
\nn\\
&\hspace*{4mm} +\!2(\ep_1^{}\!\hsm\cdot\hsm p_2^{})(\ep_2^{}\!\hsm\cdot\hsm v_3^{})
-\!2(\ep_1^{}\hsm\!\cdot\hsm v_3^{})(\ep_2^{}\!\cdot\hsm p_1^{})\!  \Big\} 
\nn\\
&= 0 + O(E^0) \,, 
\label{eq:ApApAL}
\\
\tT [A^{an_1^{}}_\pm \! A^{bn_2^{}}_\pm \! A^{cn_3^{}}_5] 
&= g f^{abc}  \big(\ep_1^{}\!\cdot\hsm\ep_2^{}\big)
(\tilde{a}_{n_1^{}n_3^{}n_2^{}}^{}\hsm M_{n_1^{}}^{}\! - 
\tilde{a}_{n_2^{}n_3^{}n_1^{}}^{}\hsm M_{n_2^{}}^{})
\nn\\
&= g f^{abc} {a}_{n_1^{}n_2^{}n_3^{}}^{}  (\ep_1^{}\!\cdot\hsm\ep_2^{})
\big(M_{n_1^{}}^2\!\!-\!M_{n_2^{}}^2\big)\hsm M_{n_3^{}}^{-1}
= 0 + O(E^0) \hs,
\label{eq:ApApA5}
\end{align}
\eeqs
where $v_3^\mu \!=\hsm\hsm \ep_{3,L}^\mu\hsm\!-\hsm p_3^\mu/{M_{n_3^{}}^{}}\! 
\!=\! O(M_{n_3^{}}^{}\hsm /\hsm E_3^{})$, 
and we have used the sum rules \eqref{app-eq:anml} 
to simplify the second KK amplitude \eqref{eq:ApApA5}.\
From the above, we see that the leading contributions of $O(E^1)$ vanishes,
and the remaining contributions are of the same $O(E^0)$ as the residual term 
$\hs\TT_v^{}\hsm =\hsm \TT[A^{an_1^{}}_\pm \! A^{bn_2^{}}_\pm \! A^{cn_3^{}}_\mu]\hs v^\mu$.\  

\vs 

We have also analyzed the 3-point scattering amplitudes
of the pure longitudinal KK gauge bosons and of their corresponding KK Goldstone bosons.\ 
The 3-point longitudinal KK gauge boson amplitude $\TT[3A_L^{an}]$ 
has the leading energy dependence
of $E^4$ by the power counting rule \eqref{eq:E-counting-AL+A5},
which arises from the trilinear KK gauge boson vertex
(depending on the first power of external momenta) 
and the three external longitudinal polarization vectors,
each of which has the high energy behavior
\beq
\label{eq:epL-epS-v-order}
\ep_L^\mu =\ep_S^\mu \hsm +\hsm v^\mu =O(E^1)\hsm +\hsm O(E^{-1})\hs,
\eeq
according to Eq.\eqref{eq:epL-epS-v},
where $\hs\ep_S^\mu\!=\!p_j^\mu\hsm /M_{n_j^{}}$ for the $j$-th external state ($j\!=\!1,2,3$).\  
The four external momenta (from either the trilinear vertex or the $\ep_S^\mu$ part
of the external longitudinal polarization vector $\ep_L^\mu$) will form two bilinear contractions
like $\,p_i^{}\!\cdot\hsm p_j^{}$ or $p_j^2$, which can be converted into squared KK-masses
due to the momentum conservation $\hs p_1^{}\!+\!p_2^{}\!+\!p_3^{}\!=\!0\hs$  
and the on-shell condition of each external momentum $p_j^2\!=\!\!-M_{n_j^{}}^2$.\
So all the leading energy dependence of $E^4$ will be reduced to $E^0$.\  The remaining contributions
arise from replacing the external longitudinal polarization vectors by one, two, and three
vectors of $v^\mu\!\!=\!O(E^{-1})$, 
which can contribute to the 3-point KK gauge boson amplitude with leading energy dependence of
$E^0$, $E^0$, and $E^{-2}$ respectively.\   
Hence, the 3-point longitudinal KK gauge boson amplitude should actually have a deduction on
its leading energy dependence,  $E^4\ito E^0$.\ 
We can show that this is consistent with the requirement of the GAET identity \eqref{eq:5D-ETI3-ALQ}.\ 
We note that the corresponding 3-point KK Goldstone amplitude vanishes at tree level
due to the absence of cubic vertex $A_5^{an}$-$A_5^{am}$-$A_5^{a\ell}$.\ 
Thus, according to the GAET identity \eqref{eq:5D-ETI3-ALQ}, the leading nonzero contribution
on its right-hand side arises from the residual terms with the trilinear vertex having
two KK Goldstone bosons and one KK gauge boson (contracting with the external vector $v^\mu$)
which is apparently of $O(E^0)$.\  Note that the second power counting rule in 
Eq.\eqref{eq:E-counting-AL+A5} gives: 
$D_E^{}\!=\!4\!-\!3\!-\!1\!=\!0\hs$, where we have inputs  
$(N,\hs N_v)\!=\!(3,\hs 1)$ and $\bar{V}_3^{\min}\!=\!0\hs$.\ 
Hence, the GAET identity \eqref{eq:5D-ETI3-ALQ} also enforces the energy deduction of 
$E^4\!\ito\hsm  E^0$ for the 3-point longitudinal KK gauge boson amplitude on its 
left-hand side.\

\vs 

Then, we explicily compute the 3-point scattering amplitude of pure longitudinal KK gauge bosons, 
in which each external longituding polarization vector can be split   
into two parts, $\ep_L^\mu =\ep_S^\mu + v^\mu$, as shown in Eq.\eqref{eq:epL-epS-v-order}.\  
Thus, we derive the following expanded scattering amplitude of 3 longitudinal KK gauge bosons:
\\[-6mm]
{\small 
\begin{align}
& \TT[A^{a n_1^{}}_L\! A^{b n_2^{}}_L\! A^{c n_3^{}}_L] 
= -\ii g f^{abc} ({M_{n_1^{}}\!M_{n_2^{}}\hsm M_{n_3^{}}})^{-1}a_{n_1^{} n_2^{} n_3^{}}^{}\times
\nn\\[-1mm]
&\hspace*{6mm}
\Big[ (p_1^{}\!\cdot\hsm p_2^{})( p_3^{}\!\cdot\hsm v_2^{}) - (p_1^{} \!\cdot\hsm p_2^{})(p_3^{}\!\cdot\hsm v_1^{}) 
- ( p_2^{}\!\cdot\hsm p_3^{})(p_1^{}\!\cdot\hsm v_2^{}) +(p_1^{}\!\cdot\hsm p_3^{})( p_2^{}\!\cdot\hsm v_1^{})
\nn\\[-1mm]
&\hspace*{4mm}
+( p_2^{}\!\cdot\hsm p_3^{})(p_1^{}\!\cdot\hsm v_3^{}) -(p_1^{}\!\cdot\hsm p_3^{})( p_2^{}\!\cdot\hsm v_3^{})
+(p_1^{}\!\cdot\hsm v_3^{})( p_2^{}\!\cdot\hsm v_1^{}) - (p_1^{}\!\cdot\hsm p_2^{})( v_1^{}\!\cdot\hsm v_3^{})
-(p_1^{}\!\cdot\hsm v_2^{}) (p_2^{}\!\cdot\hsm v_3^{}) 
\nn\\	
&\hspace*{4mm}
+(p_1^{}\!\cdot\hsm p_2^{})(v_2^{}\!\cdot\hsm v_3^{}) - (p_1^{}\!\cdot\hsm v_2^{}) (p_3^{}\!\cdot\hsm v_1^{})
+(p_1^{}\!\cdot\hsm p_3^{})(v_1^{}\!\cdot\hsm v_2^{}) + (p_1^{}\!\cdot\hsm v_3^{}) (p_3^{}\!\cdot\hsm v_2^{})
-(p_1^{}\!\cdot\hsm p_3^{})(v_2^{}\!\cdot\hsm v_3^{})
\nn\\
&\hspace*{4mm}
+(p_1^{}\!\cdot\hsm v_3^{})(v_1^{}\!\cdot\hsm v_2^{}) - (p_1^{}\!\cdot\hsm v_2^{})( v_1^{}\!\cdot\hsm v_3^{})
-(p_2^{}\!\cdot\hsm p_3^{})(v_1^{}\!\cdot\hsm v_2^{}) + (p_2^{}\!\cdot\hsm v_1) (p_3^{}\!\cdot\hsm v_2^{})
+(p_2^{}\!\cdot\hsm p_3^{}) (v_1^{}\!\cdot\hsm v_3)
\nn\\[-1mm]
&\hspace*{4mm} 
-(p_2^{}\!\cdot\hsm v_3^{})( p_3^{}\!\cdot\hsm v_1^{}) - (p_2^{}\!\cdot\hsm v_3^{})( v_1^{}\!\cdot\hsm v_2^{})
+(p_2^{}\!\cdot\hsm v_1^{})(v_2^{}\!\cdot\hsm v_3^{}) + (p_3^{}\!\cdot\hsm v_2^{}) (v_1^{}\!\cdot\hsm v_3^{})
-(p_3^{}\!\cdot\hsm v_1^{}) (v_2^{}\!\cdot\hsm v_3^{}) 
\Big].
\label{eq:Amp-3AL-v}
\end{align}
}
\\[-6mm]
\hspace*{-2mm}
In the above amplitude, we see that all the terms having leading energy dependence of $E^4$ disappear as expected,  
and all the remaining terms can be classfied into the following 4 types according to the numbers of $v_j^{}$'s
contained therein:
\begin{equation}
\label{eq:3AL-Amp-4type}
(p_i^{} \hsm\cdot\hsm p_j^{})(p_k^{} \hsm\cdot\hsm v_\ell^{}), \quad
(p_i^{} \hsm\cdot\hsm v_j^{})(p_k^{} \hsm\cdot\hsm v_\ell^{}), \quad
(p_i^{} \hsm\cdot\hsm p_j^{})(v_k^{} \hsm\cdot\hsm v_\ell^{}), \quad  
(p_i^{} \hsm\cdot\hsm v_j^{})(v_k^{} \hsm\cdot\hsm v_\ell^{}).
\end{equation}
Due to the momentum conservation
$\hs p_1^{}\hsm +\hsm p_2^{}\hsm +\hsm p_3^{}\!=\! 0\hs$  
and the on-shell condition of each external momentum, 
the bilinear momentum products can all be converted into squared KK-masses,
$\hs p_i^{} \hsm\cdot\hsm p_j^{} \!=\! \frac{1}{2} (M_{n_i}^2 \hsm\!+\! M_{n_j}^2 \hsm\!-\! M_{n_k}^2)$ 
(with $i\!\neq\! j\!\neq\! k$)
and $p_j^2\!=\!-M_{n_j^{}}^2$.\  
Since $v_j^{} \!=\! O(M_{n_j^{}}\hsm /E_j)$, 
the 3 longitudinal KK gauge boson amplitude \eqref{eq:Amp-3AL-v} 
could only have leading energy dependence of $E^0$, 
which is given by the terms of the first and second types in Eq.\eqref{eq:3AL-Amp-4type}.\ 
Thus, the GAET identity \eqref{eq:5D-ETI3-ALQ} for the 3-point tree-level scattering amplitude of longitudinal KK gauge bosons 
take the following form: 
\begin{equation}
\label{eq:3AL=3A5+Tv}
\TT[A^{a n_1^{}}_L\! A^{b n_2^{}}_L\! A^{c n_3^{}}_L] 
\hs =\hs \ii^3\hs \tT[A^{a n_1^{}}_5\! A^{b n_2^{}}_5\! A^{c n_3^{}}_5]\hsm +\hsm \TT_v^{}
\hs =\hs 0 + O(E^0) \hs ,
\end{equation}
where on the right-hand side the 3-point KK Goldstone amplitude vanishes and the residual term $\hs\TT_v\hs$ 
gives the leading nonzero contributions 
to the corresponding longitudinal KK gauge boson amplitude on the left-hand side.\
Finally, we can further deduce the $O(E^0)$ contributions to the 3-point scattering amplitude 
of pure longitudinal KK gauge bosons as follows:
{\small 
\begin{align}
& \hspace*{-8mm}
\TT_0^{}[A^{a n_1^{}}_L\! A^{b n_2^{}}_L\! A^{c n_3^{}}_L] 
= -\ii\hs g f^{abc}  
(M_{n_1^{}} M_{n_2^{}} M_{n_3^{}})^{-1} a_{n_1^{} n_2^{} n_3^{}}^{} \times
\nn\\
& \Big[ M_{13}^{}(p_2^{}\!\cdot\hsm v_1^{})\!-\! M_{12}^{}(p_3^{}\!\cdot\hsm v_1^{})\!-\! 
M_{23}^{}(p_1^{}\!\cdot\hsm v_2^{}) \!+\!M_{23}^{}(p_1^{}\!\cdot\hsm v_3^{}) \!+\!M_{12}^{}( p_3^{}\!\cdot\hsm v_2^{})
\!-\!M_{13}( p_2^{}\!\cdot\hsm v_3^{})
\nn\\
&  +(p_1^{}\!\cdot\hsm  v_3^{})( p_2^{}\!\cdot\hsm v_1^{}) \!-\!(p_1^{}\!\cdot\hsm  v_2^{}) (p_2^{}\!\cdot\hsm  v_3^{}) 
-(p_1^{}\!\cdot\hsm v_2^{})( p_3^{}\!\cdot\hsm v_1^{}) \!+\! (p_1^{}\!\cdot\hsm v_3^{})(p_3^{}\!\cdot\hsm v_2^{}) 
\nn\\
&
+(p_2^{}\!\cdot\hsm v_1^{})( p_3^{}\!\cdot\hsm v_2^{}) -(p_2^{}\!\cdot\hsm  v_3^{})(p_3^{}\!\cdot\hsm v_1^{})  
\Big],
\end{align}
}
\hspace*{-2mm}
where we have defined 
${M_{ij}^{}} \hsm\equiv\hsm p_i^{} \hsm\cdot\hsm p_j^{} \!=\! \frac{1}{2} (M_{n_i}^2 \!+\! M_{n_j}^2 \!-\! M_{n_k}^2)$ 
with $i\!\neq\! j\!\neq\! k\hs$ denoting the three external states.\

\vspace*{1.5mm}
\subsubsection{\hspace*{-2.5mm}Warped GRET for 3-Point KK Graviton Amplitudes}
\label{sec:3.1.2}
\vspace*{1.5mm}

In this subsection, we compute the 3-point 
massive scattering amplitudes of KK gravitons
and of the corresponding gravitational KK Goldstone bosons.\ 
With these we explicitly demonstrate that the KK GRETs 
\eqref{eq:KK-GRET1b} and \eqref{eq:KK-GRET-hV1f} hold at tree level. 

\vs

Using the Feynman rules of Appendix\,\ref{app:C2}, we explicitly compute the most general 
3-point on-shell scattering amplitudes of KK gravitons at tree level as given in Eq.\eqref{eq:Amp-3h-exact}.\ 
Using this, we derive the 3-point KK graviton scattering amplitude with the external states\,(1,\,2,\,3) 
having helicities $(\pm1,\,\pm1,\,\pm2)$.\ We also compute the corresponding on-shell 
3-point scattering amplitudes with external states\,(1,\hs 2) replaced by the gravitational 
KK vector Goldstone bosons having helicities $\pm1\hs$.\ Thus, we derive the following 3-point
gravitational KK scattering amplitudes: 
\beqs 
\label{eq:hhh112-VVh112}
\begin{align} 
\label{eq:hhh112} 
\hspace*{-4mm}
& \M[h^{\pm1}_{n_1}h^{\pm1}_{n_2}h^{\pm2}_{n_3}\hs] 
= {-\ka} \Big[  (\ep_1^{} \!\cdot\hsm p_2^{}) (\ep_1^L \!\cdot\hsm \ep_2^L) (\ep_2^{} \!\cdot\hsm\ep_3^{}) 
(\ep_3^{}\!\cdot\hsm p_1^{}) +(\ep_1^{}\!\cdot\hsm \ep_2^L) (\ep_1^L\!\cdot\hsm p_2^{}) (\ep_2^{}\!\cdot\hsm\ep_3^{}) (\ep_3^{}\!\cdot\hsm p_1^{}) 
\nn\\[-1mm]
\hspace*{-4mm} & \hspace*{5mm} 
+(\ep_1^{}\!\cdot\hsm p_2^{}) (\ep_1^L\!\cdot\hsm \ep_2^{}) (\ep_2^L\!\cdot\hsm \ep_3^{}) (\ep_3^{}\!\cdot\hsm p_1^{}) 
+(\ep_1^{}\!\cdot\hsm \ep_2^{}) (\ep_1^L \!\cdot\hsm p_2^{}) (\ep_2^L\!\cdot\hsm \ep_3^{}) (\ep_3^{}\!\cdot\hsm p_1^{}) 
\nn\\
\hspace*{-4mm} & \hspace*{5mm}  
+(\ep_3^{}\!\cdot\hsm\ep_1^{}) (\ep_1^L\!\cdot\hsm\ep_2^L)( \ep_2^{}\!\cdot\hsm p_3^{})( \ep_3^{}\!\cdot\hsm p_1^{})  
+(\ep_3^{}\!\cdot\hsm \ep_1^{})(\ep_1^L\!\cdot\hsm \ep_2^{})( \ep_2^L \!\cdot\hsm p_3^{})( \ep_3^{}\!\cdot\hsm p_1^{}) 
\nn\\
\hspace*{-4mm} & \hspace*{5mm}  
+(\ep_1^{}\!\cdot\hsm \ep_2^L)( \ep_3^{}\!\cdot\hsm \ep_1^L) (\ep_2^{}\!\cdot\hsm p_3^{})
(\ep_3^{}\!\cdot\hsm p_1^{}) +(\ep_1^{}\!\cdot\hsm \ep_2^{})(\ep_3^{} \!\cdot\hsm\ep_1^L)(\ep_2^L\!\cdot\hsm p_3^{})
(\ep_3^{} \!\cdot\hsm p_1^{}) 
\nn\\
\hspace*{-4mm} & \hspace*{5mm} 
+(\ep_1^{}\!\cdot\hsm\ep_2^L)( \ep_1^L\!\cdot\hsm \ep_2^{}) (\ep_3^{}\!\cdot\hsm p_1^{})^2 +(\ep_1^{}\!\cdot\hsm \ep_2^{})
( \ep_1^L\!\cdot\hsm \ep_2^L) (\ep_3^{} \!\cdot\hsm p_1^{})^2 
\\
\hspace*{-4mm} & \hspace*{5mm} 
+(\ep_3^{}\!\cdot\hsm \ep_1^{})( \ep_1^L\!\cdot\hsm p_2^{})( \ep_2^{}\!\cdot\hsm p_3^{})( \ep_2^L\!\cdot\hsm \ep_3^{}) 
+(\ep_3^{}\!\cdot\hsm \ep_1^{})( \ep_1^L \!\cdot\hsm p_2^{})( \ep_2^{} \!\cdot\hsm \ep_3^{})( \ep_2^L\!\cdot\hsm p_3^{}) 
\nn\\
\hspace*{-4mm} & \hspace*{5mm}  
+(\ep_1^{} \!\cdot\hsm p_2^{})( \ep_3^{} \!\cdot\hsm\ep_1^L)(\ep_2^{}\!\cdot\hsm p_3^{})( \ep_2^L\!\cdot\hsm \ep_3^{}) 
+(\ep_1^{} \cdot p_2^{})( \ep_3^{}\!\cdot\hsm \ep_1^L)( \ep_2^{} \!\cdot\hsm \ep_3^{})( \ep_2^L\!\cdot\hsm p_3^{}) 
\nn\\
\hspace*{-4mm} & \hspace*{5mm}   
+2 (\ep_1^{}\!\cdot\hsm p_2^{})( \ep_1^L\!\cdot\hsm p_2^{})( \ep_2^{}\!\cdot\hsm \ep_3^{})
( \ep_2^L\!\cdot\hsm \ep_3^{}) +2 (\ep_3^{}\!\cdot\hsm \ep_1^{}) (\ep_3^{}\!\cdot\hsm\ep_1^L)(\ep_2^{}\!\cdot\hsm p_3^{})
( \ep_2^L\!\cdot\hsm p_3^{}) \Big]\al_{n_1^{}n_2^{}n_3^{}}^{}\hs,
\nn\\
\label{eq:VVh112}
\hspace*{-4mm} & 
\MT[\VV^{\pm1}_{n_1}\VV^{\pm1}_{n_2}h^{\pm2}_{n_3}\hs] =
{\ka} \Big[\hsm (\ep_1^{}\!\cdot\hsm \ep_2^{})(\ep_3^{}\!\cdot\hsm p_1^{})^2\!
+\hsm (\ep_2^{}\!\cdot\hsm\ep_3^{})(\ep_1^{}\!\cdot\hsm p_2^{})(\ep_3^{}\!\cdot\hsm p_1^{}) 
+\hsm (\ep_3^{}\!\cdot\hsm\ep_1^{})(\ep_2^{}\!\cdot\hsm p_3^{})(\ep_3^{}\!\cdot\hsm p_1^{}) 
\nn\\
\hspace*{-4mm} & \hspace*{5mm} 
-(\mathbb{M}_{n_1^{}}^2\!\!+\! \mathbb{M}_{n_2^{}} \mathbb{M}_{n_3^{}} \!\!+\! \mathbb{M}_{n_3^{}}^2)(\ep_3^{}\!\cdot\hsm \ep_1^{})
(\ep_3^{}\!\cdot\hsm\ep_2^{})\Big] \tilde{\al}^{}_{n_1^{}n_2^{}n_3^{}}\hs,
\end{align}
\eeqs
where $\ep_j^{}\!=\!\ep_j^{\pm}$ denotes transverse polarization vector associated 
with the external KK graviton $j\hs (=\!1,2,3)$, 
and $\ep_{j'}^L$ ($j'\!=\!1,2$) denotes longitudinal polarization vector associated 
with the external KK graviton\,$j'$\hs.\ 
[The polarization tenors of each massive KK graviton with helicities $(\pm2,\,\pm1,\,\pm0)$ 
are defined in Eq.\eqref{eq:Gpol} of Appendix\,\ref{app:A}.]
In the above, the three-point gravitational KK coupling coefficients  
$(\al_{n_1^{}n_2^{}n_3^{}}^{},\,\tilde{\al}_{n_1^{}n_2^{}n_3^{}}^{})$ 
are defined in \eqrefe{app-eq:KKYM-RS-couplings} of Appendix\,\ref{app:C1}.\ 
Then, we make high energy expansions of the above 3-point gravitational KK scattering amplitudes 
\eqref{eq:hhh112} and \eqref{eq:VVh112}, 
and derive their leading-order scattering amplitudes as follows:
\beqs
\label{eq:GRET-LO-h1h1h2-VVh2}
\begin{align}
\label{eq:GRET-LO-h1h1h2}
\hspace*{-3mm}
\M_0^{}[h^{\pm1}_{n_1}h^{\pm1}_{n_2}h^{\pm2}_{n_3}\hs] & = 
\frac{\,\ka\big(\mathbb{M}_{n_3^{}}^2\!\!-\!\mathbb{M}_{n_1^{}}^2\!\!-\!\mathbb{M}_{n_2^{}}^2\big)\,}
{2\hs \mathbb{M}_{n_1^{}}^{}\!\mathbb{M}_{n_2^{}}^{}}
\Big[\!(\ep_1^{}\!\cdot\hsm\ep_2^{})(\ep_3^{}\!\cdot\hsm p_1^{})^2\!
+(\ep_2^{}\!\cdot\hsm\ep_3^{})(\ep_1^{}\!\cdot\hsm p_2^{})(\ep_3^{}\!\cdot\hsm p_1^{})
\nn\\
&\hspace*{5mm} 
+\hsm (\ep_3^{}\!\cdot\hsm\ep_1^{})(\ep_2^{}\!\cdot\hsm p_3^{})(\ep_3^{}\!\cdot\hsm p_1^{})\Big]
\al_{n_1^{}n_2^{}n_3^{}}^{} ,
\\
\label{eq:GRET-LO-VVh2}
\hspace*{-3mm}
\MT_0^{}[\VV^{\pm1}_{n_1}\VV^{\pm1}_{n_2}h^{\pm2}_{n_3}\hs] & =
{\ka} \Big[(\ep_1^{}\!\cdot\hsm \ep_2^{})(\ep_3^{}\!\cdot\hsm p_1^{})^2\!
+\hsm (\ep_2^{}\!\cdot\hsm\ep_3^{})(\ep_1^{}\!\cdot\hsm p_2^{})(\ep_3^{}\!\cdot\hsm p_1^{})
\nn\\
&\hspace*{5mm} 
+\hsm (\ep_3^{}\!\cdot\hsm\ep_1^{})(\ep_2^{}\!\cdot\hsm p_3^{})(\ep_3^{}\!\cdot\hsm p_1^{})\Big] \tilde{\al}^{}_{n_1^{}n_2^{}n_3^{}}.
\end{align}
\eeqs 
The equivalence between the two leading-order KK scattering amplitudes
$\M_0^{}$ and $\MT_0^{}$ in Eqs.\eqref{eq:GRET-LO-h1h1h2}-\eqref{eq:GRET-LO-VVh2} 
is realized by the identity,
\beq
\label{eq:GRET2-LLT-3pt}
\M_0^{}[h^{\pm1}_{n_1^{}}h^{\pm1}_{n_2^{}}h^{\pm2}_{n_3^{}}\hs]
\hs =\hs -\MT_0^{}[\VV^{\pm1}_{n_1^{}}\VV^{\pm1}_{n_2^{}}h^{\pm2}_{n_3^{}}\hs]\hs , 
\eeq 
where the minus sign on the right-hand side comes from the multiplicative modification factor 
$\hat{C}_{\rm{mod}}^{}\!=(-\ii)^2$ at tree level, 
and \eqrefe{eq:GRET2-LLT-3pt} enforces the following condition:
\begin{equation}
\label{eq:al-tal-3pt}
\(\mathbb{M}_{n_1^{}}^2\!\!+\!\mathbb{M}_{n_2^{}}^2\!-\!\mathbb{M}_{n_3^{}}^2\)\!\al_{n_1^{}n_2^{}n_3^{}}^{}\!
=\, 2\hs \mathbb{M}_{n_1^{}}^{}\! \mathbb{M}_{n_2^{}}^{}\tilde{\al}_{n_1^{}n_2^{}n_3^{}}^{}\hs.
\end{equation} 
We have directly proved that the above condition \eqref{eq:al-tal-3pt} 
does hold for the warped 5d KK gravity theory 
as shown in Eq.\eqref{app-eq:al-tal-general} of Appdenix\,\ref{app:D}.\
We compare the above Eq.\eqref{eq:al-tal-3pt} with Eq.\eqref{eq:a-at-3pt} which holds the 
GAET \eqref{eq:GAET-LLT-3pt}
for 3-point KK gauge boson (KK Goldstone boson) amplitude.\ 
We observe that the condition \eqref{eq:al-tal-3pt} has {\it identical structure} to 
the condition \eqref{eq:a-at-3pt} and can be inferred from Eq.\eqref{eq:a-at-3pt} 
under the following mass and coupling correspondences (replacements):
\beq
M_{n_{\hsm j}^{}}^{}\!\!\to \MG_{n_{\hsm j}}^{},~~~~
a_{n_1^{}n_2^{}n_3^{}}^{} \!\!\to \al_{n_1^{}n_2^{}n_3^{}}^{},~~~~
\tilde{a}_{n_1^{}n_2^{}n_3^{}}^{} \!\!\to \tilde{\al}_{n_1^{}n_2^{}n_3^{}}^{}.
\eeq 
As we will show in Section\,\ref{sec:4.1}, the above correspondences (replacements) are
consistent with the double-copy construction of the GRET from the GAET 
for the 3-point KK scattering amplitudes.

Besides, for the special case of $(n_1^{},\hs n_2^{},\hs n_3^{})\!=\!(n,\hs n,\hs j)$, 
the condition \eqref{eq:al-tal-3pt} can be simplified as follows: 
\begin{equation}
\label{eq:al-tal-nnj}
\(1\!-\!\Fr{1}{2}\hs \rh_j^2\)\!\al_{nnj}^{} =\, \tilde{\al}_{nnj}^{}\hs,
\end{equation}
with the defined mass ratio $\,\rh_j^{}\!=\!\MGj\hsm /\MGn\hs$.\

\vs 

Next, we derive the 3-point KK graviton scattering amplitude with external states\,(1,\,2) 
being longitudinal KK gravitons of helicity-0 and with the external state 3 having helicities $\pm2\hs$.\ 
We also compute the 3-point KK Goldstone scattering amplitude with the external states\,(1,\,2) 
replaced by the corresponding gravitational KK Goldstone bosons.\ 
Thus, using the general 3-point KK amplitudes \eqref{eq:Amp-3h-exact} of Appendix\,\ref{app:C2}, 
we infer the following 3-point gravitational KK scattering amplitudes at the leading order of high energy expansion: 
\beqs
\label{eq:GRET-3pt-LLpp}
\begin{align}
\label{eq:hhh-LLT}
\M_0^{}[h^L_{n_1^{}}\hsm h^L_{n_2^{}}\hsm h^{\pm2}_{n_3^{}}\hs] 
&= \frac{\,\ka\hs (\ep_3^{}\!\cdot\hsm p_1^{})^2\,}{\,6\hs\mathbb{M}_{n_1^{}}^{2}\!\mathbb{M}_{n_2^{}}^{2}\,}
\!\[2 \mathbb{M}_{n_1^{}}^2\!\mathbb{M}_{n_2^{}}^2 \hsm\!+\!\(\mathbb{M}_{n_1^{}}^2\!\!+\!\mathbb{M}_{n_2^{}}^2\!-\!\mathbb{M}_{n_3^{}}^2\)^{\!2}\,\] \!\hsm\al_{n_1^{}n_2^{}n_3^{}}^{} \hs,
\\[1mm]
\label{eq:pph}
\MT_0^{}[\phi_{n_1^{}}^{}\!\phi_{n_2^{}}^{}\!h_{n_3^{}}^{\pm2}\hs]
&= \ka (\ep_3^{}\!\cdot\hsm p_1^{})^2 \tilde{\be}_{n_1^{}n_2^{}n_3^{}}^{}\hs,
\end{align}
\eeqs
where the trilinear KK coupling coefficient $\tilde{\be}_{n_1^{}n_2^{}n_3^{}}^{}\!$ 
is defined in \eqrefe{app-eq:KKGR-RS-betaT-nml} of Appendix\,\ref{app:C1} and the KK Goldstone boson amplitude
\eqref{eq:pph} actually equals its full amplitude without expansion,
i.e., $\MT [\phi_{n_1^{}}^{}\!\phi_{n_2^{}}^{}\!h_{n_3^{}}^{\pm2}\hs]
\!=\hsm \ka (\ep_3^{}\!\cdot\hsm p_1^{})^2 \tilde{\be}_{n_1^{}n_2^{}n_3^{}}^{}$.\
Then, to hold the KK GRET \eqref{eq:KK-GRET} for the 3-point KK scattering amplitudes, 
\beq
\label{eq:GRET1-LLT-3pt}
\M_0^{}[h^L_{n_1}h^L_{n_2}h^{\pm2}_{n_3}\hs]  
\hs =\hs \MT_0^{}[\phi_{n_1^{}}^{}\phi_{n_2^{}}^{}h_{n_3^{}}^{\pm2}\hs]\hs , 
\eeq 
we derive the following condition to ensure 
the equivalence between the two gravitational KK scattering amplitudes \eqref{eq:hhh-LLT} and \eqref{eq:pph}:
\begin{equation}
\label{eq:al-tbe-LLT-55T}
\[\!\( \hsm \mathbb{M}_{n_1^{}}^2\hsm\!+\! \mathbb{M}_{n_2^{}}^2\hsm\!-\!\mathbb{M}_{n_3^{}}^2\)^{\hsm 2}\hsm
\!+\hsm 2\hs\mathbb{M}_{n_1}^2\!\mathbb{M}_{n_2}^2 \]\!\hsm\al_{n_1^{}n_2^{}n_3^{}}^{}  
=\,6\hs\mathbb{M}_{n_1}^2\hsm\mathbb{M}_{n_2}^2\tilde{\be}_{n_1^{}n_2^{}n_3^{}}^{}\hs.
\end{equation}
The proof of the condition \eqref{eq:al-tbe-LLT-55T} is highly nontrivial and 
we present a direct proof of it around Eq.\eqref{eq:al-tbe-LLT-55T-3pt} of Appdenix\,\ref{app:D}.\ 
For the case of $(n_1^{},\hs n_2^{},\hs n_3^{})\!=\!(n,\hs n,\hs j)$, 
the condition \eqref{eq:al-tbe-LLT-55T} can be simplified as follows: 
\begin{equation}
\label{eq:al-tbe-nnj}
(\rh_j^4\!-\!4\hs \rh_j^2\!+\!6)\al_{nnj}^{} =\, 6\hs\tilde{\be}_{nnj}^{}\hs,
\end{equation}
with the mass ratio $\,\rh_j^{}\!=\!\MGj\hsm /\MGn\hs$.\

\vs
Then, we consider a mixed 3-point scattering amplitudes including two external KK gravitons with helicities $\pm1$ and one external KK graviton with helicity zero.\  
Thus, we derive the following 3-point scattering amplitudes of KK gravitons and of the corresponding KK Goldstone bosons at the leading order of high energy expansion:
\beqs 
\label{eq:h11L-V11Phi-3pt}
\begin{align}
\label{eq:h11L-3pt}
\hspace*{-7mm}
\M_0^{}\big[h^{\pm1}_{n_1^{}} h^{\pm1}_{n_2^{}} h^L_{n_3^{}}\big] 
&= \frac{\,\ka \big[ 2\!-\!( \rh_{13}^2 \!+\! \rh_{23}^2) \!-\!(\rh_{13}^2\!-\!\rh_{23}^2)^2 \big]\,}
{\,2\sqrt{6\,}\hs \rh_{13}^{}\rh_{23}^{}\,}
(\ep_1^{}\!\cdot\hsm p_2^{})(\ep_2^{}\!\cdot\hsm p_1^{})\al_{n_1^{}n_2^{}n_3^{}} \hs, 
\\
\hspace*{-7mm}
{\MT}\big[\VV^{\pm1}_{n_1}\VV^{\pm1}_{n_2}\phi_{n_3^{}}^{}\big] &=  
{-\ka} \Bigg[\!\sqrt{\frac{3}{2}\,} 
(\ep_1^{}\!\cdot\hsm p_2^{})(\ep_2^{}\!\cdot\hsm p_1^{}) 
\!-\! \frac{\,{\mathbb{M}_{n_3}^2} \hs}{\sqrt{6\,}\hs}(\ep_1^{}\!\cdot\hsm\ep_2^{})
(\rh_{13}^{}\rh_{23}^{} \!+\! 2\hs \rh_{13}^{}\!+\! 2\hs \rh_{23}^{})\! \Bigg]\hsm
\tilde\rho_{n_1^{}n_2^{}n_3^{}}^{}
\nn\\
\hspace*{-7mm}
&= \M_0^{}\big[\VV^{\pm1}_{n_1}\VV^{\pm1}_{n_2}\phi_{n_3^{}}^{}\big] + O(E^0)\hs,
\\
\hspace*{-7mm}
\label{eq:V11Phi-3pt}
{\MT}_0^{}\big[\VV^{\pm1}_{n_1}\VV^{\pm1}_{n_2}\phi_{n_3^{}}^{}\big] &= 
{-\ka} \sqrt{\frac{3}{2}\,}\hs  
(\ep_1^{}\!\cdot\hsm p_2^{})(\ep_2^{}\!\cdot\hsm p_1^{}) 
\tilde\rho_{n_1^{}n_2^{}n_3^{}}^{}\hs, 
\end{align} 
\eeqs 
where we have defined the KK mass ratios 
$\hs \rh_{13}^{}\!=\! \mathbb{M}_{n_1^{}}^{}\!/\mathbb{M}_{n_3^{}}^{}$ and 
$\hs \rh_{23}^{}\!=\!\mathbb{M}_{n_2^{}}^{}\hsm /\mathbb{M}_{n_3^{}}^{}$,
and the $\VV_{n_1^{}}^{\pm1}$-$\VV_{n_2^{}}^{\pm1}$-$\phin$ trilinear coupling coefficient  
$\tilde\rho_{n_1^{}n_2^{}n_3^{}}^{}$ is defined in \eqrefe{app-eq:KKGR-RS-couplings} of Appendix\,\ref{app:C}.\ 
There is another helicity combination for such mixed 3-point KK scattering amplitudes, but they vanish at the
leading order of high energy expansion:
\beq
\M_0^{}\big[h^{\pm 1}_{n_1^{}} h^{\mp 1}_{n_2^{}} h^L_{n_3^{}}\big] = 0\hs, ~~~~~
{\MT}_0^{}\big[\VV^{\pm 1}_{n_1^{}} \VV^{\mp 1}_{n_2^{}} {\phi_{n_3^{}}^{}}\big] = 0\hs.
\eeq 

Using the 3-point gravitational KK scattering amplitudes \eqref{eq:h11L-3pt}-\eqref{eq:V11Phi-3pt},
we find that holding the GRET,
\beq 
\label{eq:GRET-h11L=VVphi}
\M_0^{}\big[h^{\pm1}_{n_1^{}} h^{\pm1}_{n_2^{}} h^L_{n_3^{}}\big] \,=\,
-\MT_0^{}\big[\VV^{\pm1}_{n_1}\VV^{\pm1}_{n_2}\phi_{n_3^{}}^{}\big],
\eeq 
will impose the following condition:
\begin{equation}
\label{eq:cond-h11L=VVphi}
\Big[ 2\!-\!( \rh_{13}^2 \!+\! \rh_{23}^2) \!-\! (\rh_{13}^2 \!-\! \rh_{23}^2)^2 \Big]
{\al_{n_1^{} n_2^{} n_3^{}}^{}}
\!=\hs 6\hs \rh_{13}^{}\rh_{23}^{}\hs\tilde\rho_{n_1^{}n_2^{}n_3^{}}^{}\hs,
\end{equation}
where the KK mass ratios $(\rh_{13}^{},\hs \rh_{23}^{})$ and 
the trilinear KK coupling coefficient $\tilde\rho_{n_1^{}n_2^{}n_3^{}}^{}$ are
defined below Eq.\eqref{eq:h11L-V11Phi-3pt}.\ 
The above identity \eqref{eq:cond-h11L=VVphi} can be directly proven 
as in \eqref{eq-app:cond-h11L=VVphi} of Appendix\,\ref{app:D}.\ 
Hence, we have explicitly established the GRET in its mixed form \eqref{eq:GRET-h11L=VVphi} 
for the 3-point gravitational KK scattering amplitudes.\  

\vs

Besides, we consider a specific case with  
$(n_1^{},\hs n_2^{},\hs n_3^{})\!=\!(n,\hs n,\hs j)$.\ 
So we have $\hs \rh_{13}^{}\!=\!\rh_{23}^{}\hs$ and thus 
the above GRET condition \eqref{eq:cond-h11L=VVphi} is simplified as follows:
\beq
\label{eq:cond-h11L=VVphi-nnj}
(\rh_j^2\!-\!1)\hs\al_{nnj}^{} = 3 \hs\tilde{\rho}_{nnj}^{}\hs,
\eeq 
where the mass ratio $\hs \rh_j^{}\!=\!\MGj/\MGn\hs$.\ 
For  the case of compactified flat 5d under $S^1\!/\ZZ$, 
the KK index $j$ could only take values of $j=0,2n$ in Eq.\eqref{eq:cond-h11L=VVphi-nnj}.\ 

\vs

Finally, we discuss the 3-point scattering amplitudes of pure longitudinal KK gravitons ($3h_n^L$) and
of the corresponding gravitational KK Goldstone bosons ($3\phin$).\  
We gave a brief comment on this in a separate paragraph below Eq.\eqref{eq:E-counting-hnL}.\ 
The tree-level 3-point longitudinal KK graviton amplitude $\M [3h_n^L]$ has a 
possible leading energy dependence of $E^8$ by using the power counting rule \eqref{eq:E-counting-hnL-or-hn55}
for $N\!\!=\!3\,$.\
But, we note that the 3-point KK graviton amplitude arises from the trilinear KK graviton vertex
(whose coupling factor contains products of two external momenta)
and is expected to have individual leading energy dependence of $E^8$
after including the 3 external longitudinal polarization tenors
of KK gravitons, each of which has the high energy behavior,
\beq
\label{eq:vt-munu-2}
\vep^{\mn}_L = \sqrt{\frac{2}{3}\,}\hs\vep_S^{\mn}\!+\hsm \vt^{\mn} = O(E^2) +O(E^0)\hs,
\eeq
based on Eqs.\eqref{eq:epLmunu}-\eqref{eq:v-munu}, where for the $j$-th external KK graviton the 
scalar polarization tensor 
$\vep_S^{\mn}\hsm (p_j^{})\!=\!p_j^\mu p_j^\nu/\MG_{n_j}^2$ (with $j\!=\!1,2,3$).\
Thus, each leading energy term of $O(E^8)$ is given by picking up the leading energy term $\vep_S^{\mn}$
from all 3 external longitudinal polarization tenors ($\vep^{\mn}_L\hs$) which have contractions with the
two-momentum-factors in of the trilinear KK graviton vertex.\ These momenta will form up 4 pairs of bilinear
contraction ($\,p_i^{}\hsm\cdot\hsm p_j^{}$) which can be all converted into certain combinations of 
squared KK-masses by using the momentum conservation $\hs p_1^{}\!+\!p_2^{}\!+\!p_3^{}\!=\!0\hs$  
and the on-shell condition of each external momentum $p_j^2\!=\!-M_{n_j^{}}^2$.\
Hence, all the leading energy terms of $O(E^8)$ are reduced to $O(E^0)$ in this way.\
The remaining contributions include to replace one, two, and three external longitudinal polarization tensors
($\vep^{\mn}_L\hs$) by the residual tensor $\vt^{\mn}\!\!=\!O(E^0)$ respectively.\
For those external momentum factors contracting among themselves, they will be converted into squared KK-masses
due to the momentum conservation and the on-shell conditions on all the external momenta.\ 
Hence, a given external momentum $p_j^{\mu}$ could contribute a nontrivial energy factor
$E^1$ to the 3-point KK graviton amplitude only if it contracts with a residual tensor 
$\vt^{\mn}\hsm\!=\!O(E^0)$ from Eq.\eqref{eq:vt-munu-2}.\ 
The terms having maximal number of such contractions arise from picking up two
$\vt^{\mn}$ tensors from two of the external KK graviton states via Eq.\eqref{eq:vt-munu-2}; so the
third external longitudinal polarization tensor contributes a leading momentum term of $\vep_S^{\mn}$
via Eq.\eqref{eq:vt-munu-2} and the trilinear KK graviton vertex contributes the leading coupling factors
with two external momenta; thus all the 4 external momenta may contract with the two external $\vt^{\mn}$ tensors
[via Eq.\eqref{eq:vt-munu-2}] and thus makes the leading energy contributions of $O(E^4)$.\
On the other hand, the tree-level amplitude of 3 gravitational
KK Goldstone bosons $\MT [3\phin]$ arises from trilinear KK Goldstone vertex
whose coupling factor contains products of two external momenta
$\,p_i^{}\cdot p_j^{}$.\  So the naive power counting shows that this KK Goldstone amplitude
would have leading energy dependence of $E^2$ in agreement with Eq.\eqref{eq:E-counting-hnL-or-hn55}.\
But the momentum conservation and the on-shell condition of each external state convert
all the bilinear momentum products ($\,p_i^{}\cdot p_j^{}$) in the trilinear Goldstone vertex coupling 
into squared KK-masses, so the 3-point KK Goldstone amplitude actually has energy dependence $E^0$.\ 
Then, we can apply the GRET \eqref{eq:KK-GRET} and note that on its right-hand side the corresponding 3-point
KK Goldstone amplitude has the actual leading energy dependence $E^0$.\ 
Hence, the GRET further ensures a large energy cancellation of $\hs E^4\!\ito\hsm E^0\hs$ 
in the 3-point longitudinal KK graviton amplitude.\ 

\vs 

Then, we can explicitly compute the 3-point scattering amplitudes of pure longitudinal KK gravitons 
$\M[3h_n^L]$ and
of the corresponding gravitational KK Goldstone bosons $\MT[3\phin]$.\  
Under high energy expansion, we derive
these two KK amplitudes down to $O(E^0)\hs$:
\\[-5mm]
{\small 
\beqs
\label{eq:Amp3hL-3phi-E0}
\begin{align}
\label{eq:Amp3hL-E0}
\hspace*{-5mm}
\M\big[h^{L}_{n_1^{}}\! h^{L}_{n_2^{}}\! h^L_{n_3^{}}\big] &=  
\frac{~{-\kappa}\hs\al_{n_1^{}n_2^{}n_3^{}}^{}\rh_{12}^2\rh_{13}^2\hs\MG_{n_1^{}}^2~}{24\sqrt{6\,}}
\hsm\big(\!-\!6698\!+\!7210\hs \rh_{31}^2\!-\!20400\hs \rh_{21}^2\!-\!18940\hs \rh_{21}^4 
\!+\!14830\hs \rh_{21}^2r_{31}^2
\nn\\
\hspace*{-5mm}
& \hspace*{5mm}
-\!26471\hs \rh_{31}^4\!-\!5520\hs \rh_{21}^6\!+\!7310\hs \rh_{21}^4\rh_{31}^2
\!-\!2950\hs \rh_{21}^2\rh_{31}^4\!+\!362\hs \rh_{31}^6\!-\!186\hs \rh_{21}^8
\nn\\
\hspace*{-5mm}
& \hspace*{5mm}
+\!458\hs \rh_{21}^6\rh_{31}^2\!-\! 375\hs \rh_{21}^4\rh_{31}^4
\!+\!114\hs \rh_{21}^2\rh_{31}^6 \!-\!11\hs \rh_{31}^8
\big) ,
\\
\label{eq:Amp3phi-E0}
\hspace*{-5mm}
{\MT}\big[\phi_{n_1^{}}^{}\!\phi_{n_2^{}}^{}\!\phi_{n_3^{}}^{}\big] &= 
-\sqrt{\frac{2}{3}\,} \kappa \Big[ \hs\tilde{\omega}_{n_1^{}n_2^{}n_3^{}}^{}\!\MG_{n_1^{}}^2\!\hsm 
\big(1 \!+\hsm \rh_{21}^2 \!+\hsm \rh_{31}^2 \big)\! + 
\xoverline{\omega}_{n_1^{} n_2^{} n_3^{}}^{} \!\!+\hsm \xoverline{\omega}_{n_2^{} n_3^{} n_1^{}}^{} 
\!\!+\hsm \xoverline{\omega}_{n_3^{} n_1^{} n_2^{}}^{}
\Big]  ,
\end{align}
\eeqs
}
\hspace*{-3mm}
where we have defined the mass ratio $\rh_{ij}^{} \hsm\!=\! \MG_{n_i^{}}^{} / \MG_{n_{\!j}^{}}^{}\hsm$,
the trilinear KK graviton coupling $\al_{n_1^{}n_2^{}n_3^{}}^{}$ and KK Goldstone couplings
$(\tilde{\omega}_{n_1^{}n_2^{}n_3^{}}^{}\hsm,\hs \xoverline{\omega}_{n_1^{} n_2^{} n_3^{}}^{})$ 
are defined in Eq.\eqref{app-eq:KKGR-RS-couplings}.\ 
We see that both the above KK scattering amplitudes receive nonzero contributions at $O(E^0)$.\  
Even though the power counting rule \eqref{eq:E-counting-hnL-or-hn55} suggests that the 3-point 
gravitational KK Goldstone amplitude \eqref{eq:Amp3phi-E0} would scale like $E^2$, 
we find that its actual leading energy dependence is only of $O(E^0)$ 
due to the momentum conservation and
the on-shell condition for each external momentum, which agrees with our discussion 
in the previous paragraph.\footnote{%
From the viewpoint of double-copy (cf.\ Sec.\,\ref{sec:4.1}), 
this is also related to the fact that the trilinear KK Goldstone vertex
$A_5^{an}$-$A_5^{bm}$-$A_5^{c\ell}$ vanishes identically in the compactified KK Yang-Mills gauge theories 
as shown in Eq.\eqref{eq:3AL=3A5+Tv}.}\ 
According to the GRET \eqref{eq:KK-GRET} together with Eq.\eqref{eq:Delta-vV+vvh},
the residual term contains an amplitude having two (spin-0) KK Goldstone states ($\phi_n^{}$ and $\phi_m^{}$) 
and one external KK state $v_\mu^{}\VV^\mu_\ell$ (with helicity $\pm1$), which also contributes to $O(E^0)$.\ 
This explains why the two KK amplitudes \eqref{eq:Amp3hL-E0} and \eqref{eq:Amp3phi-E0}  
do not simply equal despite that they are both of $O(E^0)$.\

\vspace*{1.5mm}
\subsection{\hspace*{-2.5mm}Warped GAET and GRET: 
from 3-Point to 4-Point Amplitudes}
\label{sec:3.2}
\vspace*{1.5mm}

In Section\,\ref{sec:3.1} we have explicitly established the GAET for the 3-point
KK gauge boson scattering amplitudes and the GRET for the 3-point KK graviton scattering amplitudes
in the high energy limit.\  
Using the GAET and GRET for the basic 3-point KK scattering amplitudes, 
we will further prove explicitly in this subsection that the GAET and GRET hold for the 4-point
KK scattering amplitudes in the high energy limit.\ This is realized by proving that the 
mass-coupling identities (sum rules) ensuring the basic 3-point-level GAET and GRET can further  
guarantee the validity of the GAET and GRET for the 4-point KK scattering amplitudes.\  
With these we also conjecture that this proof can be extended through iterations 
to the general $N$-point ($N\hsm\!\geqq\! 5$) KK scattering amplitudes in the high energy limit.\

\vspace*{1mm}
\subsubsection{\hspace*{-2.5mm}Warped GAET: 
from 3-Point to 4-Point KK Amplitudes}
\label{sec:3.2.1}
\vspace*{1mm}

In this section, we first compute the 4-point elastic scattering amplitude of longitudinal KK gauge bosons 
$\TT[A_L^{an}A_L^{bn}\ito A_L^{cn} A_L^{dn}]$ and the corresponding KK Goldstone boson scattering amplitude 
$\tT[A_5^{an} A_5^{bn}\ito A_5^{cn} A_5^{dn}]$ at tree level in the 't\,Hooft-Feynman gauge ($\xin\!=\!1$).\
Under high energy expansion, we derive the leading-order KK amplitudes as follows:
\beqs
\label{eq:T-4ALA5-LO}
\begin{align}
\label{eq:T-4AL-LO}
\TT_{0}^{}[A_L^{an}A_L^{bn}\!\ito\hsm A_L^{cn} A_L^{dn}] &= 
g^2\big(\CC_s^{}\KK_s^{0} + \CC_t^{}\KK_t^0 + \CC_u^{}\KK_u^0\big) ,
\\[1mm]
\label{eq:T-4A5-LO} 
\tT_{0}^{}[A_5^{an} A_5^{bn}\!\ito\hsm A_5^{cn} A_5^{dn}] &= 
g^2 \big(\CC_s^{}\KKt_s^0 + \CC_t^{}\KKt_t^0 + \CC_u^{}\KKt_u^0\big) , 
\end{align}
\eeqs
where the non-Abelian group factors $(\CC_s^{},\, \CC_t^{},\, \CC_u^{})$ are defined as
\begin{equation}
\big(\, \CC_s^{},\, \CC_t^{},\, \CC_u^{} \,\big)  =
\big(f^{abe}f^{cde}\hsm,\, f^{ade}f^{bce}\hsm,\,  f^{ace}f^{dbe} \big) ,
\end{equation}
and they satisfy the color Jacobi identity:
\begin{equation}
\label{eq:Cstu-Jacobi-ID}
\CC_s^{}\! + \CC_t^{}\! + \CC_u^{} = 0\,.
\end{equation}
In Eqs.\eqref{eq:T-4AL-LO}-\eqref{eq:T-4A5-LO}, both the leading-order KK scattering amplitudes
are of $O(E^0)$ under the high energy expansion, as required by the KK GAET \eqref{eq:KK-ET1-N}
which enforces the energy cancellation of $E^4\!\ito\! E^0$ for the 4-point 
longitudinal KK scattering amplitude $\TT[A_L^{an}A_L^{bn}\!\ito\hsm A_L^{cn} A_L^{dn}]$ 
as shown in Eq.\eqref{eq:E-counting-An}.\ 
The exact cancellations of $O(E^4)$ and $O(E^2)$ terms in 
longitudinal KK amplitude $\TT[A_L^{an}A_L^{bn}\!\ito\hsm A_L^{cn} A_L^{dn}]$
impose the following sum-rule-type conditions respectively:
\\[-8mm]
\beqs
\label{eq:SumRule-YM-4pt}
\begin{align}
\label{eq:SumRule1-YM-4pt}
\sum_{j=0}^{\infty}\!a_{nnj}^2 &= a_{nnnn}^{}\,, 
\\
\label{eq:SumRule2-YM-4pt}
\sum_{j=0}^{\infty}\Mjj\hs a_{nnj}^2 &= \frac{4}{\,3\,}\hs \Mnn a_{nnnn}^{}\,.
\end{align}
\eeqs
The above conditions \eqref{eq:SumRule1-YM-4pt} and \eqref{eq:SumRule2-YM-4pt} are proved respectively  
in Eq.\eqref{eq:app-a-4pt} and Eq.\eqref{app-eq:SumRule2-YM-4pt} of Appendix\,\ref{app:D}.\
Combining the above two conditions, we can eliminate the quartic coupling $a_{nnnn}^{}$ and deduce the
following sum rule on the trilinear coupling coefficient $a_{nnj}^{}$ alone:
\begin{equation}
\label{eq:SumRule-YM-3pt}
\sum_{j=0}^{\infty}\!\(\!\frac{\hs 4\hs}{3}-r_j^2\)\hsm\!a_{nnj}^2\hs =\, 0 \hs.
\end{equation}

In \eqrefe{eq:T-4ALA5-LO}, 
the leading-order longitudinal KK gauge boson amplitudes 
for each channel are given by
\begin{subequations}
\label{eq:K0stu-L} 
\begin{align}
\label{eq:K0s-L}
\hspace*{-5mm}
\KK_s^0 &=  -\frac{\ct}{\,4\,}
\Bigg[4\hs a_{nn0}^2+\sum_{j=1}^\infty\! \big(r_j^2 \!+\! 2\big)^{\!2} a^2_{nnj}\Bigg],
\\
\label{eq:K0t-L}
\hspace*{-5mm}
\KK_t^0 &= \frac{1}{\,(1\!+\!\ct)\,}\bigg\{\!
(6\!-\!19\ct\!-\!10\ctt\!-\!\cttt)\hs a_{nn0}^2
+\!\sum_{j=1}^\infty\! \Big[ \!\big(3 r_j^4\!-\!10 r_j^2\!+\!6\big) 
\!-\!\big(r_j^4\!-\!8 r_j^2\!+\!19\big)\ct   
\nn\\
\hspace*{-5mm}
& \quad\, +\hsm 2 \big(r_j^2\!-\!5\big)\ctt\!-\!\cttt \Big] a_{nnj}^2 
\!-(2\!+\!\ct\!-\!2\ctt\!-\!\cttt)\hs a_{nnnn}^{}\bigg\}\hs, 
\\ 
\label{eq:K0u-L}
\hspace*{-5mm}
\KK_u^0 &= \frac{-1}{\,(1\!-\!\ct)\,}\bigg\{\!
(6\!+\!19\ct\!-\!10\ctt\!+\!\cttt)\hs a_{nn0}^2
+\!\sum_{j=1}^\infty\! \Big[ \!\big(3 r_j^4\!-\!10 r_j^2\!+\!6\big) 
\!+\!\big(r_j^4\!-\!8 r_j^2\!+\!19\big)\ct
\nn\\
\hspace*{-5mm} 
&\quad\, +\hsm 2 \big(r_j^2\!-\!5\big)\ctt\!+\!\cttt\Big] a_{nnj}^2 
\!-(2\!-\!\ct\!-\!2\ctt\!+\!\cttt)\hs a_{nnnn}^{}\bigg\}\hs,
\end{align}
\end{subequations}
and the leading-order KK Goldstone boson amplitudes for each channel are given by 
\begin{equation}
\label{eq:K0stu-5}
\hspace*{-5mm}
\KKt_s^0 =  -\ct\!\sum_{j=0}^\infty\! \tilde{a}_{nnj}^2\hs,\quad~
\KKt_t^0 =  \frac{\,3\!-\!\ct\,}{\,1\!+\!\ct\,}\!\sum_{j=0}^\infty\! \tilde{a}_{nnj}^2\hs,\quad~
\KKt_u^0 = -\frac{\,3\!+\!\ct\,}{\,1\!-\!\ct\,}\!\sum_{j=0}^\infty\!\tilde{a}_{nnj}^2\hs,
\end{equation}
where we define the abbreviations 
$(s_{n\theta}^{},\hs c_{n\theta}^{})\!=\!(\sin\hsm n\theta,\hs \cos\hsm n\theta)$
with $\theta$ being the scattering angle.\ 
In the above the KK quartic coupling coefficient $a_{nnnn}^{}$ is defined 
in \eqrefe{app-eq:KKYM-RS-couplings} of Appendix\,\ref{app:C}.\ 

\vs

The KK GAET \eqref{eq:KK-ET1-N} requires the equivalence between the two leading-order KK scattering amplitudes \eqref{eq:T-4AL-LO} and \eqref{eq:T-4A5-LO}.\ 
To prove this explicitly, we compute the difference between the leading-order sub-amplitudes \eqref{eq:K0stu-L} and \eqref{eq:K0stu-5} for each scattering channel: 
\beqs
\label{eq:dK-dKt}
\begin{align}
\label{eq:dK-dKt-s}
\KK_s^0\! -\hsm\KKt_s^0 &=
-\frac{\,\ct\,}{4}\Bigg[\sum_{j=1}^{\infty}r_j^2\hs (7 \!+\! r_j^2)\hs a_{nnj}^2 
\!-\! \sum_{j=0}^{\infty}\! 4\hs\tilde{a}_{nnj}^2\Bigg],
\\
\label{eq:dK-dKt-t}
\KK_t^0\!-\hsm\KKt_t^0 &=\frac{-1}{\,4\hs (1\!+\!\ct)}\!
\Bigg\{\!\sum_{j=1}^{\infty}\! r_j^2\hs\big[7 \!-\! 3\hs r_j^2 \!+\! (7 \!+\! r_j^2)\hs\ct 
\!+\! 4\hs\ctt\big] \hs a_{nnj}^2 
\!+\! 4\hs (3 \!-\!\ct)\!\sum_{j=0}^{\infty}\!\tilde{a}_{nnj}^2\!\Bigg\},
\\
\label{eq:dK-dKt-u}
\KK_u^0\!-\hsm\KKt_u^0 &=\frac{1}{\,4\hs (1\!-\!\ct)}\!
\Bigg\{\!\sum_{j=1}^{\infty}\! r_j^2\hs\big[7 \!-\! 3\hs r_j^2 \!-\! (7 \!+\! r_j^2)\hs\ct 
\!+\! 4\hs\ctt\big] \hs a_{nnj}^2 
\!+\! 4\hs (3 \!+\!\ct)\!\sum_{j=0}^{\infty}\!\tilde{a}_{nnj}^2\!\Bigg\},
\end{align}
\eeqs
where we have utilized Eq.\eqref{eq:SumRule-YM-4pt} to eliminate the quartic coupling coefficient $\hs a_{nnnn}^{}\hs$
and Eq.\eqref{eq:SumRule-YM-3pt} to eliminate the constant coefficient of $a_{nnj}^{2}$.\
Inpsecting the three difference expressions \eqref{eq:dK-dKt-s}-\eqref{eq:dK-dKt-u}
and the sum rule \eqref{eq:SumRule-YM-3pt},
we find that they can be equal to each other provided the following sum rule condition
is satisfied:
\begin{equation}
\label{eq:SumRule1-YM-4pt-E0} 
\sum_{j=0}^{\infty}\!\(\!1\hsm -\hsm\frac{1}{2}r_j^2 \!\)^{\!\!2}\! a_{nnj}^2 
= \sum_{j=0}^{\infty} \tilde{a}_{nnj}^2 \hs.
\end{equation}
This condition \eqref{eq:SumRule1-YM-4pt-E0} can be inferred directly from the identity
\eqref{eq:a-at-nnj}  by squaring its two sides and then summing over KK index $j\hs$.\ 
Note that Eq.\eqref{eq:a-at-nnj} was proved to hold the GAET \eqref {eq:GAET-LLT-3pt}
[in the case of $(n_1^{},n_2^{},n_3^{})\!=\!(n,\hs n,\hs j)$] 
for the most fundamental 3-point KK scattering amplitudes at the leading order.\ 
Hence, the validity of the GAET at 4-point-level of KK scattering amplitudes is guaranteed
by the fundamental 3-point-level GAET \eqref {eq:GAET-LLT-3pt}.\  

\vs

With these we can further simplify the amplitude-difference in each channel of
Eq.\eqref{eq:dK-dKt} and derive the following:
\begin{equation}
\label{eq:K-Kt-nnnn}
\KK_i^0-\KKt_i^0 \,=\,
-2\hs\ct\!\sum_{j=1}^\infty\! r_j^2 a_{nnj}^2 \,,
\end{equation}
where the index $\hs i\!\in\! \{s,t,u\}$ denotes the kinematic channel under consideration.\ 
Hence, we sum up the contributions of three kinematic channels and apply 
the color Jacobi identity \eqref{eq:Cstu-Jacobi-ID} 
to prove the equivalence between the two leading-order KK scattering amplitudes 
\eqref{eq:T-4AL-LO} and \eqref{eq:T-4A5-LO}:
\begin{equation}
\label{eq:T0-KK-ET-nnnn}
\TT_{0}^{}[4A^{an}_L] \!-\! \tT_{0}^{}[4A^{an}_5] =
-2\hs\ct\!\sum_{j=1}^\infty\! r_j^2 a_{nnj}^2 \!\times\! 
{(\CC_s\hsm +\CC_t^{}\hsm +\CC_u)} = 0 \,.
\end{equation}

In summary, based upon the GAET \eqref{eq:GAET-LLT-3pt}
for the most fundamental 3-point KK scattering amplitudes,  
we have derived the GAET for leading-order 
4-point KK gauge scattering amplitudes \eqref{eq:T-4AL-LO}-\eqref{eq:T-4A5-LO}
as the outcome:
\begin{equation}
\label{eq:GAET1-nnnn}
\TT_{0}^{}[A_L^{an}A_L^{bn}\!\ito\hsm A_L^{cn} A_L^{dn}] 
\,=\, \tT_{0}^{}[A_5^{an} A_5^{bn}\!\ito\hsm A_5^{cn} A_5^{dn}]\,. 
\end{equation}

The above analysis demonstrates that the GAET for 3-point KK scattering amplitudes
is the most fundamental formulation of the GAET, with which we can further deduce the validity of the GAET for 4-point KK scattering amplitudes.\ Hence, we conclude that 
{\it the validity of GAET \eqref{eq:GAET1-nnnn} for 4-point KK scattering amplitudes can be reduced to 
the validity of GAET \eqref{eq:GAET-LLT-3pt} for 3-point  KK scattering amplitudes} 
(which we have already proved in Sec.\,\ref{sec:3.1.1}).\ 
Extending this result, we would conjecture that 
the GAET for $N$-point KK scattering amplitudes ($N\!\!\geqq\!5$) can be also reduced to 
the validity of the GAET for the 3-point KK scattering amplitudes.\

\vs

As a consistency check, we examine the flat 5d limit by taking the warped parameter $k\ito 0\hs$.\ 
Substituting the flat-space couplings of Table\,\ref{app-table:Flat-values} (Appendix\,\ref{app:C}) into 
the above amplitudes \eqref{eq:K0stu-L}-\eqref{eq:K0stu-5}, 
we obtain the following leading-order amplitudes:
\beqs
\begin{align}
\KK_s^0 &=  -\frac{\,11\,}{2}\ct  \,,
\hspace*{-15mm}
&\KKt_s^0 &=  -\frac{\,3\,}{2}\ct  \,,
\\[0.5mm]
\KK_t^0 &= \frac{\,5\!-\!11\ct\!-\!4\ctt\,}{2\hs (1\!+\hsm\ct)} \,,
\hspace*{-15mm}
&\KKt_t^0 &= \frac{\,3\hs (3\!-\!\ct)}{\,2\hs (1\!+\hsm\ct)\,} \,,
\\[0.5mm]
\KK_u^0 &\dis = -\frac{\,5\!+\!11\ct\!-\!4\ctt}{2(1-\ct)} \,,
\hspace*{-15mm}
& \KKt_u^0 &\dis = -\frac{\,3(3\!+\hsm\ct)\,}{\,2\hs (1\!-\hsm\ct)\,} \,,
\end{align}
\eeqs
which agree with the previous results for the compactified flat 5d 
Yang-Mills gauge theories (under $S^1\!/\ZZ$)
as given in Refs.\,\cite{Hang:2021fmp}\cite{Hang:2022rjp} 
and also Refs.\,\cite{Chivukula:2001esy}\cite{Chivukula:2003kq}.\ 
Moreover, in this flat 5d limit,
the elastic scattering amplitude of the longitudinal KK gauge bosons and 
of the corresponding KK Goldstone bosons differ by the same amount in each channel:
\begin{equation}
\KK_i^0-\KKt_i^0 \,=\hs -4\hs\ct \hs, 
\end{equation}
with $\,i\!\in\! \{s,t,u\}$.\ They have vanishing contribution to the full amplitude 
due to the color Jacobi identity \eqref{eq:Cstu-Jacobi-ID}. 

\vs

Next, we consider two inelastic scattering processes,  
$(n,n) \!\ito\! (m,m)$ with $n\!\neq\! m\!\!>\!0\hs$, and 
$(n,m) \ito (\ell,q)$ with $n\!\neq\! m\!\neq\!\ell\!\neq\! q\!>\!0\hs$.\
For the inelastic scattering process $(n,n) \!\ito\!(m,m)$, 
we compute the 4-point leading-order scattering amplitudes 
of the longitudinal KK gauge bosons and of the corresponding KK Goldstone bosons:
\beqs
\label{eq:Tnnmm}
\begin{align}
\label{eq:T0nnmm-AL}
\TT_0^{}[A_L^{an}A_L^{bn}\!\ito\! A_L^{cm}A_L^{dm}]  
\,& =\, g^2\big({\CC_s^{}}\hs\KK_{s}^{\rm{in}\hs0} \!+ {\CC_t^{}}\hs\KK_t^{\rm{in}\hs0}\!+ {\CC_u^{}}\hs\KK_u^{\rm{in}\hs0}\big),
\\[1mm]
\label{eq:T0nnmm-A5}
\tT_0^{}[A_5^{an}A_5^{bn}\!\ito\! A_5^{cm}A_5^{dm}]
\,& =\, g^2 \big({\CC_s^{}}\hs\KKt_s^{\rm{in}\hs0}\! + {\CC_t^{}}\hs\KKt_t^{\rm{in}\hs0}\! + {\CC_u^{}}\hs\KKt_u^{\rm{in}\hs0}\big).
\end{align}
\eeqs
In Eqs.\eqref{eq:T0nnmm-AL}-\eqref{eq:T0nnmm-A5}, 
both the leading-order inelastic KK scattering amplitudes
are of $O(E^0)$ under the high energy expansion, as required by the KK GAET \eqref{eq:KK-ET1-N}
which enforces the energy cancellations of $E^4\!\ito\! E^0$ for the 4-point 
longitudinal KK scattering amplitude $\TT[A_L^{an}A_L^{bn}\!\ito\hsm A_L^{cm} A_L^{dm}]$ 
as shown in Eq.\eqref{eq:E-counting-An}.\ 
The exact cancellations of $O(E^4)$ and $O(E^2)$ terms in the inelastic 
longitudinal KK amplitude $\TT[A_L^{an}A_L^{bn}\!\ito\hsm A_L^{cm} A_L^{dm}]$
impose the following sum-rule-type conditions:
\beqs
\begin{align}
\label{eq:4AL-E4-cancel}
\sum_{j=0}^{\infty} a_{nnj}a_{mmj} &\,=\, \sum_{j=0}^{\infty} a_{nmj}^2 = a_{nnmm} \,,
\\
\label{eq:4AL-E2a-cancel}
\sum_{j=1}^{\infty}\! r_j^2 a_{nmj}^2 
&\,=\, \sum_{j=0}^{\infty} \!\Big[\hsm (1\!+\!r^2) \!-\! \Fr{1}{\hs 2\hs}r_j^2\Big] a_{nnj}^{}a_{mmj}^{} \hs,
\\
\label{eq:4AL-E2b-cancel}
\sum_{j=1}^{\infty} \!\frac{\,(1\!-\!r^2)^2\,}{r_j^2} a_{nmj}^2 
&\,=\, \sum_{j=0}^{\infty}\!\Big[\hsm (1\!+\!r^2) \!-\!\Fr{\hs 3\hs}{2}\hsm r_j^2\Big] a_{nnj}^{}a_{mmj}^{} \hs,
\end{align}
\eeqs
where $(r,\,r_j^{})$ denote mass ratios and are defined as  $\hs r\!=\!\Mm/\Mn$ and $r_j^{}\!=\!\Mj/\Mn\hs$.\  
In the above, the inelastic condition \eqref{eq:4AL-E4-cancel} ensures to the energy cancellations at $O(E^4)$
and is mass-independent, whereas the inelastic conditions \eqref{eq:4AL-E2a-cancel}-\eqref{eq:4AL-E2b-cancel}
guarantees the energy cancellations at $O(E^2)$.\ 
We prove the three inelastic conditions respectively in Eqs.\eqref{eq:app-a-4pt}, \eqref{app-eq:nnmm-E2-cancel-a} 
and Eqs.\eqref{app-eq:lemma-d6}-\eqref{app-eq:3.45c-f} of Appendix\,\ref{app:D}.\ 

\vs

In \eqrefe{eq:T0nnmm-AL}, for each kinematic channel, 
the 4-point leading-order scattering amplitudes of longitudinal KK gauge bosons are given by
\beqs
\begin{align} 
\KK_s^{\rm{in}\hs0} &= -\frac{\ct}{\,4\hs r^2\,} 
\bigg\{\!2\hs (5\!-\!3\hs r^4) a_{nn0}^{}a_{mm0}^{}\hsm +\! 
\sum_{j=1}^{\infty}\!\Big[ r_j^4 \!+\! 4 r_j^2\!+\!2\hs (5\!-\!3\hs r^4) \Big] 
a_{nnj}^{}a_{mmj}^{} 
\nn\\
&\hspace*{4.2mm} 
- 2\hs (3\!-\!2\hs r^2\hsm\!-\!r^4)\hs a_{nnmm}^{} \Big\}\hs ,
\\
\KK_t^{\rm{in}\hs0} &= \frac{1}{\,8\hs r^2 (1\!+\!\ct)\,} 
\sum_{j=1}^{\infty}\!\bigg\{ 2(3\!-\!\ct) r_j^4\!+\!\Big[4(1\!+\!3\hs r^2)\ct
\!-\!(3\!+\!r^2)(5\!-\!\ctt)\hsm\Big] r_j^2
\nn\\
&\hspace*{4.2mm} +\Big[(15\!-\!6\hs r^2\!+\!3\hs r^4)\!-\!(2\!+\!32\hs r^2\!+\!4\hs r^4)\ct
\!-\!(11\!+\!10\hs r^2\!-\!r^4)\ctt\!-\!2\hs\cttt\Big]  
\nn\\
&\hspace*{4.2mm} +(1\!-\!r^2)^2 \big[(5\!-\!r^2)\!+\!8\hs\ct\!+\!(3\!+\!r^2)\ctt\big] r_j^{-2} \Big\}
\hs a_{nmj}^2
-\frac{1}{\,8\hs r^2\,} \Big[ 2(3\!-\!2\hs r^2\!-\!r^4)\ct 
\nn\\
&\hspace*{4.2mm} +4\hs (3\!-\!2\hs r^2\hsm\!-\!\ctt) \big] a_{nnmm}^{}\,,
\\
\KK_u^{\rm{in}\hs0} &= -\frac{1}{\,8\hs r^2 (1\!-\!\ct)\,} 
\sum_{j=1}^{\infty}\!\bigg\{ 2(3\!+\!\ct) r_j^4\!-\!\Big[4(1\!+\!3\hs r^2)\ct
\!+\!(3\!+\!r^2)(5\!-\!\ctt)\hsm\Big] r_j^2
\nn\\
&\hspace*{4.2mm} +\Big[(15\!-\!6\hs r^2\!+\!3\hs r^4)\!+\!(2\!+\!32\hs r^2\!+\!4\hs r^4)\ct
\!-\!(11\!+\!10\hs r^2\!-\!r^4)\ctt\!+\!2\hs\cttt\Big]  
\nn\\
&\hspace*{4.2mm} +(1\!-\!r^2)^2 \big[(5\!-\!r^2)\!-\!8\hs\ct\!+\!(3\!+\!r^2)\ctt\big] r_j^{-2} \Big\}
\hs a_{nmj}^2
+\frac{1}{\,8\hs r^2\,} \Big[ 2(3\!-\!2\hs r^2\!-\!r^4)\ct 
\nn\\
&\hspace*{4.2mm} -4\hs (3\!-\!2\hs r^2\hsm\!-\!\ctt) \big] a_{nnmm}^{}\,,
\end{align}
\eeqs
with the mass ratios $\hs r\!=\!\Mm/\Mn$ and $r_j^{}\!=\!\Mj/\Mn\hs$.\ 

\vs

For the corresponding 4-point leading-order inelastic scattering amplitude \eqref{eq:T0nnmm-A5} of KK Goldstone bosons, the sub-amplitudes in each kinematic channel are given by
\begin{align}
{\KKt_s^{\rm{in}\hs0}}\! = -\ct\!\sum_{j=0}^\infty \tilde{a}_{nnj}^{}\tilde{a}_{mmj}^{} \hs, ~~~~
{\KKt_t^{\rm{in}\hs0}}\! =
\frac{~3\!-\!\ct~}{\,1\!+\!\ct\,}\!\sum_{j=0}^\infty \tilde{a}_{nmj}^{2}\hs, ~~~~
{\KKt_u^{\rm{in}\hs0}}\! =
-\frac{~3\!+\!\ct~}{1\!-\!\ct}\!\sum_{j=0}^\infty \tilde{a}_{nmj}^{2} \hs.
\end{align}
Then, we find that in each channel, the 4-point inelastic scattering amplitudes of the longitudinal KK gauge bosons and of the corresponding KK Goldstone bosons differ by the same amount:
\begin{equation}
\label{eq:K-Kt-nnmm}
{\KK^{\rm{in}\hs0}_i - \KKt^{\rm{in}\hs0}_i} \,= -\frac{\ct}{\,2\hs r^2\,}\!
\sum_{j=0}^{\infty}\!\Big[(3\!+\!r^2)\hs r_j^2 \hsm +\hsm 2(1\!-\!r^4) \Big] 
a_{nnj}^{}a_{mmj}^{}\hs ,
\end{equation}
provided that the following sum rule conditions hold:
\beqs
\label{eq:a-at-All-nnj-nnmm}
\begin{align}
\label{eq:a-at2-nnj-nnmm}
\sum_{j=0}^{\infty} \!\tilde{a}_{nnj}^{} \tilde{a}_{mmj}^{} 
\,&=\, 
\sum_{j=0}^{\infty}\! \(\hsm 1 \!-\! \Fr{1}{2} r_j^2 \) \! 
\!\(\hsm 1 \!-\! \Fr{1}{2} \bar{r}_j^2 \) \!a_{nnj}^{}a_{mmj}^{} \hs, 
\\
\label{eq:-at2=atat-nnj-nnmm}
\sum_{j=0}^{\infty} \!\tilde{a}_{nmj}^2 \,&=\, 
\sum_{j=0}^{\infty}\!\tilde{a}_{nnj}^{} \tilde{a}_{mmj}^{}
= \tilde{a}_{nnmm}^{}\,,
\end{align}
\eeqs 
where we have defined the mass ratios $\,r_j^{}\!=\!\Mj/\Mn$ and $\,\bar{r} \!=\! \Mj/\Mm$, 
and the quartic KK Goldstone coupling coefficient 
$\tilde{a}_{nnmm}^{} \!=\!\lrb{\fft_n^{} \fft_n^{} \fft_m^{} \fft_m^{}}$.\ 
We note that the above inelastic condition \eqref{eq:a-at2-nnj-nnmm} can be directly inferred  
from Eq.\eqref{eq:a-at-nnj} by squaring its two sides followed by summing over the KK index $j\hs$.\
[{\hs}We recall that Eq.\eqref{eq:a-at-nnj} is the condition to hold the GAET \eqref{eq:GAET-LLT-3pt} 
at the level of 3-point KK amplitudes.]\ 
The condition \eqref{eq:-at2=atat-nnj-nnmm} holds because it is just the special case of the basic relations
\eqref{app-eq:a-completeness} which rely on the completeness condition \eqref{eq-app:complete-cond}.\

\vs

\eqrefe{eq:K-Kt-nnmm} demonstrates that the inelastic scattering amplitudes 
\eqref{eq:T0nnmm-AL} and \eqref{eq:T0nnmm-A5} are equal 
due to the color Jacobi identity \eqref{eq:Cstu-Jacobi-ID}.\ 
This explicitly establishes the GAET for 4-point inelastic KK scattering amplitudes:
\begin{equation}
\label{eq:GAET-nnmm}
\TT_0^{}[A_L^{an}A_L^{bn}\!\ito\! A_L^{cm}A_L^{dm}] \,=\, \tT_0^{}[A_5^{an}A_5^{bn}\!\ito\! A_5^{cm}A_5^{dm}] \,.
\end{equation}
According to our discussion below Eq.\eqref{eq:a-at-All-nnj-nnmm}, we stress that 
{\it the validity of the inelastic GAET \eqref{eq:GAET-nnmm} is guaranteed by the validity of
the most fundamental GAET \eqref{eq:GAET-LLT-3pt} at the level of 3-point KK amplitudes.}\

As a consistency check, we take the flat 5d limit $k\ito 0\hs$ 
and derive the following inelastic KK scattering amplitudes for each kinematic channel:
\beqs
\begin{align}
\KK_s^{\rm{in}\hs0} &= -\ct \!+\hsm (r^2\hsm -\!r^{-2})\ct  \,,
\hspace*{-15mm}
& \KKt_s^{\rm{in}\hs0} &=  -\ct  \,,
\\[1mm]
\KK_t^{\rm{in}\hs0} &= \frac{\,3\!-\!\ct\,}{\,1\!+\!\ct\,} +(r^2\!-\!r^{-2})\ct  \,,
\hspace*{-15mm}
& \KKt_t^{\rm{in}\hs0} &= \frac{\,3\!-\!\ct\,}{\,1\!+\!\ct\,} \,,
\\[1mm]
\KK_u^{\rm{in}\hs0} & =  -\frac{\,3\!+\!\ct\,}{\,1\!-\!\ct\,} \hsm +\hsm (r^2\!-\!r^{-2})\ct  \,,
\hspace*{-15mm}
& \KKt_u^{\rm{in}\hs0} & = -\frac{\,3\!+\!\ct\,}{\,1\!-\!\ct\,} \,.
\end{align}
\eeqs
Thus, the inelastic scattering amplitudes of the longitudinal KK gauge bosons and 
of the KK Goldstone bosons differ by the same amount in each kinematic channel:
\begin{equation}
\label{eq:Ki-tKi-nnmm}
{\KK_i^{\rm{in}\hs0}-\KKt_i^{\rm{in}\hs0} }\,=\, (r^2\!-\!r^{-2})\hs\ct \hs , 
\end{equation}
where each individual channel is marked by the kinematic index $\,i \!\in\! \{s,t,u\}$.\ Hence, the difference \eqref{eq:Ki-tKi-nnmm} gives zero contribution to the inelastic KK scattering amplitudes \eqref{eq:T0nnmm-AL}-\eqref{eq:T0nnmm-A5} due to the color Jacobi identity \eqref{eq:Cstu-Jacobi-ID}.\ 

\vs

Then, we study the general inelastic scattering process $(n,m) \!\ito\! (\ell,q)$
with all the external KK indices $(n,m,\ell,q)$ unequal.\
Under the high energy expansion,
we compute its 4-point leading-order inelastic scattering amplitudes 
of the longitudinal KK gauge bosons and of the corresponding KK Goldstone bosons,
which take the following form:
\beqs
\label{eq:T0nmlq}
\begin{align}
\label{eq:T0nmlq-AL}
\TT_0^{}[A_L^{an}A_L^{bm}\!\ito\! A_L^{c\ell}A_L^{dq}]  
\,& =\, g^2\big({\CC_s^{}}\hs\KK_{s}^{\rm{in}\hs0} \!+ {\CC_t^{}}\hs\KK_t^{\rm{in}\hs0}\!+ {\CC_u^{}}\hs\KK_u^{\rm{in}\hs0}\big),
\\[1mm]
\label{eq:T0nmlq-A5}
\tT_0^{}[A_5^{an}A_5^{bm}\!\ito\! A_5^{c\ell}A_5^{dq}]
\,& =\, g^2 \big({\CC_s^{}}\hs\KKt_s^{\rm{in}\hs0}\! + {\CC_t^{}}\hs\KKt_t^{\rm{in}\hs0}\! + {\CC_u^{}}\hs\KKt_u^{\rm{in}\hs0}\big).
\end{align}
\eeqs
In each kinematic channel, 
we derive the leading-order scattering amplitudes of longitudinal KK gauge bosons as follows:
\beqs
\label{eq:K0stu-L-nmlq}
\begin{align}
\KK_s^{\rm{in}\hs0}=&\,\frac{1}{\,4\Mn\Mm\Ml\Mq\,}  
\bigg\{\! \sum_{j=1}^{\infty}  \Big[M_j^{-2}  (\Mnn\!-\!\Mmm)(\Mll\!-\!\Mqq)(\Mnn\!+\!\Mmm\!+\!\Mll\!+\!\Mqq)
\nn\\[-1mm]
&-\ct M_j^4 \Big]a_{nmj}^{}a_{\ell qj}^{} \!+\hsm \ct\Big[ M_\ell^4 \!+\! M_q^4 \!+\!(\Mnn \!+\! \Mmm)(\Mnn \!+\! \Mmm\!+\!\Mqq)
\nn\\[-1mm]
&+\! \Mll (\Mmm \!+\! \Mnn \!+\! 2 \Mqq) \Big] a_{nm\ell q}^{}\bigg\} \hs, 
\\
\KK_t^{\rm{in}\hs0}=&\,\frac{1}{\,16\Mn\Mm\Ml\Mq\,}
\bigg\{\!\sum_{j=1}^{\infty} \Big[\!-\!4 M_j^4 \!+\! 2(5\!+\!\ct)M_j^2 (\Mnn\!+\!\Mmm \!+\!\Mll\!+\!\Mqq)
\nn\\[-1mm]
&-\hsm 2 (1\!-\!\ct)M_j^{-2} (\Mnn\!-\!\Mqq)(\Mmm\!-\!\Mll)(\Mnn\!+\!\Mmm \!+\!\Mll\!+\!\Mqq)\Big] a_{nqj}^{}a_{m\ell j}^{}
\nn\\[-1mm]
&+\frac{16}{\,1\!+\!\ct\,} \sum_{j=0}^{\infty}(\Mnn\!+\!\Mqq\!-\!M_j^2)(\Mmm\!+\!\Mll\!-\!M_j^2) a_{nqj}^{}a_{m\ell j}^{}
\nn\\[-1mm]
&-\!2 \Big[\hsm (\Mnn \!+\! \Mqq)(\Mnn \!+\! 4 \Mmm \!+\! 4 \Mll\!+\! \Mqq) \!-\hsm \ct( \Mnn \!+\! \Mmm \!+\! \Mll \!+\! \Mqq)^2
\nn\\[-1mm]
&+(\Mmm \!+\! M_{\ell}^2)^2 \Big] a_{nm\ell q}^{}
\bigg\}\hs,
\\
\KK_u^{\rm{in}\hs0}=&\,\frac{1}{\,16\Mn\Mm\Ml\Mq\,}\bigg\{\!
\sum_{j=1}^{\infty} \!\Big[4 M_j^4 \!-\! 2(5\!-\!\ct)M_j^2 (\Mnn\!+\!\Mmm \!+\!\Mll\!+\!\Mqq)
\nn\\[-1mm]
&+\hsm 2 (1\!+\!\ct) M_j^{-2}(\Mnn\!-\!\Mll) (\Mmm\!-\!\Mqq)(\Mnn\!+\!\Mmm \!+\!\Mll\!+\!\Mqq)\hsm\Big] a_{n\ell j}^{}a_{mqj}^{}
\nn\\[-1mm]
&-\frac{16}{\,1\!-\!\ct\,} \sum_{j=0}^{\infty} (\Mnn\!+\!\Mll\!-\!M_j^2)(\Mmm\!+\!\Mqq\!-\!M_j^2) a_{n\ell j}^{} a_{mqj}^{}
\nn\\[-1mm]
&+\hsm 2\Big[\hsm(\Mnn \!+\! \Mll )(\Mnn \!+\! 4 \Mmm \!+\! \Mll \!+\! 4 \Mqq) \!+\! \ct( \Mnn \!+\! \Mmm \!+\! \Mll \!+\! \Mqq)^2
\nn\\[-1mm]
&+\hsm ( \Mmm \!+\! \Mqq)^2 \Big] a_{nm\ell q}^{}\bigg\} \hs.
\end{align}
\eeqs
We further compute the corresponding 4-point leading-order scattering amplitude of KK Goldstone bosons in each channel,
\beqs
\label{eq:Kt0stu-5-nmlq}
\begin{align}
\KKt_s^{\rm{in}\hs0} &= -\ct  \sum_{j=0}^\infty \tilde{a}_{nmj}^{}\tilde{a}_{\ell qj}^{} \hs,
\\
\KKt_t^{\rm{in}\hs0} &=
\frac{~3\!-\!\ct~}{\,1\!+\!\ct\,}\!\sum_{j=0}^\infty \tilde{a}_{nqj}^{}\tilde{a}_{m\ell j}^{}\hs,
\\
\KKt_u^{\rm{in}\hs0}  &=-\frac{~3\!+\!\ct~}{1\!-\!\ct}\!\sum_{j=0}^\infty \tilde{a}_{n\ell j}^{}\tilde{a}_{mqj}^{} \hs.
\end{align}
\eeqs
Then, we demonstrate that in each kinematic channel the above inelastic scattering amplitudes of
the longitudinal KK gauge bosons and of the KK Goldstone bosons differ by the same amount:
\begin{align}
\label{eq:K-Kt-nmlq}
\hspace*{-3mm}
\KK^{\rm{in}\hs0}_i \!- \KKt^{\rm{in}\hs0}_i 
= \frac{~\Mnn\!+\!\Mmm\!+\!\Mll\!+\!\Mqq~}{4\Mn\Mm\Ml\Mq}\!
\sum_{j=1}^{\infty}\! \Mjj\hsm\Big[\hsm 
(1 \!+\!\ct) a_{nqj}^{}a_{m \ell j}^{} \!-\! (1 \!-\!\ct) a_{n\ell j}^{}a_{mqj}^{} \Big] ,
\end{align}
where we have used Eq.\eqref{eq:a-at-3pt} as the crucial condition
[which holds the GAET \eqref{eq:GAET-LLT-3pt} for the basic 3-point KK amplitudes].\ 
In the above, we have also imposed the following sum rule conditions:
\beqs
\label{eq:E2-cancelation-nmlq}
\begin{align}
\label{eq:E2-1st-nmlq}
& a_{nm\ell q}^{} =\sum_{j=0}^{\infty}a_{nmj}a_{\ell qj}^{} =\sum_{j=0}^{\infty}a_{n\ell j}^{} a_{mqj}^{}
=\sum_{j=0}^{\infty}\! a_{nqj}^{}a_{\ell mj}^{} \hs,
\\[1mm]
\label{eq:E2-2nd-nmlq}
& a_{nm\ell q}^{} = \frac{1}{~\Mnn\!+\!\Mmm\!+\!\Mll\!+\!\Mqq~}\!
\sum_{j=1}^{\infty}\!\Mjj(a_{nmj}^{}a_{\ell qj}^{}\!+\!a_{n\ell j}^{} a_{mqj}^{}\!+\! a_{nqj}^{}a_{m\ell j}^{}) \hs,
\\[1mm]
&
\label{eq:E2-cancelation-nmlq-s} (\Mnn\!-\!\Mmm)(\Mll\!-\!\Mqq)\!\sum_{j=1}^{\infty}\!\frac{\,a_{nmj}^{}a_{\ell qj}^{}\,}{\Mjj} 
\,=\, \sum_{j=0}^{\infty}\! \Mjj(a_{nqj}^{}a_{m\ell j}^{} \!-\! a_{n\ell j}^{}a_{mqj}^{}) \hs,
\\[1mm]
& (\Mnn\!-\!\Mll)(\Mmm\!-\!\Mqq)\!\sum_{j=1}^{\infty}\!\frac{\,a_{n\ell j}^{}a_{mqj}^{}\,}{\Mjj}
\,=\,\sum_{j=0}^{\infty}\!\Mjj(a_{nqj}^{}a_{m\ell j}^{}- a_{nmj}^{}a_{\ell qj}^{}) \hs,
\\[1mm]
& (\Mnn\!-\!\Mqq)(\Mll\!-\!\Mmm)\!\sum_{j=1}^{\infty}\!\frac{\,a_{nqj}^{}a_{m\ell j}^{}\,}{\Mjj}
\,=\,\sum_{j=0}^{\infty}\! \Mjj (a_{nmj}^{}a_{\ell qj}^{} \!-\! a_{n\ell j}^{}a_{mqj}^{}) \hs.
\end{align}
\eeqs
The relations \eqref{eq:E2-1st-nmlq} can be readily proved as in Eq.\eqref{app-eq:a-completeness}
which relies on the completeness condition \eqref{eq-app:complete-cond}.\ 
The above formula \eqref{eq:E2-2nd-nmlq} is derived 
as in Eq.\eqref{app-eq:justforfun} of Appendix\,\ref{app:D} and  
the last three conditions of \eqref{eq:E2-cancelation-nmlq}  
are derived around Eqs.\eqref{app-eq:justforfun3}-\eqref{app-eq:3.57-f}.\ 
With the above, we explicitly prove the GAET for the general 4-point inelastic KK scattering amplitudes 
with all external KK indices being different:
\begin{equation}
\label{eq:GAET1-nmlq} 
\TT_0^{}[A_L^{an}A_L^{bm}\!\ito\! A_L^{c\ell}A_L^{dq}] \,=\, \tT_0^{}[A_5^{an}A_5^{bm}\!\ito\! A_5^{c\ell}A_5^{dq}] \hs.
\end{equation}
Here we find again that the validity of the GAET \eqref{eq:GAET1-nmlq} 
for the 4-point inelastic KK scattering amplitudes 
is guaranteed by the validity of the fundamental 3-point-level GAET \eqref{eq:GAET-LLT-3pt}.\

Taking the flat 5d space limit $k\!\ito\! 0\hs$, we examine the inelastic KK scattering process  
$(n,2n)\ito(3n,4n)$ with $\hs n\!>\!0$ for instance.\ 
The relevant scattering amplitudes of longitudinal KK gauge bosons 
and its corresponding KK Goldstone bosons can be derived from Eqs.\eqref{eq:K0stu-L-nmlq}-\eqref{eq:Kt0stu-5-nmlq}
as follows:
\beqs
\begin{alignat}{3}
\KK^{\rm{in}\hs0}_s &= \frac{3}{32}(35\!+\!43\ct) \hs, &\qquad~~~
&\KKt^{\rm{in}\hs0}_s = -\frac{1}{2}\ct \hs,
\\[1mm]
\KK^{\rm{in}\hs0}_t &= \frac{~451\!+\!468\ct\!+\!145\ctt~}{64(1\!+\!\ct)}\hs,  &\qquad~~~
&\KKt^{\rm{in}\hs0}_t = \frac{~3\!-\!\ct~}{\,2(1\!+\!\ct)\,} \hs,
\\[1mm]
\KK^{\rm{in}\hs0}_u &=-\frac{~31\!-\!48\ct\!+\!145\ctt~}{64(1\!-\!\ct)}\hs,  &\qquad~~~
&\KKt^{\rm{in}\hs0}_u =- \frac{~3\!+\!\ct~}{\,2(1\!-\!\ct)\,} \hs.
\end{alignat}
\eeqs
From the above, we explicitly compute the difference between  $\KK^{\rm{in}\hs0}_i$ and $\KKt^{\rm{in}\hs0}_i$
which gives the same amount for all three channels:
\begin{equation}
\label{eq:K-Kt-1234n-flat}
\KK^{\rm{in}\hs0}_i - \KKt^{\rm{in}\hs0}_i = \frac{5}{\,32\,}(21\!+\!29\hs\ct)\hs,
\end{equation}
where the index $i\!\in\!\{s,t,u\}$ denotes the three kinematic channels.\ 
We have verified that the above difference \eqref{eq:K-Kt-1234n-flat} fully agrees with the flat 5d limit
($k\ito 0$) of Eq.\eqref{eq:K-Kt-nmlq} for the choice of external KK indices
$(n,m,\ell,q)=(n,2n,3n,4n)$.\  

\vs
Finally, we analyze the inelastic KK scattering channel $(0,0)\!\ito\!(n,n)$.\ 
Under high energy expansion, we compute the leading-order scattering amplitudes of the 
KK gauge bosons and of the KK Goldstone bosons as follows:
\beqs
\label{eq:ATALA5-00nn}
\begin{align}
\label{eq:ATAL-00nn}
\TT_0^{}[A^{a0}_{\pm} A^{b0}_{\mp} \!\ito\! A^{cn}_L A^{dn}_L]  
&\,=\, g^2 
\big({\CC_s^{}}\hs\KK^{\rm{in}\hs0}_s \!+ {\CC_t^{}}\hs\KK^{\rm{in}\hs0}_t \!+ {\CC_u^{}}\hs\KK^{\rm{in}\hs0}_u\big),
\\[1mm]
\label{eq:ATA5-00nn}
\tT_0^{}[A^{a0}_{\pm} A^{b0}_{\mp} \!\ito\! A^{cn}_5 A^{dn}_5] 
&\,=\, g^2 
\big({\CC_s^{}} \KKt^{\rm{in}\hs0}_s \!+ {\CC_t^{}} \KKt^{\rm{in}\hs0}_t \!+ {\CC_u^{}} \KKt^{\rm{in}\hs0}_u\big).
\end{align}
\eeqs
In the above, the sub-amplitudes $\{\KK_i^{\rm{in}\hs0}\}$ and
$\{\KKt_i^{\rm{in}\hs0}\}$ for each kinematic channel are derived as follows: 
\\[-9mm]
\beqs
\label{eq:K-Kt-00nn}
\begin{align}
\KK_s^{\rm{in}\hs0} &= -\KKt_s^{\rm{in}\hs0}=0\,, 
\\[1mm]
\KK_t^{\rm{in}\hs0} &= -\KKt_t^{\rm{in}\hs0} = -(1\!-\hsm\ct)\hs \ff_0^2 \,, 
\\[1mm]
\KK_u^{\rm{in}\hs0} &= -\KKt_u^{\rm{in}\hs0} = (1\!+\hsm\ct)\hs \ff_0^2  \,,
\end{align}
\eeqs
which establish the equality between the two leading-order inelastic KK scattering amplitudes
\eqref{eq:ATAL-00nn} and \eqref{eq:ATA5-00nn}.\  Hence, we have GAET for the 4-point inelastic
KK scattering amplitudes of $(0,0)\ito (n,n)\hs$:
\beq 
\TT_0^{}[A^{a0}_{\pm} A^{b0}_{\mp} \!\ito\! A^{cn}_L A^{dn}_L]  
\,=\, -\tT_0^{}[A^{a0}_{\pm} A^{b0}_{\mp} \!\ito\! A^{cn}_5 A^{dn}_5] \hs.
\eeq 

As a consistency check, we can take the flat 5d limit $k\ito 0\hs$ and obtain:
\beqs
\begin{align}
\KK_s^{\rm{in}\hs0} &= -\KKt_s^{\rm{in}\hs0}=0\,, 
\\[1mm]
\KK_t^{\rm{in}\hs0} &= -\KKt_t^{\rm{in}\hs0} = -(1\!-\hsm\ct)\hs, 
\\[1mm]
\KK_u^{\rm{in}\hs0} &= -\KKt_u^{\rm{in}\hs0} = (1\!+\hsm\ct)\hs. 
\end{align}
\eeqs
These agree with the previous inelastic KK scattering amplitudes derived 
for the compactified flat 5d space with orbifold $S^1\!/\ZZ$ \cite{Li:2022rel}.\

\vspace*{1mm}
\subsubsection{\hspace*{-2.5mm}Warped GRET: 
from 3-Point to 4-Point KK Amplitudes}
\label{sec:3.2.2}
\vspace*{1.5mm}

In this subsection, we first prove the wraped KK gravitational equivalence theorem (GRET) 
at leading order for the elastic process $(n,n)\ito (n,n)$ 
with all the external KK gravitons being longitudinally polarized, 
namely, $\M_0[4h^n_L] = \MT_0[4\phin]$.\ 
Then, we analyze two inelastic scattering processes $(n,n)\ito (m,m)$ and $(0,0)\ito (n,n)$  
with the external KK gravitons having helicities $\pm 1\hs$.\ 
We will prove the GRETs for these two inelastic processes respectively,  
$\M_0[2h^n_L\ito 2h^m_L] = \MT_0[2\phin\ito 2\phim]$ and 
$\M_0[2h^0_{\pm 2} \ito 2h^n_{\pm 1}] = -\MT_0[2h^0_{\pm 2}\ito 2\VV^n_{\pm 1}]$.

\vspace*{1mm}
\subsubsection*{$\blacklozenge$\,Elastic Scattering Process \boldmath{$(n,n)\ito(n,n)$}:}
\vspace*{1mm}

We first derive the 4-point elastic scattering amplitude of 
the spin-0 gravitational KK Goldstone bosons $\phin\phin\!\ito\hsm\phin\phin$
for the warped 5d GR under the $S^1\!/\ZZ$ compactification.\ 
We use the Feynman rules given in Appendix\,\ref{sec:FR-KKGR} and compute explicitly
the KK Goldstone boson amplitude to the leading order of high energy expansion:
\begin{align}
\label{eq:AmpE2-4phin}
\MT_0^{}[4\phi_n^{}] = \frac{~\ka^2s\hs (7\!+\!\ctt)^2\!\csc^2\!\theta~}{64}
\!\sum_{j=0}^\infty \!\tilde{\be}_{nnj}^{2}\,,
\end{align}
where the leading-order scattering amplitude 
$\MT_0^{}[4\phin]\!\equiv\!\MT_0^{}[\phin\phin\!\ito\hsm\phin\phin]$
is of $O(E^2)$.\
To explicitly demonstrate the GRET for 4-point KK scattering amplitudes, 
we compare the leading-order scattering amplitude
\eqref{eq:AmpE2-4phin} of the gravitational KK Goldstone bosons with 
the corresponding longitudinal KK graviton amplitude 
$\M_0^{}[4h_L^n]\!\equiv\!\M_0^{}[h_L^nh_L^n\!\ito\hsm h_L^nh_L^n]$ 
\cite{Chivukula:2020hvi},
\begin{align}
\label{eq:AmpE2-4hnL}
\M_0^{}[4h_L^n]=
\frac{~\ka^2s\hs (7\!+\!\ctt)^2\!\csc^2\!\theta~}{2304}
\sum_{j=1}^\infty\! \rh_j^2\big(\rh_j^6\!-\!12\hs \rh_j^2\!+\!11\big) \al_{nnj}^2 \,,
\end{align}
where the KK mass ratio $\rh_j^{}\!=\!\MGj/\MGn$.\ 
To hold the KK GRET \eqref{eq:KK-GRET1b} requires the equality between the two leading-order
scattering amplitudes \eqref{eq:AmpE2-4hnL} and \eqref{eq:AmpE2-4phin},
\begin{equation}
\label{eq:GRET-LO-4pt}	
\M_0^{}[4h_L^n] \,=\, \MT_0^{}[4\phin]\,,
\end{equation}
which gives the following nontrivial sum-rule condition:
\begin{equation}
\label{eq:SumR-4pt-al-be}
\sum_{j=1}^\infty\! \rh_j^2\big(\rh_j^6\!-\!12\hs \rh_j^2\!+\!11\big) \al_{nnj}^2
= 36\sum_{j=0}^\infty \! \tilde{\be}_{nnj}^{2} \,.
\end{equation} 

To prove the above condition \eqref{eq:SumR-4pt-al-be} for the 4-point GRET \eqref{eq:GRET-LO-4pt}, 	 
we make use of the important 
fundamental 3-point GRET condition \eqref{eq:al-tbe-nnj} to replace the coupling 
$\tilde{\be}_{nnj}^{}$ by $\al_{nnj}^{}$ and convert Eq.\eqref{eq:SumR-4pt-al-be} 
to the following form:
\begin{equation}
\label{eq:SumR-4pt-al2-nnj}
\sum_{j=1}^{\infty}\!\big(8\hs \rh_j^6\!-\!40\hs \rh_j^4 \!+\! 59\hs \rh_j^2 \!-\!36\big)\al_{nnj}^2 
=\hs 0 \hs.
\end{equation}
Then, the rest of the proof is to justify the condition \eqref{eq:SumR-4pt-al2-nnj}
which can be achieved by using the properties of the 5d wavefunctions and the resultant relations.\ 
To prove the condition \eqref{eq:SumR-4pt-al2-nnj},
we first derive the following sum rule identities:
\\[-6mm]  
\beqs
\label{eq:SumRule-GR-4pt}
\begin{align}
\label{eq:SumRule-GR-4pt-1a}
\sum_{j=0}^{\infty}\!\al_{nnj}^2 &= {\al_{nnnn}^{}}\hs, \qquad
\\
\label{eq:SumRule-GR-4pt-1b}
\sum_{j=0}^{\infty}\!\rh_j^2\al_{nnj}^2 &= \frac{\,4\,}{3}\al_{nnnn}^{}\hs,
\end{align}
\eeqs 
where the quartic KK coupling coefficient $\al_{nnnn}^{}$ is defined 
in Eq.\eqref{app-eq:KKGR-RS-couplings-4pt} of Appendix\,\ref{app:D}.\ 
The identities \eqref{eq:SumRule-GR-4pt-1a} and \eqref{eq:SumRule-GR-4pt-1b} are proved in 
Eqs.\eqref{eq:app-al-4pt} and \eqref{app-eq:alpha2nnj=M2nnnn} respectively.\
Combining the identities \eqref{eq:SumRule-GR-4pt-1a} and \eqref{eq:SumRule-GR-4pt-1b}, we eliminate the 
quartic KK coupling coefficient $\al_{nnnn}^{}$ and obtain: 
\beq 
\label{eq:al-nnj-new}
\sum_{j=0}^{\infty}\!\(\!1\!-\!\frac{3}{\,4\,}\rh_j^2\!\)\!\al_{nnj}^2 = 0 \hs .
\eeq  
Using Eq.\eqref{eq:al-nnj-new}, 
we can further simplify the identity \eqref{eq:SumR-4pt-al2-nnj} as follows:
\begin{align} 
\label{eq:SumRule-GR-4pt-2}
\sum_{j=0}^{\infty}\! \(\!\rh_j^6\hsm -\hsm 5\hs\rh_j^4\hsm +\!\frac{\,16\,}{3}\!\)\!\al_{nnj}^2 
= 0 \hs, 
\end{align}
which is proved in Eq.\eqref{app-eq:3.72a} of Appendix\,\ref{app:D} 
by using the equations of motion \eqref{Aeq:uv}-\eqref{Aeq:wv}.\ 
This completes the proof of the identity \eqref{eq:SumR-4pt-al2-nnj} and thus establishes the GRET 
\eqref{eq:GRET-LO-4pt} for the 4-point gravitational KK scattering amplitudes.\

\vspace*{1mm}
\subsubsection*{$\blacklozenge$\,Inelastic Scattering Process \boldmath{$(n,n)\ito(m,m)$}:}
\vspace*{1mm}

In this part, we compute the inelastic amplitudes for the longitudinal KK graviton scattering  
process $h^n_L \hs h^n_L \ito h^m_L \hs h^m_L$
and for corresponding gravitational KK Goldstone boson scattering 
$\phi_n \hs \phi_n \ito \phim \hs \phim$,
with which we explicitly prove the corresponding GRET.

\vs 

Using the Feynman rules given in Appendix\,\ref{sec:FR-KKGR}, we compute explicitly
the KK gravitational Goldstone amplitude to the leading order (LO) of high energy expansion: 
\begin{align}
\label{eq:LO-Amp-4phi-nnmm}
\MT_0[\phin\hs\phin \ito \phim\hs\phim] &=\frac{\,\ka^2\hs\sz\,}{\,128\sin^2\!\theta\,}
\sum_{j=0}^\infty\!
\[128 \tilde{\be}_{nmj}^2 \!-\! (29\!-\!28\ctt\!-\!\ctf)\tilde{\be}_{nnj}^{}\tilde{\be}_{mmj}^{}\]
\nn\\[0mm]
& =\frac{~\ka^2(7\!+\!\ctt)^2\sz\,}{\,64\sin^2\!\theta\,}\tilde{\be}_{nnmm}^{} \,.
\end{align}
Then, we compute the corresponding KK graviton scattering amplitude at the leading order 
of high energy expansion:
\begin{align}
\label{eq:LO-amp-nnmm} 
& 
\M_0^{}[h^n_L \hs h^n_L\ito h^m_L \hs h^m_L] 
=  \frac{\ka^2 s_0^{}}{~6912s_{\theta}^2r^4(1\!+\hsm \rh^2)~} 
\bigg\{\!\!-\!\Big\{513(1\!+\hsm\rh^8)\!+\!577\rh^2(1\!+\hsm\rh^4)+429\hs\rh^4 
\nn\\
&+\! 2\!\[87(1\!+\hsm \rh^8)\!+\!118\rh^2(1\!+\hsm \rh^4) \!+\!96\rh^4\]\!\ctt  
\!+\!15\!\[\hsm 1\!+\hsm \rh^8\!+\hsm 3\rh^2(1\!+\hsm \rh^2\!+\hsm \rh^4)\]
\!c_{4\theta}^{} \Big\}\!
\sum_{j=1}^{\infty}\!r_j^2\al_{nnj}^{} \al_{mmj}^{} 
\nn\\ 
&-\!\frac{1}{~3456s_{\theta}^2r^4(1\!+\hsm\rh^2)~}
\Big\{223(1\!+\hsm\rh^8)\!+\!7\rh^2(1\!+\hsm\rh^2) \!-\!131\rh^4 
\!+\!2\!\[31(1\!+\hsm\rh^8)\!-\!14\hs\rh^2(1\!+\hsm\rh^4)\!-\!56\hs\rh^4\]\!c_{2\theta}^{} 
\nn \\
& 
+\!3\!\[\hsm 1\!+\hsm\rh^8\!+\hsm7\rh^2(1\!+\hsm\rh^4)\!+\! 17\rh^4\]\!c_{4\theta}^{}\Big\} \!\sum_{j=1}^{\infty}\!\rh_j^2\al_{nmj}^2
\!+\!\frac{\,(1\!+\hsm\rh^2 \!+\hsm \rh^4)(7\!+\hsm \ctt)^2\,}{576s_{\theta}^2\rh^4}\!
\sum_{j=1}^{\infty}\!\rh_j^4\al_{nmj}^2 
\nn\\
& +\!\frac{\,(1\!+\hsm3\ctt)\,}{3456\hs\rh^4} \!\sum_{j=1}^{\infty}\!\rh_j^8\al_{nnj}^{} \al_{mmj}^{} 
\!-\!\frac{\,(37\!+\hsm11\ctt)\,}{1728\hs s_{\theta}^2\hs\rh^4} 
\!\sum_{j=1}^{\infty}\rh_j^8\al_{nmj}^2  \bigg\} ,
\end{align}
where we have defined the KK mass ratios $\rh=\MGm/\MGn$ and $\rh_j = \MGj/\MGn$.

\vs

To prove GRET for the inelastic scattering amplitudes 
\eqref{eq:LO-amp-nnmm} and \eqref{eq:LO-Amp-4phi-nnmm}, 
we use the fundamental 3-point GRET condition \eqref{eq:al-tbe-LLT-55T} 
to convert the trilinear KK coupling $\tilde{\be}_{nmj}^{}$ to $\al_{nmj}^{}$
as follows:
\\[-5mm]
\begin{equation}
\[(1\!+\hsm \rh^2 \hsm\!-\hsm \rh_j^2)^2 \!+\! 2\hs\rh^2\]\!\al_{nmj}^{} = 6\hs\rh^2 \tilde{\be}_{nmj}^{} \,. 
\end{equation}
Then, we derive the following sum rules which ensure the energy cancellations in the 
longitudinal KK graviton scattering amplitude $\M [h^n_L \hs h^n_L\ito h^m_L \hs h^m_L]$
from $O(E^{10})$ to $O(E^2)$ under high energy expansion.\ These sum rules include
%
\begin{align}
& \al_{nnmm}^{} = \sum_{j=0}^\infty\! \al_{nnj}^{} \al_{mmj}^{} 
= \sum_{j=0}^\infty \!\al_{nmj}^2 \,, 
\end{align}
for the  $O(E^{10})$ to $O(E^8)$ cancellations,
\begin{align}
2(1\!+\hsm\rh^2)\al_{nnmm}^{} = 
\sum_{j=1}^{\infty}\!\rh_j^2 (\al_{nnj}^{}\al_{mmj}^{} \!+\!2\al_{nmj}^2) \hs, 
\end{align}
for the  $O(E^{8})$ to $O(E^6)$ cancellations,
\begin{align}
\sum_{j=1}^{\infty}\!\rh_j^4(\al_{nnj}^{}\al_{mmj}^{}\!-\!\al_{nmj}^2) 
=\[2(1\!+\hsm\rh^2)\!+\!\frac{(1\!-\hsm\rh^2)^2}{\,2(1\!+\hsm\rh^2)\,}\]\! 
\sum_{j=1}^{\infty}\!\rh_j^2(\al_{nnj}^{}\al_{mmj}^{}\!-\!\al_{nmj}^2) \hs,
\end{align}
for the  $O(E^{6})$ to $O(E^4)$ cancellations, and 
\beqs 
\begin{align}
\sum_{j=1}^{\infty}\!\rh_j^6\al_{nnj}^{}\al_{mmj}^{} 
&=\frac{\hs 5\hs}{2} (1\!+\hsm\rh^2)\!\sum_{j=1}\!\rh_j^4\al_{nnj}^{}\al_{mmj}^{} 
\!+\!4\hs\rh^2(1\!+\hsm\rh^2)\al_{nnmm}^{} 
\nn\\
& \hspace{4mm} 
-6\hs\rh^2\sum_{j=1}^{\infty}\!\rh_j^2\al_{nnj}^{}\al_{mmj}^{} 
\!-\!4\rh^2\!\sum_{j=1}^{\infty}\!\rh_j^2\al_{nmj}^2 \,,
\\
\sum_{j=1}^{\infty}\!\rh_j^6\al_{nmj}^2 
&= \frac{\hs 5\hs}{2} (1\!+\hsm\rh^2) \!\sum_{j=1}\!\rh_j^4\al_{nmj}^2 
\!+\! \frac12 (1\!+\hsm\rh^2)(1\!+\hsm6\rh^2\!+\hsm\rh^4)\al_{nnmm}^{} 
\nn \\
&\hspace{4mm}
-(1\!+\hsm\rh^4)\!\sum_{j=1}^{\infty}\!\rh_j^2\al_{nnj}^{}\al_{mmj}^{} \!-\!2(1\!+\hsm\rh^2)^2\!\sum_{j=1}^{\infty}\!\rh_j^2\al_{nmj}^2 \hs,
\end{align}
\eeqs
for the  $O(E^{4})$ to $O(E^2)$ cancellations.\ 
We prove these sum rule relations in Appendices\,\ref{app:D} and \ref{app:E},
as given by Eqs.\eqref{eq:app-al-4pt}, \eqref{app-eq:GRET-nnmm-Mj2}, \eqref{app-eq:GRET-nnmm-Mj4-c} and \eqref{app-eq:GRET-nnmm-Mj6}.\ 
With these, we can establish the GRET for the inelastic scattering channel as follows:
\begin{equation}
\label{eq:GRET-nnmm}
\M_0^{}[h^n_L \hs h^n_L\ito h^m_L \hs h^m_L] = \MT_0[\phin\hs\phin \!\ito \phim\hs\phim]\,.
\end{equation}

\vspace*{1mm}
\subsubsection*{$\blacklozenge$\,Inelastic Scattering Process \boldmath{$(0,0)\ito(n,n)$}:}
\vspace*{1mm}
\vs 

Next, we consider a mixed inelastic KK graviton scattering process 
$h^{+2}_0 \hs h^{-2}_0 \ito h^{+1}_n \hs h^{-1}_n$
and its corresponding KK Goldstone boson scattering channel 
$h^{+2}_0 \hs h^{-2}_0 \!\ito \VV^{+1}_n \hs \VV^{-1}_n$, 
where the initial state includes two massless zero-mode gravitons with helicities $\pm 2\hs$, and
both the final state KK gravitons and KK vector Goldstones have helicities $\pm 1\hs$.\ 
By explicit calculation, we derive the following KK scattering amplitudes:
\beqs
\label{eq:00nn-hh-VV}
\begin{align}
\label{eq:00nn-hh}
& \M[h^{+2}_0h^{-2}_0 \!\ito h^{+1}_n h^{-1}_n] = 
-\frac{~\ka^2 (1\!+\!\ct)(1\!-\!\ct)^3 (12\Mnn \!+\!\sz )\sz~}
{16(4\hs c_\theta^2\Mnn \!+\! s_\theta^2\sz)} \al_{000}^{} \al_{nn0}^{} \hs,
\\
\label{eq:00nn-VV}
& \MT[h^{+2}_0 \hs h^{-2}_0 \ito \VV^{+1}_n \hs \VV^{-1}_n] 
= \frac{~\ka^2 (1\!+\!\ct)(1\!-\!\ct)^3 (3\sz\!+\!2\Mnn) (\sz\!-\!4\Mnn)~}
{48(4\hs c_\theta^2\Mnn \!+\! s_\theta^2\sz)} \al_{000}^{}\tilde{\al}_{nn0}^{} \hs.
\end{align}
\eeqs
%
Making high energy expansion, we derive the leading-order KK scattering amplitudes:
\beqs
\label{eq:LO-00nn-hh-VV}
\begin{align}
\label{eq:LO-00nn-hh}	
\M_0^{} [h^{+2}_0 h^{-2}_0 \!\ito h^{+1}_n h^{-1}_n] 
& = -\frac{\,\ka^2\sz\,}{\,16\,}(1\!-\!\ct)^2 \al_{000}^{} \al_{nn0}^{}  \hs, 
\\
\label{eq:LO-00nn-VV}
\MT_0^{} [h^{+2}_0 h^{-2}_0 \!\ito \VV^{+1}_n \VV^{-1}_n] 
& = +\frac{\,\ka^2\sz\,}{\,16\,} (1\!-\!\ct)^2 \al_{000}^{}\tilde{\al}_{nn0}^{} \hs,
\end{align}
\eeqs
where $\hs\al_{000}^{} \!=\! \al_{nn0}^{} \!=\! \tilde{\al}_{nn0}^{} \!=\! \uu_0^{}$ 
and $\uu_0^{}$ is given by Eq.\eqref{Aeq:u0-def}.\
We see that the above KK graviton scattering amplitude equals  
the corresponding KK Goldstone scattering amplitude at the leading order
after including an overall factor 
$(-\ii)^2\!=\!-1$ from the (tree-level) modification factor \eqref{eq:Cmod-GB} 
on its right-hand side,
\beq 
\label{eq:GRET2-00nn}
\M_0^{} [h^{+2}_0 h^{-2}_0 \!\ito h^{+1}_n h^{-1}_n]  
= -\MT_0^{} [h^{+2}_0 h^{-2}_0 \!\ito \VV^{+1}_n \VV^{-1}_n] \hs.
\eeq 
This agrees with the GRET \eqref{eq:KK-GRET-hV1f}.

\vs

In summary, we stress that in the proof of the condition \eqref{eq:SumR-4pt-al-be}
for the 4-point GRET \eqref{eq:GRET-LO-4pt}, the basic 3-point 
GRET condition \eqref{eq:al-tbe-nnj} plays a key role.\
Hence, we conclude that our proof of the most basic 3-point GRET identity \eqref{eq:GRET1-LLT-3pt} 
or its equivalent sum rule condition \eqref{eq:al-tbe-nnj} does automatically guarantee the validity
of the 4-point GRET identity \eqref{eq:GRET-LO-4pt}	or its equivalent sum rule condition
\eqref{eq:SumR-4pt-al-be}.\ 
We stress that 
{a major goal of the present work} is to demonstrate that the validities of the GAET and GRET 
for the 4-point KK scattering amplitudes can be proved from the corresponding GAET and GRET
identities (sum rule conditions) at the level of the 3-point KK scattering amplitudes.\  

\vs

Finally, it is natural to conjecture that the validity of general $N$-point ($N\!\!\geqq\!4$) GAET and GRET 
can be all reduced to that of the 3-point GAET and GRET.\
This is because all the 4-point gauge (gravity) couplings can be reduced to 
the summed products of 3-point gauge (gravity) couplings as shown in Eq.\eqref{app-eq:a-at-completeness} 
(due to the completeness conditions of 5d wavefunctions), 
and there is no more independent new gauge (gravity) couplings for $N\!\!\geqq\!4\hs$.\footnote{%
It is also known that all the $N$-point ($N\!\!\geqq\!4$) nonlinear gravitational self-couplings 
are reducible and can be induced precisely as having 3-point trilinear graviton couplings\,\cite{Deser}.}\  
Hence, we would expect that for the general $N$-point ($N\!\!\geqq\!4$) KK amplitudes the GAET and GRET 
cannot impose extra new conditions on the 3-point KK couplings (to avoid over-constraint and inconsistency), 
so that the validities of the $N$-point GAET and GRET should be reduced to that of the 3-point GAET and GRET.\
We note that a rigorous general proof on this reduction is more nontrivial, 
especially because each massive scattering amplitude of longitudinal KK gauge bosons (KK gravitons) 
involves large energy cancellations.\ This interesting task is beyond the current scope 
and we would leave it to our future work.

\section{\hspace*{-2.5mm}Extended Double-Copy for Warped KK 
Equivalence Theorems}
\label{sec:4}
\vspace*{1mm}

In this section, we study the double-copy construction of massive gravitational KK scattering amplitudes 
at tree level from the corresponding massive KK gauge scattering amplitudes 
under the warped 5d compactification of $S^1\!/\ZZ\hs$.\  
This is nontrivial and challenging
because the warped 5d compactification leads to highly nonlinear KK mass-spectrum 
and the scattering amplitude contains in each kinematic channel the sum over 
infinite number of KK-pole-exchanges,\ which is in contrast to the flat 5d case.\ 
It was proved\,\cite{Li:2022rel} 
that the double-copy construction for $N$-point ($N\!\!\geqq\hsm\!4$) massive gauge/gravity 
KK scattering amplitudes can directly work out 
only for toroidal compactification with flat extra dimensions,
which has linear mass-spectrum proportional to KK index and 
exhibits single-pole exchange in each kinematic channel.\ 
Because the KK gauge/gravity theories under warped 5d compactification do not satisfy these conditions, 
they cannot hold double-copy for $N$-point ($N\!\!\geqq\hsm\!4$) KK amplitudes 
exactly or at the sub-leading orders.\  
Nevertheless, we will demonstrate in Section\,\ref{sec:4.1} 
that the double-copy can be realized for 3-point ($N\!\!=\!3$) 
KK gauge/gravity scatttering amplitudes without high energy expansion.\ 
Then, in Section\,\ref{sec:4.2} we construct the double-copy 
for 4-point KK gauge/gravity scatttering amplitudes 
at the leading order (LO) of high energy expansion
which is the key to derive the GRET from GAET.\   
We further study in Section\,\ref{sec:4.3} the extension to 
$N$-point ($\!N\!\! \geqq\!4\hs$) KK gauge/gravity scattering amplitudes 
at the LO and their LO double-copy construction.\
Finally, we give a prescription for realizing the warped double-copy 
at the LO of high energy expansion.

\vspace*{1mm}
\subsection{\hspace*{-2.5mm}Double-Copy Construction of 3-Point Amplitudes and Warped GRET}
\label{sec:4.1}
\vspace*{1.5mm}

We have computed the most general 3-point on-shell scattering amplitudes of KK gauge bosons in Eq.\eqref{eq:AAA-n1n2n3},
which takes the following form:
\beqs 
\label{eq:AAA-3pt}
\begin{align}
\TT[\{\ep_j^{}\}] 
& = g\hs f^{abc}\NN [\{\ep_j^{}\}]  \hs,
\\
\NN [\{\ep_j^{}\}] &= -\ii\hs 2\hs a_{n_1^{}n_2^{}n_3^{}}^{}\!
\[ (\ep_1^{}\!\cdot\hsm\ep_2^{})(\ep_3^{}\!\cdot\hsm p_1^{}) \hsm +\hsm 
(\ep_2^{}\!\cdot\hsm\ep_3^{})(\ep_1^{}\!\cdot\hsm p_2^{})\hsm +\hsm 
(\ep_3^{}\!\cdot\hsm\ep_1^{})(\ep_2^{}\!\cdot\hsm p_3^{}) \]\hsm,
\end{align}
\eeqs 
where $\{\ep_j^{}\}\!=\!\{\ep_{\lam_j^{}}^{\mu}\}$  denotes the polarization vectors for the three external
KK gauge bosons $\{A^{a\hs n_j^{}}_{\lam_j^{}}\}$ with $j\!=\!1,2,3$.\ 
In parallel, we can reexpress the most relevant 3-point scattering amplitudes \eqref{eq:55T} of KK Goldstone bosons 
with gauge bosons:
\begin{equation}
\label{eq:55T-dc}
\tT[A_5^{an_1^{}}\!A_5^{bn_2^{}}\!A_{\pm}^{cn_3^{}}] 
= g f^{abc}\NNt [\ep_3^{}]  \hs,~~~~~
\NNt [\ep_3^{}] = -\ii\hs 2\hs (\ep_3^{}\!\cdot\hsm p_1^{})\hs {\tilde{a}_{n_1^{}n_2^{}n_3^{}}^{}} \hs,
\end{equation}
where the cubic KK Goldstone-gauge coupling coefficient $\hs\tilde{a}^{}_{n_1^{}n_2^{}n_3^{}}\!$ 
is given by \eqrefe{app-eq:KKYM-RS-couplings} of Appendix\,\ref{app:C1}.\ 

\vs

The conventional BCJ double-copy method\,\cite{BCJ}\cite{BCJ-Rev} applies to $N$-point scattering amplitudes
of massless gauge bosons and gravitons with $N\!\!\geqq\!4\hs$.\  
For the case of  $N\!\!=\!3\hs$, the massless double-copy can be directly realized 
after imposing on-shell conditions on the 3-point gauge/gravity amplitudes.\  
For the present study, we first extend it to the 3-point ($N\!=\!3$) massive KK scattering amplitudes of
the warped gauge/gravity theories.\ 
This is much more nontrivial as the mass spectra of the KK gauge bosons and
KK gravitons also differ from each other.\ 
Since each on-shell 3-point gauge amplitude \eqref{eq:AAA-3pt} or \eqref{eq:55T-dc}
only has a single color factor and numerator without internal propagator, 
we may formulate the 3-point CK correspondence by using the following double-copy relations 
and the gauge/gravity coupling correspondence as follows:
\begin{equation}
\label{eq:CK-3pt}
f^{abc} {\longrightarrow\,} \NN [\{\ep_j^{}\}]\hs,  \hspace*{6mm}
f^{abc} {\longrightarrow\,} \NNt [\ep_3^{}]\hs,  \hspace*{6mm}
g {\,\longrightarrow\, -\frac{\,\ka\,}{4}}\hs.
\end{equation}
Moreover, at each KK level-$n$, we set up the KK-mass correspondence (replacement),
$\Mn\ito \mathbb{M}_n\hs$,
which means to replace each mass-eigenvalue $\Mn$ of KK gauge bosons
[determined by \eqrefe{Aeq:YM-Mn}] by mass-eigenvalue $\MGn$ of KK gravitons
[determined by \eqrefe{Aeq:GR-Mn}].\ 

\vs

With the above, we make the following double-copy construction of 
the on-shell 3-point KK graviton scattering amplitude from the product of 
the corresponding on-shell 3-point KK gauge boson scattering amplitudes \eqref{eq:AAA-3pt}:
\begin{equation}
\label{eq:M=TxT-3pt}
\M^{\rm{dc}}[\{e_j^{},\hs\ep_{j}^{}\}] =  
-\frac{\,\ka\hs \al_{n_1^{}n_2^{}n_3^{}}^{}\,}{~4\hs a_{n_1^{}n_2^{}n_3^{}}^{2}\,} 
\frac{1}{8}\Big( \NN[\{e_j^{}\}] \!\times\! \NN[\{\ep_{j}^{}\}] + \{e_j^{} \!\leftrightarrow\! \ep_{j}^{}\} \Big),
\end{equation}
where  
$\{e_j^{}\}\!=\!\{e_{\lam_j}^{\mu}\}$ denotes another set of polarization vectors in parallel to the polarization vecors
$\{\ep_{j}^{}\}\!=\!\{\ep_{\lam_{j}'}^{\mu}\!\}$, 
and the overall factor $1\hsm /8\!=\!1\hsm /2^3$ arises from symmetrizing the two
polarization vectors associated with each external KK graviton polarization.\ 
The two polarization vectors $\{e_j^{},\hs\ep_{j}^{}\}$ should compose the
polarization tensor $\vep_j^{\mn}\!=\!\vep_{\hat\lam_j}^{\mn}$ of the external KK graviton-$j\hs$ according to Eq.\eqref{eq:Gpol}.\ 
Also, because the graviton polarization tensor is traceless $\vep_{\hat\lam_j}^{\mn}\eta_{\mn}^{}\!=\!0\hs$,
we have the corresponding products vanish:
\beq 
\hs e_j^{\pm}\!\cdot\hsm\ep_{j}^{\pm} \!=\!0\hs, ~~~~\hs 
e_j^{\pm}\!\cdot\hsm\ep_{j}^{L} \!=\!0\hs, ~~~~~
e_j^{\pm}\!\cdot\hsm\ep_{j}^{\mp}+e_j^{L}\!\cdot\hsm\ep_{j}^{L} \!=\!0\hs. 
\eeq 
This helps to simplify our explicit results of double-copy.\
Using the polarization tensors \eqrefe{eq:Gpol} for the three external KK graviton states, we can express the
double-copy formula of the 3-point KK graviton scattering amplitude as follows:
\begin{align}
\label{eq:M=sumTxT-3pt} 
& \M^{\rm{dc}}\hsm\big[h_{n_1^{}}^{\hat\lam_1}h_{n_2^{}}^{\hat\lam_2}h_{n_3^{}}^{\hat\lam_3}\big]
= \sum_{\lam^{}_j,\lam'_j}\!\!\hsm
\Big(\!\prod_j\!\hsm C_{\hsm\lam_j^{}\lam'_j}^{\hat\lam_j^{}}\Big)\M\big[\{e_i^{}\}^{\lam_j}\!,\{\ep_i^{}\}^{\lam'_j}\big]
\nn\\
& \hspace*{4mm}
= -\frac{~\ka\hs \al_{n_1^{}n_2^{}n_3^{}}^{}\,}{~4\hs a_{n_1^{}n_2^{}n_3^{}}^{2}\,}\!\!
\sum_{\lam_j,\lam'_j}\!\!
\Big(\!\prod_j\!\hsm C_{\hsm\lam_j^{}\lam'_j}^{\hat\lam_j^{}}\Big)
\hs \NN[\{\hsm e_{\lam_j}^{}\!\}] \hsm\!\times\!\hsm \NN[\{\hsm\ep_{\hsm\lam_j'}^{}\!\}] 
\nn\\
&\hspace*{4mm}
\equiv \frac{~\ka\hs \al_{n_1^{}n_2^{}n_3^{}}^{}\,}{~4~}\!\!
\sum_{\lam_j,\lam'_j}\!\!
\Big(\!\prod_j\!\hsm C_{\hsm\lam_j^{}\lam'_j}^{\hat\lam_j^{}}\Big) \xoverline{\M}^{\rm{dc}} \,,
\end{align}
where the helicity index
$\,\hat\lam_j^{}\!=\!\{\pm 2,\pm 1,0\}
\!=\!\{\pm 2,\pm 1,L\}\,$
labels the 5 helicity states of each external massive KK graviton.\
The polarization tensors of the external KK graviton states are defined as follows:
\beqs \label{eq:plz-tensor-combination}
\begin{align}
&\vep^{\mn}_{\hs\hat\lam_{\hsm j}^{}} =
\sum_{\lam^{}_j,\lam'_j} \!\!
C_{\hsm\lam_j^{}\lam'_j}^{\hs\hat\lam_{\hsm j}^{}}\hs
\ep_{\lam_j^{}}^\mu\hs\ep_{\lam'_j}^\nu \,, 
\\
& C_{\pm 1,\pm 1}^{\pm 2} \!=\! 1 \hs, \quad
C_{\pm 1,0}^{\pm 1} \!= C_{0,\pm 1}^{\pm1}\hsm \! =\!\frac{1}{\sqrt{2\,}\,}\hs, \quad
C_{\pm 1,\mp 1}^{\hs 0} \!= \!\frac{1}{\sqrt{6\,}\,} \hs, \quad
C_{0,0}^{\hs 0}\!=\! \sqrt{\frac{2}{3}\hs} \hs.
\end{align}
\eeqs

\vs

Then, we use the double-copy method to compute the on-shell 3-point scattering amplitudes 
of KK gravitons and derive the following formula:
\begin{align}
\label{eq:DC-hhh-3pt}
& \xoverline{\M}^{\rm{dc}}
=(e^{}_1\ep^{}_3 ) (e^{}_2 p^{}_3 ) (e^{}_3 p^{}_2 ) (\ep^{}_1\ep^{}_2 )+(e^{}_1e^{}_3 ) (e^{}_2 p^{}_3 ) (p^{}_2\ep^{}_3 ) (\ep^{}_1\ep^{}_2 )+(e^{}_1 p^{}_3 ) (e^{}_2 p^{}_3 ) (e^{}_3\ep^{}_2 ) (\ep^{}_1\ep^{}_3 )
\nn\\
&+(e^{}_1\ep^{}_2 ) (e^{}_2 p^{}_3 ) (e^{}_3 p^{}_2 ) (\ep^{}_1\ep^{}_3 )+(e^{}_1 p^{}_3 ) (e^{}_2 p^{}_3 ) (e^{}_3\ep^{}_1 ) (\ep^{}_2\ep^{}_3 )+(e^{}_1e^{}_3 ) (e^{}_2 p^{}_3 ) (p^{}_3\ep^{}_1 ) (\ep^{}_2\ep^{}_3 )
\nn\\
&+(e^{}_1\ep^{}_2 ) (e^{}_2 p^{}_3 ) (e^{}_3\ep^{}_1 ) (p^{}_2\ep^{}_3 )-2 (e^{}_1\ep^{}_2 ) (e^{}_2\ep^{}_1 ) (e^{}_3 p^{}_2 ) (p^{}_2\ep^{}_3 )-2 (e^{}_1e^{}_2 ) (e^{}_3 p^{}_2 ) (p^{}_2\ep^{}_3 ) (\ep^{}_1\ep^{}_2 )
\nn\\
&+(e^{}_1e^{}_3 ) (e^{}_2\ep^{}_1 ) (p^{}_3\ep^{}_2 ) (p^{}_2\ep^{}_3 )+(e^{}_1e^{}_2 ) (e^{}_3\ep^{}_1 ) (p^{}_3\ep^{}_2 ) (p^{}_2\ep^{}_3 )-(e^{}_1 p^{}_3 ) (e^{}_2\ep^{}_3 ) (e^{}_3 p^{}_2 ) (\ep^{}_1\ep^{}_2 )
\nn\\
&-(e^{}_1 p^{}_3 ) (e^{}_2\ep^{}_1 ) (e^{}_3 p^{}_2 ) (\ep^{}_2\ep^{}_3 )-(e^{}_1 p^{}_3 ) (e^{}_2\ep^{}_1 ) (e^{}_3\ep^{}_2 ) (p^{}_2\ep^{}_3 )-(e^{}_1 p^{}_3 ) (e^{}_2e^{}_3 ) (p^{}_2\ep^{}_3 ) (\ep^{}_1\ep^{}_2 )
\nn\\
&+(e^{}_1\ep^{}_3 ) (e^{}_2 p^{}_3 ) (e^{}_3\ep^{}_2 ) (p^{}_3\ep^{}_1 )-(e^{}_1\ep^{}_2 ) (e^{}_2\ep^{}_3 ) (e^{}_3 p^{}_2 ) (p^{}_3\ep^{}_1 )-(e^{}_1e^{}_2 ) (e^{}_3 p^{}_2 ) (p^{}_3\ep^{}_1 ) (\ep^{}_2\ep^{}_3 )
\nn\\
&-(e^{}_1\ep^{}_2 ) (e^{}_2e^{}_3 ) (p^{}_3\ep^{}_1 ) (p^{}_2\ep^{}_3 )-(e^{}_1e^{}_2 ) (e^{}_3\ep^{}_2 ) (p^{}_3\ep^{}_1 ) (p^{}_2\ep^{}_3 )-2 (e^{}_1 p^{}_3 ) (e^{}_2\ep^{}_3 ) (e^{}_3\ep^{}_2 ) (p^{}_3\ep^{}_1 )
\nn\\
&-2 (e^{}_1 p^{}_3 ) (e^{}_2e^{}_3 ) (p^{}_3\ep^{}_1 ) (\ep^{}_2\ep^{}_3 )+(e^{}_1 p^{}_3 ) (e^{}_2\ep^{}_3 ) (e^{}_3\ep^{}_1 ) (p^{}_3\ep^{}_2 )+(e^{}_1 p^{}_3 ) (e^{}_2e^{}_3 ) (p^{}_3\ep^{}_2 ) (\ep^{}_1\ep^{}_3 )
\nn\\
&+(e^{}_1e^{}_2 ) (e^{}_3 p^{}_2 ) (p^{}_3\ep^{}_2 ) (\ep^{}_1\ep^{}_3 )+(e^{}_1\ep^{}_3 ) (e^{}_2\ep^{}_1 ) (e^{}_3 p^{}_2 ) (p^{}_3\ep^{}_2 )-2 (e^{}_1\ep^{}_3 ) (e^{}_2 p^{}_3 ) (e^{}_3\ep^{}_1 ) (p^{}_3\ep^{}_2 )
\nn\\
&-\! 2 (e^{}_1e^{}_3 ) (e^{}_2 p^{}_3 ) (p^{}_3\ep^{}_2 ) (\ep^{}_1\ep^{}_3 )\!+\!(e^{}_1\ep^{}_3 ) (e^{}_2e^{}_3 ) (p^{}_3\ep^{}_1 ) (p^{}_3\ep^{}_2 )\!+\!(e^{}_1e^{}_3 ) (e^{}_2\ep^{}_3 ) (p^{}_3\ep^{}_1 ) (p^{}_3\ep^{}_2 )\,,
\end{align}
where we have used the shorthand notation $(ab)\!\equiv\! (a \hsm\cdot\hsm b)$, and
we also make the replacement $M_{n}^{}\!\to\hsm \MG_{n}^{}\hs$ for each KK mass.\

We compare the above double-copy formula \eqref{eq:DC-hhh-3pt} of the general 3-point KK graviton scattering amplitude
with the explicit calculation result of the same KK graviton scattering amplitude \eqref{eq:Amp-3h-exact} in
Appendix\,\ref{app:C2}.\ We find that they fully agree with each other after 
imposing the correspondence \eqref{eq:CK-3pt}.\

\vs

With the above, we then analyze the 3-point on-shell scattering amplitude of KK gravitons 
at the leading order of high energy expansion
and establish the double-copy construction of the KK gravitational equivalence theorem (GRET) 
from the KK gauge-theory equivalence theorem (GAET) 
at the level of the most fundamental 3-point KK gauge/gravity scattering amplitudes.\ 
We consider a 3-point KK graviton scattering amplitude 
containing two external KK gravitons with helicities $\pm1$
and one external KK gravition with helicities $\pm2$, 
or, including two external longitudinal polarization states (with helicity-0) 
plus one transverse polarization state (with helicities $\pm2$).\ 
Using the KK gauge boson scattering amplitudes \eqref{eq:AAA-3pt} and \eqref{eq:M=sumTxT-3pt}, 
we can construct two types of 3-point KK graviton amplitudes at the leading order of high energy expansion.\
One type of the KK graviton scattering amplitudes have helicities $(\pm1,\hs \pm1,\hs \pm2)$,  
and we derive the following:
{\small 
\beqs
\label{eq:DC-3pt-112All}
\begin{align}
\label{eq:DC-3pt-112}
\hspace*{-2mm}
&{\M^{\rm{dc}}} [h^{\pm1}_{n_1}h^{\pm1}_{n_2}h^{\pm2}_{n_3}\hs]  
={-\frac{~c_3^{}\hs\ka\,}{~4}}\hsm  
\Big(\hsm\NN [A^{\pm}_{n_1}A^{\pm}_{n_2}A^{\pm}_{n_3}]\, \NN [A^{L}_{n_1}A^{L}_{n_2}A^{\pm}_{n_3}]  
+ \!\NN [A^{\pm}_{n_1}A^{L}_{n_2}A^{\pm}_{n_3}]\NN [A^{L}_{n_1}A^{\pm}_{n_2}A^{\pm}_{n_3}] \hsm\Big)
\nn \\
\hspace*{-2mm}
&= \M_0^{\rm{dc}}[h^{\pm1}_{n_1}h^{\pm1}_{n_2}h^{\pm2}_{n_3}\hs] +O(E^0) \hs, 
\\[1mm]
\label{eq:DC-LO-3pt-112}
\hspace*{-2mm}
&{\M_0^{\rm{dc}}}[h^{\pm1}_{n_1}h^{\pm1}_{n_2}h^{\pm2}_{n_3}\hs]  
={-\frac{~c_3^{}\hs\ka\,}{~4}}
\NN_0^{}[A^{\pm}_{n_1}A^{\pm}_{n_2}A^{\pm}_{n_3}] \NN_0^{}[A^{L}_{n_1}A^{L}_{n_2}A^{\pm}_{n_3}] 
\nn\\[1mm]
\hspace*{-2mm}
&= {c_3^{}\hs a_{n_1^{}n_2^{}n_3^{}}^{2}}\!\!
\frac{\,\ka\hs (1\!-\!\rh_{13}^2\!-\!\rh_{23}^2)\,}{\,2\hs \rh_{13}^{}\hs \rh_{23}^{}\,} \!
\Big[\hsm (\ep_1^{}\!\hsm\cdot\hsmx \ep_2^{})(\ep_3^{}\!\hsm\cdot\hsmx p_1^{})^2
\!+\!(\ep_2^{}\!\hsm\cdot\hsmx\ep_3^{})(\ep_1^{}\!\hsm\cdot\hsmx p_2^{})(\ep_3^{}\!\hsm\cdot\hsmx p_1^{}) 
\!+\!(\ep_3^{}\!\hsm\cdot\hsmx\ep_1^{})(\ep_2^{}\!\hsm\cdot\hsmx p_3^{})(\ep_3^{}\!\hsm\cdot\hsmx p_1^{})\hsm\Big],
\end{align}
\eeqs 
}
\hspace*{-1.5mm}where $c_3^{}$ is a normalization factor of couplings and will determined shortly.\ 
As before we denote the mass ratio $\rh_{ij}^{}\!=\!\MG_{n_i^{}}^{}/\MG_{n_j^{}}^{}$.\ 
We note that on the right-hand side of Eq.\eqref{eq:DC-3pt-112} 
the second amplitude product has only subleading contributions
of $O(E^0)$, instead of the leading-order contributions of $O(E^2)\hs$.\ 

\vs

Another type of KK graviton scattering amplitudes have helicities $(L,\hs L,\hs \pm2)$,
where $L$ stands for longitudinal polarization with helicity-0$\hs$.\ 
We construct its exact tree-level
scattering amplitude $\M [h^L_{n_1} h^L_{n_2} h^{\pm2}_{n_3}\hs]$  
and derive its leading-order amplitude 
$\M_0^{}[h^L_{n_1} h^L_{n_2} h^{\pm2}_{n_3}\hs]$ as follows:
{\small 
\beqs 
\label{eq:DC-hhh-LL2}
\begin{align} 
\hspace*{-3mm}
{\M^{\rm{dc}}}[h^L_{n_1} h^L_{n_2} h^{\pm2}_{n_3}\hs] &= 
{-\frac{~c_3^{}\hs\ka\,}{~4}} \frac{1}{3} 
\Big\{ 2(\NN_0^{}[A^L_{n_1^{}}\! A^L_{n_2^{}}\! A^{\pm}_{n_3^{}}])^2 + 
\NN_0^{}[A^{+}_{n_1^{}}\! A^{-}_{n_2^{}}\! A^{\pm}_{n_3^{}}] \NN_0^{}[A^{-}_{n_1^{}}\! A^{+}_{n_2^{}}\! A^{\pm}_{n_3^{}}]  
\nn \\
\hspace*{-3mm}
&\hspace*{0.5cm}  
+\NN_0^{}[A^{+}_{n_1^{}}\! A^{+}_{n_2^{}}\! A^{\pm}_{n_3^{}}] \hs\NN_0^{}[A^{-}_{n_1^{}}\! A^{-}_{n_2^{}}\! A^{\pm}_{n_3^{}}] +
2\NN_0^{}[A^{+}_{n_1^{}}\! A^{L}_{n_2^{}}\! A^{\pm}_{n_3^{}}] \hs\NN_0^{}[A^{-}_{n_1^{}}\! A^{L}_{n_2^{}}\! A^{\pm}_{n_3^{}}] \Big\}\hs,  
\label{eq:DC-hhh-LL2full}
\\[1mm]
\hspace*{-3mm}
{\M^{\rm{dc}}_0}[h^L_{n_1} h^L_{n_2} h^{\pm2}_{n_3}\hs] &= 
{-\frac{~c_3^{}\hs\ka\,}{~12}}\!\LB\hsm 
2(\NN_0^{}[A^L_{n_1^{}}\! A^L_{n_2^{}}\! A^{\pm}_{n_3^{}}])^2 \!+\!   
\NN_0^{}[A^{+}_{n_1^{}}\! A^{-}_{n_2^{}}\! A^{\pm}_{n_3^{}}] 
\NN_0^{}[A^{-}_{n_1^{}}\! A^{+}_{n_2^{}}\! A^{\pm}_{n_3^{}}]\hsm\RB
\nn\\[1mm]
\hspace*{-3mm}
&= {c_3^{}\hs a_{n_1^{}n_2^{}n_3^{}}^{2}}
\frac{\,\ka\hs (\ep_3^{}\!\hsm\cdot\hsmx p_1^{})^2\,}{6}  \!\!\[ 2 +\!
\(\!\!\frac{\,1 \!-\! \rh_{13}^2 \!-\! \rh_{23}^2\,}{\rh_{13}^{}\hs \rh_{23}^{}}\!\!\)^{\!\!\!2}\, \]\! ,
\label{eq:DC-hhh-LL2-LO}
\end{align}
\eeqs 
}
\hspace*{-1.5mm}where the KK mass ratios are defined as\, 
$\hs \rh_{13}^{}\!=\!\mathbb{M}_{n_1^{}}^{}\!/\mathbb{M}_{n_3^{}}^{}$ and 
$\hs \rh_{23}^{}\!=\!\mathbb{M}_{n_2^{}}^{}\hsm /\mathbb{M}_{n_3^{}}^{}$.\ 
We note that under the high energy expansion and on the righ-hand side of Eq.\eqref{eq:DC-hhh-LL2full},
the first two terms in the braces give leading-order contributions and the last two terms are
subleading.\ 

\vs
Then, for the 3-point leading-order KK gravitational amplitudes \eqref{eq:GRET-LO-VVh2} and \eqref{eq:pph},
we can construct them by double-copy of the corresponding two 3-point KK gauge boson amplitudes: 
\beqs
\label{eq:DC-3pt-GB}
\begin{align}
{\MT_0^{\rm{dc}}}[\VV^{\pm1}_{n_1^{}}\VV^{\pm1}_{n_2^{}}h^{\pm2}_{n_3^{}}\hs]
&= {-\frac{~ c_3^{}\hs\ka\,}{~4}} \NN_0^{}[A^{\pm}_{n_1}A^{\pm}_{n_2}A^{\pm}_{n_3}]\, 
\NNt_0^{}[A^{5}_{n_1}A^{5}_{n_2}A^{\pm}_{n_3}]
\nn\\
&= {\ka\hs c_3^{}\hs a_{n_1^{}n_2^{}n_3^{}}^{}\hsm\tilde{a}_{n_1^{}n_2^{}n_3^{}}^{} } 
\Big[\hsm (\ep_1^{}\!\cdot\hsm \ep_2^{})(\ep_3^{}\!\cdot\hsm p_1^{})^2
\!+\!(\ep_2^{}\!\cdot\hsm \ep_3^{})(\ep_1^{}\!\cdot\hsm p_2^{})(\ep_3^{}\!\cdot\hsm p_1^{})
\nn\\
& \hspace*{3.4cm}+\hsm (\ep_3^{}\!\cdot\hsm \ep_1^{})(\ep_2^{}\!\cdot\hsm p_3^{})
(\ep_3^{}\!\cdot\hsm p_1^{})\Big] \hs,
\label{eq:LO-DC-3pt-GB112}
\\[1mm]
{\MT_0^{\rm{dc}}}[\phi_{n_1^{}}^{}\phi_{n_2^{}}^{}h_{n_3^{}}^{\pm2}\hs]
&= {-\frac{~c_3^{}\hs\ka\,}{~4} (\NNt_0^{}[A^5_{n_1^{}}A^5_{n_2^{}}A^{\pm}_{n_3^{}}])^2 }
= {\ka\hs c_3^{}\hs\tilde{a}_{n_1^{}n_2^{}n_3^{}}^{2}} \(\ep_3^{}\!\cdot\hsm p_1^{}\)^2 .
\label{eq:LO-DC-3pt-GB002}
\end{align}
\eeqs

By comparing the gravitational KK scattering amplitudes \eqref{eq:DC-3pt-112All}-\eqref{eq:DC-3pt-GB} 
(obtained via double-copy construction) 
with the same type of amplitudes \eqref{eq:GRET-LO-h1h1h2}-\eqref{eq:GRET-LO-VVh2}
and \eqref{eq:hhh-LLT}-\eqref{eq:pph}
(obtained via explicit Feynman-diagram approach), 
we find that the two sets of KK scattering amplitudes share exactly {\it the same kinematic structure} and 
can equal to each other so long as we require that 
the KK gauge boson mass be replaced by the corresponding KK graviton mass at each KK level and
the KK gauge coupling factor are replaced by the corresponding KK gravitational coupling factor 
[cf.\ Eq.\eqref{eq:CK-3pt}] together with the following replacements of the trilinear coupling factors:
\begin{align}
\label{eq:Conversion-C3}
c_3^{}\, a_{n_1^{}n_2^{}n_3^{}}^{2}\!\!\to \al_{n_1^{}n_2^{}n_3^{}}^{},\quad
c_3^{}\, a_{n_1^{}n_2^{}n_3^{}}^{} \tilde{a}_{n_1^{}n_2^{}n_3^{}}^{} \!\!\to \tilde{\al}_{n_1^{}n_2^{}n_3^{}}^{},\quad
c_3^{}\, \tilde{a}_{n_1^{}n_2^{}n_3^{}}^2 \!\!\to \tilde{\be}_{n_1^{}n_2^{}n_3^{}}^{} .
\end{align}
%
In general, such replacements are nontrivial because the coupling raitos 
(such as the ratio $\al_{n_1^{}n_2^{}n_3^{}}^{}\hsm /a_{n_1^{}n_2^{}n_3^{}}^{2}$) do not equal one.\
This means that for the double-copy of KK gauge/gravity theories with warped 5d space, the coupling
replacements \eqref{eq:Conversion-C3} are necessary and nontrivial.\  
We note that in the flat 5d limit of $\,k\ito 0\hs$, these ratios take simple values (cf.~{Table\,\ref{app-table:Flat-values}}).\   
For instance, choosing $\hs (n_1^{},n_2^{},n_3^{})\!=\!(n,\hs n,\hs j)\,$ with $n\!>\!0$, 
we have 
\begin{equation}
c_3^{}=\frac{\,\al_{nnj}^{}\,}{a^2_{nnj}}
= \frac{\tilde{\al}_{nnj}}{\,a^{}_{nnj}\tilde{a}_{nnj}}
=\frac{\,\tilde{\be}_{nnj}^{}\,}{\tilde{a}^2_{nnj}\,}
=\LB\!\!
\begin{array}{cl}
1\hs,  & \quad j\!=\!0 \,, 
\\[1mm] 
\frac{1}{\sqrt{2\,}\,}\hs, &\quad j\!=\!2\hs n \,,
\end{array}
\right.
\end{equation}
whereas for the case of $\hs (n_1^{},n_2^{},n_3^{})\!=\!(n,\hs m,\hs |n\hsm\pm\hsm m|)\,$ with $n\!\neq\!m$
in the flat 5d limit, we have 
\begin{equation}
c_3^{}=\frac{\,a^2_{nm|n\pm m|}\,}{\,\al_{nm|n\pm m|}^{}\,} 
= \frac{\,a_{nm|n\pm m|}^{}\tilde{a}_{nm|n\pm m|}^{}\,}{\tilde{\al}_{nm|n\pm m|}^{}}
=\frac{\,\tilde{a}^2_{nm|n\pm m|}\,}{\,\tilde{\be}_{nm|n\pm m|}^{}\,}
=\frac{1}{\sqrt{2\,}\,} \,.
\end{equation}
In fact, as we have checked, for the flat 5d space under the $S^1$ toroidal compactification
without orbifold\,\cite{Li:2022rel}, all the above coupling ratios will become $c_3^{}\!=\!1$. 

\vs 

Finally, using the coupling replacements \eqref{eq:Conversion-C3}, 
we compare the leading-order KK graviton amplitudes 
\eqref{eq:DC-LO-3pt-112} and \eqref{eq:DC-hhh-LL2-LO} 
with the corresponding gravitational KK Goldstone boson amplitudes
\eqref{eq:LO-DC-3pt-GB112} and \eqref{eq:LO-DC-3pt-GB002} respectively.\
We find that they share exactly the same kinematic structures and obey the GRET,
\beqs 
\begin{align}
\label{eq:GRET112-3pt-DC}
{\M_0^{\rm{dc}}}[h^{\pm1}_{n_1}h^{\pm1}_{n_2}h^{\pm2}_{n_3}\hs] &= -{\MT_0^{\rm{dc}}}[\VV^{\pm1}_{n_1^{}}\VV^{\pm1}_{n_2^{}}h^{\pm2}_{n_3^{}}\hs]\hs, 
\\
\label{eq:GRET002-3pt-DC}
{\M^{\rm{dc}}_0}[h^L_{n_1} h^L_{n_2} h^{\pm2}_{n_3}\hs] &= 
{\MT_0^{\rm{dc}}}[\phi_{n_1^{}}^{}\phi_{n_2^{}}^{}h_{n_3^{}}^{\pm2}\hs]\hs,
\end{align}
\eeqs 
provided the following conditions hold respectively,
\beqs 
\label{eq:3pt-dcCond-h112+h002}
\begin{align}
\label{eq:3pt-dcCond-h112}
(\rh_{13}^2\!+\!\rh_{23}^2\!-\!1)\al_{n_1^{}n_2^{}n_3^{}}^{} &\!=\hs 
2\hs\rh_{13}^{}\rh_{23}^{}\tilde{\al}_{n_1^{}n_2^{}n_3^{}}^{} ,
\\
\label{eq:3pt-dcCond-h002}
\[\hsm (1\!-\!\rh_{13}^2\!-\!\rh_{23}^2)^2\!+\!2\hs\rh_{13}^{2}\rh_{23}^{2}\]\!\hsm\al_{n_1^{}n_2^{}n_3^{}}^{}
&\!=\hs 6\hs\rh_{13}^{2}\rh_{23}^{2} \tilde{\be}_{n_1^{}n_2^{}n_3^{}}^{} ,
\end{align}
\eeqs 
where the mass ratios 
$\hs \rh_{13}^{}\!=\!\mathbb{M}_{n_1^{}}^{}\!/\mathbb{M}_{n_3^{}}^{}$ and 
$\hs \rh_{23}^{}\!=\!\mathbb{M}_{n_2^{}}^{}\hsm /\mathbb{M}_{n_3^{}}^{}$.\ 
Comparing the above conditions \eqref{eq:3pt-dcCond-h112}-\eqref{eq:3pt-dcCond-h002} 
with the conditions  \eqref{eq:al-tal-3pt} and \eqref{eq:al-tbe-LLT-55T}
(which we obtained by explicit Feynman diagram approach), we find that the
identities \eqref{eq:3pt-dcCond-h112} and \eqref{eq:al-tal-3pt} are identical, and
the same is true for the two identities \eqref{eq:3pt-dcCond-h002} versus \eqref{eq:al-tbe-LLT-55T}.\ 
This establishes our successful double-copy constructions of the GRET \eqref{eq:GRET112-3pt-DC} and
\eqref{eq:GRET002-3pt-DC} (in warped KK gravity theory) from the GAET 
\eqref{eq:GAET-LLT-3pt} (in the warped KK gauge theory). 

\vspace*{1mm}

In summary, we have established the double-copy construction from the GAET to GRET
for the 3-point leading-order (LO) KK scattering amplitudes:
\begin{equation}
\label{eq:T3AL-M3hL-outline}
\begin{aligned}
\M_0^{(3)}[h^n_L] &~\xlongequal{~\scalebox{.9}{\text{GRET}}~}~ \MT_0^{(3)}[\phin]
\\[.5mm]
\xuparrow{.5cm}\scalebox{.9}{$\text{DC}_3$}\hspace*{-0.1cm}
&\hspace*{2.1cm} \scalebox{.9}{$\text{DC}_3$}\xuparrow{.5cm}
\\[.5mm]
\TT_0^{(3)}[A^n_L] &~\xlongequal{~\scalebox{.9}{\text{GAET}}~}~ 
\tT_0^{(3)}[A^n_5]
\end{aligned}
\end{equation}
In Eq.\eqref{eq:T3AL-M3hL-outline}, along the horizontal directions,
the lower (upper) equality presents the GAET (GRET) of 3-point scattering amplitudes
of KK gauge bosons (KK gravitons).\ 
Along the direction of a given vertical arrow, each correspondence is established 
by the double-copy (DC$_3^{}$) under the extended massive color-kinematics dualities \eqref{eq:CK-3pt} and \eqref{eq:Conversion-C3}.\
The left vertical arrow represents the double-copy (DC$_3^{}$) 
construction of the 3-point LO KK graviton
amplitude $\M_0^{(3)}[h^n_L]$ from the corresponding LO KK gauge boson amplitude $\TT_0^{(3)}[A^n_L]$.\  
The right vertical arrow represents the double-copy (DC$_3^{}$) 
construction of the 3-point LO gravitational 
KK Goldstone boson amplitude $\MT_0^{(3)}[\phin]$ from the corresponding 
(gauge theory) LO KK Goldstone boson amplitude $\tT_0^{(3)}[A^n_5]$.\ 
Finally, this picture also holds if we replace the two amplitudes
$\M_0^{(3)}[h^n_L]$ and $\MT_0^{(3)}[\phin]$ in the top row by
$\M_0^{(3)}[h^n_{\pm1}]$ and $\MT_0^{(3)}[\VV_{\pm1}^{n}]$ respectively.\

\vspace*{2mm}
\subsection{\hspace*{-2.5mm}Double-Copy Construction of 4-Point Amplitudes and Warped GRET}
\label{sec:4.2}
\vspace*{1.5mm}

In this subsection, we study double-copy construction for the 4-point KK gauge/gravity scattering amplitudes
and for their corresponding 4-point KK Goldstone boson scattering amplitudes
at the leading order (LO) of high energy expansion.\ 
As we will demonstrate, under high energy expansion, the leading-order KK amplitudes no longer contain
summation of infinite number of KK mass-poles.\ 
By construction they can exhibit an effective single massless-pole in each channel
and the associated numerators obey the kinematic Jacobi identity, which enable the realization of double-copy 
at the leading order.\
In the subsections\,\ref{sec:4.2.1} and \ref{sec:4.2.2}, we study the double-copy constructions for
the elastic and inelastic scattering amplitudes of KK gravitons respectively.

\vspace*{1.5mm}
\subsubsection{\hspace*{-2.5mm}Construction of 4-Point Elastic KK Graviton Amplitudes}
\label{sec:4.2.1}
\vspace*{1mm}

At the leading order of high energy expansion, we can reexpress the 4-point scattering amplitudes 
\eqref{eq:T-4AL-LO}-\eqref{eq:T-4A5-LO}
of the longitudinal KK gauge bosons and of the KK Goldstone bosons as follows: 
\beqs
\label{Amp-ALA5-nnnn}
\begin{align}
\label{Amp-AL-nnnn}
\TT_0^{}[4A^{n}_{L}]
& \,=\, g^2 \! \(\! \frac{\,\CC_s^{} \NN_s^0\,}{\sz}
+ \frac{\,\CC_t^{} \NN_t^0\,}{\tz} + \frac{\,\CC_u^{} \NN_u^0 \,}{\uz}  \!\)  \!,
\\[1mm]
\label{Amp-A5-nnnn}
\tT_0^{}[4A^{n}_5]
& \,=\, g^2 \! \(\! \frac{\,\CC_s^{} \NNt_s^0\,}{\sz} + \frac{\,\CC_t^{} \NNt_t^0\,}{\tz}
+ \frac{\,\CC_u^{} \NNt_u^0\,}{\uz}  \!\)  \!,
\end{align}
\eeqs
where the Mandelstam variables $(\sz,\tz,\uz)$ are defined in Eq.\eqref{eq:s0-t0-u0} 
of Appendix\,\ref{app:A} and obey the condition
$\sz\!+\!\tz\!+\!\uz\!=\!0\hs$.\
In the above, the kinematic numerators $\big(\NN_s^{0},\, \NN_t^{0},\, \NN_u^{0}\big)$
and $\big(\NNt_s^{0},\, \NNt_t^{0},\, \NNt_u^{0}\big)$  are connected to the
subamplitudes of Eqs.\eqref{eq:T-4AL-LO}-\eqref{eq:T-4A5-LO} and Eqs.\eqref{eq:K0stu-L}-\eqref{eq:K0stu-5}
via the folllowing relations:
\beqs
\label{eq:Nj-Ntj-def-4n}
\begin{align}
\label{eq:Nj-def-4n}
\big(\NN_s^{0},\, \NN_t^{0},\, \NN_u^{0}\big) &=
\big(\sz\hs\KK_s^0,\, \tz\hs\KK_t^0,\, \uz\hs\KK_u^0\big) ,
\\
\label{eq:Ntj-def-4n}
\big(\NNt_s^{0},\, \NNt_t^{0},\, \NNt_u^{0}\big) &=\big(\sz\hs\KKt_s^0,\,\tz\hs\KKt_t^0,\,\uz\hs\KKt_u^0\big),
\end{align}
\eeqs
Using the definitions \eqref{eq:Nj-Ntj-def-4n} and the difference relation \eqref{eq:K-Kt-nnnn}, we derive the generalized gauge transformations between the numerators of elastic scattering amplitudes of KK gauge bosons and of the KK Goldstone bosons:
\begin{equation}
\label{eq:GGT-NjNtj-nnnn}
\NN^0_j \,=\, \NNt_j^0 + s_{0j}^{}\Theta \,, \hspace*{10mm} 
\Theta =\! -2\hs\ct\!\sum_{i=0}^{\infty}\! r_{i}^2 a_{nni}^2 \,,
\end{equation}
where $s_{0j}^{}\!\in\!\{\sz,\hs \tz,\hs \uz\}$ denotes the massless Mandelstam variables.\
Inspecting the leading-order numerators \eqref{eq:Nj-def-4n} and \eqref{eq:Ntj-def-4n}
for elastic scattering amplitudes of KK gauge bosons and of KK Goldstone bosons, 
we find that they satisfy the kinematic Jacobi identities:
%
\beqs
\label{eq:Jacobi-4pt}
\begin{align}
\label{eq:KJacobi-N0}
\sum_j \NN_j^0  &\hs =\hs
\sum_j \NNt_j^0 + \Theta\sum_j\! s_{0j}^{} = 0\hs,
\\
\label{eq:KJacobi-N50}
\sum_j \NNt_j^0 &\hs =\hs 0 \hs,
\end{align}
\eeqs
where the index $j\!\in\!\{s,t,u\}$.\  We can readily prove the above second identity \eqref{eq:KJacobi-N50} by using 
the formulas \eqref{eq:Ntj-def-4n} and \eqref{eq:K0stu-5},
whereas the first identity \eqref{eq:KJacobi-N0} holds because of  $\sz\hsm +\hsm\tz\hsm +\hsm\uz\hsm =0\hs$.

\vs

We extend the conventional color-kinematics duality 
for massless gauge/gravity scattering amplitudes\,\cite{BCJ}\cite{BCJ-Rev} 
to the massive scattering amplitudes of the longitudinal KK gauge bosons 
and of the corresponding KK Goldstone bosons in \eqrefe{Amp-ALA5-nnnn}.\ 
We formulate the massive color-kinematics duality by using the following double-copy relations:
\begin{equation}
\label{eq:CK-4pt}
\CC_j^{} {\,\longrightarrow\,} \NN_j^{0}\hs, \hspace*{8mm}
\CC_j^{} {\,\longrightarrow\,} \NNt_j^{0}\hs, \hspace*{8mm}
g^2 {\,\longrightarrow\,} -\! (\ka /4)^2 \hs. 
\end{equation}
Thus, we construct the corresponding scattering amplitudes of the longitudinal KK gravitons
and of the gravitational KK Goldstone bosons, to the leading-order contributions of $\mO(E^2)$ under the high energy expansion.\  At each KK level-$n$, we also set up the KK-mass correspondence (replacement)
from KK gauge bosons to KK gravitons, $\Mn\!\to \mathbb{M}_n\hs$.

\vs

Next, we make double-copy construction of the 4-point scattering amplitudes of (helicity-zero) longitudinal KK gravitons and
of the KK gravitational Goldstone bosons from the corresponding KK gauge/Goldstone boson scattering amplitudes.\   
Applying the double-copy correspondence \eqref{eq:CK-4pt} to the 4-point scattering amplitudes 
\eqref{Amp-AL-nnnn}-\eqref{Amp-A5-nnnn} of the longitudinal KK gauge bosons 
and of the corresponding KK Goldstone bosons at the leading order of high energy expansion, 
we derive the following 4-point scattering amplitudes of 
the longitudinal KK gravitons and of the KK gravitational Goldstone bosons:
\beqs 
\label{eq:DC-hL-phi-4n}
\begin{align}
\label{eq:DC-hL-4n}
\M_0^{\rm{dc}}[4h_L^n] &= -\frac{\,c_4^{}\ka^2\,}{\,16\,}\hsm 
\Bigg[\!\frac{\,(\NN_s^0)^2\,}{\sz}+\frac{\,(\NN_t^0)^2\,}{\tz}
+\frac{\,(\NN_u^0)^2\,}{\uz}\Bigg],
\\
\label{eq:DC-phi-4n}
\MT_0^{\rm{dc}}[4\phin] &= -\frac{\,c_4^{}\ka^2\,}{\,16\,}\hsm 
\Bigg[\!\frac{\,(\NNt_s^0)^2\,}{\sz}+\frac{\,(\NNt_t^0)^2\,}{\tz}
+\frac{\,(\NNt_u^0)^2\,}{\uz}\Bigg],
\end{align}
\eeqs 
where the normalization factor $c_4^{}$ will be determined by taking the flat 5d limit ($k\ito 0$).\ 
We already established\,\cite{Li:2022rel} the exact double-copy formula 
for the flat 5d space under orbifold compactification of $S^1\!/\ZZ\hs$,
which leads to $\hs c_4^{} \!=\! 2/3\hs$ \cite{Li:2022rel}.\ 

\vs 

Applying the generalized gauge transformation \eqref{eq:GGT-NjNtj-nnnn} to
the numerators of the longitudinal KK graviton scattering amplitude
\eqref{eq:DC-hL-4n} and making use of the Jacobi identity \eqref{eq:KJacobi-N50}, 
we can directly prove the equivalence between the leading-order scattering 
amplitudes of the longitudinal KK gravitons and of the gravitational KK Goldstone bosons:
\begin{equation}
\label{eq:GRET-DC-4n}
\M_0^{\rm{dc}}[4h_L^n] \,=\, \MT_0^{\rm{dc}}[4\phin] \,.
\end{equation}
This is a proof of the KK gravitational equivalence theorem (GRET) from the 
KK gauge-theory equivalence theorem (GAET) by using the double-copy approach 
at the level of 4-point scattering amplitudes. 

\vs

Based on the GRET relation \eqref{eq:GRET-DC-4n}, we can compute its either side to obtain the same 
4-point leading-order gravitational KK scattering amplitude.\ Since the KK Goldstone numerators 
$\{\NNt_j^0\}$ [as in Eqs.\eqref{eq:Ntj-def-4n}\eqref{eq:K0stu-5}] are much simpler than 
the KK graviton numerators $\{\NN_j^0\}$ [as in Eqs.\eqref{eq:Nj-def-4n}\eqref{eq:K0stu-L}],
we can use the KK Goldstone numerators $\{\NNt_j^0\}$ to explicitly compute the leading-order
amplitudes \eqref{eq:DC-hL-phi-4n} as follows:
\begin{align}
\M_0^{\rm{dc}}[4h_L^n] & = \MT_0^{\rm{dc}}[4\phi_n^{}] 
= -\frac{\,c_4^{}\ka^2\hs}{\,16\,\hs}\hsm 
\Bigg[\!\frac{\,(\NNt_s^0)^2\,}{\sz}+\frac{\,(\NNt_t^0)^2\,}{\tz}
+\frac{\,(\NNt_u^0)^2\,}{\uz}\!\Bigg]
\nn\\
&=\frac{~c_4^{}\ka^2\sz(7\!+\!\ctt)^2\hsm\csc^2\!\theta~}{64} 
\bigg(\!\sum_{j=0}^\infty\tilde{a}_{nnj}^2\!\bigg)^{\!\!2}  
\nn\\
&=\frac{~c_4^{}\ka^2\sz(7\!+\!\ctt)^2\hsm\csc^2\!\theta~}{64} \tilde{a}^2_{nnnn} \,,
\label{eq:M5DC-LO}
\end{align}
where in the last step we have made use of the sum rule \eqref{eq:app-alT-4pt} 
for converting the trilinear Goldstone coupling ($\tilde{a}_{nnj}^2$) 
to the quartic Goldstone coupling ($\tilde{a}_{nnnn}^{}$).\ 
Then, using the explicit Feynman diagram calculations shown in 
Eqs.\eqref{eq:AmpE2-4phin}-\eqref{eq:AmpE2-4hnL},
we have the following 4-point elastic scattering amplitudes of longitudinal KK gravitons and
of KK gravitational Goldstone bosons at the leading order of high energy expansion, which are equal to
each other due to the GRET \eqref{eq:GRET-LO-4pt}:	
\begin{align}
\M_0^{}[4h_L^n] &= \MT_0^{}[4\phi_n^{}]
=\frac{~\ka^2\sz(7\!+\!\ctt)^2\hsm\csc^2\!\theta~}{2304}\!
\sum_{j=1}^\infty\! r_j^2\big(r_j^6\!-\!12\hs r_j^2\!+\!11\big) \al_{nnj}^2
\nn\\
&=\frac{~\ka^2\sz(7\!+\!\ctt)^2\!\csc^2\!\theta~}{64} \sum_{j=0}^\infty\!\tilde{\be}_{nnj}^{2}
\nn\\ 
&=\frac{~\ka^2\sz(7\!+\!\ctt)^2\!\csc^2\!\theta~}{64} \tilde{\be}_{nnnn}^{} \,,
\label{eq:AmpE2-4hL-2}
\end{align} 
where the mass ratio $\hs r_j^{}\!=\!\MGj/\MGn$, and in the last step above we have used
the sum rule \eqref{eq:app-beT-4pt} with the trilinear and quartic gravitational Goldstone couplings 
$\tilde{\be}_{nnj}^{}$ and $\tilde{\be}_{nnnn}^{}$ 
defined respectively in Eqs.\eqref{app-eq:KKGR-RS-betaT-nml} and \eqref{app-eq:KKGR-RS-couplings-4pt-bet}.\ 

\vs

With the above, we compare the double-copied scattering amplitudes of KK gravitons (KK Goldstone bosons) 
in Eq.\eqref{eq:M5DC-LO} with the corresponding KK graviton 
(KK Goldstone boson) amplitudes by explicit Feynman diagram calculations in Eq.\eqref{eq:AmpE2-4hL-2}.\ 
They have exactly the same kinematic structure except the difference 
between the two types of quartic coupling coefficients:
\begin{equation}
\label{eq:Ratio-4n-hLphi}
\frac{~\M_0^{\rm{dc}}[4h_L^n]~}{\M_0^{}[4h_L^n]\,} \hs =\hs  
\frac{~\MT_0^{\rm{dc}}[4\phi_n^{}]~}{\MT_0^{}[4\phi_n^{}]} \hs =\hs  
\frac{~c_4^{}\tilde{a}_{nnnn}^2\,}{~\tilde{\be}_{nnnn}^{}\,} \,.
\end{equation}
As a consistency check, we take the flat 5d space limit $k\ito 0$ 
and find that the above ratio of KK amplitudes approaches one as expected:
\begin{equation}
\label{eq:Ratio-4n-flat5d}
\lim_{k\to 0}\!\frac{~\M_0^{\rm{dc}}[4h_L^n]~}{\M_0^{}[4h_L^n]\,} \hs =\hs  
\lim_{k\to 0}\!\frac{~\MT_0^{\rm{dc}}[4\phi_n^{}]~}{\MT_0^{}[4\phi_n^{}]} \hs =\hs  
\lim_{k\to 0}\!\frac{~c_4^{}\tilde{a}_{nnnn}^2\,}{~\tilde{\be}_{nnnn}^{}\,} 
= \frac{~2/3\!\times\hsm\!(3\hsm /2)^2~}{3\hsm /2}  =\hsm 1 \hs,
\end{equation}
where the flat-space coupling coefficients are given by 
$\tilde{a}_{nnnn}^{}\!=\!\tilde{\be}_{nnnn}^{}\!=\!3/2\hs$, 
as summarized in {Table\,\ref{app-table:Flat-values}}.\ 

\vs

Hence, according to Eqs.\eqref{eq:Ratio-4n-hLphi}-\eqref{eq:Ratio-4n-flat5d}, we complete our
double-copy construction of the 4-point leading-order KK gauge/gravity scattering amplitudes
by further imposing the following replacements,
\begin{equation} 
\label{eq:CK2-4pt}
c_4^{}\tilde{a}_{nnnn}^2 {\,\longrightarrow\,} \tilde{\be}_{nnnn}^{} \,, 
\end{equation}
in addition to the operations in Eq.\eqref{eq:CK-4pt}.\ 
With these, we can reexpress the leading-order double-copied gravitational KK amplitudes 
\eqref{eq:M5DC-LO} as follows:
\begin{align}
\M_0^{\rm{dc}}[4h_L^n] & = \MT_0^{\rm{dc}}[4\phi_n^{}] 
= -\frac{\,\ka^2\tilde{\be}_{nnnn}^{}\,}{\,16\hs\tilde{\al}_{nnnn}^2\,}\hsm 
\Bigg[\!\frac{\,(\NNt_s^0)^2\,}{\sz}+\frac{\,(\NNt_t^0)^2\,}{\tz}
+\frac{\,(\NNt_u^0)^2\,}{\uz}\!\Bigg]
\nn\\
&=\frac{~\ka^2\sz(7\!+\!\ctt)^2\hsm\csc^2\!\theta~}{64} \tilde{\be}_{nnnn}^{}  \,,
\label{eq:M5DC-LO-Fin}
\end{align}
which agrees with Eq.\eqref{eq:AmpE2-4hL-2}.

\vspace*{1.5mm}
\subsubsection{\hspace*{-2.5mm}Construction of 4-Point Inelastic KK Graviton Amplitudes}
\label{sec:4.2.2}
\vspace*{1.5mm}

In this subsection, we study double-copy construction of the leading-order gravitational KK scattering amplitudes
for inelastic processes. This includes the inelastic channels $(n,n)\ito(m,m)$ (with $n\!\neq\! m$), 
$(0,0)\ito(n,n)$, and $(n,m)\ito(\ell,q)$ (with unequal KK indices of $n,m,\ell,q$).\

\vspace*{1mm}
\subsubsection*{$\blacklozenge$\,Inelastic Scattering Process \boldmath{$(n,n)\ito(m,m)\hs$}:}
\vspace*{1mm}

We first consider the inelastic scattering process $(n,n)\ito(m,m)$ with $\hs n\!\neq\! m\hs$ and 
$n,m\!\in\!\mathbb{Z}^+$.\ 
Making high energy expansion, we can reexpress the leading-order inelastic KK scattering amplitudes 
\eqref{eq:T0nnmm-AL}-\eqref{eq:T0nnmm-A5} as follows:
\beqs
\label{Amp-ALA5-nnmm}
\begin{align}
\label{Amp-AL-nnmm}
\TT_0^{}[2A^n_L2A_L^m]
&\,=\, g^2 \! \(\! \frac{\,\CC_s \NN_s^{\rm{in}\hs0}\,}{\sz}
+ \frac{\,\CC_t \NN_t^{\rm{in}\hs0}\,}{\tz} + \frac{\,\CC_u \NN_u^{\rm{in}\hs0}\,}{\uz}  \!\)  \!,
\\[1mm]
\label{Amp-A5-nnmm}
\tT_0^{}[2A^n_5 2A_5^m]
&\,=\, g^2 \! \(\! \frac{\,\CC_s \NNt_s^{\rm{in}\hs0}\,}{\sz} + \frac{\,\CC_t \NNt_t^{\rm{in}\hs0}\,}{\tz}
+ \frac{\,\CC_u \NNt_u^{\rm{in}\hs0}\,}{\uz}  \!\)  \!,
\end{align}
\eeqs
where the leading-order kinematic numerators 
$\big(\NN_s^{\rm{in}\hs0},\, \NN_t^{\rm{in}\hs0},\, \NN_u^{\rm{in}\hs0}\big)$
and $\big(\NNt_s^{\rm{in}\hs0},\, \NNt_t^{\rm{in}\hs0},\, \NNt_u^{\rm{in}\hs0}\big)$  are  given by
\beqs
\label{eq:Nj-Ntj-def-nnmm}
\begin{align}
\label{eq:Nj-def-nnmm}
\big(\NN_s^{\rm{in}\hs0},\, \NN_t^{\rm{in}\hs0},\, \NN_u^{\rm{in}\hs0}\big) &=
\big(\sz\hs\KK_s^{\rm{in}\hs0},\, \tz\hs\KK_t^{\rm{in}\hs0},\, \uz\hs\KK_u^{\rm{in}\hs0}\big),
\\
\label{eq:Ntj-def-nnmm}
\big(\NNt_s^{\rm{in}\hs0},\, \NNt_t^{\rm{in}\hs0},\, \NNt_u^{\rm{in}\hs0}\big) &=
\big(\sz\hs\KKt_s^{\rm{in}\hs0},\,\tz\hs\KKt_t^{\rm{in}\hs0},\,\uz\hs\KKt_u^{\rm{in}\hs0}\big).
\end{align}
\eeqs
Using \eqrefe{eq:Nj-Ntj-def-nnmm} and the difference relation
\eqref{eq:K-Kt-nnmm}, we derive the generalized gauge transformations between
the kinematic numerators of inelastic scattering amplitudes, analogous to \eqrefe{eq:K-Kt-nnnn}:
\\[-7mm]
\beqs
\begin{align}
\label{eq:GGT-NjNtj-nnmm}
\NN^{\rm{in}\hs0}_j &= \NNt_j^{\rm{in}\hs0} + s_{0j}^{}\Theta_{\rm{in}}^{} \,, 
\\
\Theta_{\rm{in}}^{} &= -\frac{\ct}{\,2\hs r^2\,}\!
\sum_{i=0}^{\infty}\!\Big[\hsm (3\!+\!r^2)\hs r_i^2 \hsm +\hsm 2(1\!-\!r^4) \Big] 
a_{nni}^{}\hs a_{mmi}^{}\hs ,
\end{align}
\eeqs
where the mass ratios $\hs r \!=\! \Mm/\Mn$ and $r_i^{}\!=\!M_i^{}/\Mn$.\ 
Then, we sum up the leading-order kinematic numerators \eqref{eq:Nj-def-nnmm} and \eqref{eq:Ntj-def-nnmm}
respectively and deduce the following: 
\beqs
\label{eq:Jacobi-4pt-nnmm}
\begin{align}
\label{eq:KJacobi-N0-nnmm}
\sum_{j} \NN_j^{\rm{in}\hs0} &\hs =\hs \sum_{j} \NNt_j^{\rm{in}\hs0} \!+\!
(\sz\!+\!\tz\!+\!\uz)\Theta_{\rm{in}}^{} \!= 0 \,,
\\[-1mm]
\label{eq:KJacobi-N50-nnmm}
\sum_{j}\NNt_j^{\rm{in}\hs0} &\hs =\hs \sz \ct 
\sum_{i=0}^{\infty} (\tilde{a}_{nmi}^2 \!-\hsm \tilde{a}_{nni}^{}\tilde{a}_{mmi}^{}) =0 \,,
\end{align}
\eeqs
where the index $\hs j\!\in\!\{s,t,u\}$, and we have used the kinematic identity $\sz\!+\!\tz\!+\!\uz\hsm =\hsm 0\hs$
and the sum rule condition \eqref{eq:app-alT-4pt}.\ 
Eqs.\eqref{eq:KJacobi-N0-nnmm} and \eqref{eq:KJacobi-N50-nnmm}
demonstrate that the two sets of numerators $\{\NN_j^{}\}$ and $\{\NNt_j^{}\}$ satisfy 
the kinematic Jacobi identities respectively.\

\vs

Thus, with the massive CK duality based on the kinematic Jacobi identities 
\eqref{eq:KJacobi-N0-nnmm}-\eqref{eq:KJacobi-N50-nnmm},
we formulate the following double-copy correspondence for
the inelastic scattering amplitudes of the longitudinal KK gauge bosons and 
of the KK Goldstone bosons in \eqrefe{Amp-ALA5-nnmm}: 
\begin{equation}
\label{eq:CK-4pt-nnmm}
\CC_j^{} {\,\longrightarrow\,} \NN_j^{\rm{in}\hs0}\hs, \hspace*{7mm}
\CC_j^{} {\,\longrightarrow\,} \NNt_j^{\rm{in}\hs0}\hs, \hspace*{7mm}
g^2 {\,\longrightarrow\,} -(\ka /4)^2 \hs. 
\end{equation}
Applying the double-copy correspondence \eqref{eq:CK-4pt-nnmm} 
to the inelastic KK scattering amplitudes 
\eqref{Amp-AL-nnmm}-\eqref{Amp-A5-nnmm} of the longitudinal KK gauge bosons 
and of the KK Goldstone bosons at the leading order of high energy expansion, 
we derive the following:
\beqs 
\label{eq:DC-hL-phi-nnmm}
\begin{align}
\label{eq:DC-hL-nnmm}
&\M_0^{\rm{dc}}[2h_L^n2h_L^m] = -\frac{\,c_4^{\rm{in}}\ka^2\,}{\,16\,}\hsm 
\Bigg[\frac{\,(\NN_s^{\rm{in}\hs0})^2\,}{\sz}+\frac{\,(\NN_t^{\rm{in}\hs0})^2\,}{\tz}
+\frac{\,(\NN_u^{\rm{in}\hs0})^2\,}{\uz}\Bigg],
\\
\label{eq:DC-phi-nnmm}
&\MT_0^{\rm{dc}}[2\phin2\phim] = -\frac{\,c_4^{\rm{in}}\ka^2\,}{\,16\,}\hsm 
\Bigg[\frac{\,(\NNt_s^{\rm{in}\hs0})^2\,}{\sz}+\frac{\,(\NNt_t^{\rm{in}\hs0})^2\,}{\tz}
+\frac{\,(\NNt_u^{\rm{in}\hs0})^2\,}{\uz}\Bigg],
\end{align}
\eeqs 
where the normalization factor $c_4^{\rm{in}}$ can be determined 
by taking the flat 5d limit $k\ito 0\hs$.\  
For the flat 5d compactification with orbifold $S^1\!/\ZZ\hs$, this factor is given by 
$\hs c_4^{\rm{in}} \!=\! 1\hs$ \cite{Li:2021yfk}\cite{Li:2022rel}.

\vs

Since the kinematic numerators of KK Goldstone bosons $\{\NNt_j^{\rm{in}\hs0}\}$ are much simpler than that of the longitudinal KK gauge bosons $\{\NN_j^{\rm{in}\hs0}\}$, with the GRET relation \eqref{eq:GRET-DC-4n} we can use the kinematic numerators of KK Goldstone bosons to explicitly compute the leading-order KK scattering amplitudes:
\begin{align}
\label{eq:M5DC-LO-nnmm}
&\M_0^{\rm{dc}}[2h^n_L 2h^m_L] = \MT_0^{\rm{dc}}[2\phin2\phim] 
= -\frac{\,c_4^{\rm{in}}\ka^2\hs}{\,16\,\hs}\hsm 
\Bigg[\!\frac{\,(\NNt_s^{\rm{in}\hs0})^2\,}{\sz}+\frac{\,(\NNt_t^{\rm{in}\hs0})^2\,}{\tz}
+\frac{\,(\NNt_u^{\rm{in}\hs0})^2\,}{\uz}\!\Bigg]
\nn\\
&= -\frac{~c_4^{\rm{in}}\ka^2\sz~}{\,32\,}\!
\[ 2c_{\theta}^2  \hsm -\! \frac{\,(3\!-\!\ct)^2 \,}{\,1\!+\!\ct\,} 
\!-\! \frac{\,(3\!+\!\ct)^2 \,}{\,1\!-\!\ct\,}\!\] \!\tilde{a}_{nnmm}^2
\nn\\
&=\frac{~c_4^{\rm{in}}\ka^2(7\!+\!\ctt)^2\sz~}{64 \sin^2\!\theta}\hs\tilde{a}_{nnmm}^2\,,
\end{align}
where in the second line we have utilized the sum rule \eqref{eq:app-alT-4pt} 
to convert the trilinear coupling into the quartic coupling.\ 
In the above, the normalization factor $c_4^{\rm{in}}$ 
may be determined by taking the flat 5d space limit ($k\ito 0$)
which gives $c_4^{\rm{in}}\!=\!1$ \cite{Li:2021yfk}\cite{Li:2022rel}.\ 
Moreover, using the explicit Feynman diagram approach, we derive the 4-point 
inelastic scattering amplitudes of longitudinal KK gravitons and
of KK gravitational Goldstone bosons at the leading order of high energy expansion.\ 
They are equal to each other due to the GRET \eqref{eq:GRET-nnmm} and we present them as follows: 
\begin{align}
\label{eq:AmpE2-4hL-2-nnmm}
\M_0^{}[2h^n_L 2h^m_L] = \MT_0^{}[2\phin2\phim] 
=\frac{~\ka^2(7\!+\!\ctt)^2\sz\,}{\,64\sin^2\!\theta\,}\tilde{\be}_{nnmm}^{} \,,
\end{align} 
where the sum rule \eqref{eq:app-beT-4pt} is used in the last step.

\vs

With these, we compare the double-copied amplitudes of KK gravitons (KK Goldstone bosons)
in Eq.\eqref{eq:M5DC-LO-nnmm} with the corresponding KK graviton (KK Goldstone bosons) amplitudes 
by explicit Feynman diagram calculations in Eq.\eqref{eq:AmpE2-4hL-2-nnmm}.\ 
We find that they have exactly the same kinematic structure except the difference 
between the two types of quartic coupling coefficients:
\begin{equation}
\label{eq:Ratio-nnmm-hLphi}
\frac{~\M_0^{\rm{dc}}[2h_L^n2h_L^m]~}{\M_0^{}[2h_L^n2h_L^m]\,} \hs =\hs  
\frac{~\MT_0^{\rm{dc}}[2\phin2\phim]~}{\MT_0^{}[2\phin2\phim]} \hs =\hs  
\frac{~c_4^{\rm{in}}\tilde{a}_{nnmm}^2\,}{~\tilde{\be}_{nnmm}^{}\,} \,.
\end{equation}
Taking the flat 5d space limit ($k\ito 0\hs$), 
the above ratio of KK amplitudes approaches one as expected:
\begin{equation}
\label{eq:Ratio-nnmm-flat5d}
\lim_{k\to 0}\!\frac{~\M_0^{\rm{dc}}[2h_L^n2h_L^m]~}{\M_0^{}[2h_L^n2h_L^m]\,} \hs =\hs  
\lim_{k\to 0}\!\frac{~\MT_0^{\rm{dc}}[2\phin2\phim]~}{\MT_0^{}[2\phin2\phim]} \hs =\hs  
\lim_{k\to 0}\!\frac{~c_4^{\rm{in}}\tilde{a}_{nnmm}^2\,}{~\tilde{\be}_{nnmm}^{}\,} 
= \frac{~1\!\times\! 1^2\,}{1}= 1 \hs,
\end{equation}
where values of the flat 5d coupling coefficients 
$\tilde{a}_{nnmm}^{}$ and $\tilde{\be}_{nnmm}^{}$ are given in {Table\,\ref{app-table:Flat-values}}
of Appendix\,\ref{app:C}.\ 

\vs
Accordingly, based on Eqs.\eqref{eq:Ratio-nnmm-hLphi}-\eqref{eq:Ratio-nnmm-flat5d}, 
we impose the following correspondence for the 
double-copy construction of leading-order inelastic KK gauge/gravity amplitudes,
\begin{equation}
\label{eq:CKin-4pt}
c_4^{\rm{in}}\tilde{a}_{nnmm}^2 {\,\longrightarrow\hs\,} \tilde{\be}_{nnmm}^{} \,,
\end{equation}
alongside the operations in Eq.\eqref{eq:CK-4pt-nnmm}.\ 
With these, we can reexpress the double-copied leading-order gravitational KK amplitudes 
\eqref{eq:M5DC-LO-nnmm} as follows:
\begin{align}
\M_0^{\rm{dc}}[2h_L^n2h_L^m] & = \MT_0^{\rm{dc}}[2\phin2\phim] 
= -\frac{\,\ka^2\tilde{\be}_{nnmm}^{}\,}{\,16\hs\tilde{a}_{nnmm}^2\,}\hsm 
\Bigg[\!\frac{\,(\NNt_s^{\rm{in}\hs0})^2\,}{\sz}+\frac{\,(\NNt_t^{\rm{in}\hs0})^2\,}{\tz}
+\frac{\,(\NNt_u^{\rm{in}\hs0})^2\,}{\uz}\!\Bigg]
\nn\\
&=\frac{~\ka^2(7\!+\!\ctt)^2\sz~}{64\sin^2\!\theta} \tilde{\be}_{nnmm}^{}  \,,
\label{eq:M5DC-LO-Fin-nnmm}
\end{align}
which agrees with Eq.\eqref{eq:AmpE2-4hL-2-nnmm}.

\vspace*{1mm}
\subsubsection*{$\blacklozenge$\,Inelastic Scattering Process \boldmath{$(n,m)\ito(\ell,q)\hs$}:}
\vspace*{1mm}

Consider the general inelastic scattering of the longitudinal KK gauge bosons $A^n_LA^m_L\ito A^\ell_LA^q_L$ 
and of the corresponding KK Goldstone bosons $A^n_5A^m_5\ito A^\ell_5A^q_5\hs$.\ 
Under high energy expansion, we can reexpress the leading-order inelastic KK scattering amplitudes 
\eqref{eq:T0nmlq-AL} and \eqref{eq:T0nmlq-A5} as follows:
\beqs
\label{Amp-ALA5-nmlq}
\begin{align}
\label{Amp-AL-nmlq}
\TT_0^{}[A^n_LA_L^mA_L^{\ell}A_L^q]
&\,=\, g^2 \! \(\! \frac{\,\CC_s^{} \NN_s^{\rm{in}\hs0}\,}{\sz}
+ \frac{\,\CC_t^{} \NN_t^{\rm{in}\hs0}\,}{\tz} + \frac{\,\CC_u^{} \NN_u^{\rm{in}\hs0}\,}{\uz}  \!\)  \!,
\\[1mm]
\label{Amp-A5-nmlq}
\tT_0^{}[A^n_LA_L^mA_L^{\ell}A_L^q]
&\,=\, g^2 \! \(\! \frac{\,\CC_s^{} \NNt_s^{\rm{in}\hs0}\,}{\sz} + \frac{\,\CC_t^{} \NNt_t^{\rm{in}\hs0}\,}{\tz}
+ \frac{\,\CC_u^{} \NNt_u^{\rm{in}\hs0}\,}{\uz}  \!\)  \!,
\end{align}
\eeqs
where the kinematic numerators 
$\big(\NN_s^{\rm{in}\hs0},\, \NN_t^{\rm{in}\hs0},\, \NN_u^{\rm{in}\hs0}\big)$
and $\big(\NNt_s^{\rm{in}\hs0},\, \NNt_t^{\rm{in}\hs0},\, \NNt_u^{\rm{in}\hs0}\big)$  are  given by
\beqs
\label{eq:Nj-Ntj-def-nmlq}
\begin{align}
\label{eq:Nj-def-nmlq}
\big(\NN_s^{\rm{in}\hs0}\hsm,\, \NN_t^{\rm{in}\hs0}\hsm,\, \NN_u^{\rm{in}\hs0}\big) &=
\big(\sz\hs\KK_s^{\rm{in}\hs0}\hsm,\, \tz\hs\KK_t^{\rm{in}\hs0}\hsm,\, \uz\hs\KK_u^{\rm{in}\hs0}\big),
\\[1mm]
\label{eq:Ntj-def-nmlq}
\big(\NNt_s^{\rm{in}\hs0}\hsm,\, \NNt_t^{\rm{in}\hs0}\hsm,\, \NNt_u^{\rm{in}\hs0}\big) &=
\big(\sz\hs\KKt_s^{\rm{in}\hs0}\hsm,\,\tz\hs\KKt_t^{\rm{in}\hs0}\hsm,\,\uz\hs\KKt_u^{\rm{in}\hs0}\big).
\end{align}
\eeqs
Then, we derive the generalized gauge transformations which connect the numerators of KK gauge bosons 
to that of the KK Goldstone bosons:
\beqs
\begin{align}
\NN_j^{\rm{in} \hs 0} &= \NNt_j^{\rm{in}\hs0} + s_{0j}^{} \Theta_{\rm{in}} \,,
\\[1mm]
\Theta_{\rm{in}} &= \frac{~M_n^2\!+\!M_m^2\!+\!M_{\ell}^2\!+\!M_q^2~}{4 \Mn \Mm \Ml \Mq}\!
\sum_{i=1}^{\infty}\! M_i^2\hsm\Big[\hsm (1\!+\!\ct) a_{nqi}^{}a_{m \ell i}^{} 
\!-\! (1\!-\!\ct) a_{n\ell i}^{}a_{mqi}^{} \Big].
\end{align}
\eeqs
With these, we compute the sum of each set of the leading-order kinematic numerators
$\{\NN_j^{\rm{in}\hs0}\}$ and $\{\NNt_j^{\rm{in}\hs0}\}$ as follows:
\beqs 
\label{eq:KJacobi-nmlq}
\begin{align}
\sum_j\NNt_j^{\rm{in}\hs0} 
& = -\frac{\,\sz\,}{2}\!\sum_{j=0}^{\infty}\! 
\[ 2\ct \tilde{a}_{nmj}^{} \tilde{a}_{\ell q j}^{} \!+\! (3\!-\!\ct) \tilde{a}_{nqj}^{}\tilde{a}_{m\ell j}^{} 
\!-\!(3\!+\!\ct) \tilde{a}_{n\ell j}^{} \tilde{a}_{mqj}^{} \]
\nn\\[0mm]
&=-\frac{\sz}{2}\[ 2\ct+(3-\ct)\!-\!(3\!+\!\ct)\]\tilde{a}_{nm\ell q}^{} = 0 \,,
\\[1.5mm]
\sum_{j} \NN_j^{\rm{in}\hs0}  &=\sum_{j} \NNt_j^{\rm{in}\hs0} \!+\!
(\sz\!+\!\tz\!+\!\uz)\Theta_{\rm{in}}^{} \!= 0 \,.
\end{align}
\eeqs 
This proves the kinematic Jacobi identities for the leading-order inelastic KK scattering amplitudes:
\begin{equation}
\label{eq:KJacobi1-nmlq}
\NN_s^{\rm{in}\hs0} \!+\hsm \NN_t^{\rm{in}\hs0} \!+\hsm \NN_u^{\rm{in}\hs0} = 0 \hs,~~~~~
\NNt_s^{\rm{in}\hs0} \!+\hsm \NNt_t^{\rm{in}\hs0} \!+\hsm \NNt_u^{\rm{in}\hs0} = 0 \hs.
\end{equation} 

Next, using the extended massive double-copy we construct the inelastic scattering amplitudes 
of KK gravitons and of gravitational KK Goldstone bosons at the leading order of 
high energy expansion, which also obey the GRET:
\begin{align}
\label{eq:DC-LO-4h-nmlq}
& \M_0^{\rm{dc}}[h^n_Lh^m_Lh^\ell_Lh^q_L] = \MT_0^{\rm{dc}}[\phin\phim\phil\phi_q^{}] 
= -\frac{\,c_4^{\rm{in}}\ka^2\,}{\,16\,}\hsm
\Bigg[\!\frac{\,(\NNt_s^{\rm{in}\hs0})^2\,}{\sz}\!+\!\frac{\,(\NNt_t^{\rm{in}\hs0})^2\,}{\tz}
\!+\!\frac{\,(\NNt_u^{\rm{in}\hs0})^2\,}{\uz}\!\Bigg]
\nn\\ 
&= -\frac{~c_4^{\rm{in}}\ka^2\sz~}{\,32\,}\hsm
\[ 2c_{\theta}^2  \hsm -\! \frac{\,(3\!-\!\ct)^2 \,}{\,1\!+\!\ct\,} 
\!-\! \frac{\,(3\!+\!\ct)^2 \,}{\,1\!-\!\ct\,}\!\]\!\tilde{a}_{nm\ell q}^{2}
\nn\\
&=\frac{~c_4^{\rm{in}}\ka^2(7\!+\!\ctt)^2\sz~}{64\sin^2\!\theta}\tilde{a}_{nm\ell q}^{2} 
= \frac{~\ka^2(7\!+\!\ctt)^2\sz~}{64\sin^2\!\theta}\tilde{\be}_{nm\ell q}^{}
\,,
\end{align}
where in the second equality the completeness relation \eqref{app-eq:at-completeness} is applied 
and in the last equality we have used the conversion relation \eqref{eq:CKnmlq-4pt}
which is established at the end of this paragraph.\ 
In the above, the normalization factor $c_4^{\rm{in}}$ 
may be determined by taking the flat 5d limit ($k\ito 0$)
which gives $c_4^{\rm{in}}\!=\!2\hs$ \cite{Li:2021yfk}\cite{Li:2022rel}.\  
On the other hand, using the explicit Feynman diagram approach, we compute the inelastic scattering amplitudes 
of gravitational KK Goldstone bosons (KK gravitons) at the leading order of high energy expansion:
\begin{align}
\hspace*{-2mm}
&\M_0^{}[h^n_Lh^m_Lh^\ell_Lh^q_L] = \MT_0^{}[\phin\phim\phil\phi_q^{}]
\nn\\
\hspace*{-2mm}
&= - \frac{\,\ka^2\st}{32} \sum_{j=0}^{\infty}\!\[\!
(15\!+\!\ctt)\tilde{\be}_{nmj}^{}\tilde{\be}_{\ell qj}^{}
\!-\!\frac{\,4(5\!+\!2\ct\!+\!\ctt)\,}{1\!+\hsm\ct}\tilde{\be}_{nq j}^{}\tilde{\be}_{m\ell j}^{}
\!-\!\frac{\,4(8\!+\!7\ct\!+\!\cttt)\,}{1\!-\hsm\ctt}\tilde{\be}_{n\ell j}^{}\tilde{\be}_{mqj}^{}\hsm\]
\nn\\
\hspace*{-2mm}
&=\frac{~\ka^2(7\!+\!\ctt)^2\sz~}{64\sin^2\!\theta}\tilde{\be}_{nm\ell q}^{}  \,.
\label{eq:Fey-LO-4h-nmlq}
\end{align}
%
As a consistency check, we take the limit of flat 5d space ($k\ito 0$) 
and find that the ratio of KK scattering amplitudes approaches one:
\begin{equation}
\label{eq:c4-alT-beT-nmlq}
\lim_{k\to 0}\!\frac{~c_4^{\rm{in}}\tilde{a}_{nm\ell q}^2\,}{~\tilde{\be}_{nm\ell q}^{}\,}
= \frac{~2\!\times\!(1\hsm /2)^2~}{1\hsm /2}  = 1 \hs,
\end{equation}
where the values of couplings $\tilde{a}_{nm\ell q}^{}$ and $\tilde{\be}_{nm\ell q}^{}$ 
in the flat space limit are summarized in 
{Table\,\ref{app-table:Flat-values}}.
Hence, we set up the following double-copy correspondence for 
the above leading-order inelastic KK gauge/gravity scattering amplitudes,
\begin{equation}
\label{eq:CKnmlq-4pt} 
c_4^{\rm{in}}\hs \tilde{a}_{nm\ell q}^2 {\,\hs\longrightarrow~} \tilde{\be}_{nm\ell q}^{} \,,
\end{equation}
together with Eq.\eqref{eq:CK-4pt-nnmm}.\

\vspace*{1mm}
\subsubsection*{$\blacklozenge$\,Inelastic Scattering Process \boldmath{$(0,0)\ito(n,n)\hs$}:}
\vspace*{1mm}

Finally, we analyze the mixed 4-point inelastic scattering amplitudes of $\hs (0,\hs 0)\hsm\ito (n,\hs n)\hs$.\ 
According to \eqrefe{eq:K-Kt-00nn}, we verify that the kinematic Jacobi identities for the numerators 
$\{\NN_i^{\rm{in}\hs0}\}$ and $\{\NNt_i^{\rm{in}\hs0}\}$ hold.\ 
Applying the massive double-copy correspondence \eqref{eq:CK-4pt-nnmm}, we construct the  
double-copied gravitational KK scattering amplitudes at the leading order:
\begin{equation}
\label{eq:h00nn-DC}
\M_0^{\rm{dc}}[2h^0_{\pm2} 2h^n_L] =\MT_0^{\rm{dc}}[2h^0_{\pm2} 2\phin] 
= \frac{\,c_4^{\rm{in}}\ka^2}{32}\sz (1\!-\!\ctt) \hs a_{0nn}^4 \hs,
\end{equation}
where $a_{0nn}^{}\!\!=\! \ff_0^{}$ is given by \eqrefe{eq:ann0=f0}.\  
On the other hand, using the explicit Feynman diagram approach and the GRET identity \eqref{eq:GRET-LO-4pt}, 
we compute the 4-point inelastic gravitational KK scattering amplitudes at the leading order of high energy expansion:	 
\begin{equation}
\label{eq:eq:AmpE2-4hL-2-00nn}
\M_0^{}[2h^0_{\pm2} 2h^n_L] = \MT_0^{}[2h^0_{\pm2} 2\phin] 
=\frac{\,\ka^2}{\,32\,}\sz (1\!-\!\ctt)\hs \al_{0nn}^2 \,,
\end{equation}
where $\hs \al_{0nn}^{}\!\!=\! \uu_0^{}$ is given by \eqrefe{eq:alpha-nn0=u0}.\ 
We see that the two types of KK scattering amplitudes \eqref{eq:h00nn-DC} and \eqref{eq:eq:AmpE2-4hL-2-00nn}
exhibit the same kinematic structure, and thus the mixed double-copy construction \eqref{eq:h00nn-DC} is
successful.\  As a consistency check, we take the flat 5d limit ($k\ito 0$) and find that  
the ratio of the two KK amplitudes approaches unit as expected:
\begin{equation}
\label{eq:Ratio-00nn-flat5d}
\lim_{k\to 0}\!\frac{~\M_0^{\rm{dc}}[2h^0_{\pm2} 2h^n_L]~}{\M_0^{}[2h^0_{\pm2} 2h^n_L]\,} \hs =\hs  
\lim_{k\to 0}\!\frac{~\MT_0^{\rm{dc}}[2h^0_{\pm2} 2\phin]~}{\MT_0^{}[2h^0_{\pm2} 2\phin]} \hs =\hs  
\lim_{k\to 0}\!\frac{~c_4^{\rm{in}}\hs a_{0nn}^4\,}{~\al_{0nn}^{2}\,} 
=\frac{1 \times 1^4}{1^2}= 1 \hs,
\end{equation}
where the values of the coupling coefficients $a_{0nn}^{}$ and $\al_{0nn}^{}$ in the flat 5d limit 
are summarized in Table\,\ref{app-table:Flat-values}.\
Hence, for the mixed inelastic KK scattering process of $\hs (0,\hs 0)\hsm\ito (n,\hs n)\hs$, 
we set up the following double-copy correspondence for the leading-order inelastic KK gauge/gravity 
scattering amplitudes,
\begin{equation}
\label{eq:CK00nn-4pt} 
c_4^{\rm{in}}\hs a_{0nn}^4 {\,\hs\longrightarrow~} \al_{0nn}^{2} \,,
\end{equation}
together with Eq.\eqref{eq:CK-4pt-nnmm}.\

\vs

For a further demonstration, we study the double-copy constructions of the gravitational 
KK scattering amplitudes for another type of mixed processes 
$h^{+2}_0 \hs h^{-2}_0 \!\ito\! h^{+1}_n \hs h^{-1}_n$ and 
$h^{+2}_0 \hs h^{-2}_0 \!\ito\! \VV^{+1}_n \hs \VV^{-1}_n$.\
For this purpose, we compute the following 4-point mixed amplitudes 
for the scattering of two zero-mode gauge bosons into two final-state KK gauge bosons 
(with helicities $\pm 1$ or $0$) or into two final-state KK Goldstone bosons at the leading order 
of high energy expansion:
\beqs
\label{eq:AmpLO-4A}
\begin{align}
\label{eq:Amp0-A+A-A+A-}
\TT_0^{}[A^{a0}_{+} A^{b0}_{-} \!\ito\! A^{cn}_+ A^{dn}_-] 
&= g^2 \! \(\! \frac{~{\CC_s^{}\hs\NN_s^{\pp\rm{in}\hs0}}\,}{\sz}\hsm +\hsm \frac{~{\CC_t^{}\hs\NN_t^{\pp\rm{in}\hs0}}~}{\tz} 
\hsm +\hsm \frac{~{\CC_u^{}\hs \NN_u^{\pp\rm{in}\hs0}}~}{\uz}  \!\)  
\nn\\
&= -g^2 \sz\! \[\! \frac{\, (1\!-\!\ct)^2\hs \CC_t^{}\,}{2\hs \tz} - \frac{\,(1\!-\!\ct)^2\hs \CC_u^{} \,}{2\hs \uz}  \!\]\!
a_{000}^{}\hs a_{nn0}^{} \,,
\\[1mm]
\label{eq:Amp0-A+A-ALAL}
\TT_0^{}[A^{a0}_{+} A^{b0}_{-} \!\ito\! A^{cn}_L A^{dn}_L] &=  g^2 \! \(\! \frac{\,\CC_s^{} \NN_s^{\rm{in}\hs0}\,}{\sz} + \frac{\,\CC_t^{} \NN_t^{\rm{in}\hs0}\,}{\tz} + \frac{\,\CC_u^{} \NN_u^{\rm{in}\hs0}\,}{\uz}  \!\) 
\nn\\
&=  g^2\sz \! \[\! -\frac{\, (1\!-\!\cct)\hs \CC_t^{} \,}{2\hs \tz} + \frac{\,(1\!-\!\cct)\hs \CC_u^{}\,}{2\hs \uz} \!\]\!
a_{000}^{}\hs a_{nn0}^{} \,,	
\\[1mm]
\label{eq:Amp0-A+A-A5A5}
\tT_0^{}[A^{a0}_{+} A^{b0}_{-} \!\ito\! A^{cn}_5 A^{dn}_5] &= g^2 \! \(\! \frac{\,\CC_s^{} \NNt_s^{\rm{in}\hs0}\,}{\sz} + \frac{\,\CC_t^{} \NNt_t^{\rm{in}\hs0}\,}{\tz} + \frac{\,\CC_u^{} \NNt_u^{\rm{in}\hs0}\,}{\uz}  \!\)  
\nn\\
&= g^2\sz \! \[\! \frac{\, (1\!-\!\cct)\hs \CC_t^{} \,}{2\hs \tz} - \frac{\,(1\!-\!\cct)\hs \CC_u^{}\,}{2\hs \uz} \!\]\! 
a_{000}^{}\hs \tilde{a}_{nn0}^{} \,,
\end{align}
\eeqs
where the coupling coefficients 
$a_{000}^{} \!=\! a_{nn0}^{} \!=\!\tilde{a}_{nn0}^{}\!=\!\ff_0^{}$
according to Eq.\eqref{eq:alpha3-a3-eg}.\  
We see that the above leading-order KK scattering amplitudes 
\eqref{eq:Amp0-A+A-ALAL} and \eqref{eq:Amp0-A+A-A5A5} obey 
the GAET \eqref{eq:KK-ET1-N} as expected.

\vs

With the mixed leading-order gauge boson scattering amplitudes \eqref{eq:AmpLO-4A}, 
we construct the double-copied graviton scattering amplitude and the corresponding 
scattering amplitude with gravitational KK Goldstone bosons:
\beqs 
\begin{align}
\label{eq:DC-h0h0-hn1hn1} 
\M^{\rm{dc}}_0 [h^{+2}_0 \hs h^{-2}_0 \!\ito\! h^{+1}_n \hs h^{-1}_n]
&= -\frac{\,c_4^{\rm{in}} \ka^2\,}{16} \! \(\!\! \frac{~\NN_s^{\rm{in}\hs0} \NN_s^{\pp\rm{in}\hs0}\,}{\sz}
\hsm +\hsm \frac{~\NN_t^{\rm{in}\hs0} \NN_t^{\pp\rm{in}\hs0}\,}{\tz}
\hsm +\hsm \frac{\,\NN_u^{\rm{in}\hs0} \NN_u^{\pp\rm{in}\hs0}\,}{\uz}  \!\) 
\nn\\
&= -\frac{~c_4^{\rm{in}} \ka^2\sz~}{16} (1\!-\!\ct)^2 a_{000}^2 \hs a_{nn0}^{2} \,, 
\\[1.mm]
\label{eq:DC-h0h0-V+V-}
\MT^{\rm{dc}}_0 [h^{+2}_0 \hs h^{-2}_0 \!\ito\! \VV^{+1}_n \hs \VV^{-1}_n] 
&= -\frac{\,c_4^{\rm{in}} \ka^2\,}{16} \! \(\!\! \frac{~\NNt_s^{\rm{in}\hs0} {\NN_s^{\pp\rm{in}\hs0}}\,}{\sz} 
\hsm +\hsm \frac{~\NNt_t^{\rm{in}\hs0} {\NN_t^{\pp\rm{in}\hs0}}\,}{\tz} 
\hsm +\hsm \frac{\,\NNt_u^{\rm{in}\hs0} {\NN_u^{\pp\rm{in}\hs0}}\,}{\uz}  \!\)  
\nn\\
&= \frac{~c_4^{\rm{in}} \ka^2\sz~}{16} (1\!-\!\ct)^2 
a_{000}^2 \hs a_{nn0}^{} \tilde{a}_{nn0}^{} \,, 
\end{align}
\eeqs
where the coupling coefficients 
$a_{000}^{} \!=\! a_{nn0}^{} \!=\!\tilde{a}_{nn0}^{}\!=\!\ff_0^{}$
according to Eq.\eqref{eq:alpha3-a3-eg}.\ 
From the above, we can verify the equality between the two scattering amplitudes 
\eqref{eq:DC-h0h0-hn1hn1} and \eqref{eq:DC-h0h0-V+V-}: 
\begin{equation}
\label{eq:GRET-DC-00nn-2211}
\M^{\rm{dc}}_0 [h^{+2}_0 \hs h^{-2}_0 \!\ito\! h^{+1}_n \hs h^{-1}_n] 
\,=\, 
-\MT^{\rm{dc}}_0 [h^{+2}_0 \hs h^{-2}_0 \!\ito\! \VV^{+1}_n \hs \VV^{-1}_n] \,.
\end{equation}
Then, we explicitly compute the above leading-order amplitudes by using the Feynman diagram approach:
\\[-7mm]
\beqs 
\begin{align}
\label{eq:Fey-h0h0-hn1hn1} 
\M_0^{} [h^{+2}_0 \hs h^{-2}_0 \!\ito\! h^{+1}_n \hs h^{-1}_n]
&= -\frac{~\ka^2\sz~}{16}  (1\!-\!\ct)^2 \al_{000}^{}\hs \al_{nn0}^{} \,,
\\	
\label{eq:Fey-h0h0-V+V-} 
\MT_0^{} [h^{+2}_0 \hs h^{-2}_0 \!\ito\! \VV^{+1}_n \hs \VV^{-1}_n] 
&= \frac{~\ka^2\sz~}{16}  (1\!-\!\ct)^2 \al_{000}^{}\hs \tilde{\al}_{nn0}^{} \,,
\end{align}
\eeqs
where the coupling coefficients are given by 
$\hs\al_{000}^{} \!=\! \al_{nn0}^{} \!=\!{\tilde{\al}_{nn0}^{}}\!=\!\uu_0^{}\hs$ 
according to Eq.\eqref{eq:alpha3-a3-eg}.\ 
This also establishes the equivalence between the leading-order scattering amplitudes 
\eqref{eq:Fey-h0h0-hn1hn1} and \eqref{eq:Fey-h0h0-V+V-}: 
\begin{equation}
\label{eq:GRET-00nn-2211}
\M_0^{} [h^{+2}_0 \hs h^{-2}_0 \!\ito\! h^{+1}_n \hs h^{-1}_n]
\,=\,-\MT_0^{} [h^{+2}_0 \hs h^{-2}_0 \!\ito\! \VV^{+1}_n \hs \VV^{-1}_n] \,,
\end{equation}
in full agreement with the double-copied GRET \eqref{eq:GRET-DC-00nn-2211}.\ 
As a consistency check, we take the flat 5d limit ($k\ito 0$) and find that  
the ratio of the two gravitational amplitudes \eqref{eq:DC-h0h0-hn1hn1} and \eqref{eq:Fey-h0h0-hn1hn1}  
[or, \eqref{eq:DC-h0h0-V+V-} and \eqref{eq:Fey-h0h0-V+V-}]  
approaches one as expected:
\begin{equation}
\label{eq:Ratio-h0h0VV-flat5d}
\lim_{k\to 0}\!\frac{~\M_0^{\rm{dc}}[2h^0_{\pm2} 2h^n_{\pm1}]~}{\M_0^{}[2h^0_{\pm2} 2h^n_{\pm1}]\,} \hs =\hs  
\lim_{k\to 0}\!\frac{~\MT_0^{\rm{dc}}[2h^0_{\pm2} 2\VV^n_{\pm1}]~}{\MT_0^{}[2h^0_{\pm2} 2\VV^n_{\pm1}]} \hs =\hs  
\lim_{k\to 0}\!\frac{~c_4^{\rm{in}}\hs a_{000}^2 \hs a_{nn0}^{} \tilde{a}_{nn0}^{}\,}{~\al_{000}^{}\hs \tilde{\al}_{nn0}^{}\,} = 1 \hs,
\end{equation}
where the values of all coupling coefficients in the flat 5d limit 
are summarized in Table\,\ref{app-table:Flat-values}.\ 
Comparing the leading-order scattering amplitudes 
\eqref{eq:DC-h0h0-hn1hn1}-\eqref{eq:DC-h0h0-V+V-} 
and \eqref{eq:Fey-h0h0-hn1hn1}-\eqref{eq:Fey-h0h0-V+V-}, 
we see that they share exactly the same kinematic structure and can be equal under the following 
correspondence:
\beqs 
\vspace*{-3.5mm}
\label{eq:CKh0h0-h1h1-VV-4pt} 
\begin{align}
\label{eq:CKh0h0-h1h1-4pt} 
c_4^{\rm{in}}\hs a_{000}^2 \hs a_{nn0}^{2}  
&{\,\,\longrightarrow~} 
\al_{000}^{}\hs {\al}_{nn0}^{} \,,
\\[0.7mm]
\label{eq:CKh0h0-VV-4pt} 
c_4^{\rm{in}}\hs a_{000}^2 \hs a_{nn0}^{} \tilde{a}_{nn0}^{} 
&{\,\,\longrightarrow~} 
\al_{000}^{}\hs \tilde{\al}_{nn0}^{} \,.
\end{align}
\eeqs 

In summary, we have established the double-copy construction from the GAET to the GRET
for the 4-point leading-order (LO) KK scattering amplitudes:
\begin{equation}
\label{eq:T4AL-M4hL-outline}
\begin{aligned}
\M_0^{(4)}[h^n_L] &~\xlongequal{~\scalebox{.9}{\text{GRET}}~}~ \MT_0^{(4)}[\phin]
\\[.5mm]
\xuparrow{.5cm}\scalebox{.9}{$\text{DC}_4$}\hspace*{-0.1cm}
&\hspace*{2.1cm} \scalebox{.9}{$\text{DC}_4$}\xuparrow{.5cm}
\\[.5mm]
\TT_0^{(4)}[A^n_L] &~\xlongequal{~\scalebox{.9}{\text{GAET}}~}~ 
\tT_0^{(4)}[A^n_5]
\end{aligned}
\end{equation}
In parallel to \eqrefe{eq:T3AL-M3hL-outline}, we discuss the meaning of 
the above Eq.\eqref{eq:T4AL-M4hL-outline}.\   Along the horizontal directions,
the lower (upper) equality presents the GAET (GRET) at the level of 4-point KK scattering amplitudes.\ 
On the other hand, along the direction of each vertical arrow, the correspondence 
is established by the double-copy (DC) under the extended massive color-kinematics dualities
and the KK gauge/gravity coupling correspondences 
for both the elastic scattering [as shown in Eqs.\eqref{eq:CK-4pt}\eqref{eq:CK2-4pt}] 
and the inelastic scattering [as shown in Eqs.\eqref{eq:CK-4pt-nnmm}\eqref{eq:CKin-4pt}\eqref{eq:CKnmlq-4pt}\eqref{eq:CK00nn-4pt}\eqref{eq:CKh0h0-h1h1-VV-4pt}].\  
The left vertical arrow represents the double-copy (DC$_4^{}$) 
construction of the 4-point LO KK graviton
amplitude $\M_0^{(4)}[h^n_L]$ from the corresponding LO KK gauge boson amplitude $\TT_0^{(4)}[A^n_L]$.\  
The right vertical arrow represents the double-copy (DC$_4^{}$) 
construction of the 4-point LO gravitational 
KK Goldstone boson amplitude $\MT_0^{(4)}[\phin]$ from the corresponding 
(gauge theory) LO KK Goldstone boson amplitude $\tT_0^{(4)}[A^n_5]$.\ 
Finally, this picture also holds if we replace the two KK scattering amplitudes
$\M_0^{(4)}[h^n_L]$ and $\MT_0^{(4)}[\phin]$ in the top row by
the other two KK scattering amplitudes 
$\M_0^{(4)}[h^n_{\pm1}]$ and $\MT_0^{(4)}[\VV_{\pm1}^{n}]$ respectively.\

\vspace*{2mm}
\subsection{\hspace*{-2.5mm}Extension of Warped LO Double-Copy with
\boldmath{$N\!\hsm \geqq\!4\hs$}}
\label{sec:4.3}
\vspace*{1.5mm}

So far, we have explicitly demonstrated that the warped double-copy can be realized 
for 3-point and 4-point KK scattering amplitudes at the leading order of high energy expansion.\  
In this subsection, we discuss the extension of the warped LO double-copy to the 
$N$-point KK gauge/gravity amplitudes with $N\!\hsm \geqq\!4\hs$.\ 
Using the GAET [cf.\,\eqrefe{eq:5D-ETI3-ALQ}] and GRETs [cf.\,Eqs.\eqref{eq:KK-GRET-hV} and \eqref{eq:KK-GRET}], we find that it is more convenient to analyze the 
LO KK Goldstone boson amplitudes.\ 

\vs

\begin{figure}
\centering
\includegraphics[width=10cm]{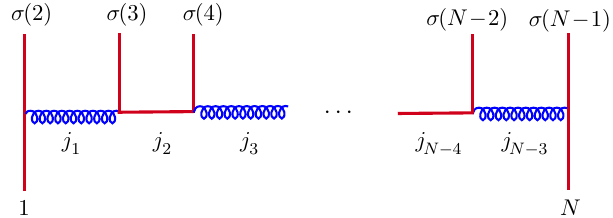}
\vspace*{-2mm}
\caption{\small
$N$-point half-ladder (multi-peripheral) diagrams for the scalar KK Goldstone boson scattering at tree level 
($N\!\in\!2\hs\mathbb{Z}\,)$ which contribute the leading energy-dependence  
in the compactified 5d gauge theory.\ 
Here the external states 1 and $N$ are fixed, and 
$\si\!=\!\{\si(2),\cdots\!,\si(N\!-\!1)\}\!\in\! S_{N-2}^{}$ represents 
a permutation of the remaining $(N\hsm\!-\!2)$ external states.\ 
}
\label{fig:2}
\end{figure}

For a given $S$-matrix element $\mathbb{S}$ with $N$ external states in (3+1)d spacetime, 
its mass-dimension is determined by the spacetime dimension $d\!=\!4$ and the number of 
external states $N$ \cite{Weinberg:1978kz}\cite{Hang:2021fmp}:
\begin{equation}
\label{eq:DS}
D_{\mathbb{S}}^{} \,=\,4 - N\hs .
\end{equation}
Then, we first consider the pure KK Goldstone amplitude having even number of external states,
with $N\hsm\!\!\geqq\! 4$ and $N\!\!\in\!\! 2\hs\mathbb{Z}\,$.\
For the KK gauge theory, the leading energy-power dependence of the $N$-point KK Goldstone boson 
$(A^{an}_5)$ amplitude $\,\tT\,$ is given in the second equation of \eqrefe{eq:E-counting-AL+A5} 
with $\bar{V}_3^{\min}\hsm\!=\!N_v^{}\hsm\!=\!0\hs$.\ 
Thus, the mass-dimension of the amplitude $\,\tT\,$ and 
its leading energy-power dependence are equal:
\begin{equation}
\label{eq:DtT=DE-NA5}
D_{\tT}(N\!A_5^{an}) = D_E^{}(N\!A_5^{an}) =\hs 4\!-\!N  \hs.
\end{equation}
Hence, \eqrefe{eq:DtT=DE-NA5} indicates that we can express the LO scattering amplitudes of 
KK Goldstone bosons as follows:
\begin{equation}
\label{eq:tT0-N}
\tT_0^{}\big[A^{a_1^{}n_1^{}}_{5}, \cdots\!, A^{a_N^{}n_N^{}}_{5}\big]
\,=\, \tilde{a}_{n_1^{}\cdots n_N^{}}^{}\hs g^{N-2}\!
\sum_{i=1}^{(2N-5)!!} \!\!\!\frac{~\CC_i^{}\, \NNt_i^{0}\,}{\D_i^{}}\,,
\end{equation}
where $\D_i^{}$ denotes the products of denominators of propagators.\  
The effective $N$-point coupling coefficients 
$\tilde{a}_{n_1^{}\cdots n_N^{}}^{}$ 
is dimensionless because the 4d gauge coupling is dimensionless, $[g]\!=\!0\hs$.\ 
Moreover, by taking the Del\,Duca-Dixon-Maltoni (DDM) color decomposition\,\cite{DDM}, 
we can express scattering amplitude 
\eqref{eq:tT0-N} in terms of Kleiss-Kuijf basis\,\cite{Kleiss:1988ne},
which corresponds to diagrams with the half-ladder (multi-peripheral) structure 
shown in Fig.\,\ref{fig:2} and contributes the leading energy-dependence.\  
Thus, we can readily write down the $N$-point KK coupling coefficient 
$\tilde{a}_{n_1^{}\cdots n_N^{}}^{}$ as follows:
\begin{align}
\tilde{a}_{n_1^{}\cdots n_N^{}}^{} 
&= \sum_{j_1^{},\ldots,j_{N-3}^{}}^{}\hspace*{-4mm} \tilde{a}_{n_1^{}n_{\si(2)}j_1^{}}^{}\tilde{a}_{j_2^{}n_{\si(3)}j_1^{}}^{}
{\tilde{a}_{j_2^{}n_{\si(4)}j_3^{}}^{}}
\!\cdots\hs\tilde{a}_{j_{N-4}^{}n_{\si(N-2)}j_{N-3}^{}}^{}\tilde{a}_{n_{\si(N-1)}n_{N}^{}j_{N-3}^{}}^{}
\nn\\[1mm]
&= \lrb{\tilde{f}_{n_1^{}}^{}\! \cdots \tilde{f}_{n_N^{}}^{}} \hs,
\end{align}
where the completeness relation \eqref{eq-app:completeness} 
has been imposed, and the brackets $\lrb{\cdots}$ are defined in 
Eq.\eqref{app-eq:def-[[]]} of Appendix\,\ref{app:D}.\  
In the above, the external states 1 and $N$ are fixed, and
$\si\!=\!\{\si(2),\cdots\!,\si(N\! -\! 1)\}\!\in\! S_{N-2}^{}$  
represents a permutation of the remaining $(N\hsm\!-\!2)$ external states.\ 
Every color-ordered amplitude shares the same KK coupling coefficient, 
because $\tilde{a}_{n_1^{}\!\cdots\hs n_N^{}}^{}\!$ is just the integration of 5d wavefunctions $\{\tilde{f}_{n_i^{}}^{}\hsm\}$ associated with each external state $\{A^{a_i^{}n_i^{}}_{5}\hsm\}$   
and thus the KK indices of $\tilde{a}_{n_1^{}\cdots n_N^{}}^{}\!$ 
are symmetric and independent of the color ordering.\
We also note that the LO KK Goldstone numerators $\{\NNt_i^{0}\}$ in Eq.\eqref{eq:tT0-N}
should still obey the kinematic Jacobi identities for $N$-point amplitude since they
are fully determined by kinematics (similar to the flat 5d case) 
and the effect of the warped 5d compactification only
affects the KK coupling coefficients $\tilde{a}_{n_1^{}\cdots n_N^{}}^{}$.\

\vs

As for the KK gravity theory, we first consider the scattering amplitudes of the gravitational 
KK scalar Goldstone bosons ($\phin$).\ 
Based on the extended CK duality and using \eqrefe{eq:tT0-N}, we can construct  
the $N$-point LO amplitudes of gravitational KK scalar Goldstone bosons  
via double-copy (DC) as follows: 
\begin{equation}
\label{eq:tM0-N}
\MT_0^{\rm{dc}}\big[\phi_{n_1^{}}^{}\!\cdots\phi_{n_N^{}}^{}\big] 
= -\,c_N^{}\tilde{a}_{n_1^{}\!\cdots n_N^{}}^{2}
\!\!\(\frac{\ka}{4}\)^{\!\!N-2} 
\sum_{i=1}^{(2N-5)!!} \!\!\frac{~(\NNt_i^0)^{2}\,}{\D_i^{}}\hs.
\end{equation}
Then, we can derive the leading energy-power dependence of the KK gravitational Goldstone amplitude 
\eqref{eq:tM0-N}:
\begin{equation}
\label{eq:E-counting-hn55}
D_E^{}(N\!A_5^{an}) ~\xrightarrow{~\rm{DC}~}~ D_E^{}(N\!\phin)  
= 2(4\!-\!N)+2(N\!-\!3) = 2\,.
\end{equation}
This confirms our power counting rule \eqrefe{eq:E-counting-hnL-or-hn55} at tree level 
($L\!=\!0$).\ 
Moreover, based on the double-copy correspondence, we make the following replacement:
\begin{equation}
c_N^{}\tilde{a}_{n_1^{}\cdots n_N^{}}^2 \,\longrightarrow\,~ 
\tilde{\be}_{n_1^{}\cdots n_N^{}}^{} .
\end{equation}	
Thus, the $N$-point KK scalar gravitational Goldstone scattering amplitude can be expressed as follows:
\begin{equation}
\label{eq:tM0-N-2}
\MT_0^{}\big[\phi_{n_1^{}}^{}\ldots\phi_{n_N^{}}^{}\big] 
= -\tilde{\be}_{n_1^{}\cdots n_N^{}}^{}\!\hsm 
\(\frac{\ka}{4}\)^{\!\!N-2} \sum_{\ell=1}^{(2N-5)!!}\!\! \frac{\,(\NNt_\ell^0)^{2}\,}{\D_\ell^{}}\,,
\end{equation}
where the KK Goldstone coupling coefficient $\tilde{\be}_{n_1^{}\cdots n_N^{}}^{}\!$ 
is dimensionless.\ To understand the dimensionlessness of 
$\tilde{\be}_{n_1^{}\cdots n_N^{}}^{}\hsm$, we compute the difference between 
the mass-dimension of the KK Goldstone amplitude $\MT_0^{}(N\!\phin)$ and its leading energy-dependence 
via Eqs.\eqref{eq:DS} and \eqref{eq:E-counting-hn55}, which yields:
\begin{equation}
\label{eq:deltaD-MT-2Z}
\Delta D(\MT_0^{}) = D_{\MT}(N\!\phin)\hsm -\hsm D_E^{}(N\!\phin) =  2\hsm -\! N = [\ka^{N-2}] \hs.
\end{equation}
Eq.\eqref{eq:deltaD-MT-2Z} shows that the residual mass-dimension of the LO amplitude $\MT_0^{}$ 
fully comes from the gravitational coupling constant $\ka$ 
which has mass-dimension $[\ka]\!=\!-1$.\ 
The LO KK gravitational Goldstone amplitude \eqref{eq:tM0-N-2} is mass-independent
(resembling a ``massless'' amplitude) and is contributed by the trilinear vertices of 
$\phi_{n_1^{}}^{}\!$-$\hs\phi^{}_{n_2^{}}\!$-$\hs h^{\mn}_{n_3^{}}$ 
type (containing two partial derivatives).\ 
Hence, we deduce that the KK coupling coefficient $\tilde{\be}_{n_1^{}\cdots n_N^{}}^{}\!$ 
is dimensionless and takes the following form:
\begin{equation}
\tilde{\be}_{n_1^{}\cdots n_N^{}}^{} = 
\sum_{j_1^{},\cdots,j_{N-3}^{}=0}^{\infty}\!\!\!\! \tilde{\be}_{n_1^{}n_{2}^{}j_1^{}}^{}\cdots\tilde{\be}_{n_{N-1}^{}n_{N}^{}j_{N-3}^{}}^{}\!
=\lrb{\ww_{n_1^{}}^{} \cdots \ww_{n_N^{}}^{}} .
\end{equation}
This shows that the coupling coefficient $\tilde{\be}_{n_1^{}\cdots n_N^{}}^{}$ only depends on the 
5d wavefunction $\{\ww_{n_i^{}}^{}\}$ associated with the external states $\{\phi_{n_i^{}}^{}\}\hs$.

\vs

Then, we consider the case for odd number of external states 
with $N\!\hsm\geqq\!4\hs$ and $N\!\in\! 2\hs\mathbb{Z} +\hsm 1$, 
the mass-dimension of $N$-point KK Goldstone amplitude in the KK gauge theory 
and its leading energy-power dependence are given by
\begin{equation}
D_{\tT}^{}(N\!A_5^n) = 4\!-\!N , \qquad 
D_E^{}(N\!A_5^n)  = 3\!-\!N .
\end{equation}
Thus, the difference $D_{\tT}^{}\hsm -\hsm D_E^{}\!=\!1\hs$ 
for the odd number $N$.\  
Actually, this difference is due to the fact that the $N$-point KK Goldstone amplitude (with odd $N$)
must contain a non-derivative cubic vertex $A_5^{an_1^{}}$-$A_\mu^{bn_2^{}}$-$A_\nu^{cn_3^{}}$ 
in Eq.\eqref{eq-app:A-A-A5}.\ 
Hence, its KK coupling coefficient has mass-dimension  
$\,[\tilde{a}_{n_1^{}\cdots n_N^{}}^{}] \!=\! [(M_{n_2}^2\!\!-\!M_{n_3}^2)/M_{n_1}^{}]\!=\!1\hs$,  
which can make up the full mass-dimension 
of the KK Goldstone amplitude, $D_{\tT}^{} \!=\! 4\!-\!N\hs$.\ 
This means thar for the $N$-point KK scattering amplitude (with odd $N$), 
the leading KK Goldstone amplitude contains one power of KK mass 
and has its leading energy dependence  
$D_{\!E}^{} \! =\hsm 3\hs -\hsm N$, which is lower than the leading energy dependence of 
the mass-independent KK Goldstone amplitude with even number $N$.\  
For such massive amplitudes with odd number $N$, there is no systematic way to realize 
the BCJ-type double-copy for the warped KK gauge/gravity theories.

\vs

Next, we consider the double-copy construction for the $N$-point scattering amplitudes of 
KK vector Goldstone bosons ($\VV_n^\mu$).\ The dimensional analysis including the counting
of leading energy dependence is similar to the aforementioned conclusion 
on the $N$-point amplitudes of the KK scalar Goldstone bosons ($\phin$).\ 
For the $N$-point KK scattering amplitudes (with even number $N$), 
we obtain the following scattering amplitude:
\begin{align}
\label{eq:tM0-NNt}
\hspace*{-8mm}
\MT_0^{}\big[\VV_{n_1^{}}^{\pm1},\cdots\!,\VV_{n_N^{}}^{\pm1}\big] 
&= -\hs\tilde{\al}_{n_1^{}\cdots n_N^{}}^{} \!\!\(\!\frac{\ka}{4}\)^{\!\!N-2}\! \sum_{\ell=1}^{\,(2N-5)!!}\!\!\!\!\frac{1}{\,\D_\ell^{}\,}
\sum_{j=0}^{N}\sum_{i=0}^{\binom{N}{j}}\! 
\NNt_{\ell,i}^0[(N\!\!-\!j)A_5^{an};jA_{\pm}^{an}]
\nn\\
\hspace*{-8mm}
&\quad~\times\NNt_{\ell,i}^0[\hs (N\!\!-\!j)A_{\pm}^{an};jA_5^{an}] \hs,
\end{align}
where for every possible product of two numerators in the sums 
we have replaced the relevant product of KK gauge coupling coefficients 
by the corresponding common gravitational KK coupling coefficient:
\begin{equation}
\label{eq:N-LODC-replace}
c_N^{} \lrb{\cdots\ff_{n_k^{}}^{}\!\!\!\cdots\fft_{n_{k'}^{}}^{}\!\!\!\cdots}\hsm
\lrb{\cdots\fft_{n_k^{}}^{}\!\!\!\cdots\ff_{n_{k'}^{}}^{}\!\!\!\cdots}
\,\longrightarrow~ 
\tilde{\al}_{n_1^{}\hsm\cdots\hsx n_N^{}}^{} \,, 
\end{equation}
where the $N$-point KK gauge coupling coefficient $\lrb{\cdots\ff_{n_k^{}}^{}\!\!\!\cdots\fft_{n_{k^\pp}^{}}^{}\!\!\cdots}$ 
contains $(N\!-\!j)$ wavefunctions associated with $A_5^{an}$ 
and $j$ wavefunctions associated with $A_\pm^{an}$, 
whereas the $N$-point KK gauge coupling coefficient $\lrb{\cdots\fft_{n_k^{}}^{}\!\!\!\cdots\ff_{n_{k'}^{}}^{}\!\!\!\cdots}$
includes $j$ wavefunctions associated with $A_5^{an}$ and $(N\!-\!j)$ 
wavefunctions associated with $A_\pm^{an}\hs$.\ 
The bracket notation $\lrb{\cdots}$ for defining the KK coupling coefficient 
as an integral of 5d wavefunctions over the 5th coordinate 
is given in Eq.\eqref{app-eq:def-[[]]} of Appendix\,\ref{app:D}.\ 
In Eq.\eqref{eq:N-LODC-replace}, $c_N^{}$ is a normalization factor and can be determined 
by taking the flat 5d limit ($k\ito 0$).\ 
The effective $N$-point coupling coefficient 
$\tilde{\al}_{n_1^{}\cdots\hs n_N^{}}^{}\!$ is dimensionless and can be expressed as follows:  
%
\begin{align}
\label{eq:tal-N}
\tilde{\al}_{n_1^{}\cdots\hs n_N^{}}^{} &= 
\sum_{j_1^{},\ldots,j_{N-3}^{}=0}^{\infty}\hspace*{-4mm} 
\tilde{\rho}_{n_1^{}n_{2}^{}j_1^{}}^{}
\tilde{\rho}_{j_2^{}n_{3}^{}j_1^{}}^{}\!
\tilde{\al}_{j_2^{}n_{4}^{}j_3^{}}^{}\!\cdots
{\tilde{\al}_{j_{N-4}^{}n_{N-2}^{}j_{N-3}^{}}^{}}
\!\tilde{\al}_{n_{N-1}^{}n_{N}^{}j_{N-3}^{}}^{}\!
\nn\\[1mm]
&= \lrb{\vv_{n_1^{}}^{} \!\!\cdots \vv_{n_N^{}}^{}}\hs.
\end{align}
In the above, we have chosen the case with 2 $\tilde{\rho}_{nm\ell}^{}$'s and $(N\!\!-\hsm 4)$ $\tilde{\al}_{nm\ell}^{}$'s as an example, whereas the results computed for other combinations of $\tilde{\rho}_{nm\ell}^{}$ 
and $\tilde{\al}_{nm\ell}^{}$ will give the same coupling coefficient 
$\lrb{\vv_{n_1^{}}^{} \!\!\cdots \vv_{n_N^{}}^{}}$
as shown in \eqrefe{eq:tal-N}.\ 
This is because, regardless of the numbers and positions of $\tilde{\rho}_{nm\ell}^{}$ and $\tilde{\al}_{nm\ell}^{}$, the internal lines (indices) 
can be summed over by using the completeness relation  \eqref{eq-app:completeness}.\ 
The final result only depends on the 5d eigenfunctions 
$\{\vv_{n_i^{}}^{}\hsm\}$ associated with the external states $\{\VV_{n_i^{}}^\mu\hsm\}$.\ 
Similarly, for the scattering amplitudes containing $N$ external states (with odd number $N$), 
the KK scattering amplitudes will be proportional to KK masses 
and thus have lower energy-power dependence.\ 
So there is no clear evidence that the double-copy construction would hold 
for these massive amplitudes.

\vs 

We can further prove the $N$-point LO double-copy relation by utilizing 
the Cachazo-He-Yuan (CHY) formalism\,\cite{CHY}.\ 
We start by considering the $(4 +\delta)$-dimensional spacetime with
$\delta$ extra spatial dimensions under the toroidal compactification.\
(Since we will study the KK scattering amplitudes at the leading order (LO) only, 
the current discussion under toroidal compactification 
will give the same results for the case of
orbifold compactification of $S^1\!/\ZZ$ at the LO of high energy expansion.)\
For a momentum $\hat{p}^{}_a$ living in the
6d spacetime, we can decompose it in the following form:
\begin{equation}
\hat{p}_{a}^{M}=
\( p_{a}^{\mu}\hs; M_{n_a^{}}^{}\!,\, 0 \), \qquad
a \!\in\! \{1 , \cdots\hsm, N\}  \hs,
\end{equation}
where $\,p^{\mu}_a\,$ denotes the momentum in 
the $(3\!+\!1)$d spacetime and $M_{n_a^{}}^{}\!$ is the momentum-component in  
the 5th dimension (realized as discretized KK mass-eigenvalues).\ 
For each polarization vector, we have two choices, depending on 
whether it is restricted to the extra dimension or not.\ 
Thus, we can define the polarizations for each KK gauge boson and 
its corresponding KK Goldstone boson as follows:
\begin{align}
\ep^M_{a,\rm{g}} = (\ep^\mu_a\hs;\, 0\hs,\, 0)\hs, \qquad~~
\ep^M_{a,\rm{s}} = (\vec{0}\hs;\, 0\hs,\,1)\hs,
\end{align}
where the subscripts ``g'' and ``s'' stand for 
the KK guage boson and KK Goldstone boson respectively.\
The LO scattering amplitude $\A_N^{0}$ in either gauge theory or gravity theory 
can be re-written as the expression of a contour integral:
\begin{equation}
\label{eq:CHY-AN}
\A_N^{} = \oint_\Ga^{}\! z_{ij}^2z_{jk}^2z_{ki}^2\!\!\! 
\prod_{\substack{a=1 \\ (a \neq i,j,k)}}^N\!\!\!\!\!
\frac{\,\td z_a^{}\,}{\mathbb{E}_a^{}}\hs I_N^{} \hs,  \qquad
\mathbb{E}_a^{} = \!\!\sum_{\substack{b=1 \\ (b \neq a)}}^N\!\!
\!\frac{~p_a^{} \!\cdot\hsm p_b^{}\!+\!M_{n_a^{}}^{}\!M_{n_b^{}}^{}\,}
{z_a^{} \!-\! z_b^{}} \,,
\end{equation}
where the contour $\Ga$ encloses all the solutions 
for the scattering equations of $\mathbb{E}_a^{}$ 
and $I_N^{}$ is the theory-dependent integrand.\
Under high energy expansion, the above formulations 
can be further derived at the leading order:
\begin{equation}
\label{eq:CHY-A0N}
\A_N^{0} = \oint_\Ga^{}\! z_{ij}^2z_{jk}^2z_{ki}^2\!\!\! 
\prod_{\substack{a=1 \\ (a \neq i,j,k)}}^N \!\!\!\!\!
\frac{\,\td z_a^{}\,}{\mathbb{E}_a^{0}}\, I_N^{0} \,,  \qquad
\mathbb{E}_a^0 = \!\!\sum_{\substack{b=1 \\ (b \neq a)}}^N\!\!
\frac{~p_a^{} \!\cdot\! p_b^{}~}{z_a^{} \!-\! z_b^{}} \,.
\end{equation}

We note that in the high energy limit and for the KK Goldstone boson sector, 
the 5d KK Yang-Mills gauge theory is reduced to a 4d KK Yang-Mills-Scalar (YMS) theory, 
whereas the 5d KK Einstein gravity theory is reduced to 
a 4d KK Einstein-Maxwell-Scalar (EMS) theory.\ 
According to the dimensional compactification of the CHY formalism\,\cite{Cachazo:2014xea},  
we can express the scalar integrands 
for the YMS (single flavor) amplitude and the EMS amplitude as follows:
\beqs
\label{eq:I-YMS-EMS}
\begin{align}
\label{eq:I-YMS}
I_{N}^{0,\rm{YMS:s}} &=\, 
\tilde{a}_{n_1^{}\cdots n_N^{}}^{} {\CC[\al]}\hs \rm{PT}[\al]
(\rm{Pf}\hs{{\mathbb{X}}_N^{}})(\rm{Pf}'\hsmx {{\mathbb{A}}_{\hsm N}^{}}) \hs,
\\[.5mm]
\label{eq:I-EMS}
I_{N}^{0,\rm{EMS:s}} &=\, {c_N^{}} \tilde{a}_{n_1^{}\cdots n_N^{}}^{2}
(\rm{Pf}\hs{{\mathbb{X}}_N^{}})^2(\rm{Pf}'\hsmx {{\mathbb{A}}_N^{}})^2 \hs,
\end{align}
\eeqs
where the symbols Pf and $\rm{Pf}'$ denote the Pfaffian and reduced Pfaffian respectively.
The matrix quantities ${\mathbb{X}}_N^{}$ and ${\mathbb{A}}_N^{}$ are given by 
\begin{equation}
({\mathbb{X}}_N^{})_{ab}^{} = \left\{\!\! 
\begin{aligned} 
\frac{1}{\,z_{a}^{}\!-\!z_{b}^{}\,}\hs, 
& \hspace*{4mm} (\rm{for}~a \!\neq\! b),
\\[1mm]
0\hs,~  \hspace*{3mm}
& \hspace*{4mm}  (\rm{for}~a\!=\!b),
\end{aligned}\right.
\hspace*{10mm}
({\mathbb{A}}_N^{})_{ab}^{} = \left\{\!\!
\begin{aligned} 
\frac{p_{a}^{}\!\cdot\hsm p_{b}^{}}{\,z_{a}^{}\!-\!z_{b}^{}\,}\hs, 
& \hspace*{4mm}  (\rm{for}~a \!\neq\! b),
\\[1mm]
0\hs,~  \hspace*{3mm}
& \hspace*{4mm}  (\rm{for}~a\!=\!b).
\end{aligned}\right.
\end{equation}
In Eq.\eqref{eq:I-YMS}, the color factor $\hs\CC[\al]$ is decomposed into
the DDM basis and contains $(N\!-\hsm 2)!\hs$ elements, as defined below:
\begin{equation}
\CC [\al] \,= \sum_{e_1^{},\cdots\hsm ,\hs e_{N-3}^{}}
\hspace*{-4mm}
f^{a_1^{} a_{\al(2)} e_1^{}} f^{e_1^{} a_{\al(3)} e_2^{}} \cdots
f^{e_{N-3}^{} a_{\al(N-1)} a_N^{}}
\,,
\end{equation}
where the 1st and $N$-th labels are fixed in the color ordering $\al\hs$, i.e.,
$\al\!=\hsm\![1,\al(2),\cdots\!,\al(N\hsm\!-\hsm\!1),N]
\!\hsm\in\! S_{N-2}^{}\hs$.\ 
The symbol $\rm{PT}[\al]$ stands for the Parke-Taylor factor and is given by 
\begin{equation}
\rm{PT}[\al] = 
\frac{1}{~(z_1^{} \!-\! z_{\al(2)}) \cdots   (z_{\al(N-1)} \!-\! z_N^{})(z_N^{} \!-\! z_1^{})~} \hs.
\end{equation}
The above CHY formulation works for even number $N$, whereas for the odd number $N$ 
the Pfaffian is trivially zero.\ 
This is consistent with the conclusions we reached earlier in this subsection.\ Namely, 
in the case of $N$-point scattering (with even number $N$), 
the leading-order double-copy holds.\  
But, for the $N$-point scattering (with odd number $N$),
the KK GAET and GRET (which are connected by double-copy) will take the trivial form of  
$0\hsm =\hsm 0\hs$ at the leading order of high energy expansion.\ 
Note that the scattering equations are the same for both the KK YMS theory and the KK EMS theory.\ 
So the above LO double-copy relation \eqref{eq:I-EMS} holds under the replacement
$\,c_N^{}\tilde{a}_{n_1^{}\cdots\hs n_N^{}}^2\!\!\!\!\longrightarrow\!
\tilde{\be}_{n_1^{}\cdots\hs n_N^{}}^{}$.\ 
Using Eqs.\eqref{eq:CHY-A0N}-\eqref{eq:I-YMS-EMS} and the fundamental BCJ relations, 
we derive the extended KLT-type double-copy formula for the $N$-point 
LO scattering amplitude of gravitational KK Goldstone bosons:
\begin{align}
\MT_0^{\rm{dc}}\big[\phi_{n_1^{}}^{}\!,\hsm\cdots\!,\phi_{n_N^{}}^{}\big] = -\,c_N^{}\tilde{a}_{n_1^{}\!\cdots\hs n_N^{}}^{2}
\!\(\!\frac{\,\ka\,}{4}\)^{\!\!N-2}\!\!\!\!\!\!\!
\!\!\sum_{\{\al,\be\} \in S_{N \!-\!3}^{}}\hspace*{-4mm}
\tT_{0}^{}\big[\hsm A^{a_{\al(i)}n_{\al(i)}}_{5}\big] 
\mathcal{K}[\al|\be]
\, \tT_{0}^{}\big[\hsm A^{a_{\be(i)}n_{\be(i)}}_{5}\big]\hs,
\end{align}
where $\mathcal{K}[\al|\be]$ is the KLT kernel.\ 
Finally, we note that the CHY double-copy construction of the $N$-point LO scattering amplitude 
of the vector-type gravitational KK Goldstone bosons  
$\MT_0^{\rm{dc}}\big[\VV_{n_1^{}}^{\pm1},\!\cdots\!,\VV_{n_N^{}}^{\pm1}\big]$ 
can be similarly realized.

\vspace*{1mm}
\subsection{\hspace*{-2.5mm}Prescription for Warped Double-Copy Construction at Leading Order}
\label{sec:4.4}
\vspace*{1mm}

Our previous work\,\cite{Li:2022rel} proved that for the compactified KK gauge/gravity theories,
only the toroidal compactification of flat extra dimensions can satisfy the mass spectrum condition
and directly realize extended double-copy of massive KK gauge/gravity scattering amplitudes.\
It was also proved\,\cite{Li:2021yfk} that
under the toroidal compactification of flat extra dimensions such massive KK double-copies can be
derived as the field theory limit of the massive KLT relations of KK open/closed string amplitudes.\
But we note that the compactified warped 5d gauge/gravity theories
with orbifold $S^1\! /\ZZ$ (such as the RS1 model\,\cite{RS1})
clearly do not meet this criterion.\
This is because such warped KK gauge/gravity theories have highly nonlinear mass spectra
and the mass spectrum of KK gauge bosons differ from that of the KK gravitons due to the 5d warped metric,
whereas coupling coefficients of the KK gauge bosons and of the KK gravitons also differ significantly.\
Moreover, the $S^1\hsm /\ZZ$ orbifold compactification spoils the KK number conservation and
the KK scattering amplitudes do not exhibit single-pole structure in each channel.\
Nevertheless, as demonstrated in Sections\,\ref{sec:4.1}-\ref{sec:4.2}, for tree-level KK gauge/gravity
scattering amplitudes, we find proper ways (or restriction) to evade these problems.\ 
These include:\ 
{\bf (i)}\,the 3-point double-copied KK graviton scattering amplitudes (from that of the KK gauge bosons)
exhibit exactly the same kinematic structure as the original 3-point KK grviton amplitudes even without
high energy expansion except that we need to set up a proper prescription for the correspondence
between the double-copied trilinear KK gauge couplings and the original trilinear KK graviton couplings,
whereas under such prescription
the double-copy works for the corresponding 3-point gravitational KK Goldstone
amplitudes only at the leading-order of high energy expansion;\footnote{%
We found previously\,\cite{Li:2022rel} that for the gauge/gravity
KK Goldstone scattering amplitudes with toroidal compactification of flat extra dimensions,
the double-copy construction also works only at the leading order of high energy expansion.}\
{\bf (ii)}\,at the leading-order of high energy expansion,
the numerators of 4-point scattering amplitudes of KK gauge bosons and of KK Goldstone bosons
obey the kinematic Jacobi identities;
{\bf (iii)}\,the double-copied leading-order KK graviton scattering amplitudes
(constructed from that of the KK gauge bosons)
exhibit exactly the same kinematic structure as the original leading-order amplitudes of
KK grvitons except that we need to set up a proper prescription for the correspondence
between double-copied KK gauge couplings and the original KK graviton couplings.\

\vs

Based upon the above observation and the double-copy analysis shown in Sections\,\ref{sec:4.1}-\ref{sec:4.2},
we summarize the prescriptions for successful double-copy constructions of the 3-point and 4-point KK
graviton scattering amplitudes in the warped 5d gauge/gravity theories under orbifold compactification
of $S^1\!/\ZZ\hs$:
\begin{enumerate}
	
\item[{\bf 1).}]
For the present double-copy construction of $N$-point warped KK graviton scattering amplitudes from the corresponding $N$-point warped KK gauge boson scattering amplitudes, we first replace the 4d gauge coupling $g$ by the 4d gravitational coupling $\ka$ as follows:
\begin{equation}
\label{eq:DC-g-kappa}
g^{N\hsm -2} ~\longrightarrow~ \!\hsm -\hsmx\(\frac{\ka}{4}\)^{\!\!N\hsm -2} \,.
\end{equation}
For instance, this gives $\hs g\ito- \ka/4\,$ for $N\!=\!3$ 
and $\hs g^2\ito -\ka^2/16$ for $N\!=\!4\hs$.\
This double-copy replacement \eqref{eq:DC-g-kappa} is the similar to what we established for the compactified flat 5d gauge/gravity theories\,\cite{Li:2021yfk}\cite{Li:2022rel}.\ 
The difference is that in the present analysis, due to an overall factor of $\hsx\ii\hs$ of the 
$N$-point (for $N\!=\hs$odd) gauge amplitude, the corresponding double-copied gravitational 
amplitude has an overall factor of $-1\hs$.\ Thus, the factor $(-)^{N+1}$ in Refs.\cite{Li:2021yfk}\cite{Li:2022rel} will be replaced by $-1$ in this study.

\item[{\bf 2).}]
Then, for the scattering amplitudes of KK gauge bosons (KK Goldstone bosons), 
we apply the extended massive color-kinematics duality 
and make the following group factor replacements (double-copy relations)
as in Eq.\eqref{eq:CK-3pt} for the 3-point KK amplitudes, or as in Eqs.\eqref{eq:CK-4pt}\eqref{eq:CK-4pt-nnmm}
for the 4-point KK amplitudes:
\beq
\label{eq:CKG-3pt}
\begin{array}{rlrll}
f^{abc}   & {\longrightarrow~} \NN [\{\ep_j^{}\}]\hs,  \hspace*{6mm}
& f^{abc} & {\longrightarrow~} \NNt [\ep_3^{}]\hs,     \hspace*{6mm}
& \text{(3-point~amplitudes),}
\\[1mm]
\CC_j^{} & {\longrightarrow~} \NN_j^{0}\hs,   \hspace*{6mm}
& \CC_j^{} & {\longrightarrow~} \NNt_j^{0}\hs,  \hspace*{6mm}
& \text{(4-point~elastic~amplitudes),}
\\[1mm]
\CC_j^{} & {\longrightarrow~} \NN_j^{\rm{in}\hs 0}\hs,   \hspace*{6mm}
&\CC_j^{} & {\longrightarrow~}\NNt_j^{\rm{in}\hs 0}\hs,  \hspace*{6mm}
& \text{(4-point~inelastic~amplitudes).}
\end{array}
\eeq

\item[{\bf 3).}]
At each KK level-$n$, we replace the mass-eigenvalue $\Mn$ of KK gauge bosons
[determined by \eqrefe{Aeq:YM-Mn}] by mass-eigenvalue $\MGn$ of KK gravitons
[determined by \eqrefe{Aeq:GR-Mn}], i.e., $\Mn\!\to\hsm \mathbb{M}_n\hs$.\

\item[{\bf 4).}]
At each KK level-$n$, we further replace the involved KK gauge couplings
by the corresponding KK gravitational couplings as in Eq.\eqref{eq:Conversion-C3} for the 3-point KK amplitudes
and as in Eqs.\eqref{eq:CK2-4pt}\eqref{eq:CKin-4pt}\eqref{eq:CK00nn-4pt} for the 4-point KK amplitudes:
\begin{align}
&\hspace*{-9mm}
c_3^{} a_{n_1^{}n_2^{}n_3^{}}^{2}\!\!\to \al_{n_1^{}n_2^{}n_3^{}}^{},\quad
c_3^{} a_{n_1^{}n_2^{}n_3^{}}^{} \tilde{a}_{n_1^{}n_2^{}n_3^{}}^{} \!\!\to \tilde{\al}_{n_1^{}n_2^{}n_3^{}}^{},\quad
c_3^{} \tilde{a}_{n_1^{}n_2^{}n_3^{}}^2 \!\!\to \tilde{\be}_{n_1^{}n_2^{}n_3^{}}^{} ,
\nn\\[1mm]
&\hspace*{-9mm}
c_4^{}\tilde{a}_{nnnn}^2 \!\to \tilde{\be}_{nnnn}^{} \hs,~~~~
c_4^{\rm{in}}\tilde{a}_{nnmm}^2 \!\to \tilde{\be}_{nnmm}^{} \hs,~~~~
c_4^{\rm{in}}\hs a_{0nn}^4 \!\to \al_{0nn}^{2} \hs,
\nn\\[1mm]
&\hspace*{-9mm}c_4^{\rm{in}} a_{000}^2 a_{nn0}^2 \!\to \al_{000}^{} \al_{nn0}^{} \hs,~~~~
c_4^{\rm{in}} a_{000}^2 a_{nn0}^{} \tilde{a}_{nn0}^{} \!\to \al_{000}^{} \tilde{\al}_{nn0}^{} \hs, 
\end{align}
where the overall coefficient $c_3^{}$ or $c_4^{}$ is the relevant normalization factor and
its determination is given in Sections\,\ref{sec:4.1}-\ref{sec:4.2}.\ 
Finally, in Section\,\ref{sec:4.3}, we studied the double-copy construction 
for the $N$-point ($N\!\!\geqq\!4$) LO KK Goldstone boson amplitudes.\ 
Thus for the $N$-point LO double-copy, we impose the following correspondence (replacement) 
between the KK gauge and gravity coupling coefficients: 
\beqs
\begin{align}
&c_N^{} \lrb{\cdots\ff_{n_k^{}}^{}\!\!\!\cdots\fft_{n_{k'}^{}}^{}\!\!\!\cdots}\hsm
\lrb{\cdots\fft_{n_k^{}}^{}\!\!\!\cdots\ff_{n_{k'}^{}}^{}\!\!\!\cdots}
\,\longrightarrow~
\tilde{\al}_{n_1^{}\hsmx\cdots\, n_N^{}}^{} \hs, 
\\
&c_N^{}\hs\tilde{a}_{n_1^{}\!\cdots\, n_N^{}}^2 \,\longrightarrow~ 
\tilde{\be}_{n_1^{}\cdots n_N^{}}^{} .
\end{align}
\eeqs

\end{enumerate}

\section{\hspace*{-2.5mm}Conclusions}
\label{sec:5}

\begin{figure}[t]
\centering
\includegraphics[width=13cm]{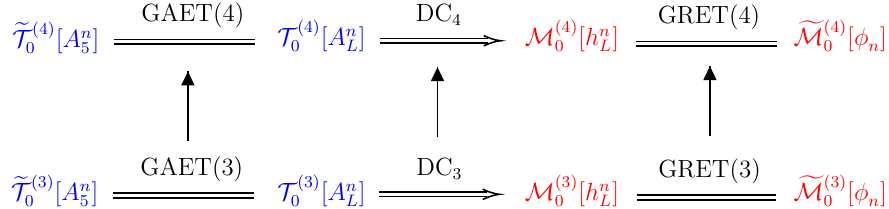}
\caption{\small 
Schematic Summary of the present analyses: Equivalence Theorems and Double-Copy Correspondences from 3-point scattering amplitudes to 4-point scattering amplitudes and from massive KK gauge scattering amplitudes to massive KK gravitational scattering amplitudes at the leading order of high energy expansion.}
\label{fig:4}
\end{figure}

In this work, we conducted comprehensive analyses on the structure of scattering amplitudes of massive Kaluza-Klein (KK) states in the compactified 5-dimensional warped gauge and gravity theories.\ 
We presented systematic formulations of the gauge theory equivalence theorem (GAET) and the gravitational 
equivalence theorem (GRET) within the $\Rxi$ gauge and up to loop level.\ 
Under high energy expansion, the GAET quantitatively connects the scattering amplitudes of longitudinal KK gauge bosons to that of the corresponding KK Goldstone bosons, whereas the GRET connects
the scattering amplitudes of massive KK gravitons of helicity-0 (helicity-1) to that of the corresponding
gravitational KK scalar (vector) Goldstone bosons.\ 
A key point of our work is {\it to take the GAET of KK Yang-Mills gauge theories as the
truly fundamental ET formulation,} from which {\it the GRET of  
the corresponding KK gravity theories can be reconstructed} by using double-copy 
at the leading order (LO) of high energy expansion.\ 
We systematically studied the extended double-copy construction of 
3-point and 4-point massive KK gauge/gravity scattering amplitudes at tree level.\  
We proved that under color-kinematics correspondence and gauge-gravity coupling
correspondence, the double-copy of 3-point physical KK gauge-boson/graviton scattering amplitudes 
can be realized without high energy expansion, 
and for the corresponding 3-point gauge/gravity KK Goldstone 
scattering amplitudes the double-copy can be achieved at the leading order (LO)
of high energy expansion.\
Moreover, we demonstrated that the double-copy can be realized for 4-point massive 
scattering amplitudes of KK gauge bosons and of KK gravitons 
(as well as their corresponding KK Goldstone bosons) to the
leading order of high energy expansion.\
In addition, we can reduce the validity of 4-point-level GAET and GRET
to the validity of 3-point-level GAET and GRET.\ Hence, the GAET and GRET formulations
at the level of 3-point KK scattering amplitudes are the most fundamental formulations, 
from which the GAET and GRET formulations at 4-point-level of KK scattering amplitudes 
can be inferred.\ 
We derived the GRET from GAET by LO double-copy constructions for
3-point and 4-point KK scattering amplitudes.\ 
We further extended this LO double-copy construction 
to the general $N$-point gauge/gravity KK scattering amplitudes with $N\!\hsm \geqq\!4\hs$.\ 
A more elaborated summary of our results and conclusions in each sections are presented as below,
which is followed by a schematic summary shown in the last paragraph of this section 
together with Fig.\,\ref{fig:4}.

\vs 

In Section\,\ref{sec:2}, we proved the formulations of the GAET and GRET 
within the $\Rxi$ gauge and up to loop level.\ 
The GAET formulation was presented in Eqs.\eqref{eq:5D-ETI3-ALQ}-\eqref{eq:5D-GAET-Cmod} and
Eq.\eqref{eq:KK-ET1-N}, whereas the GRET formulations were given 
in Eqs.\eqref{eq:KK-GRET}\eqref{eq:KK-GRET1b} for KK gravitions with helicity-0 (called type-I)
and in Eqs.\eqref{eq:KK-GRET-hV}\eqref{eq:KK-GRET-hV1f} 
for KK gravitions with helicity-1 (called type-II).\ 
In essence, the GAET and GRET reflect the geometric ``Higgs'' mechanism through the KK compactifications,
which hold not only for the compactified flat extra dimensions\,\cite{Chivukula:2001esy,Chivukula:2002ej}\cite{Hang:2021fmp,Hang:2022rjp}, 
but also for the compactified warped extra dimensions as proved in Sections\,\ref{sec:2.2}-\ref{sec:2.3}.\ 
They determine the high energy behaviors of massive KK graviton scattering amplitudes and 
ensure the nontrivial interlancing cancellations among contributions of different KK levels 
as happened in the scattering amplitudes of KK gauge bosons and of KK gravitons.\

In Section\,\ref{sec:3}, we analyzed the structure of 3-point and 4-point massive scattering amplitudes of 
KK gauge bosons and of KK gravitons (as well as the corresponding scattering amplitudes of 
KK Goldstone bosons), including the interconnections between the 3-point and 4-point KK amplitudes.\ 
In Section\,\ref{sec:3.1},  
we explicitly proved the warped GAET and GRET 
for the fundamental 3-point massive scattering amplitudes of KK gauge bosons and of KK gravitons.\ 
We found that the nontrivial realization of GAET is given by the 3-point KK gauge boson amplitude
with two longitudinal external states and one transverse external state (of $LLT$ type),
as shown in Eq.\eqref{eq:GAET-LLT-3pt}.\ We proved that to hold this 3-point-level GAET \eqref{eq:GAET-LLT-3pt}
requires the condition \eqref{eq:a-at-3pt} which is directly proved in Eq.\eqref{app-eq:a-ta-general} 
of Appendix\,\ref{app:D}.\ Then, we computed the 3-point KK graviton amplitude\footnote{%
We derived the most general 3-point scattering amplitude of KK gravitons in Eq.\eqref{eq:Amp-3h-exact} 
of Appendix\,\ref{app:C}.} 
of helicities $(\pm1,\pm1,\pm2)$ and its corresponding KK Goldstone amplitude 
in Eqs.\eqref{eq:hhh112}-\eqref{eq:VVh112}
and Eqs.\eqref{eq:GRET-LO-h1h1h2}-\eqref{eq:GRET-LO-VVh2}.\ 
To hold the 3-point-level GRET \eqref{eq:GRET2-LLT-3pt} imposes the condition \eqref{eq:al-tal-3pt}   
which we proved in Eq.\eqref{app-eq:al-tal-general} of Appdenix\,\ref{app:D}.\
We further computed the 3-point KK graviton amplitude of helicities 
$(0,\hs 0,\hs\pm2)$ and its corresponding KK Goldstone amplitude in 
Eqs.\eqref{eq:hhh-LLT}-\eqref{eq:pph}.\ 
To hold the 3-point-level GRET \eqref{eq:GRET1-LLT-3pt} leads to the nontrivial condition 
\eqref{eq:al-tbe-LLT-55T} which we proved in Eq.\eqref{eq:al-tbe-LLT-55T-3pt} of Appdenix\,\ref{app:D}.\
We also computed the mixed 3-point KK graviton amplitude of helicities $(\pm1,\pm1,\hs 0)$
and its corresponding KK Goldstone amplitude in Eq.\eqref{eq:h11L-V11Phi-3pt}.\
To hold the GRET \eqref{eq:GRET-h11L=VVphi} imposes the condition \eqref{eq:cond-h11L=VVphi}
which was proved in \eqref{eq-app:cond-h11L=VVphi} of Appendix\,\ref{app:D}.\

In Section\,\ref{sec:3.2}, we further demonstrated that the validity of the warped GAET and GRET for 
4-point massive KK scattering amplitudes can be effectively reduced to the validity of these theorems 
for the 3-point massive KK scattering amplitudes.\ 
For the 4-point massive elastic and inelastic scattering amplitudes
of KK gauge bosons and KK Goldstone bosons, 
we explicitly established the warped GAET as in Eqs.\eqref{eq:GAET1-nnnn},
\eqref{eq:GAET-nnmm}, and \eqref{eq:GAET1-nmlq} respectively.\ We proved that the validity of the 
4-point-level GAET in these cases is ensured by the validity of the fundamental 
3-point-level GAET \eqref{eq:GAET-LLT-3pt}.\   
Then, we analyzed the 4-point massive scattering amplitudes
of KK gravitons and gravitational KK Goldstone bosons.\ 
We explicitly established the warped GRET as in Eq.\eqref{eq:GRET-LO-4pt}	 
for the 4-body elastic scattering channel.\ 
The validity of the 4-point-level GRET \eqref{eq:GRET-LO-4pt} relies on the 
sum rule condition \eqref{eq:SumR-4pt-al-be}, while the proof of Eq.\eqref{eq:SumR-4pt-al-be} 
requires the sum rule condition \eqref{eq:al-tbe-nnj} to play the key role, where 
the condition \eqref{eq:al-tbe-nnj} just ensures the validity of the 3-point-level
GRET \eqref{eq:GRET1-LLT-3pt} in the case of $(n_1{},n_2^{},n_3^{})\!=\!(n,\hs n,\hs j)$.\ 
This shows that validity of the 4-point-level GRET \eqref{eq:GRET-LO-4pt} is reduced to the validity 
of the 3-point-level GRET \eqref{eq:GRET1-LLT-3pt}.\ 
We further computed the mixed 4-point inelastic scattering amplitudes 
of KK gravitons and of KK Goldstone bosons 
as in Eqs.\eqref{eq:00nn-hh}-\eqref{eq:00nn-VV}.\ 
Under high energy expansion, we found that the leading-order amplitudes 
\eqref{eq:LO-00nn-hh}-\eqref{eq:LO-00nn-VV} are simple enough in this case.\  
Thus, we explicitly established the 4-point-level GRET \eqref{eq:GRET2-00nn} 
without a need of additional sum rule condition.\

In Section\,\ref{sec:4}, we studied the double-copy construction of the massive gravitational 
KK scattering amplitudes from the corresponding massive KK gauge scattering amplitudes 
for the warped 5d gauge and gravity theories with the orbifold compactification of $S^1\!/\ZZ\hs$.\ 
This is highly nontrivial and challenging since it was proved\,\cite{Li:2022rel}   
that the direct construction of double-copy for $N$-point massive gauge/gravity KK 
scattering amplitudes (with $N\!\!\geqq\hsm\!4$) can directly work out 
only for toroidal compactifications with flat extra dimensions.\
Nevertheless, we could realize double-copy construction 
for the warped massive gauge/gravity theories
with proper restrictions and prescriptions so as to evade the previous conditions\,\cite{Li:2022rel}.\ 
In Section\,\ref{sec:4.1}, we newly proved that the double-copy can be constructed for 
the 3-point full scattering amplitudes of KK gravitons at tree level 
for warped massive gauge/gravity theories.\
We set up the 3-point color-kinematics (CK) correspondence and 
the gauge-gravity coupling correspondence as in Eq.\eqref{eq:CK-3pt},  
and the KK mass replacement $\Mn\!\to\hsm \mathbb{M}_n\hs$ at each KK level.\   
With these, we first presented the general 3-point doubel-copy formulas (with any  
polarization tensors of physical KK graviton states) as in 
Eqs.\eqref{eq:M=sumTxT-3pt}-\eqref{eq:DC-hhh-3pt}.\ 
Then, we explicitly constructed the 3-point KK graviton amplitudes with helicities
$(\pm1,\pm1,\pm2)$ and $(0,\hs 0,\hs \pm2)$ in 
Eq.\eqref{eq:DC-3pt-112All} and Eq.\eqref{eq:DC-hhh-LL2} respectively.\ 
We further derived their corresponding gravitational KK Goldstone boson amplitudes
in Eqs.\eqref{eq:LO-DC-3pt-GB112}-\eqref{eq:LO-DC-3pt-GB002}.\
The required conversions of gauge-gravity coupling constants are given in Eq.\eqref{eq:Conversion-C3}.\  
With these we established 
extended massive double-copy constructions of the GRET \eqref{eq:GRET112-3pt-DC} and
\eqref{eq:GRET002-3pt-DC} (in the warped KK gravity theory) from the GAET
\eqref{eq:GAET-LLT-3pt} (in the warped KK gauge theory) 
at the level of 3-point KK scattering amplitudes.\ 

In Section\,\ref{sec:4.2}, 
we demonstrated that the extended double-copy construction can be achieved 
for the 4-point KK gauge/gravity scatttering amplitudes 
at the leading-order (LO) of high energy expansion.\   
For the 4-point elastic scattering $(n,n)\ito (n,n)\hs$, the LO amplitudes of KK gauge bosons and 
of KK Goldstone bosons have their effective numerators connected by the 
generalized gauge transformations \eqref{eq:GGT-NjNtj-nnnn} and obey the kinematic Jacobi identities
\eqref{eq:Jacobi-4pt}.\  The double-copies of these two LO amplitudes of KK gauge bosons and of
KK Goldstone bosons give the corresponding LO amplitudes of KK gravitons 
and of the gravitational KK Goldstone bosons as in 
Eqs.\eqref{eq:DC-hL-phi-4n} and \eqref{eq:M5DC-LO}.\  
Using the coupling conversion \eqref{eq:CK2-4pt}, we derived the double-copied LO gravitational
KK amplitudes \eqref{eq:M5DC-LO-Fin} which agree with the same amplitudes \eqref{eq:AmpE2-4hL-2} 
(by explicit Feynman diagram approach).\ 
In parallel, we studied the double-copy constructions for the LO inelastic KK gauge/gravity
scattering amplitudes, including the processes $(n,n)\ito (m,m)$ (with $n\!\neq\!m$),
$(n,m)\ito (\ell, q)$ (with $n\!\neq\!m\!\neq\!\ell\!\neq\!q$), and $(0,0)\ito (n,n)$
(with $n\!>\!0$).\ We found that their effective numerators satisfy the kinematic Jacobi identity 
respectively, as shown in Eqs.\eqref{eq:Jacobi-4pt-nnmm}, \eqref{eq:KJacobi-nmlq}-\eqref{eq:KJacobi1-nmlq},
and so on.\ For these inelastic processes, we presented the double-copied inelastic amplitudes 
of KK gravitons and of gravitational KK Goldstone bosons 
as in Eqs.\eqref{eq:M5DC-LO-Fin-nnmm}, \eqref{eq:DC-LO-4h-nmlq}, and
\eqref{eq:h00nn-DC}\eqref{eq:CK00nn-4pt}.\ 
In Section\,\ref{sec:4.3},  we further established that this LO double-copy construction 
can be extended to the general $N$-point KK scattering amplitudes with $N\!\! \geqq\!4\hs$.\ 
In Section\,\ref{sec:4.4}, we summarized a set of well-defined prescriptions for 
realizing the extended double-copy constructions in the warped massive KK gauge/gravity theories,
which include the double-copy constructions for 3-point full KK gauge/gravity amplitudes
and the double-copy constructions for the 4-point LO KK gauge/gravity amplitudes.\

\vspace*{0.5mm}

Finally, we present a schematic summary of the present analyses as in Fig.\,\ref{fig:4}.\
In this figure, we start from the horizontal bottom row in which all entries describe the 
basic 3-point KK scattering amplitudes.\ From the left to right, the long equality sign linking 
the first two entries gives the ``GAET(3)'' which connects the 3-point LO longitudinal KK 
gauge boson amplitude $\TT_0^{(3)}[A_L^n]$  
to the corresponding LO KK Goldstone boson amplitude $\tT_0^{(3)}[A_5^n]$;  then the long equality sign
linking the last two entries gives the ``GRET(3)'' which connects 
the 3-point LO longitudinal KK graviton amplitude 
$\M_0^{(3)}[h_L^n]$  to the corresponding LO gravitational KK Goldstone boson amplitude 
$\MT_0^{(3)}[\phin]$; in the middle, the horizontal arrow indicates the extended LO double-copy ``DC$_3$''
which constructs the 3-point LO longitudinal KK graviton (KK Goldstone) amplitude from the
3-point LO longitudinal KK gauge boson (KK Goldstone) amplitude.\ 
After this, we see that all the entries and equality signs (or arrow) in the top row are doing the
same jobs as those in the bottom row except that all entries in the top row deal with 
the 4-point KK gauge/gravity scattering amplitudes.\   
Finally, the vertical arrows from bottom to top indicate the 4-point-level ``GAET(4)'' and ``GRET(4)''
can be reduced to (reconstructed from) the fundamental ``GAET(3)'' and ``GRET(3)'' for the basic 3-point KK
gauge (gravity) amplitudes.\ Furthermore, along the horizontal top row 
we can construct the ``GRET(4)'' from ``GAET(4)'' through the extended LO double-copy ``DC$_4$''.

\vspace*{5mm}
\noindent
{\bf\large Acknowledgments}
\\
This research was supported in part by the National NSF of China 
under grants 12175136 and 12435005.\ 
YH is supported in part by the Northwestern University Amplitudes and Insight group, the Department of Physics and Astronomy, and Weinberg College.\ 

\vspace*{8mm}
\noindent
{\Large\bf Appendix}
\vspace*{-2mm}

\appendix

\section{\hspace*{-2mm}Kinematics of KK Particle Scattering}
\label{app:A}

In this Appendix we present the kinematics of 3 and 4 KK particle scattering
processes in the (3+1)d spacetime.\ We choose the 4d Minkowski metric tensor 
$\,\eta^{\mn} \!=\! \eta_{\mn}^{} \!=\! \diag(-1,1,1,1)$.

\vs
For the 3 KK particle scattering, we define the 4-momenta of the external particles as follows:
\begin{align}
\begin{alignedat}{3}
p_1^{\mu} &= (E_1^{},\hs k\st,\hs 0,\hs k\ct) \hs, \qquad 
&& E_1^{} = \sqrt{M_{1}^2 \hsm +\hsm k^2\,}\,, 
\\
p_2^{\mu} &= (E_2^{},\hs -k\st,\hs 0,\hs k\ct) \hs, \qquad 
&& E_2^{} = \sqrt{M_{2}^2\hsm +\hsm k^2\,}\,, 
\\
p_3^{\mu} &= -(E_3^{},\hs 0,\hs 0,\hs 2k\ct)\hs, \qquad 
&&E_3 = \sqrt{M_{3}^2 \hsm +\hsm 4k^2 \cct\,}\,,
\end{alignedat}
\end{align}
where $k\!=\!|\vec{p}\hs|$, $(\st,\hs\ct)=(\sin\hsm\theta ,\hs \cos\hsm\theta)$,  $p_j^2 \!=\!- M_{j}^2$ (with $j\!=\!1,2,3\hs$),
and all momenta are outgoing by convention.\  

Using the energy conservation condition $E_1^{}\!+\!E_2^{}\!=\!E_3^{}$, we can solve the magnitude of 3-momentum
$k\!=\!|\vec{p}\hs|$ as a function of the scattering angle $\theta\,$:
\beq 
\hspace*{-2mm}
k = \hsm\frac{1}{\,\sqrt{2\,}s_{2\theta}^{}}\hsm\bigg\{\!\!
\sqrt{\!M_3^4\!+\!4c_{\theta}^2\!\left[\hsm  
M_{+}^2\hsm (M_{+}^2\!\!-\!M_3^2)\!-\!2M_1^2M_2^2(\hsm 1\!-\!c_{2\theta}^{})\right]\hs}
+\hsm s_{2\theta}^{}\hsm
\Big(\! M_3^2\hsm \cot\hsm 2\theta\!-\! M_{+}^2\!\cot\hsm\theta
\Big)\!\hsm\bigg\}^{\!\!\frac{1}{2}}\!,~~~
\eeq 
where we have defined $M_{+}^2\!\!=\!M_1^2\!+\!M_2^2$ and
$(s_{2\theta}^{},c_{2\theta}^{})\!=\!(\sin 2\theta,\cos2\theta)$.\
Alternatively, we can express $\theta$ as a function of $k$\,:
\begin{equation}
\label{eq:cos-theta}
\cos\hsm\theta = \frac{1}{\,2k\,}\!\hsm
\[ 2k^2 \!+\!M_1^2 \!+\!M_2^2 \!+\!
2\sqrt{\!(k^2\!+\!M_1^2)(k^2\!+\!M_2^2)\,} \!-\!M_3^2\]^{\!\frac{1}{2}} \!.
\end{equation}
In high energy limit, we take $\hs k\ito \infty$ and $\theta\ito 0\hs$.\ 
Thus, we can make high energy expansion in terms of $1/k$ and $\theta$ as follows:
\beqs
\label{eq:k-theta}
\begin{align}
k &= \frac{\,[M_3^2\!-\!2(M_1^2\!+\!M_2^2)]^{1/2}\,}{2\hs\st} + 
\frac{\st (M_1^2\!-\!M_2^2)^2}{\,4[M_3^2\!-\!2(M_1^2\!+\!M_2^2)]^{3/2}\,} 
+ O(s_\theta^3) \,, 
\\
\cos\theta &= 1 - \frac{\,M_3^2\!-\!2(M_1^2\!+\!M_2^2)\,}{8\hs k^2} + O(k^{-4}) \,, 
\\
\sin\theta &= 
\frac{\,[M_3^2\!-\!2(M_1^2\!+\!M_2^2)]^{1/2}\,}{2\hs k} + O(k^{-3}) \,.
\end{align}
\eeqs
Inspecting Eq.\eqref{eq:k-theta}, 
we see that to have real solutions of $(k,\hs\theta)$ requires
$M_3 \!\geqq\! \sqrt{2(M_1^2 \!+\! M_2^2)\,}$.\
Moreover, using the exact formula \eqref{eq:cos-theta} and imposing the condition
$\cos\theta\!\leqq\!1$, we derive a general condition:
\beq
\label{eq:KCond-M123}
M_3^2 \geqq M_1^2 \!+\! M_2^2 \!+\! 2\sqrt{\!(k^2\!+\!M_1^2)(k^2\!+\!M_2^2)\,} \hsm -\hsm 2k^2 \,,
\eeq 
where the right-hand side is a monotonically increasing function of $k\hs$.\ 
Thus, taking $\hs k\ito 0$ and $k\ito \infty\hs$ respectively, we derive the
following conditions:
\beqs
\begin{align}
\label{eq:M3-M12-a}
M_3^{} &\geqq M_1^{} \hsm +\hsm M_2^{} \,, \\[1mm]
\label{eq:M3-M12-b}
M_3^{} &\geqq \sqrt{2(M_1^2 \!+\! M_2^2)\,} \,.
\end{align}
\eeqs
It is clear that the condition \eqref{eq:M3-M12-b} is stronger than the condition 
\eqref{eq:M3-M12-a}, and they become equivalent if and only if 
$M_1^{}\hsm\!=\!M_2^{}\,$.\

\vs

For the four-body KK scattering process  
$\,X_{1}^{} X_{2}^{} \hsm\ito\hsm X_{3}^{} X_{4}^{}\,$, 
the 4-momenta in the center-of-mass frame are defined as follows:
\begin{align}
\label{app-eq:4-momenta}
\begin{alignedat}{3}
p_1^{\mu} =& -\!(E_1^{}, 0, 0, k ),
&& \hspace*{10mm}
p_2^{\mu} =  -(E_2^{}, 0, 0, -k ),
\\
p_3^{\mu} =& \,(E_3^{}, k'\hsm\st, 0, k'\hsm\ct ),
&& \hspace*{10mm}
p_4^{\mu} =  (E_4^{}, -k'\hsm\st, 0, -k'\hsm\ct ),
\end{alignedat}
\end{align}
where we define the following Mandelstam variables:
\begin{equation}
s =-( p_1^{} \!+\hsm p_2^{} )^{2} , \qquad
t =-( p_1^{} \!+\hsm p_4^{} )^{2} ,\qquad
u =-( p_1^{} \!+\hsm p_3^{} )^{2} ,
\end{equation} 
from which we have $\hs s +t + u =\!\dis\sum_{j=1}^{4}\!M_{j}^2\hs$. 
In addition, the momenta $k$ and $k'$ in \eqrefe{app-eq:4-momenta} are defined as follows:
\begin{align}
\begin{aligned}
k &= \frac{1}{\,2\sqrt{s\,}\,}\!
\(\hsm\big[s\hsm -\!(M_{1}^{}\!+\!M_{2}^{})^2\big]\big[s\hsm -\!(M_{1}^{}\!-\!M_{2}^{})^2\big]\hsm\)^{\!\frac{1}{2}}\!,
\\
k' &=\frac{1}{\,2\sqrt{s\,}\,}\!
\(\hsm [s\hsm -\!(M_{3}^{}\!+\!M_{4}^{})^2]
[s\hsm -\! (M_{3}^{}\!-\!M_{4}^{})^2]\hsm\)^{\!\frac{1}{2}} \!.
\end{aligned}
\end{align}
We further define the massless Mandelstam variables as $(\sz,\tz,\uz)\!=\!(s,t,u)|_{M_j=0}^{}$, and thus we have:
\begin{equation}
\label{eq:s0-t0-u0}
\sz = 4k^2 \hs, \quad~~
\tz = -\frac{\,\sz\,}{2}(1\!+\hsm\ct)\hs, \quad~~
\uz  = -\frac{\,\sz\,}{2}(1\!-\hsm\ct)\hs,
\end{equation} 
where $j\!=\!1,2,3,4$ and the sum of these Mandelstam variables is given by $\hs \sz+\tz+\uz=0\hs$.

\vs

A massive KK graviton $h_n^{\mn}$ in 4d 
has 5 physical helicity states, and their polarization tensors are represented by
\begin{equation}
\label{eq:Gpol}
\vep^{\mn}_{\pm 2} = \ep_{\pm}^{\mu} \ep_{\pm}^{\nu}\hs, \quad
\vep_{\pm 1}^{\mn} =\!\frac{1}{\sqrt{2\,}\,} (\ep_{\pm}^{\mu}\ep_{L}^{\nu}\!+\!\ep_{L}^{\mu}\ep_{\pm}^{\nu})\hs, \quad
\vep^{\mn}_L  \!= \frac{1}{\sqrt{6\,}\,}
(\ep^\mu_{+}\ep^\nu_{-} \!+\hsm \ep^\mu_{-}\ep^\nu_{+} \!+\hsm 2\hs\ep^\mu_L\ep^\nu_L) \hs,
\end{equation}
where $\ep_{\pm}^\mu$ and $\ep_{L}^\mu$ are the spin-1 polarization vectors:
\begin{equation}
\ep_{\pm}^\mu  \hsm = \pm \frac{e^{\mp\ii\phi}}{\sqrt{2\,}\,} 
( 0, \ct c_{\phi}^{} \pm\!\ii\hs s_{\phi}^{},\hs \ct s_{\phi}^{} \!\mp\!\ii\hs c_{\phi}^{}, -\st ) \hs,\quad~
\ep_L^\mu \!=\hsm \frac{1}{\,\MG_{n}^{}\,}
(k , E_n^{}\st c_{\phi}^{}, E_n^{}\st s_{\phi}^{}, E_n^{}\ct )\,.
\end{equation}
In the above, the KK graviton $h_n^{\mn}$ moves in an arbitrary direction with polar and azimuthal angles $(\theta,\hs \phi)$.\  
For instance, consider the 4-body elastic scattering at the KK level-$n$, we have 
$\,E_{n_j}^{}\!\!=\hsm E\,$ and $\MG_{n_j}^{}\!\!=\hsm M\hs$.\ 
Then, we can define the following polarization vectors in the center-of-mass frame:
\begin{equation}
\begin{alignedat}{3}
\ep^\mu_{1,\hs\pm} &= \frac{1}{\sqrt{2\,}\,} (0,\mp 1, \ii,0)\hs,
\hspace*{8mm}
&& \ep^\mu_{1,L} =  -\frac{E}{M}(\be,0,0,1)\hs,
\\
\ep^\mu_{2,\hs\pm} &= \frac{1}{\sqrt{2\,}\,}(0,\pm 1,  \ii,0)\hs,
\hspace*{8mm}
&&\ep^\mu_{2, L} =- \frac{E}{M}(\be,0,0,-1) \hs,
\\
\ep^\mu_{3,\hs\pm} &= \frac{1}{\sqrt{2\,}\,}(0, \pm \ct, -\ii , \mp\st )\hs,
\hspace*{8mm}
&&\ep^\mu_{3,L} = \frac{E}{M} (\be, \st, 0, \ct ) \hs,
\\
\ep^\mu_{4,\hs\pm} &=\frac{1}{\sqrt{2\,}\,}(0, \mp \ii\hs\ct, -\ii, \pm \st)\hs,
\hspace*{8mm}
&&\ep^\mu_{4,L} = \frac{E}{M} (\be, -\st, 0, -\ct ) \hs,
\end{alignedat}
\end{equation}
where $\hs\be\!=\!(1\!-\!M^2\hsm /E^2)^{1/2}$.\ The polarizations for the inelastic scattering processes can be derived 
in a similar way.

\vspace*{1.5mm}
\section{\hspace*{-2mm}BRST Quantization for GRET Formulation}
\label{app:B}
\vspace*{1mm}

In this Appendix, we provide detailed derivations to support the formulation of 
warped GAET in Section\,\ref{sec:2.2} and the formulation of 
warped GRET in Section\,\ref{sec:2.3}.\  
This includes the BRST quantizations 
used in Sections\,\ref{sec:2.2} and \ref{sec:2.3.1}, the formulation
of the warped GRET type-I presented in Section\,\ref{sec:2.3.3}, 
and the formulation of the warped GRET type-II 
given in Section\,\ref{sec:2.3.2}.\ 

\vspace*{1.5mm}
\subsection{\hspace*{-2mm}BRST Quantization for Warped 5d Gauge Theory}
\label{appx:B.1}
\vspace*{1.5mm}

In this sub-Appendix, we present the detailed derivations of the BRST quantization 
for warped 5d gauge theory, which are needed for the warped GAET formulation 
in Section\,\ref{sec:2.2}.\ 

\vs

We first expand the Lagrangian \eqref{eq:5dYMLa-1} to the quadratic order:
\begin{equation}
\label{Aeq:5dYMLa-2}
\hat{\La}_{\rm{YM}}^{(2)} = - \frac{\,e^{A(z)}\hs}{4} \!\LB\! 
\Big(\pd_{\mu}\hA_{\nu}^a\!-\pd_{\nu}\hA_{\mu}^a\Big)^{\!2}\!\hsm 
+2\!\[\!\Big(\pd_{z} \hA_{\mu}^a\Big)^{\!2}\!
+ \!\Big(\pd_\mu \hA_5^a\Big)^{\!2}\]\!
-4\hs\pd_z^{}\hA_\mu^a \pd^\mu\!\hA_5^a \RB\hsm .
\end{equation}
The mixing term between the 5d components $\hA_\mu^a$ and $\hA_5^a$ 
in Eq.\eqref{Aeq:5dYMLa-2} can be eliminated by constructing 
an $R_\xi^{}$ gauge-fixing term as follows:
\begin{equation}
\label{Aeq:LGF-gauge-5d}
\hat{\La}_{\rm{GF}}^{} = 
-\frac{~e^{A(z)}\hs}{2\hs\xi}\big(\hat\FF^a\big)^{\!2},
~~~~~
\hat\FF^a=\pd^\mu\hsm \hA_\mu^a \!+\hsm \xi ( A'\hsm\!+\hsm\pd_z)\hA_5^a \,,
\end{equation}
where $\hs\xi\hs$ is the gauge-fixing parameter and 
$A'\!\hsm =\hsm\pd_z A(z)\hs$.\ 
Using Eqs.\eqref{Aeq:5dYMLa-2}-\eqref{Aeq:LGF-gauge-5d}, we derive the 5d 
equation of motion (EOM) for the free fields $(\hA_\mu^a,\hA_5^a$):
%
\beqs
\label{Aeq:5dEOM-A} 
\begin{align}
\label{Aeq:5dEOM-Amu} 
[\pd_\nu^2 + (A'\!+\pd_z^{})\pd_z^{}]\hA_\mu^a
-(1\!-\xi^{-1})\pd_\mu^{}\pd_\nu^{}\hA^{a\nu} &=\,0 \hs,
\\
\label{Aeq:5dEOM-A5} 
\pd_\mu^2\hA_5^a+\xi \pd_z^{}(A'\!+\pd_z^{})\hA_5^a 
&=\,0 \hs.
\end{align} 
\eeqs

Then, we compactify the warped 5d space under $S^1\!/\ZZ$
orbifold and impose the Neumann and Dirichlet boundary conditions on
the 5d gauge field components
$\hA_{\mu}^a$ and $\hA_5^a$ respectively, as given in \eqrefe{eq:BC-Amu-A5} of the main text.\
Thus, substituting the KK expansions \eqref{eq:AA5Exp}
into the 5d EOM \eqref{Aeq:5dEOM-A} and using the 4d EOM for free KK gauge fields,
\beqs 
\label{Aeq:4d-EOM-AmuA5}
\begin{align}
(\pd_\nu^2\hsm -\!M_n^2){A_\mu^{an}}
-(1\!-\hsm{\xi^{-1}_n})\pd_\mu^{}\pd_\nu^{}A_n^{a\nu} &=\,0\hs , 	
\\[0.6mm]
(\pd_\nu^2\hsm -\!\Mtt_n^2){A_n^{a5}} &=\, 0\hs , 
\end{align}
\eeqs 
we deduce the following EOMs for the
5d eigenfunctions $\ff_n^{}(z)$ and $\fft_n^{}(z)\hs$:
\beqs
\label{Aeq:EOM-5d-fnfnT}
\begin{align}
\label{Aeq:EOM-fn-Amu}
(A'\hsm +\pd_z) \pd_z \ff_n(z)&= -M_n^2\, \ff_n^{}(z)\hs,
\\[.5mm]
\label{Aeq:EOM-fnT-A5}
\pd_z (A'\hsm +\pd_z) \fft_n^{}(z)&= -{\xi^{-1}_n}\!\Mtt_n^2\, \fft_n^{}(z)\hs,
\end{align}
\eeqs
where $M_n$ and $\Mtt_n$ are the mass eigenvalue of KK level-$n\hs$.\
Applying the partial derivative $\pd_z^{}$
on both sides of Eq.\eqref{Aeq:EOM-fn-Amu}
and comparing it with Eq.\eqref{Aeq:EOM-fnT-A5}, we find that
$\hs\pd_z^{}\ff_n^{}\!\propto\hsm\fft_n^{}$ and
$\Mtt_n^2\!=\!\xin M_n^2$.\
So, with proper normalization we infer the following equations
from the above Eq.\eqref{Aeq:EOM-fn-Amu}-\eqref{Aeq:EOM-fnT-A5}:
\\[-10mm]
\beqs
\begin{align}
\pd_z \ff_n^{}(z) &= -\Mn \hs\fft_n^{}(z) \hs,
\\[.5mm]
(A'\!+\pd_z)\hs \fft_n^{}(z) &= \Mn \ff_n^{}(z) \hs,
\end{align}
\eeqs
which connect 5d wavefunctions $\ff_n^{}(z)$ and $\fft_n^{}(z)$.\
Substituting the KK expansions \eqref{eq:AA5Exp} into Eq.\eqref{eq:BC-Amu-A5}, 
we obtain the boundary conditions imposed on  $(\ff_n^{},\fft_n^{})\hs$:
\begin{equation}
\label{Aeq:BC-YM}
\pd_z^{} \ff_n^{}(z)\big|_{z=0,L}=0\,,  \qquad
\fft_n^{}(z)\big|_{z=0,L}=0 \,.
\end{equation}
They also obey the orthonormal conditions:
\begin{equation}
\label{Aeq:Normalize-YM}
\frac{1}{L\,}\!\hsmx\int_0^{L} \!\!\!\td z \,
e^{A(z)}\hs\ff_n^{}(z)\hs\ff_m^{}(z) \hs =\hs \delta_{nm}\hs,
\hspace*{6mm}
\frac{1}{L\,}\!\hsmx\int_0^{L}\!\!\!\td z \,
e^{A(z)}\hs\fft_n^{}(z)\hs\fft_m^{}(z)\hs =\hs \delta_{nm} \,.
\end{equation}

The 5d wavefunctions $\ff_n^{}(z)$ and $\fft_n^{}(z)$ can be solved 
from the equation of motion \eqrefe{Aeq:EOM-5d-fnfnT}
in terms of Bessel functions of order $\,\nu\hsm =\!1,0\,$:
\beqs
\label{Aeq:fn-fnT}
\begin{align}
\label{Aeq:fn}	
\ff_n^{}(z) &=  \frac{\,e^{-A(z)}\,}{N_n^{}}\!
\[\hsm J_1^{}\!\hsmx\(\!\frac{\Mn}{k}e^{-A(z)}\!\)\! + 
b_{n0}^{}\hs  Y_1^{}\!\hsmx\(\!\frac{\Mn}{k}e^{-A(z)}\!\)\!\]\!,
\\[1mm]
\label{Aeq:fnT}	
\fft_n^{}(z) &=  \frac{\,e^{-A(z)}\,}{{N}_n^{}}\!
\[\hsm J_0^{}\!\hsm\(\!\frac{\Mn}{k}e^{-A(z)}\!\)\! + 
b_{n0}^{} Y_0^{}\!\(\!\frac{\Mn}{k}e^{-A(z)}\!\)\!\]\!,
\end{align}
\eeqs
where the solution $\ff_n^{}(z)$ agrees with  Refs.\,\cite{RS-gauge}\cite{RS-gauge2}.\ 
In the above, the normalization factors $N_n^{}$ 
can be fixed by the orthonormal conditions \eqref{Aeq:Normalize-YM}, 
whereas the coefficient $b_{n0}^{}$ is derived as follows:
\begin{equation}
\label{Aeq:bn0}	
b_{n0}^{}=-\frac{~J_0^{}\!\(\!\frac{\Mn}{k}\!\)~}
{~Y_0^{}\!\(\!\frac{\Mn}{k}\!\)~}\,.
\end{equation}
The KK mass $\Mn$ is determined by roots of the eigenvalue equation:
\begin{equation}
\label{Aeq:YM-Mn}
J_0^{}\hsm\!\(\!\!\frac{\,\Mn}{k} e^{-A(L)}\!\) + b_{n0}^{}\hs  Y_0^{}\hsm\!\(\!\!\frac{\,\Mn}{k}e^{-A(L)}\!\) = 0 \,.
\end{equation}
For the massless zero-mode wavefunction $\ff_0^{}\hs$, we can solve it from 
the eigenvalue equation \eqref{Aeq:EOM-fn-Amu}  and obtain:
\begin{equation}
\ff_0^{} \,=\hs \sqrt{\!\frac{~1\!-\hsm e^{-A(L)}~}{A(L)}}\,.
\end{equation}

Next, given the 5d gauge-fixing \eqref{Aeq:LGF-gauge-5d}, we write down the following 
5d Faddeev-Popov ghost term:
\begin{equation}
\label{Aeq:FP-gauge-5d}
\hat{\La}_{\rm{FP}}^{} = e^{A(z)}\hat{\bar{c}}^a \widehat{\tt s}\hat{\FF}^a \,,
\end{equation}
where $\hs\sss\hs$ is the Becchi-Rouet-Stora-Tyutin (BRST) operator and the 5d BRST transformations
take the form:
\begin{equation}
\label{Aeq:BRST-5d}
\widehat{\tt s}\hat{A}^{aN}\!= D_b^{aN}\!(\hat{A})\hat{c}^b,
~~~~~
\sss\hat{c}^a \!= \Fr{1}{2}\hat{g}_5^{}f^{abc}\hat{c}^b\hat{c}^c,
~~~~~
\sss\hat{\bar{c}}^a \!= -\xi^{-1}\hsm\hat\FF^a .
\end{equation}
We make KK expansions for the 5d ghost fields,
\beqs
\label{Aeq:ghost-exp}
\begin{align}
\hat{c}^a(x, z) &\,=\,
\frac{1}{\sqrt{L\,}\,}
\sum_{n=0}^{\infty} c^{a}_{n}(x)\hs\ff_n^{}(z) \,,
\\
\hat{\bar{c}}^a (x,z) &\,= \,\frac{1}{\sqrt{L\,}\,}
\sum_{n=1}^{\infty}\bar{c}^a_{n}(x) \hs\fft_n^{}(z)\,,
\end{align}
\eeqs
and derive the BRST transformations for the KK fields,
\begin{align}
\begin{alignedat}{3}
\label{Aeq:BRST-KK} 
\widehat{\tt s}{A}^{a\mu}_n &= D^{ab,\mu}_{nm}{c}_m^b\,,
~~~~~
& \widehat{\tt s}{A}^{a5}_n &= D^{ab,5}_{nm}{c}_m^b\,,
\\[1mm]
\sss{c}_n^a &= \Fr{1}{2}{g}\hs f^{abc}
\mathfrak{D}^{ab,m\ell}_{n}{c}_m^b{c}_{\ell}^c\,,
~~~~~
& \sss{\bar{c}}_n^a &= -\xi_n^{-1}\hsm\FF_n^a \,,
\end{alignedat}
\end{align}
where the following notations are defined,
\begin{align}
D^{ab,\mu}_{nm} &= -\delta^{ab}\delta_{nm}^{}\pd^\mu +
\(\!\!\frac{1}{\,L\,}\!\int_0^L\!\!\td z\,e^{2A}\hs
\ff_n^{}\ff_m^{}\ff_{\ell}^{}\!\)\!
g\hs f^{abc}A^{c\mu}_{\ell} \,,
\nn\\
\label{eq:DDD-KKghosts}
D^{ab,5}_{nm} & = -\delta^{ab}\delta_{nm}^{}\Mm +
\(\!\!\frac{1}{\,L\,}\!\int_0^L\!\!\td z\,e^{2A}\hs
\fft_n^{}\ff_m^{}\fft_{\ell}^{}\!\)\!
g\hs f^{abc}A^{c5}_{\ell} \,,
\\
\mathfrak{D}^{ab,m\ell}_{n} &=
\frac{1}{\,L\,}\!\int_0^L\!\!\td z\,e^{2A}\hs
\ff_n^{}\ff_m^{}\ff_{\ell}^{} \,.
\nn
\end{align}

\vspace*{1.5mm}
\subsection{\hspace*{-2mm}BRST Quantization for Warped 5d Gravity}
\label{sec:B.1}
\label{appx:B.2}
\vspace*{1.5mm}

In this sub-Appendix, we present the detailed derivations of the BRST quantization for
the warped 5d gravity theory, which are needed for the GRET formulation 
presented in Section\,\ref{sec:2.3} of the main text.\ 

\vs 

We derive the warped 5d GR Lagrangian to the quadratic order which takes the following form: 
\\[-7mm]
\begin{align}
\label{Aeq:LagLOExp}
\hat\La_{\rm{RS}}^{(2)} = &~ e^{3A(z)}\bigg\{\! -\!\frac{1}{\,4\,}\hh^{\mn}\Big\{
\!\big(\eta_{\mu\al}^{}\pd_\nu^{}\pd_\be^{}+\eta_{\mu\be}^{}\pd_\nu^{}\pd_\al^{}
+\eta_{\nu\al}^{}\pd_\mu^{}\pd_\be^{} 
+\eta_{\nu\be}^{}\pd_\mu^{}\pd_\al^{}\big)\!
-2\big(\eta_{\mn}^{}\pd_\al^{}\pd_\be^{}
\nn\\
&\, +\eta_{\ab}^{}\pd_\mu^{}\pd_\nu^{}\big)\!
-(\eta_{\mu\al}^{} \eta_{\nu\be}^{}+\eta_{\mu\be}^{}\eta_{\nu\al}^{}
-2 \eta_{\mn}^{}\eta_{\ab}^{})\Big[\pd^2\!+\!(3A'\!+\pd_z^{})\pd_z^{}\Big]\Big\} \hh^{\ab}
\nn\\
&\, + \hat{\VV}^\mu\big(\eta_{\mn}^{} \pd^2\!-\pd_\mu^{}\pd_\nu^{}\big) \hat{\VV}^\nu\!
+\!\frac{3}{\,4\,}\hat{\phi}\Big[\pd^2\!-2(A'\!+\hsm\pd_z^{})(2A'\!+\hsm\pd_z^{})\Big]\hat{\phi} 
\nn\\
&- \!\frac{1}{\,2\,}\Big[\hh^{\mn}(\eta_{\mu\al}^{}\pd_\nu^{}\hsm +\eta_{\nu\al}^{} \pd_\mu^{}\!
-2\eta_{\mn}^{}\pd_\al^{})(3A'\!+\pd_z^{})\hat{\VV}^\al\! 
+\hsm \hat{\VV}^\mu(\eta_{\mu\al}^{}\pd_\be^{}\!+\hsm\eta_{\mu\be}^{} \pd_\al^{}
\nn\\
& -2\hs \eta_{\ab}^{}\pd_\mu^{}) \pd_z^{}\hh^{\ab}\Big]
+\frac{3}{\,4\,} \Big[\hh(3 A'\!+\hsm\pd_z^{})(2A'\!+\hsm\pd_z^{}) \hat{\phi}
+\hat{\phi}(A'\!+\hsm\pd_z^{}) \pd_z^{} \hh\Big]
\nn\\
& -\!\frac{3}{\,2\,}\Big[\hat{\VV}^\mu (2 A'\!+\pd_z^{})\pd_\mu^{}\hat{\phi}
+\hat{\phi}(A'\!+\hsm\pd_z^{})\pd_\mu^{}\hat{\VV}^\mu\Big] \bigg\}\hs .
\end{align}
This agrees with the literature\,\cite{Chacko:2003yp,Callin:2004zm,Gherghetta:2005se,Lim:2008}.\
To eliminate the mixing terms appearing in Eq.\eqref{Aeq:LagLOExp}, 
we choose the gauge-fixing terms as follows:
\begin{equation}
\label{Aeq:L-GF-KKgravity} 
\hat{\La}_{\rm{GF}}^{} = 
-\frac{\,e^{3A(z)}\,}{\xi} \big(\hat{\FF}_\mu^2 + \hat{\FF}_5^2\,\big)\hs,
\end{equation}
where $\,\xi\,$ is a gauge-fixing parameter and the gauge-fixing functions 
$\hat{\FF}_\mu^{}$ and $\hat{\FF}_5^{}$ take the following form:
\\[-7mm]
\beqs
\label{Aeq:FmuF5-5d}
\begin{align}
\hat{\FF}_\mu^{} &=\pd^\nu \hh_{\mn}
\!-\!\(\!1\!-\!\frac{1}{\,2\xi\,}\)\!\pd_\mu \hh
+\xi (3A'\!+\hsm\pd_z^{}) \hat{\VV}_{\mu} \,,
\\[1mm]
\hat{\FF}_5^{} &=\frac{1}{\,2\,}\! 
\[\pd_z^{} \hh-3\hs\xi (2A'\!+\pd_z) \hat{\phi}-\hsm 2\pd_\mu^{}\hsm \hat{\VV}^{\mu}\]\!.
\end{align}
\eeqs
In the above, we denote the gravitational gauge-fixing parameter 
by $\,\xi\,$ for notational convenience, 
but it is independent of the gauge-fixing parameter $\,\xi\,$ 
for gauge theories as defined in Eq.\eqref{Aeq:LGF-gauge-5d}.\

For the BRST quantization approach, the ghost fields $(\hat{c}^M\!,\hs\hat{\bar{c}}^M)$ 
are introduced in the path intergral formulation.\
The 5d Faddeev-Popov ghost Lagrangian for the 5d GR with warped metric takes the following form:
\begin{equation}
\hat{\La}_{\rm{FP}}^{} \,=\, e^{3A(z)}\, \hat{\bar{c}}^M \sss \hat{\FF}_M^{} \hs,
\end{equation}
where the 5d gauge-fixing functions $\hat{\FF}_M^{}\!\!=\hsm\!(\hat{\FF}_\mu^{},\hat{\FF}_5^{})$ 
are given by \eqrefe{eq:FmuF5-5d}.\
The BRST transformations for 5d graviton, ghost and anti-ghost fields take the following form:
\begin{align}
\sss\hat{h}_{MN}^{} &= -\pd_M^{} \hat{c}_N^{} - \pd_N^{} \hat{c}_M^{} + \hka\!\(\hh_{MN}^{} \pd_P^{} - \hh_{MP}^{} \pd_N^{} - \hh_{NP}^{} \pd_M^{} \)\!\hat{c}^P ,
\nn\\[1mm]
\sss\hat{c}_M^{} &= \hka\hs\hat{c}_N^{} \pd^N \hat{c}_M^{} \,, \qquad
\sss\hat{\bar{c}}_M^{} = -2\hs\xi^{-1} \hat{\FF}_M^{} \,.
\end{align}
The BRST transformations exhibit nilpotency and obey $\sss^2\!=0\,$.

\vs

Using the KK decompositions \eqref{eq:5d-BC-hVphi} in the main text, 
we derive the EOMs for the 5d gravitational KK wavefunctions as follows:\footnote{%
Note that our analyses below and elsewhere in this work are self-contained and
well understood.\ It does not invoke any additional symmetries such as SUSY and alike.}\
\beqs
\label{Aeq:EOM-GR}
\begin{align}
\label{Aeq:EOM-GR-un}
(3A'\!+\hsm \pd_z^{}) \pd_z \uu_n^{}(z) &= -\MGnn \,\uu_n^{}(z)\hs,
\\[.5mm]
\label{Aeq:EOM-GR-vn-1}
\pd_z^{}(3A'\!+\hsm\pd_z^{})\vv_n^{}(z) &=-\MGnn \,\vv_n^{}(z)\hs,
\\[.5mm]
\label{Aeq:EOM-GR-vn-2}
(2 A'\!+\hsm\pd_z^{})(A'+\pd_z) \vv_n^{}(z) &=  -\MGnn \,\vv_n^{}(z)\hs,
\\[.5mm]
\label{Aeq:EOM-GR-wn}
(A'\!+\hsm\pd_z^{})(2 A'\!+\hsm\pd_z^{}) \ww_n^{}(z) &=  -\MGnn \,\ww_n^{}(z)\hs,
\end{align}
\eeqs
where the 5d wavefunctions satisfy the following orthonormal conditions, 
\begin{align}
\label{Aeq:Normalization-GR}
\frac{1}{\,L\,}\!\int_0^{L}\!\!\!\td z \, e^{3A(z)}\hs\chi_n^{}(z)\,\chi_m^{}(z)
= \delta_{nm}^{}\,,
\end{align}
with the notation 
$\chi_n^{}\!=\hsm\{\uu_n^{},\hs \vv_n^{},\hs \ww_n^{}\}\hs$.\
In the above and hereafter,
we use the notation $\MGn$ to denote the mass eigenvalue of KK gravitons at level-$n\hs$,
which differs from the KK gauge boson mass eigenvalue $\Mn$ of level-$n\hs$
and is in contrast to the case of compactified flat 5d extra dimension.\
In addition, Eqs.\eqref{Aeq:EOM-GR-vn-1}-\eqref{Aeq:EOM-GR-vn-2} are actually equivalent
and we give both formulas for convenience of later analysis.\

\vs

Moreover, the 5d wavefunctions $\uu_n^{}(z)$ and $\vv_n^{}(z)$ are connected to each other
by the following equations,
\\[-9mm]
\beqs
\label{Aeq:uv}
\begin{align}
\label{Aeq:uv-1}
\pd_z^{} \uu_n^{}(z) &= -\MGn \vv_n^{}(z) \,,
\\[.5mm]
\label{Aeq:uv-2}
(3A'\!+\pd_z^{})\vv_n^{}(z) &= \MGn \uu_n^{}(z) \,,
\end{align}
\eeqs
whereas the 5d wavefunctions $\ww_n^{}(z)$ and $\vv_n^{}(z)$ are connected by the equations,
\beqs
\label{Aeq:wv}
\begin{align}
\label{Aeq:wv-1}
(2A'\!+\pd_z^{})\ww_n^{}(z) &= -\MGn \vv_n^{}(z) \hs,
\\[0.5mm]
\label{Aeq:wv-2}
(A'\!+\pd_z^{})\vv_n^{}(z) &= \MGn \ww_n^{}(z) \hs.
\end{align}
\eeqs
By using Eqs.\eqref{Aeq:uv-2} and \eqref{Aeq:wv-2},
we can derive the following identity connecting the three types of wavefunctions:
\\[-3mm]
\begin{equation}
\label{Aeq:un=wn+vn}
\uu_n^{}(z) = \ww_n^{}(z) + \frac{\,2A'\,}{\MGn}\vv_n^{}(z)\,.
\end{equation}
In passing, the relation of Eq.\eqref{Aeq:un=wn+vn} to the
quantum-mechanical SUSY algebra was discussed in the literature\,\cite{SUSY}.

\vs

Solving the equation of motion \eqref{Aeq:EOM-GR}, we can express 
the 5d wavefunctions $\uu_n^{}(z),\hs\vv_n^{}(z)$ and $\ww_n^{}(z)$
in terms of the Bessel functions of order
$\nu\hsm =\hsm 2,1,0\hs$ respectively:
\beqs
\label{Aeq:KK-uvw} 
\begin{align}
\uu_n^{}(z) &\,=\,
\frac{\,e^{-2A(z)\,}}{N_n^\pp}\!\[J_2^{}\!\(\!\frac{\,\MGn}{k}e^{-A(z)}\!\)\! +
b_{n1}^{} Y_2^{}\!\(\frac{\MGn}{k}e^{-A(z)}\!\)\] \!,
\\[1mm]
\vv_n^{}(z) &\,=\,
\frac{e^{-2A(z)}}{N_n^\pp}\[J_1^{}\!\(\!\frac{\,\MGn}{k}e^{-A(z)}\!\)\! +
b_{n1}^{} Y_1^{}\!\(\frac{\,\MGn}{k}e^{-A(z)}\!\)\]\!,
\\[1mm]
\ww_n^{}(z) &\,=\,
\frac{e^{-2A(z)}}{N_n^\pp}\[J_0^{}\!\(\,\frac{\MGn}{k}e^{-A(z)}\!\)\! +
b_{n1}^{}Y_0^{}\!\(\frac{\,\MGn}{k}e^{-A(z)}\!\)\]\!,
\end{align}
\eeqs
where $N_n^\pp$ is the normalization factor and $b_{n1}^{}$ is given by
\begin{equation}
\label{Aeq:KK-bn1}
b_{n1}^{}\,=\hs -\frac{\,J_1^{}\!\hsm\(\!\frac{\,\MGn\,}{k}\!\)}
{~Y_1^{}\!\hsm\(\!\frac{\,\MGn\,}{k}\!\)\,} \,.
\end{equation}
The KK mass eigenvalue $\MGn$ is determined by solving the following equation:
\begin{equation}
\label{Aeq:GR-Mn}
J_1^{}\!\hsm\(\!\frac{\,\MGn}{k} e^{-A(L)}\!\)\! +
b_{n1}^{} Y_1^{}\!\hsm\(\!\frac{\,\MGn}{k}e^{-A(L)}\!\) = 0 \,.
\end{equation}
We note that the KK graviton mass $\MGn$ solved from the above eigenvalue equation \eqref{Aeq:GR-Mn}
differs from the KK gauge boson mass $\Mn$ solved from Eq.\eqref{Aeq:YM-Mn}.\
Then, the zero-mode wavefunctions $\uu_0^{}(z)$ and $\ww_0^{}(z)$
can be solved as follows:
\beqs
\label{Aeq:u0-w0}
\begin{align}
\label{Aeq:u0-def}
\uu_0^{} &\,= \sqrt{2\,}\hsm\big[e^{A(L)}\!+e^{2A(L)}\big]^{\hsm -\frac{1}{2}}\,,
\\
\label{Aeq:w0-def}
\ww_0^{}(z) &\,= \sqrt{2\,}\hs e^{-2A(z)}\hsm\big[1\!+e^{-A(L)}\big]^{\hsm -\frac{1}{2}},
\end{align}
\eeqs
where $\uu_0^{}$ has no dependence on $z\hs$.\

\vs

Then, we make KK expansions for the 5d Lagrangian \eqref{Aeq:LagLOExp}  
and integrate $z$ over the interval $[\hs 0,\hs L\hs]$, 
with which we deduce the 4d KK Lagrangian to quadratic order as follows: 
\begin{align}
\label{Aeq:LagLOKK}
\La^{(2)}_{\rm{RS}}
\,=\, & \sum_{n=0}^{\infty}\bigg[
\fr{1}{\,2\,}(\pd_\mu h_n^{})^2\!
+\! \fr{1}{\,2\,} \MGnn(h_n^{})^2\!
-\!\fr{1}{\,2\,}(\pd^\rho h^{\mn}_n)^2\!
- \!\fr{1}{\,2\,} \MGnn(h^{\mn}_n)^2\!
-\!(\pd^\mu \VV^{\nu}_n)^2\!
\nn\\[0mm]
& \hspace*{8.6mm}
+\! (\pd_\mu^{} \VV_n^{\mu})^2\! 
-\!\frac{3}{\,4\,}(\pd_{\mu}^{}\phi_n^{})^2
+\frac{3}{2} \MGnn\phi_n^2
-\pd_\mu^{} h^{\mn}_n\pd_\nu^{} h_n^{}\! 
+\pd_\mu^{} h^{\mu\rho}_n\pd^\nu h_{\nu\rho,n}^{}
\nn\\[0mm]
& \hspace*{8.6mm}
+2 \MGn h_n^{} \pd_\mu^{} \VV^{\mu}_n\!
- 2 \MGn  h^{\mn}_n \pd_\mu \VV_{\nu,n}^{}\!
- \!\frac{3}{\,2\,} \MGnn h_n^{} \phi_n^{}\!
-3 \MGn \pd_\mu^{} \VV^{\mu}_n\phi_n^{}\bigg] .
\end{align}
From the above, we can construct the following 4d $R_{\xi}^{}$ gauge-fixing terms 
for the KK Lagrangian:
\begin{equation}
\label{Aeq:GF-4d}
\La_{\rm{GF}}^{} = \int_0^L \!\!\td z\, \hat{\La}_{\rm{GF}}^{} 
\hs =\hs -\sum_{n=0}^{\infty}\!\frac{1}{~\xin \hs}\!
\[(\FF_n^\mu)^2\!+\!(\FF_n^5)^2\] \!,
\end{equation}
where $\hs\xin$ is the gauge-fixing parameter for each KK level-$n\hs$, and 
the gauge-fixing functions $\FF_n^\mu$ and $\FF_n^5$ take the following forms:
\\[-6mm]
\beqs
\label{eq:Fmun-F5n}
\begin{align}
\label{eq:Fmun}
\FF_n^\mu &\,=\,\pd_{\nu} h_n^{\mn}\! 
-\!\(\!1\!-\!{\frac{1}{\,2\hs\xin\,}}\!\)\!\pd^{\mu} h_n^{}
\!+\hsm \xin \MGn \VV^{\mu}_n  \,,
\\[0mm]
\label{eq:F5n}
\FF_n^5 &\,=\hs {\frac{1}{2}}\! \(\MGn h_n^{}\!-\! 3\xin\MGn\phin\!+\! 2\pd_{\mu}^{} \VV^\mu_n\)\!.
\end{align}
\eeqs
They take the same forms as what we obtained 
for the flat 5d compactification\,\cite{Hang:2021fmp}\cite{Hang:2022rjp}.\
The above gauge-fixing terms are designed 
to cancel the quadratic mixing terms between the KK graviton fields
$h_n^{\mn}$ and the KK gravitational Goldstone fields $\VV_n^\mu$ and $\phin\hs$.\ 
To ensure proper normalization of the kinetic terms of $\VV_n^\mu$ and $\phin$,
we make the following rescaling,
\\[-3.5mm]
\begin{align}
\label{Aeq:rescaling-Vn-phin}
\VV^{\mu}_n \to \frac{1}{\sqrt{2\,}\,} \VV^{\mu}_n  \,, \quad~~~
\phin  \to \sqrt{\frac{2}{3}\,}\phin  \,,
\end{align}
under which the gauge-fixing functions \eqref{eq:Fmun}-\eqref{eq:F5n} become,
\beqs
\label{Aeq:Fmun-F5n2}
\begin{align}
\label{Aeq:Fmun2}
\FF_n^\mu &\,=\,\pd_{\nu} h_n^{\mn}\! -\!\(\!1\!-\!
{\frac{1}{\,2\hs\xin}\,}\)\!\pd^{\mu} h_n^{}
\!+\hsm\frac{1}{\sqrt{2\,}\,} \xin \MGn \VV^{\mu}_n  \,,
\\[0mm]
\label{Aeq:F5n2}
\FF_n^5 &\,=\hs {\frac{1}{\hs 2\hs}}\,
\!\MGn h_n^{}\!-\! {\sqrt{\frac{3}{2}}}\, 
\xin\MGn\phin \!+\!{\frac{1}{\sqrt{2\,}\,}}\, \pd_{\mu}^{}\VV^\mu_n\,.
\end{align}
\eeqs
Thus, we can derive the propagators of KK fields $(h_n^{\mn},\hs \VV_n^\mu,\hs\phi_n^{})$
at tree level (as in \cite{Hang:2021fmp}) which take the simple diagonal forms 
in the 't\,Hooft-Feynman gauge
($\xi_n^{}\!=\hsm 1$)\hs:
\beqs
\label{Aeq:KKpropagator-FHooft} 
\begin{align}
\D_{h,nm}^{\mn\ab}(k)
&= -\frac{\,\ii\dnm\,}{2}
\frac{\,\eta^{\mu \al}\eta^{\nu \be}\!+\!\eta^{\mu \be}\eta^{\nu\al}\!-\!\eta^{\mu\nu}\eta^{\al\be}\,}{\,k^{2}\!+\!\MGnn\,} \,,
\label{Aeq:Prophh-xi=1} 
\\
\label{Aeq:Prophmu5-xi=1}
\D^{\mn}_{\VV,nm}(k) &=
-\frac{~\ii\eta^{\mn}\dnm~}{\,k^{2}\!+\!\MGnn\,} \,,
\\
\label{Aeq:Prop55-xi=1}
\D_{\!\phi,nm}^{}(k) &=
-\frac{\,\ii \dnm\,}{~k^2 \!+\! \MGnn~} \,.
\end{align}
\eeqs

Next, we make KK expansions for the 5d ghost and anti-ghost fields as follows:
\beqs
\begin{align}
\hat{c}^\mu (x,z) &= \frac{1}{\sqrt{L\,}\,} \!\sum_{n=0}^{\infty} \!c^\mu_n(x)\hs\uu_n(z) \hs,\qquad
\hat{\bar{c}}^\mu (x,z) =
\frac{1}{\sqrt{L\,}\,}\!\sum_{n=0}^{\infty} \!\bar{c}_n^\mu(x)\hs\uu_n(z)\hs,
\\
\hat{c}^5 (x, z) &=
\frac{1}{\sqrt{L\,}\,}\!\sum_{n=1}^{\infty} \!c_n^5(x)\hs\vv_n(z) \hs,\qquad
\hat{\bar{c}}^5 (x, z)=
\frac{1}{\sqrt{L\,}\,}\!\sum_{n=1}^{\infty} \!\bar{c}_n^5 (x)\hs\vv_n(z) \hs,
\end{align}
\eeqs
where we use the fact that 
the 5d ghost fields $(\hat{c}^{\hs\mu},\hat{c}^{\hs 5})$ 
share the same $\ZZ$ parities as that of the gravitational fields 
$(\hat{h}^{\mn},\hs \hat{\VV}^{\mu})$ respectively, 
as shown by Eqs.\eqref{eq:5d-BC-hVphi} and \eqref{eq:5d-BC-cmu-c5} of the main text.\  
Using the KK expansions \eqref{eq:KK-Exp-GR}, 
we derive the BRST transformations for the KK fields in 4d spacetime:
\beqs
\label{Aeq:BRST-KK-Fields}
\begin{align}
\label{Aeq:BRST-h}
\sss h_n^{\mn} &= -\,\pd^\mu c^\nu_n \!-\!\frac{1}{2}\eta^{\mn}\MGn c^5_n
\!+\!\frac{\ka}{2}\lrb{\uu_n^{}\uu_m^{}\uu_\ell^{}} (h^{\mn}_m\pd^\al \!-\! 2h^{\mu\al}_{m}\pd^\nu)c_{\al}^\ell
\!+\!\frac{\ka}{2}\lrb{\uu_n^{}\uu_m^{}\vv_\ell^{}}\MGl h^{\mn}_m c^5_\ell
\nn\\
&\quad\,+\!\frac{\ka}{\,2\sqrt{2\,}\,}\lrb{\uu_n^{}\vv_m^{}\uu_\ell^{}}\eta^{\mn}\MGl\VV^\al_m c_{\al}^{\ell}
\!-\!\frac{\ka}{\sqrt{2\,}\,}\lrb{\uu_n^{}\vv_m^{}\vv_\ell^{}}\VV^\mu_m \pd^\nu c^5_\ell
\!+\!\frac{\ka}{\sqrt{6\,}\,}\big(\lrb{\uu_n^{}\ww_m^{}\uu_\ell^{}}\phim\pd^\mu c^\nu_\ell
\nn\\
&\quad\, -\lrb{\uu_n^{}\ww_m^{}\vv_\ell^{}}\eta^{\mn}\MGl\phim c^5_\ell\big) + (\mu \leftrightarrow \nu )  \hs,
\\[1mm]
\label{Aeq:BRST-V}
\sss\VV_n^\mu &= -\sqrt{2}\hs\pd^\mu c^5_n
\hsm +\hsm\sqrt{2\,}\hs\MGn c_n^\mu \!+\hsm \sqrt{2\,}\hs\ka\lrb{\vv_n^{}\uu_m^{}\uu_\ell^{}}\MGl h^{\mn}_m c_{\nu}^\ell
\!+\hsm\ka \lrb{\vv_n^{}\vv_m^{}\uu_\ell^{}} \big(\VV_m^\mu \pd^\nu \!-\hsm \VV_m^\nu \pd^\mu \big)c_{\nu}^\ell
\nn\\
&\quad\,
-\hsm\frac{\ka}{\sqrt{3\,}\,}(\lrb{\vv_n^{}\ww_m^{}\uu_\ell^{}} \MGl\phim c^\mu_\ell
\!+\hsm 2\lrb{\vv_n^{}\ww_m^{}\vv_\ell^{}} \phim \pd^\mu c^5_\ell)\hs,
\\[1mm]
\label{Aeq:BRST-phi}
\sss\phin &= -\sqrt{6\,}\hs\MGn c^5_n
\!+\!\sqrt{3\,}\hs\ka \lrb{\ww_n^{}\vv_m^{}\uu_\ell^{}}\MGl \VV_{m}^{\mu} c_\mu^\ell
\!+\hsm\ka\lrb{\ww_n^{}\ww_m^{}\uu_\ell^{}} \phim \pd_\mu^{} c^\mu_\ell
\nn\\
&\quad\,-\ka\lrb{\ww_n^{}\ww_m^{}\vv_\ell^{}}\MGl\phim c^5_\ell  \,,
\\[1mm]
\label{Aeq:BRST-ghost}
\sss c^\mu_n &= \ka\lrb{\uu_n^{}\uu_m^{}\uu_\ell^{}} c^\nu_m \pd_\nu^{} c^\mu_\ell
-\ka\lrb{\uu_n^{}\vv_m^{}\uu_\ell^{}}\MGl c^5_m c^\mu_\ell \,,
\\[1mm]
\sss c^5_n &= \ka\lrb{\vv_n^{}\uu_m^{}\vv_\ell^{}} c^\mu_m \pd_\mu^{} c^5_\ell
+\ka\lrb{\vv_n^{}\vv_m^{}\vv_\ell^{}}\MGl c^5_m c^5_\ell \,,
\\[1mm]
\label{Aeq:BRST-antighost}
\sss\bar{c}^\mu_n &= -\frac{2}{\,\xin\,} \FF_n^\mu \,, \qquad
\sss \bar{c}^5_n = -\frac{2}{\,\xin\,} \FF_n^5 \,,
\end{align}
\eeqs
where $\hs\ka=\hka/\hsm\sqrt{L}$ and the brackets $\lrb{\cdots}$ are defined in 
Eq.\eqref{app-eq:def-[[]]} of Appendix\,\ref{app:D}.

\vspace*{2mm}
\subsection{\hspace*{-2mm}Gravitational ET of Type-I}
\label{app:B.2}
\label{appx:B.3}
\vspace*{1.5mm}

In the Sections\,\ref{sec:2.3.2}$-$\ref{sec:2.3.3} of the main text, 
we have formulated the GRET which quantitatively connects 
each scattering amplitude of KK gravitons $h^{\mn}_n$ 
with helicities $(0,\hs\pm1)$ to that of the corresponding 
gravitational KK Goldstone bosons $(\phin,\hs\VV_n^{\pm 1})$ respectively.\ 
In this and next sub-Appendices, 
we provide detailed derivations to support the analyses presented in the main text.\   

\vs

We study a combination of gauge-fixing functions,  
$\pd_\mu^{}\FF_n^\mu \!-\xin\MGn\FF_n^5\,$,  
which eliminates the KK vector-Goldstone field $\,\VV_n^{\mu}\,$.\   
For this we derive the following formula in the momentum space:
\begin{equation}
\label{eq:app-Fmu-F5-0}
-\ii k_\mu^{}\FF_n^\mu \!-\xin\MGn\FF_n^5 \,=\,
-k_\mu^{}k_\nu^{} h_n^{\mn}
\!+\Fr{1}{2}\!\[\!(2\!-{\xi_n^{-1}})k^2\!-\xin\MGnn\]\!h_n^{}
\!+\!\sqrt{\!\Fr{3}{2}\,}{\xi_n^2}\MGnn\phin \,,
\end{equation}
where the rescaling $\phin\hsm\ito\hsm\sqrt{\!\Fr{2}{3}}\phin$ is made according to Eq.\eqref{Aeq:rescaling-Vn-phin}.\ 
Applying the on-shell condition $k^2\!=\!-\MGnn\hs$, 
we further derive Eq.\eqref{eq:app-Fmu-F5-0} in the following form:
\beqs 
\label{eq:app-Fmu-F5-1}
\begin{align}
\label{eq:app-Fmu-F5-1a}
&\ii k_\mu^{}\FF_n^\mu \!+ \xin\MGn\FF_n^5 
\,=\sqrt{\!\Fr{3}{2}\,}\MGnn\hs\FT_n^{}\,,
\\
\label{eq:app-Fmu-F5-1b}
& \FT_n^{} = \sqrt{\!\Fr{2}{3}\,}h_n^{S}
+\Fr{1}{\sqrt{6\,}\,}\big(2\hsm +\hsm\xi_n^{}\!-\xi_n^{-1}\big)h_n^{}-\xi_n^2\phin
=\mathbf{K}^T_n\mathbf{H}_n^{}
\\
\label{eq:app-Fmu-F5-1c}
&\hspace*{4.1mm}
= \sqrt{\!\Fr{2}{3}\,}\big(h_n^{S}
+h_n^{}\big)\!-\phin\hs, \hspace*{7mm}
(\text{for}~\xin\!=\! 1)\hs,
\\
& \mathbf{K}^{}_n\!= \Big(\!\sqrt{\!\Fr{2}{3}\,}\vep^S_{\mn}
\!+\!\Fr{1}{\sqrt{6\,}\,}\big(2\hsm +\hsm\xi_n^{}\!-\xi_n^{-1}\big)\eta_{\mn}^{},\,-\xi_n^2
\Big)^{\!T}\hsm\!,~~~~ \mathbf{H}_n^{} = (h_n^{\mn},\,\phi_n^{})^{T},
\\
&\hspace*{4.1mm}
= \Big(\!\sqrt{\!\Fr{2}{3}\,}\!\big(\vep^S_{\mn}\hsm +\eta_{\mn}^{}\big),\hs -1\hsm\Big)^{\!T} \!, 
\hspace*{7mm}
(\text{for}~\xin\!=\! 1)\hs.
\end{align}
\eeqs
We then derive a BRST identity involving the gauge-fixing function $\FT_n^{}$ in \eqrefe{eq:app-Fmu-F5-1b}:
%
\begin{equation}
\bla0\big|T\, \FT_n^{}\bH_m^{T} \big|0\bra (k) = -
\frac{\xin}{\sqrt{6\,}\hs\MGnn} 
\bla 0\big|T\, \sss\bH_m^{T} (\ii k_\mu\bar{c}^\mu_n\hsm +\hsm\xin\MGn\bar{c}_n^5) \big| 0\bra(k) \,,
\end{equation}
and it can be further rewritten in the following form by utilizing \eqrefe{eq:app-Fmu-F5-1}$\hs$:
\begin{equation}
\label{Aeq:app-KD-X-2} 
\bK_n^T\bD_{nm}^{}(k)=-\bX_{nm}^T(k)\,.
\end{equation}
In the above, we adopt for simplicity the 't\,Hooft-Feynman gauge condition $(\xin\hsm\!=\hsm\!1)$.\ 
Thus, in \eqrefe{Aeq:app-KD-X-2}, we define the following notations:
%
\begin{align}
\bD_{nm}^{}(k) &= \bla0\big|T \bH_n^{} \bH_m^{T} \big|0\bra(k) \,,\quad
\bX_{nm}^{}(k) = \underline{\bX}_{mj}^{}(k) \SS_{jn}^{}(k)\hs,
\nn\\[1mm]
\underline{\bX}_{mj}(k)
&=\frac{1}{2\sqrt{6\hs}\hs{\MGjj}\,}\hsm\!
\begin{pmatrix}
\bla 0\big|T\,\sss h^{\mn}_m\big|\bar{\mathsf{c}}_j^{}\bra(k)
\\[1mm]
\bla 0\big|T\,\sss \phim\big|\bar{\mathsf{c}}_j^{}\bra(k)
\end{pmatrix}
\!=\!{\frac{1}{2\sqrt{6\hs}\,}}\hsm\!
\begin{pmatrix}\!
(2k^\mu k^\nu\!/{\MGjj}\!-\!\eta^{\mn})[\delta_{mj}^{}\!+\!\Delta^{\!(3)}_{mj}(k^2)]
\\[1mm]
-\sqrt{6\,}[\delta_{mj}^{}\!+\!\widetilde{\Delta}_{mj}^{\!(4)}(k^2)]
\end{pmatrix}\hsm\!,
\nn\\[1mm]
\SS_{jn}^{}(k) &= \bla 0\big|T {\mathsf{c}_j^{} \bar{\mathsf{c}}_n^{}} \big| 0\bra(k)\hs, 
\label{Aeq:app-X-Delta34}
\end{align}
%
with $\,{\mathsf{c}}_n^{}\!\equiv\!\ii\hs \ep_\mu^S{c}^\mu_n\hsm +\hsm {c}_n^5\hs$,  
$\hs\bar{\mathsf{c}}_n^{}\!\hsm\equiv\hsm\ii\hs \ep_\mu^S\bar{c}^\mu_n\hsm + \bar{c}_n^5\hs$, 
and $\ep_\mu^S\hsm\!=\!k_\mu^{}/\MGn\hs$.\
In the above, each external momentum is chosen to be incoming.\  
The formulas \eqref{Aeq:app-KD-X-2}-\eqref{Aeq:app-X-Delta34} 
just give Eq.\eqref{eq:KD-X-2} in the main text.\ 
In addition, we see that the quantities $\Delta^{\!(3)}_{mj}(k^2)$ and $\widetilde{\Delta}^{\!(4)}_{mj}(k^2)$ 
are of loop order and are generated by the non-linear terms of the BRST transformations 
\eqref{Aeq:BRST-h} and \eqref{Aeq:BRST-phi} of $h_n^{\mn}$ and $\phin\hs$.\ 

\vs 

Next, we use the identity \eqref{eq:F-identityP} and Eq.\eqref{eq:app-Fmu-F5-1} to deduce an identity 
containing the external state $\FT_n^{}(k)\hs$: 
\begin{align}
\label{Aeq:app-Fn-KD-X}
0 &= \bla 0\hs |\,\FT_n^{}(k)\cdots {\Phi}\,|\hs 0\bra 
= \mathbf{K}^T_n \bD_{nm}^{}(k)
\bla 0\hs |\,\under{\mathbf{H}}_m^{}(k)\cdots {\Phi}\,|\hs 0\bra 
\nn\\
& = -\bX_{nm}^{T}(k)\hs
\M\hsm\big[\under{\mathbf{H}}_m^{}(k),\cdots\!,{\Phi}\hs\big], 
\end{align}
which leads to the following identity,
\begin{align}
\label{Aeq:app-Fn-AmpLSZ}
\M\hsm\big[\under{\FT}_n^{}\hsm (k),\cdots\!,{\Phi}\hs\big]= 0 \,,
\end{align}
with the amputated external state 
$\under{\FT}_n^{}\hsm (k)$ given by\,\footnote{%
We note that after the LSZ amputation in general $\Rxi$ gauge at tree level, 
the coefficient of $h_n^{}$ in $\underline{\FT}$ 
should be $-1/2$ and corrects the coefficient of $h_n^{}$ 
as given in Refs.\,\cite{Hang:2021fmp}\cite{Hang:2022rjp},
but this does not affect all the conclusions therein\,\cite{Hang:2021fmp}\cite{Hang:2022rjp}.}
\beqs 
\label{eq:app-Fn-2}
\begin{align}
\label{Aeq:app-Fn-Gfin}
\under{\FT}_n^{} &= 
\sqrt{\!\Fr{2}{3}\,}(h_n^S\!-\!\Fr{1}{2}h_n^{})\hsm -\hsm C_{nm}^{}\phim 
= h_n^L- \Omega_n^{} \hs,
\\[1mm]
\label{Aeq:app-Omega-Gfin}
\Omega_n^{} & = C_{nm}^{}\hs\phi_m^{}\!+\hsm\tilde{\Delta}_n^{}\hs,~~~~
\tilde{\Delta}_n^{} = \Fr{1}{\sqrt{6\,}\,}h_n^{}\!+\hsm\vt_n^{} \hs,~~~~
\vt_n^{}\!=v_{\mn}^{}h_n^{\mn} \hs.
\end{align}
\eeqs 
In the above, $C_{nm}^{}$ is a multiplicative modification factor
induced at loop level:
\begin{equation}
\label{Aeq-app:Cmod-GRET-1}
C_{nm}^{}(k^2) = \[\!\frac{~\mathbf{1}\!+\!\boldsymbol{\widetilde{\Delta}}^{\!(4)}(k^2)~}{\,\mathbf{1}\!+\!\boldsymbol{\Delta}^{\!(3)}(k^2)}\!\]_{mn}\!=\, \delta_{nm}^{}+O(\rm{loop}) \,,
\end{equation}
where the matrix form is used such that 
$(\boldsymbol{\Delta}^{\!(3)}_{})_{jj'}^{}\!\!=\hsm\!{\Delta}^{\!(3)}_{jj'}$ and
$(\boldsymbol{\widetilde{\Delta}}^{\!(4)})_{jj'}^{}\!\!=\!\!\widetilde{\Delta}^{\!(4)}_{jj'}$
with the matrix elements $({\Delta}^{\!(3)}_{jj'},\hs\widetilde{\Delta}^{\!(4)}_{jj'})$
from Eq.\eqref{Aeq:app-X-Delta34}.\ 
The above Eqs.\eqref{Aeq:app-Fn-KD-X}-\eqref{Aeq-app:Cmod-GRET-1} just reproduce the formulas 
\eqref{eq:Fn-ampLSZ-X}-\eqref{eq:Cmod-GRET-1} in the main text.\ 

\vs 

Thus, from Eq.\eqref{Aeq:app-Fn-AmpLSZ} we deduce another general identity for gravitational equivalence theorem (GRET):
\begin{equation}
\label{eq:app-GRET-hL-Omega}
\M\hsm\big[\hs\under{\FT}_{n_1^{}}^{}\!(k_1^{}),\cdots\!,\under{\FT}_{n_{\!N}^{}}^{}\!(k_N^{}), 
{\Phi}\hs\big] = 0 \,.
\end{equation}
From this, we further derive the following GRET identity:
\begin{align}
\label{eq:app-GRET-Fn}
\M\hsm\big[h_{n_1^{}}^L\hsm\!(k_1^{}),\cdots\!,h_{n_{\!N}^{}}^L\hsm\!(k_{N}^{}),{\Phi}\hs\big]
\hs =\hs 
\M\hsm\big[\Omega_{n_1^{}}^{}\hsm\!(k_1^{}),\cdots\!, 
\Omega_{n_{\!N}^{}}^{}\hsm\!(k_{N}^{}),{\Phi}\hs\big] \,.
\end{align}
We can directly prove this identity by expanding the right-hand side of Eq.\eqref{eq:app-GRET-Fn}
with each external state replacded by 
$\Omega_n^{}\!=\!h_n^L\!-\under{\FT}_{n}^{}$ 
and further using Eq.\eqref{eq:app-GRET-hL-Omega}.\  
This just reproduces the GRET identity \eqref{eq:GRET-Fn}
in the main text.\

\vs

We further note that at tree level the LSZ reduction may be implemented directly.\ 
Adopting the 't\,Hooft-Feynman gauge $(\xin\hsm\!=\! 1)$ 
and using the identity \eqref{eq:F-identityP}, 
we can derive a new identity involving the external line 
$\widetilde{\mathbb{F}}_n^{}(k)\hs$ as follows: 
\begin{align}
0\, &= \bla 0\hs |\,\widetilde{\mathbb{F}}_n^{}(k)\cdots \over{\Phi}\,|\hs 0\bra
= \sqrt{\!\Fr{2}{3}\,}\big(\varepsilon_{\mn}^{S}\!+\!\eta_{\mn}^{}\big)
\bla 0\hs |\,h_n^{\mn}(k)\cdots \Phi\,|\hs 0\bra
- \bla 0\hs |\,\phin(k)\cdots \Phi\,|\hs 0\bra
\nn\\[1mm]
&= \sqrt{\!\Fr{2}{3}\,}\big(\varepsilon_{\mn}^{S}\!+\!\eta_{\mn}^{}\big)
\D^{\mn\ab}_{h,nm}(k)
\hs\M\hsm\big[h_m^{\ab}(k),\cdots\!, \Phi\hs\big]
-\D^{}_{\phi,nm}(k)\hs\M\hsm\big[\phi_m^{}(k),\cdots\!,\Phi\hs\big]
\nn\\[1mm]
&= \D^{}_{\phi,nm}(k)\hs\M\hsm\big[
\sqrt{\!\Fr{2}{3}\,}(h_m^S\!-\!\Fr{1}{2}h_m^{})\hsm-\hsm\phim,
\cdots\!,\Phi\hs\big]
\nn\\[1mm]
&\equiv \over{\D}^{}_{\phi}(k)\hs
\M\hsm\big[\under{\widetilde{\mathbb{F}}}_n^{}(k),\cdots\!,\Phi\hs\big] ,
\label{eq:Fn-AmpLSZ-x}
\end{align}
where $\D^{}_{\phi,nm}(k)\hsm\hsm =\hsm\hsm\delta_{nm}^{}\over{\D}^{}_{\phi}(k)$ and
the LSZ-amputated external state $\under{\widetilde{\mathbb{F}}}_n^{}(k)$ 
is just given by \eqrefe{eq:app-Fn-2} with $C_{nm}^{}\!=\!1\hs$.\ 
We have also used the tree-level propagators 
\eqref{Aeq:Prophh-xi=1} and \eqref{Aeq:Prop55-xi=1} 
for $\xin\!=\!1\hs$.\
In addition, we can extend the above derivation to the general $\Rxi$ gauge.\
Using the $h_n^{\mn}$ and $\phin$ propagators in the $\Rxi$ gauge as given by 
Eqs.\eqref{eq-app:D-hh-Rxi}-\eqref{eq-app:D-VV-Rxi},
%
%
we derive the following identity of propagators at tree level:  
\begin{align}
\big[\varepsilon_{\mn}^{S}\!+\!\Fr{1}{2}(\xi_n^{}\!-\hsm\xi_n^{-1})\eta_{\mn}^{}\big]\D^{\mn\ab}_{h,nm}(k) 
= \xi_n^2\D_{nm}^{}(k) \big(\varepsilon^{\al\be}_{S}\hsm\!-\!\Fr{1}{2}\eta^{\al\be}\big)\hs.
\end{align}
Using this identity, we can derive the following tree-level identity 
with amputated external state $\widetilde{\mathbb{F}}_n^{}(k)\hs$:
\begin{align}
0\, &= \bla 0\hs |\,\widetilde{\mathbb{F}}_n^{}(k)\cdots \over{\Phi}\,|\hs 0\bra
\nn\\[1mm]
&= \sqrt{\!\Fr{2}{3}\,}\big[\varepsilon_{\mn}^{S}\!+\!\Fr{1}{2}(\xi_n^{}\!-\hsm\xi_n^{-1})\eta_{\mn}^{}\big]
\D^{\mn\ab}_{h,nm}(k)
\hs\M\hsm\big[h_m^{\ab}(k),\cdots\!, \Phi\hs\big]
\!-\hsm \xi_n^2\D^{}_{\phi,nm}(k)\hs\M\hsm\big[\phi_m^{}(k),\cdots\!,\Phi\hs\big]
\nn\\[1mm]
& = \xi_n^2\D_{nm}^{}(k) 
\M\hsm\Big[\!\sqrt{\!\Fr{2}{3}\,}\!
\big(h_m^S\!-\!\Fr{1}{2}h_m^{}\big) \!-\hsm \phim,\cdots\!,\Phi\Big].
\end{align}
Thus, we have amputated external state
$\under{\FT}_n^{}\hsm (k)$ obey the identity:
\beq 
\label{eq-app:Fn-RxiTree}
\M\hsm\Big[\under{\FT}_n^{}\hsm (k),\cdots\!,\Phi\Big] = 0\hs, 
\eeq 
where we have defined the following quantities at tree level,
\beqs
\begin{align}
\label{eq-app:Fn-Gfin}
\under{\FT}_n^{} &=
\sqrt{\!\Fr{2}{3}\,}(h_n^S\!-\!\Fr{1}{2}h_n^{})\hsm -\hsm \phim
= h_n^L- \Omega_n^{} \hs,
\\[1mm]
\label{eq-app:Omega-Gfin}
\Omega_n^{} & = \phi_m^{}\!+\hsm\tilde{\Delta}_n^{}\hs,~~~~
\tilde{\Delta}_n^{} = \Fr{1}{\sqrt{6\,}\,}h_n^{}\!+\hsm\vt_n^{} \hs,~~~~
\vt_n^{}\!=v_{\mn}^{}h_n^{\mn} \hs.
\end{align}
\label{eq-app:Fn-2}
\eeqs
\hspace*{-3.3mm}
We can readily extend the identity \eqref{eq-app:Fn-RxiTree} to the case
with $N$ external states $\under{\FT}_n^{}$,
which then reproduces the form of the GRET \eqref{eq:app-GRET-hL-Omega}
for $R_\xi^{}$ gauge and at tree level.\ 

\vspace*{2mm}
\subsection{\hspace*{-2mm}Gravitational ET of Type-II}
\label{app:B.3}
\label{appx:B.4}
\vspace*{1.5mm}

In Section\,\ref{sec:2.3.2} of the main text, 
we have formulated the KK GRET type-II 
which connects the scattering amplitude of KK gravitons $h^{\mn}_n$ 
with helicity $\pm1$ to that of the corresponding gravitational KK vector 
Goldstone bosons $\VV_n^\mu$ with the same helicity.\
In this sub-Appendix, we will provide detailed derivations to support the main text.

\vs
We start with the general Slavnov-Taylor (ST) type identity \eqref{eq:F-identityP} 
for the gravitational gauge-fixing functions given in \eqrefe{eq:Fmun2}-\eqref{eq:F5n2}.\ 
Then, we reexpress the gauge-fixing function \eqref{eq:Fmun2} in the following matrix notation:
\beqs
\label{Aeq:app-Fmu=KH}
\begin{align}
\FF_n^\mu &= -\frac{\,\ii\MGn\,}{\sqrt{2\,}\,} \mathbb{F}_n^{\mu} \,, \quad
\mathbb{F}_n^{\mu} =  \bK_n^T \bH_n^{} \,,
\\
\bK_n^{}\hsm &= \hsm\(\!\Fr{1}{\,\sqrt{2\,}\MGn}\!\big[2k_\be^{}\tensor{\eta}{_\al^\mu}\!
- k^\mu(2\!-\hsm\xi_n^{-1})\eta_{\ab}^{}\big],\, \ii\xin \hsm\)^{\!T} \!,
\quad
\bH_n^{}\!=\hsm \big(h_n^{\ab},\,  \VV_{n}^\mu \big)^{\!T} ,
\end{align}
\eeqs
where we have also assumed the external momentum $k^\mu$ to be incoming.\
Therefore, the ST-type identity associated with the gauge function $\mathbb{F}_n^\mu$ is given by
\begin{equation}
\label{Aeq:app-TFH-hV}
\bla0\big| T\, \mathbb{F}_n^\mu\, \bH_m^{T}\big|0\bra(k) = -\frac{\ii\xin}{\sqrt{2\,}\MGn}\bla 0\big|T\, \sss \bH_m^{T}\,\bar{c}^\mu_n\big|0\bra (k)\,.
\end{equation}
By taking \eqrefe{Aeq:app-Fmu=KH}, we can further rewrite \eqrefe{Aeq:app-TFH-hV} as follows:
\begin{equation}
\label{Aeq:app-KD-X}
\bK_n^T \bD_{nm}^{}(k)=-\bX^T_{nm}(k)\,,
\end{equation}
where we have defined the following notations:
\begin{align}
\bD_{nm}^{}(k) &= \bla0\big|T\hs\bH_n^{} \bH_m^{T}\big|0\bra\hsm (k) \hs, \qquad
\bX_{nm}^{}\hsm (k) = \underline{\bX}_{mj}^{}(k)\SS_{jn}^{}\hsm (k)\hs,
\nn\\[1mm]
\underline{\bX}_{mj}^{}(k)
&=\frac{\,\ii\hs\bar{\eta}^{\lam\mu}}{\,\sqrt{2\,}\MGj\,}\!\!
\begin{pmatrix}\!
\bla 0\big|\hs T\, \sss h^{\si\rho}_m\big|\bar{c}^{}_{j,\lam}\hs\bra\hsm (k)
\nn\\[1mm]
\bla 0\big|\hs T\,\sss \VV_m^{\nu}\big|\bar{c}^{}_{j,\lam}\hs\bra\hsm (k)
\end{pmatrix}\!
=\!\frac{1}{\sqrt{2\,}\MGj}\!\!
\begin{pmatrix}\!
(k^\si\bar{\eta}^{\rho\mu}\!+\hsm k^\rho\bar{\eta}^{\si\mu})
[\hs \delta_{mj}^{}\!+\!\Delta^{\!(1)}_{mj}(k^2)\hs]
\\[1mm]
\ii\sqrt{2\hs}\hs{\MGm}\bar{\eta}^{\mn}[\hs \delta_{mj}^{} \!+\! 
\widetilde{\Delta}^{\!(2)}_{mj}(k^2)\hs]
\end{pmatrix}\hsm\!, 
\label{Aeq:app-DX-KD=X-2}
\nn\\[1mm]
\SS_{jn}^{}(k) \hs\bar{\eta}^{\lam\mu} &= 
\bla 0\big|T{c_j^{\lam} \bar{c}_n^{\hs\mu}}\big| 0\bra(k)\,, \quad  
\bar{\eta}^{\lam\mu}=\eta^{\lam\mu} - 
\frac{\,k^{\lam} k^{\mu} (1\!-\hsm \xin)\,}{k^{2}+\xi_n^2\MGnn} \,,
\end{align}
which $\Delta^{(1)}_{mj}(k^2)$ and $\widetilde{\Delta}^{(2)}_{mj}(k^2)$ are the loop-level quantities.\ 
The above Eq.\eqref{Aeq:app-KD-X} 
just gives Eqs.\eqref{eq:KD-X-1} in the text.\ 
Moreover, according to the BRST transformations \eqref{Aeq:BRST-h}-\eqref{Aeq:BRST-V}, 
each loop-level quantity incorporates distinct contributions from the non-linear terms of the BRST transformations \eqref{Aeq:BRST-h}-\eqref{Aeq:BRST-V}:
\beqs
\label{Aeq:Delta-12}
\begin{align}
\Delta^{(1)}_{mj}(k^2) &= \big[\Delta_{h,h}^{(1)}(k^2) + \Delta_{h,\VV}^{(1)}(k^2)+ \Delta_{h,\phi}^{(1)}(k^2)\big]_{mj} \,,
\\[1.2mm]
\widetilde{\Delta}^{(2)}_{mj}(k^2) &= \big[\widetilde{\Delta}_{\VV,h}^{(2)}(k^2)+\widetilde{\Delta}_{\VV,\VV}^{(2)}(k^2) + \widetilde{\Delta}_{\VV,\phi}^{(2)}(k^2)\big]_{mj} \,.
\end{align}
\eeqs

Next, we consider the identity \eqref{eq:F-identityP}  with an external state given 
by the combination of gauge-fixing functions as in \eqrefe{Aeq:app-Fmu=KH}.\ 
With this, we can directly amputate the external state as follows:
\\[-8mm]
\begin{align}
\label{Aeq:app-Fn-mu-LSZ}
0 &= \bla 0\hs |\,\mathbb{F}_n^{\hs\mu}(k)\cdots {{\Phi}}\,|\hs 0\bra 
= \mathbf{K}^T_n \bD_{nm}^{}(k)
\bla 0\hs |\,{\under{\mathbf{H}}_m^{}(k)}\cdots {{\Phi}}\,|\hs 0\bra 
\nn\\
& =-\bX_{nm}^{T}(k)\hs\M\hsm\big[\under{\mathbf{H}}_m^{}(k),\cdots\!,{{\Phi}}\hs\big],
\end{align}
which leads to the following identity:
\begin{align}
\label{Aeq:app-Fn-mu-ID1}
\M\hsm\big[\under{\mathbb{F}}_n^{\hs\mu}\hsm (k),\cdots\!,{\Phi}\hs\big] = 0 \,,
\end{align}
where we have used the BRST identity \eqref{Aeq:app-KD-X} and
$\M[\cdots]$ denotes the amputated scattering amplitude.\ 
In \eqrefe{Aeq:app-Fn-mu-ID1}, we derive the following amputated external state 
$\under{\mathbb{F}}_n^{\hs\mu}\hs$:
\begin{equation}
\label{Aeq:app-Fnmu-Amp1}
\under{\mathbb{F}}_n^{\hs\mu}(k) =
\frac{\sqrt{2\,}\,}{\MGn}\hs k_{\nu}^{} 
h_{n}^{\mn} -{\Ch}^{\hs nm}_{\rm{mod}}\hs{\eta}^{\hs\mn}\VV^{}_{m,\nu} \hs,
\end{equation}
with the loop-induced modification factor given by
\begin{equation}
\label{Aeq-app:Cmod-GRET-2}
{\Ch}^{\hs nm}_{\rm{mod}}(k^2) = -\frac{\,\ii\MGm\,}{\MGn} 
\!\!\[\!\frac{\,{\mathbf{1}}\!+\!\widetilde{\boldsymbol{\Delta}}^{\!(2)}(k^2)\,}
{\,{\mathbf{1}}\!+\!\boldsymbol{\Delta}^{\!(1)}(k^2)\,}\!\]_{mn}
\!=-\ii\hs\delta_{nm}^{}+O(\rm{loop}) \hs.
\end{equation}
The matrix form presented above represents loop-level quantities $(\boldsymbol{\Delta}^{\!(1)}_{})_{jj'}^{}\!\!=\!\hsm {\Delta}^{\!(1)}_{jj'}$ and
$(\boldsymbol{\widetilde{\Delta}}^{\!(2)})_{jj'}^{}\!=\!\widetilde{\Delta}^{\!(2)}_{jj'}\hs$. 
We provide the above loop-level formulation for completeness, 
since our present focus in the main text 
is to analyze the KK scattering amplitudes at tree level.\ 
The above Eqs.\eqref{Aeq:app-Fn-mu-LSZ}-\eqref{Aeq-app:Cmod-GRET-2}
just give the Eqs.\eqref{eq:Fn-mu-LSZ}-\eqref{eq:Cmod-GRET-2} in the main text. 

\vs

By utilizing a transverse polarization vector $\ep^\mu_{\pm}$, we contract it with
$\under{\mathbb{F}}_n^{\hs\mu}(k)$ in \eqrefe{Aeq:app-Fnmu-Amp1}, yielding the formula:
\begin{align} 
\label{eq:app-Fn=hn1-Omega}
\under{\mathbb{F}}_n^{}\hsm (k) =h_n^{\pm1}\! -\Theta_n^{}\hs,~~~~\Theta_n^{} 
\!= {\Ch}_{\rm{mod}}^{\hs nm}\VV_m^{\pm1}	\!+ v_n^{\pm1}\hs, 
\end{align} 
where $v_{\pm1}^{\mn}\!=\hsm O(\MGn/\hsm E_n^{})$.
Then, we can generalize \eqrefe{Aeq:app-Fn-mu-ID1} 
to incorporate $N$ externally-amputated gauge-fixing functions:
\begin{equation}
\label{Aeq:app-GRET2-Fn1}
\M\hsm\big[\under{\mathbb{F}}_{n_1^{}}^{}\hsm\!(k_1^{}),\cdots\!,\under{\mathbb{F}}_{n_{\!N}^{}}^{}\hsm\!(k_{\!N}^{}),
{\Phi}\hs\big]= 0 \,. 
\end{equation} 
Using this identity, we can further derive the gravitational equivalence theorem (GRET) identity 
which connects the $h_n^{\pm1}$ amplitude to the amplitude of $\Theta_{n}^{}$:
\begin{align}
\label{Aeq:app-GRET2-hn1=Theta}
\M\hsm\big[h_{n_1^{}}^{\pm1}\hsm (k_1^{}),\cdots\!,h_{n_{\!N}^{}}^{\pm1}\hsm (k_{N}^{}), 
{\Phi}\hs\big]\hs =\hs 
\M\hsm\big[\Theta_{n_1^{}}^{}\hsm\!(k_1^{}),\cdots\!, \Theta_{n_{\!N}^{}}^{}\hsm\!(k_{N}^{}),
{\Phi}\hs\big] \hs.
\end{align}
We can prove the above identity by computing the amplitude on its right-hand side.\ 
For this, we expand $\Theta_n^{}$ in terms of 
$\,\Theta_n^{}\!=\hsm h_n^{\pm1}\!-\under{\mathbb{F}}_n^{}$
for each external state of Eq.\eqref{Aeq:app-GRET2-hn1=Theta} and thus deduce 
the left-hand side of Eq.\eqref{Aeq:app-GRET2-hn1=Theta} after applying the 
identity Eq.\eqref{Aeq:app-GRET2-Fn1} to eliminate each extenal state of $\under{\mathbb{F}}_n^{}$.\  
The above Eq.\eqref{Aeq:app-GRET2-hn1=Theta} 
just reproduce the GRET identity \eqref{eq:GRET2-hn1=Theta} 
in the main text.\

\section{\hspace*{-2mm}Feynman Rules for Warped KK Gauge and Gravity Theories}
\label{app:C}

In this Appendix, we present the relevant Feynman rules for the warped KK 
gauge and gravity theories under the 5d compactification of $S^1\!/\ZZ\hs$,
which are needed for explicitly computing the 3-point and 4-point 
KK scattering amplitudes in the main text. 

\vspace*{1.5mm}
\subsection{\hspace*{-2.5mm}Feynman Rules for Warped KK Gauge Theory}
\label{app:C1}
\label{sec:FR-KKGA}
\vspace*{1.5mm}

In this sub-Appendix, we present the relevant Feynman rules for the warped KK 
gauge theory.\ 
The propagators for KK gauge boson and the KK Goldstone boson take the following forms
in the $R_\xi^{}$ gauge:
\beqs
\begin{align}
\D_{nm}^{\mn}(p) &= - \frac{\ii\hs\delta_{nm}}{\,p^2 \!+\! \MGnn\,} 
\!\[\eta^{\mn} \!+ (\xin\!-\!1)\frac{p^\mu p^\nu}{\,p^2\hsm +\xin \MGnn\,}\] \!,
\\
\D_{nm}(p) &= -\frac{\ii\hs\delta_{nm}}{~p^2 + \xin\MGnn~}\,.
\end{align}
\eeqs
The trilinear vertices and quartic vertices for KK gauge bosons (KK Goldstone bosons)
are derived as follows:
\beqs
\label{app-eq:Feyn-3KK-4KK} 
\begin{align}
\raisebox{0.em}{\hbox{\begin{minipage}{3.5cm}
\hspace*{-4.5mm}
\centering
\includegraphics[width=3.8cm]{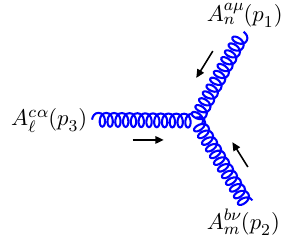}
\end{minipage}}}
\hspace{0.1em}
& \begin{aligned}
&= - g\hs a_{nm\ell}^{}\hs f^{abc}\big[\hs \eta^{\mn} (p_1^{} \!-\hsm p_2^{})^{\al}
\!+\hsm\eta^{\nu\al}(p_{2}^{} \!-\hsm p_{3}^{})^{\mu}
\\
&\hspace*{26mm}+\eta^{\al\mu} (p_{3}^{} \!-\hsm p_{1}^{})^{\nu}\hs\big]\,,
\end{aligned}
\\[1mm]
\raisebox{0.em}{\hbox{\begin{minipage}{3.5cm}
\centering
\includegraphics[width=3.5cm]{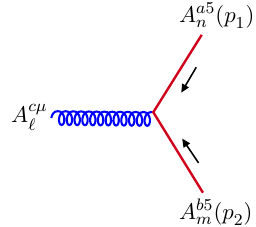}
\end{minipage}}}
\hspace{0.1em}
&= - g\hs\tilde{a}_{nm\ell}^{}\hs f^{abc}(p_1^{}\!-\hsm p_2^{})^\mu\,,
\\[1mm]
\label{eq-app:A-A-A5}
\raisebox{0.em}{\hbox{\begin{minipage}{3.5cm}
\centering
\includegraphics[width=3.2cm]{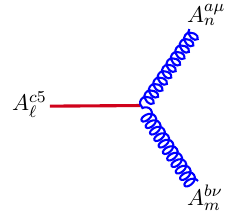}
\end{minipage}}}
\hspace{0.1em}
& = \ii g\hs a_{nm\ell}^{}\hs f^{abc}\eta^{\mn}(\Mnn\!-\!\Mmm)M_\ell^{-1}\,,
\\[1mm]
\raisebox{0.em}{\hbox{\begin{minipage}{3.5cm}
\centering
\includegraphics[width=3.2cm]{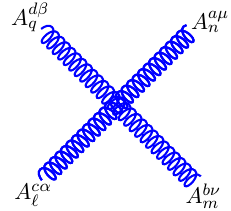}
\end{minipage}}}
\hspace{0.1em}
& \begin{array}{ll}
& =  \ii\hs g^2\hs a_{nm\ell q}^{} \big[\hs f^{abe} f^{cde}
(\eta^{\mu\al}\eta^{\nu \be}\!-\hsm\eta^{\mu\be}\eta^{\nu\al} )
\\
&\hspace*{20mm}+ f^{a c e} f^{dbe}
(\eta^{\mu \be}\eta^{\nu\al} \!-\hsm\eta^{\mn}\eta^{\ab} )
\\
&\hspace*{20mm}+f^{ade} f^{bce}
(\eta^{\mn}\eta^{\ab}\!-\hsm\eta^{\mu\al}\eta^{\nu\be} )\hs\big]  ,
\end{array}
\end{align}
\eeqs
where the effective cubic and quartic KK coupling coefficients 
$(a_{nm\ell}^{},\,\tilde{a}_{nm\ell}^{},\,a_{nm\ell q}^{},$ $\tilde{a}_{nm\ell q}^{})$ 
are defined as follows:
\\[-8mm]
\beqs
\label{app-eq:KKYM-RS-couplings}
\begin{align}
\label{eq:anml}
a_{nm\ell}^{} &= 
\frac{1}{\,L\,}\int_0^L \td z\, e^{A(z)}\, \ff_n^{}(z)\ff_m^{}(z)\ff_\ell^{}(z) \,,
\\
\label{eq:tanml}
\tilde{a}_{nm\ell}^{} &= 
\frac{1}{\,L\,}\int_0^L \td z\, e^{A(z)} \, \fft_n^{}(z)\fft_m^{}(z)\ff_\ell^{}(z)\,,
\\
\label{eq:anmlq}
a_{nm\ell q}^{} &= \frac{1}{\,L\,}\int_0^L \td z\, e^{A(z)} \,\ff_n^{}(z)\ff_m^{}(z)\ff_\ell^{}(z)\ff_q^{}(z)\,,
\\
\label{eq:atnmlq}
\tilde{a}_{nm\ell q}^{} &= 
\frac{1}{\,L\,}\int_0^L \td z\, e^{A(z)} \,\fft_n^{}(z) \fft_m^{}(z) \fft_\ell^{}(z) \fft_q^{}(z)\,.
\end{align}
\eeqs
These coupling coefficients will be used for the double-copy construction 
in Sections\,\ref{sec:4.1}-\ref{sec:4.2} and for deriving the gravitational sum rules 
in Appendix\,\ref{app:D}.
In the above Feynman rules \eqref{app-eq:Feyn-3KK-4KK}, the coupling 
$g$ is the 4d gauge coupling and is connected to the 5d gauge coupling $\hat{g}_5^{}$ via 
$\,g\hsm =\hsm \hat{g}_5^{}/\hsm\sqrt{L\,}$.\   
For the 5d physical coordinate $y$ (instead of the 5d conformal coordinate $z$), the 5d length is
$\bar{L}\!=\!\pi r_c^{}$ and the corresponding 4d gauge coupling is 
$\bar{g} \!=\! \hat{g}_5^{}/\sqrt{\bar{L}\,}$, which is related to  the gauge coupling $g$ 
via $\,\bar{g} \!=\! {g}\sqrt{\bar{L}/L\,}$.\ Thus, the cubic KK gauge coupling coefficient
$\bar{a}_{nm\ell}^{}$ (defined in 5d physical coordinate) is connected to the above
cubic KK gauge coupling coefficient ${a}_{nm\ell}^{}$ (defined in 5d conformal coordinate)
by $\bar{a}_{nm\ell}^{}\!=\!{a}_{nm\ell}^{}\sqrt{\bar{L}/L\,}$.\  
Similarly, for the quartic KK gauge coupling coefficient, we have the relation, 
$\bar{a}_{nm\ell q}^{}\!=\!(\bar{L}/L)\hs{a}_{nm\ell q}^{}\hs$.\  
%


\vspace*{1mm}
\subsection{\hspace*{-2.5mm}Feynman Rules for Warped KK Gravity Theory}
\label{app:C2}
\label{sec:FR-KKGR}
\vspace*{1.5mm}

The propagators for KK gravitons and gravitational KK Goldstone bosons 
in a general $R_\xi^{}$ gauge \eqref{eq:L-GF-KKgravity}-\eqref{eq:FmuF5-5d} 
are given by the following\,\cite{Hang:2021fmp}:
\beqs
\begin{align}
\label{eq-app:D-hh-Rxi}
&\hspace*{-4mm}
\D_{nm}^{\mn\ab}(p) =-\frac{\,\ii\hs\dnm\,}{2}\LB
\frac{\,\eta^{\mu\al}\eta^{\nu\be}\hsm\!+\hsm\eta^{\mu\be}\eta^{\nu\al}
\hsm\!-\!\eta^{\mn}\eta^{\al\be}}{p^2\!+\hsm\MGnn}
+\frac{1}{3}\!\[\frac{1}{\,p^2\!+\hsm\MGnn\,}
-\frac{1}{\,p^2\!+\!(3\hs\xin\hsm\!-\!2)\MGnn\,}\]\right.
\nn\\[1mm]
& \times\!
\(\!\eta^{\mn}\!\!-\!\frac{\,2p^{\mu}p^{\nu}\,}{\MGnn} \!\)\!\!
\(\!\eta^{\ab}\!\!-\!\frac{\,2p^{\al}p^{\be}\,}{\MGnn} \!\)
\hsm\!+\!\frac{1}{\Mnn}\!\[\frac{1}{\,p^2\!+\!\MGnn\,} -\frac{1}{\,p^2\!+\hsm\xin\MGnn\,}\]\!\!
\(\eta^{\mu\al}p^{\nu}p^{\be}\hsm\!+\hsm\eta^{\mu\be}p^{\nu}p^{\al}
\right.\nn\\[1mm]
&\left.\left.
+\,\eta^{\nu\al}p^{\mu}p^{\be}\hsm\!+\!\eta^{\nu\be}p^{\mu}p^{\al}\hsm\)
\!+\!\frac{~4\hs p^\mu p^\nu p^\al p^\be~}{\xin M_n^4}\!
\(\frac{1}{~p^2\!+\hsm\xi_n^2\MGnn~} -\frac{1}{~p^2\!+\hsm\xin\MGnn~} \)\! \RB,
\\[1mm]
\label{eq-app:D-VV-Rxi}
&\hspace*{-4mm}
\D^{\mn}_{nm}(p)  = - \frac{\ii\hs\delta_{nm}^{}}{\,p^2\!+\hsm\xin \MGnn\,}\!
\left[\eta^{\mn} \!\!-\! \frac{~p^{\mu} p^{\nu} (1\!-\!\xi_n)~}
{\,p^2\!+\xi_n^2 \MGnn~} \right]\!,
\\[1mm]
\label{eq-app:D-phiphi-Rxi}
&\hspace*{-4mm} 
\D_{nm}^{}(p) = -\frac{\ii\hs\delta_{nm}^{}}{\,p^2\hsm +\hsm (3 \xin \!\!-\!2)\MGnn\,} \,.
\end{align}
\eeqs

Then, we derive the relevant trilinear gravitational KK vertices as follows:
\beqs
\label{app-eq:3-vertices}
\begin{align}
\label{app-eq:FR-hhh}
\raisebox{0.em}{\hbox{\begin{minipage}{3.5cm}
\centering
\includegraphics[width=4.15cm]{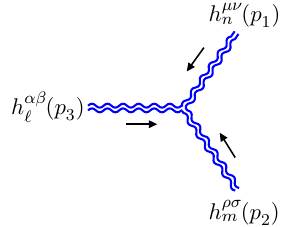}
\end{minipage}}}
&\hspace{.6cm}
= \ii\hs\ka\hs\al_{nm\ell}^{}\hs\Ga^{\mn\rho\si\ab}_{nm\ell}(p_1^{},p_2^{},p_3^{})\hs,
\\[1mm]
\label{app-eq:FR-VVh}
\raisebox{0.em}{\hbox{\begin{minipage}{3.5cm}
\hspace*{.58cm}
\centering
\includegraphics[width=3.5cm]{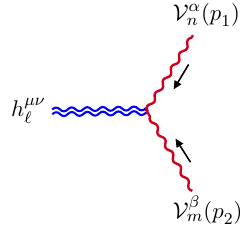}
\end{minipage}}}
&\hspace*{.62cm}
= \frac{\,\ii\hs\ka\,}{2}\tilde{\al}_{nm\ell}^{}\hs\Ga^{\ab\mn}_{nm\ell}(p_1^{},p_2^{}) \hs,
\\[1mm]
\label{app-eq:FR-pph}
\raisebox{0.em}{\hbox{\begin{minipage}{3.5cm}
\hspace*{.58cm}
\centering
\includegraphics[width=3.5cm]{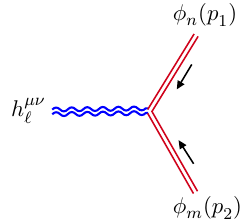}
\end{minipage}}}
&\hspace*{.62cm}
\begin{aligned} 
& = \frac{\,\ii\hs\ka\,}{2}\tilde{\be}_{nm\ell}^{} \bigg\{\hsm 2(p_1^\mu p_1^\nu+p_2^\mu p_2^\nu)
\hsm +\hsm p_1^\mu p_2^\nu \hsm +\hsm p_1^\nu p_2^\mu \hsm - \!\left[2(p_1^2 \!+\hsm p_2^2)~~
\right. 
\\
&\hspace*{21mm} 
+ 3 p_1^{} \!\cdot\hsm p_2^{} \hsm - \hsm \MG_{\ell}^2\Big]\eta^{\mn}\!\bigg\} \hs,
\end{aligned}
\\[1mm]
\label{app-eq:FR-VVp}
\raisebox{0.em}{\hbox{\begin{minipage}{3.5cm}
\hspace*{1cm}
\centering
\includegraphics[width=3.1cm]{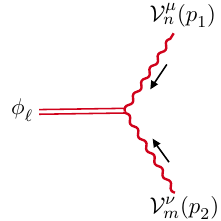}
\end{minipage}}}
&\hspace*{.62cm}
\begin{aligned}
& =\frac{\ii\hs\ka}{\sqrt{6\,}\,}\hs\tilde{\rho}_{nm\ell}^{} 
\bigg\{\!\!\hsm -\!2\hs (p_1^\mu p_1^\nu \!+\hsm p_2^\mu p_2^\nu) \!-\! 3\hs p_2^\mu p_1^\nu 
\!+\!\hsm\left[2(p_1^2 \!+\! p_2^2) \right.  
\\
& \hspace*{22mm} 
+\!3\hs p_1^{}\!\cdot\hsm p_2^{}\hsm\left]\hs \eta^{\mn} \right.\!\!\Big\} \hs,
\end{aligned}
\\[1mm]
\label{app-eq:FR-ppp}
\raisebox{0.em}{\hbox{\begin{minipage}{3.5cm}
\hspace*{4mm}
\centering
\includegraphics[width=3.75cm]{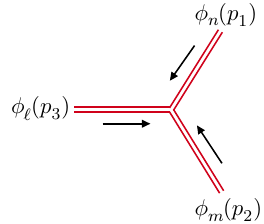}
\end{minipage}}}
&\hspace*{.62cm}
\begin{aligned}
& = \ii\hs 2\sqrt{\frac{2}{3}\,} \hs\ka \Big[ \hs\tilde{\omega}_{nm\ell}^{}
\big( p_1^{} \!\cdot\! p_2^{} \!+ p_1^{} \!\cdot\! p_3^{} \!+ p_2^{} \!\cdot\! p_3^{}\big)
\\
&\qquad\hspace*{11mm} 
+ \xoverline{\omega}_{nm\ell}^{} \hsm +\hsm \xoverline{\omega}_{n\ell m}^{} 
\hsm +\hsm \xoverline{\omega}_{m\ell n}^{} \Big] .
\end{aligned}
\end{align}
\eeqs
In the above Feynman rules \eqref{app-eq:3-vertices}, the coupling 
$\ka$ is the 4d gravitational coupling and is connected to the 5d gravitational coupling $\hat{\ka}_5^{}$ via 
$\,\ka\! =\! \hat{\ka}_5^{}/\hsm\sqrt{L\,}$.\   
For the 5d physical coordinate $y$ (instead of the 5d conformal coordinate $z$), the 5d length is
$\bar{L}\!=\!\pi r_c^{}$ and the corresponding 4d gravitational coupling is 
$\bar{\ka} \!=\! \hat{\ka}_5^{}/\sqrt{\bar{L}\,}$, which is related to  the gravitational coupling $\ka$ 
via $\,\bar{\ka} \!=\! {\ka}\sqrt{\bar{L}/L\,}$.\ Thus, the cubic KK gravitational coupling coefficient
$\bar{\al}_{nm\ell}^{}$ (defined in 5d physical coordinate) is connected to the above
cubic KK gravitational coupling coefficient ${\al}_{nm\ell}^{}$ (defined in 5d conformal coordinate)
by $\bar{\al}_{nm\ell}^{}\!=\!{\al}_{nm\ell}^{}\sqrt{\bar{L}/L\,}$.

\vs 

In the above \eqrefe{app-eq:FR-hhh}, the trilinear gravitational vertex function
$\Ga^{\mn\rho\si\ab}_{nm\ell}$ is defined as follows:
\\[-5mm]
\begin{equation}
\label{app-eq:FR-pph-Ga}
\Ga^{\mn\rho\si\ab}_{nm\ell}(p_1^{},p_2^{},p_3^{}) =  \frac{1}{\,8\,}
\big[( \MGnn\hs F_1^{} \!-\hsm p_1^2 \hs F_2^{} \hsm +\hsm F_3^{})  
\!+\! ( 1 \!\leftrightarrow\! 2) \!+\! ( 1 \!\leftrightarrow\! 3 )\big] \,,
\end{equation}
with the functions $(F_1^{},\,F_2^{},\,F_3^{})$ given by
\begin{align}
\label{app-eq:FR-pph-Ga-F1F2F3}
F_1^{} =& -\eta^{\al\be}\eta^{\mu\si} \eta^{\nu\rho}-\eta^{\al\be} \eta^{\mu\rho} \eta^{\nu\si}+2 \eta^{\al\be} \eta^{\mu\nu} \eta^{\rho\si}+\eta^{\al\mu} \eta^{\be\si} \eta^{\nu\rho}+\eta^{\al\mu} \eta^{\be\rho} \eta^{\nu\si}-\eta^{\al\mu} \eta^{\be\nu} \eta^{\rho\si}
\nn\\
& +\eta^{\al\si} \eta^{\be\mu} \eta^{\nu\rho} +\eta^{\al\rho} \eta^{\be\mu} \eta^{\nu\si}+\eta^{\al\nu} \eta^{\be\si} \eta^{\mu\rho}+\eta^{\al\nu} \eta^{\be\rho} \eta^{\mu\si}-\eta^{\al\nu} \eta^{\be\mu} \eta^{\rho\si}+\eta^{\al\si} \eta^{\be\nu} \eta^{\mu\rho}
\nn\\
&+\eta^{\al\rho} \eta^{\be\nu} \eta^{\mu\si} -3 \eta^{\al\rho} \eta^{\be\si} \eta^{\mu\nu}-3 \eta^{\al\si} \eta^{\be\rho} \eta^{\mu\nu}\,,
\nn\\[1mm]
F_2^{} =&~ 3 \eta^{\al\be} \eta^{\mu\si} \eta^{\nu\rho}+3 \eta^{\al\be} \eta^{\mu\rho} \eta^{\nu\si}-4 \eta^{\al\be} \eta^{\mu\nu} \eta^{\rho\si}-2 \eta^{\al\mu} \eta^{\be\si} \eta^{\nu\rho}-2 \eta^{\al\mu} \eta^{\be\rho} \eta^{\nu\si}+3 \eta^{\al\mu} \eta^{\be\nu} \eta^{\rho\si}
\nn\\
&+3 \eta^{\al\nu} \eta^{\be\mu} \eta^{\rho\si}-2 \eta^{\al\nu} \eta^{\be\si} \eta^{\mu\rho}-2 \eta^{\al\nu} \eta^{\be\rho} \eta^{\mu\si}+4 \eta^{\al\si} \eta^{\be\rho} \eta^{\mu\nu}+4 \eta^{\al\rho} \eta^{\be\si} \eta^{\mu\nu}-2 \eta^{\al\rho} \eta^{\be\nu} \eta^{\mu\si}
\nn\\
&-2 \eta^{\al\rho} \eta^{\be\mu} \eta^{\nu\si}-2 \eta^{\al\si} \eta^{\be\nu} \eta^{\mu\rho}-2 \eta^{\al\si} \eta^{\be\mu} \eta^{\nu\rho} \,,
\nn\\[1mm]
F_3^{} =&~ 2\eta^{\mu\nu}\eta^{\rho\si}p_1^{\al}p_2^{\be}+\eta^{\mu\nu}\eta^{\rho\si} p_1^{\al}p_3^{\be}+\eta^{\nu\rho}\eta^{\be\si}p_1^{\al}p_2^{\mu}+\eta^{\be\rho} \eta^{\nu\si}p_1^{\al}p_2^{\mu}+\eta^{\nu\rho}\eta^{\be\si}p_1^{\al}p_3^{\mu}+\eta^{\be\rho}\eta^{\nu\si}p_1^{\al}p_3^{\mu}
\nn\\
&+\eta^{\mu\rho}\eta^{\be\si}p_1^{\al}p_2^{\nu}+\eta^{\be\rho}\eta^{\mu\si}p_1^{\al}p_2^{\nu}+\eta^{\mu\rho} \eta^{\be\si}p_1^{\al}p_3^{\nu}+\eta^{\be\rho}\eta^{\mu\si}p_1^{\al}p_3^{\nu}-\eta^{\be\si}\eta^{\mu\nu}p_1^{\al}p_2^{\rho}-\eta^{\be\si}\eta^{\mu\nu}p_1^{\al}p_3^{\rho}
\nn\\
&+\eta^{\be\nu} \eta^{\mu\si}  p_1^{\al}p_3^{\rho}+\eta^{\be\mu} \eta^{\nu\si}  p_1^{\al}p_3^{\rho}- \eta^{\mu\si} \eta^{\nu\rho} p_1^{\al}p_2^{\be}- \eta^{\mu\si} \eta^{\nu\rho} p_1^{\al}p_3^{\be}-\eta^{\be\rho} \eta^{\mu\nu}p_1^{\al}p_2^{\si}-\eta^{\be\rho}\eta^{\mu\nu}p_1^{\al}p_3^{\si}
\nn\\
&+\eta^{\be\nu} \eta^{\mu\rho}p_1^{\al}p_3^{\si} +\eta^{\be\mu} \eta^{\nu\rho} p_1^{\al}p_3^{\si}-\eta^{\mu\rho} \eta^{\nu\si} p_1^{\al}p_2^{\be}- \eta^{\mu\rho} \eta^{\nu\si} p_1^{\al}p_3^{\be}-\eta^{\be\nu} \eta^{\rho\si} p_1^{\al}p_2^{\mu}-\eta^{\be\nu}\eta^{\rho\si} p_1^{\al}p_3^{\mu}
\nn\\
&-\eta^{\be\mu}\eta^{\rho\si} p_1^{\al} p_2^{\nu}-\eta^{\be\mu}\eta^{\rho\si} p_1^{\al}p_3^{\nu}+2 \eta^{\mu\nu} \eta^{\rho\si}p_1^{\be} p_2^{\al}+\eta^{\mu\nu} \eta^{\rho\si}p_1^{\be}p_3^{\al}+\eta^{\nu\rho} \eta^{\al\si} p_1^{\be}p_2^{\mu}+\eta^{\al\rho}\eta^{\nu\si}p_1^{\be}p_2^{\mu}
\nn\\
&+\eta^{\nu\rho} \eta^{\al\si}  p_1^{\be}p_3^{\mu}+\eta^{\al\rho} \eta^{\nu\si}  p_1^{\be}p_3^{\mu}+\eta^{\mu\rho} \eta^{\al\si}  p_1^{\be}p_2^{\nu}+\eta^{\al\rho} \eta^{\mu\si}  p_1^{\be}p_2^{\nu}+\eta^{\mu\rho} \eta^{\al\si}  p_1^{\be}p_3^{\nu}+\eta^{\al\rho} \eta^{\mu\si}  p_1^{\be}p_3^{\nu}
\nn\\
&-\eta^{\al\si} \eta^{\mu\nu}  p_1^{\be}p_2^{\rho}-\eta^{\al\si} \eta^{\mu\nu}  p_1^{\be}p_3^{\rho}+\eta^{\al\nu} \eta^{\mu\si}  p_1^{\be}p_3^{\rho} +\eta^{\al\mu} \eta^{\nu\si}  p_1^{\be}p_3^{\rho}- \eta^{\mu\si} \eta^{\nu\rho} p_1^{\be}p_3^{\al}-\eta^{\al\rho} \eta^{\mu\nu}  p_1^{\be}p_2^{\si}
\nn\\
&-\eta^{\al\rho} \eta^{\mu\nu}  p_1^{\be}p_3^{\si}+\eta^{\al\nu} \eta^{\mu\rho}  p_1^{\be}p_3^{\si}+\eta^{\al\mu} \eta^{\nu\rho}  p_1^{\be}p_3^{\si}- \eta^{\mu\rho} \eta^{\nu\si} p_1^{\be}p_2^{\al}- \eta^{\mu\rho} \eta^{\nu\si} p_1^{\be}p_3^{\al}-\eta^{\al\nu}  \eta^{\rho\si} p_1^{\be}p_2^{\mu}
\nn\\
&-\eta^{\al\nu}  \eta^{\rho\si} p_1^{\be}p_3^{\mu}-\eta^{\al\mu}  \eta^{\rho\si} p_1^{\be}p_2^{\nu}-\eta^{\al\mu}  \eta^{\rho\si} p_1^{\be}p_3^{\nu}-\eta^{\nu\rho} \eta^{\mu\si}  p_1^{\be}p_2^{\al}-\eta^{\al\si} \eta^{\be\rho}  p_1^{\mu}p_2^{\nu}-\eta^{\al\rho} \eta^{\be\si}  p_1^{\mu}p_2^{\nu}
\nn\\
&+\eta^{\al\be} \eta^{\rho\si}  p_1^{\mu}p_2^{\nu}-\eta^{\al\si} \eta^{\be\rho}  p_1^{\mu}p_3^{\nu}-\eta^{\al\rho} \eta^{\be\si}  p_1^{\mu}p_3^{\nu}+\eta^{\al\be} \eta^{\rho\si}  p_1^{\mu}p_3^{\nu}-\eta^{\al\be} \eta^{\nu\si}  p_1^{\mu}p_3^{\rho}-\eta^{\al\be}  \eta^{\nu\rho} p_1^{\mu}p_3^{\si}
\nn\\
&-\eta^{\al\nu}  \eta^{\rho\si} p_1^{\mu}p_2^{\be}- \eta^{\be\nu} \eta^{\rho\si} p_1^{\mu}p_2^{\al}-\eta^{\al\si} \eta^{\be\rho}  p_1^{\nu}p_2^{\mu}-\eta^{\al\rho} \eta^{\be\si}  p_1^{\nu}p_2^{\mu}+\eta^{\al\be} \eta^{\rho\si}  p_1^{\nu}p_2^{\mu}-\eta^{\al\si} \eta^{\be\rho}  p_1^{\nu}p_3^{\mu}
\nn\\
&-\eta^{\al\rho} \eta^{\be\si} p_1^{\nu}p_3^{\mu} +\eta^{\al\be} \eta^{\rho\si}  p_1^{\nu}p_3^{\mu}-\eta^{\al\be} \eta^{\mu\si}  p_1^{\nu}p_3^{\rho}-\eta^{\al\be} \eta^{\mu\rho}  p_1^{\nu}p_3^{\si}-\eta^{\al\mu}  \eta^{\rho\si} p_1^{\nu}p_2^{\be}- \eta^{\be\mu} \eta^{\rho\si} p_1^{\nu}p_2^{\al}
\nn\\
&+\eta^{\be\nu} \eta^{\mu\si}  p_1^{\rho}p_2^{\al}+\eta^{\be\mu} \eta^{\nu\si}  p_1^{\rho}p_2^{\al}+\eta^{\al\nu} \eta^{\mu\si}  p_1^{\rho}p_2^{\be}+\eta^{\al\mu} \eta^{\nu\si}  p_1^{\rho}p_2^{\be}+\eta^{\be\nu} \eta^{\al\si}  p_1^{\rho}p_2^{\mu}+\eta^{\al\nu} \eta^{\be\si}  p_1^{\rho}p_2^{\mu}
\nn\\
&+\eta^{\be\nu} \eta^{\al\si}  p_1^{\rho}p_3^{\mu}+\eta^{\al\nu} \eta^{\be\si}  p_1^{\rho}p_3^{\mu}+\eta^{\be\mu} \eta^{\al\si}  p_1^{\rho}p_2^{\nu}+\eta^{\al\mu} \eta^{\be\si}  p_1^{\rho}p_2^{\nu}-\eta^{\al\be} \eta^{\mu\si}  p_1^{\rho}p_2^{\nu}+\eta^{\be\mu} \eta^{\al\si}  p_1^{\rho}p_3^{\nu}
\nn\\
&+\eta^{\al\mu} \eta^{\be\si}  p_1^{\rho}p_3^{\nu}-\eta^{\al\be} \eta^{\mu\si}  p_1^{\rho}p_3^{\nu}-\eta^{\al\si}  \eta^{\mu\nu} p_1^{\rho}p_2^{\be}-\eta^{\al\si}  \eta^{\mu\nu} p_1^{\rho}p_3^{\be}- \eta^{\be\si} \eta^{\mu\nu} p_1^{\rho}p_2^{\al}- \eta^{\be\si} \eta^{\mu\nu} p_1^{\rho}p_3^{\al}
\nn\\
&-\eta^{\al\nu} \eta^{\be\mu}  p_1^{\rho}p_2^{\si}-\eta^{\al\mu} \eta^{\be\nu}  p_1^{\rho}p_2^{\si}+\eta^{\al\be} \eta^{\mu\nu}  p_1^{\rho}p_2^{\si}-\eta^{\al\nu} \eta^{\be\mu}  p_1^{\rho}p_3^{\si}-\eta^{\al\mu} \eta^{\be\nu}  p_1^{\rho}p_3^{\si}+2 \eta^{\al\be} \eta^{\mu\nu}  p_1^{\rho}p_3^{\si}
\nn\\
&-\eta^{\al\be}  \eta^{\nu\si} p_1^{\rho}p_2^{\mu}-\eta^{\al\be}  \eta^{\nu\si} p_1^{\rho}p_3^{\mu}+\eta^{\be\nu} \eta^{\mu\rho}  p_1^{\si}p_2^{\al}+\eta^{\be\mu} \eta^{\nu\rho}  p_1^{\si}p_2^{\al}+\eta^{\al\nu} \eta^{\mu\rho}  p_1^{\si}p_2^{\be}+\eta^{\al\mu} \eta^{\nu\rho}  p_1^{\si}p_2^{\be}
\nn\\
&+\eta^{\be\nu} \eta^{\al\rho}  p_1^{\si}p_2^{\mu}+\eta^{\al\nu} \eta^{\be\rho}  p_1^{\si}p_2^{\mu}+\eta^{\be\nu} \eta^{\al\rho}  p_1^{\si}p_3^{\mu}+\eta^{\al\nu} \eta^{\be\rho}  p_1^{\si}p_3^{\mu}+\eta^{\be\mu} \eta^{\al\rho}  p_1^{\si}p_2^{\nu}+\eta^{\al\mu} \eta^{\be\rho}  p_1^{\si}p_2^{\nu}
\nn\\
&-\eta^{\al\be} \eta^{\mu\rho}  p_1^{\si}p_2^{\nu}+\eta^{\be\mu} \eta^{\al\rho}  p_1^{\si}p_3^{\nu}+\eta^{\al\mu} \eta^{\be\rho}  p_1^{\si}p_3^{\nu}-\eta^{\al\be} \eta^{\mu\rho}  p_1^{\si}p_3^{\nu}-\eta^{\al\rho}  \eta^{\mu\nu} p_1^{\si}p_2^{\be}-\eta^{\al\rho}  \eta^{\mu\nu} p_1^{\si}p_3^{\be}
\nn\\
&- \eta^{\be\rho} \eta^{\mu\nu} p_1^{\si}p_2^{\al}- \eta^{\be\rho} \eta^{\mu\nu} p_1^{\si}p_3^{\al}-\eta^{\al\nu} \eta^{\be\mu}  p_1^{\si}p_2^{\rho}-\eta^{\al\mu} \eta^{\be\nu}  p_1^{\si}p_2^{\rho}+\eta^{\al\be} \eta^{\mu\nu}  p_1^{\si}p_2^{\rho}-\eta^{\al\nu} \eta^{\be\mu}  p_1^{\si}p_3^{\rho}
\nn\\
&-\eta^{\al\mu} \eta^{\be\nu}  p_1^{\si}p_3^{\rho}+2 \eta^{\al\be} \eta^{\mu\nu}  p_1^{\si}p_3^{\rho}-\eta^{\al\be}  \eta^{\nu\rho} p_1^{\si}p_2^{\mu}-\eta^{\al\be}  \eta^{\nu\rho} p_1^{\si}p_3^{\mu}\,.
\end{align}
In \eqrefe{app-eq:FR-VVh}, the trilinear gravitational vertex function
$\Ga^{\ab\mn}_{nm\ell}$ is defined as follows:
\begin{align}
\Ga^{\ab\mn}_{nm\ell}(p_1^{},p_2^{}) =& -2  \eta^{\ab} p_1^\mu p_1^\nu + 2 \eta^{\ab} \eta^{\mn}p_1^2 +  \eta^{\al\mu} p_1^\be p_1^\nu +  \eta^{\al\nu}p_1^\be p_1^\mu +   \eta^{\be\mu} p_1^\al p_1^\nu - \eta^{\al\nu} \eta^{\be\mu} p_1^2
\nn\\
&+\eta^{\be\nu} p_1^\al p_1^\mu -\eta^{\al\mu} \eta^{\be\nu}  p_1^2 - 2\eta^{\mn}p_1^\al p_1^\be -\eta^{\ab} p_1^\nu p_2^\mu - \eta^{\ab}p_1^\mu p_2^\nu+ 3 \eta^{\ab} \eta^{\mn} (p_1^{}\!\cdot\! p_2^{})
\nn\\
&+ \eta^{\al\mu} p_1^\nu p_2^\be \!+  \eta^{\al\nu} p_1^\mu  p_2^\be +\eta^{\be\mu}p_1^\al p_2^\nu - \eta^{\al\nu} \eta^{\be\mu}(p_1^{}\!\cdot\! p_2^{}) \!+ \eta^{\be\nu}  p_1^\al p_2^\mu -  \eta^{\al\mu} \eta^{\be\nu}(p_1^{}\!\cdot\! p_2^{})
\nn\\
&-\eta^{\mn}p_1^\be p_2^\al - 2\eta^{\mn}p_1^\al p_2^\be - 2 \eta^{\ab}p_2^\mu p_2^\nu+ 2\eta^{\ab} \eta^{\mn}p_2^2 + \eta^{\al\mu}p_2^\be p_2^\nu + \eta^{\al\nu} p_2^\be p_2^\mu 
\nn\\
&+ \eta^{\be\mu}p_2^\al p_2^\nu - \eta^{\al\nu} \eta^{\be\mu} p_2^2 +  \eta^{\be\nu}p_2^\al p_2^\mu -  \eta^{\al\mu} \eta^{\be\nu} p_2^2 - 2  \eta^{\mn}p_2^\al p_2^\be \,.
\end{align}

Moreover, the trilinear coupling coefficients 
$(\al_{nm\ell}^{},\hs\tilde{\al}_{nm\ell}^{},\hs\tilde{\be}_{nm\ell}^{},\hs\tilde{\rho}_{nm\ell}^{},\hs 
\tilde{\omega}_{nm\ell}, \hs \xoverline{\omega}_{nm\ell})$
in \eqrefe{app-eq:3-vertices} are defined as bellow:
\beqs
\label{app-eq:KKGR-RS-couplings}
\begin{align}
\label{app-eq:KKGR-RS-al_nml}
\al_{nm\ell}^{} &= \frac{1}{\,L\,}\int_0^L\!\!\td z \, e^{3A(z)} \uu_n^{}(z)\uu_m^{}(z)\uu_\ell^{}(z) \hs,
\\
\label{app-eq:KKGR-RS-alT_nml}
\tilde{\al}_{nm\ell}^{} &= \frac{1}{\,L\,}\int_0^L\!\!\td z \, e^{3A(z)} \vv_n^{}(z)\vv_m^{}(z)\uu_\ell^{}(z) \hs,
\\
\label{app-eq:KKGR-RS-betaT-nml}
\tilde{\be}_{nm\ell}^{} &= \frac{1}{\,L\,}\int_0^L\!\!\td z \, e^{3A(z)} \ww_n^{}(z)\ww_m^{}(z)\uu_\ell^{}(z) \hs,
\\
\label{app-eq:KKGR-RS-rhoT-nml}
\tilde{\rho}_{nm\ell}^{} &= \frac{1}{\,L\,}\int_0^L\!\!\td z \, e^{3A(z)} \vv_n^{}(z)\vv_m^{}(z)\ww_\ell^{}(z) \hs,
\\
\label{app-eq:KKGR-RS-omega-nml}
\tilde{\omega}_{nm\ell}^{} &= \frac{1}{\,L\,}\int_0^L\!\!\td z \, e^{3A(z)} \ww_n^{}(z)\ww_m^{}(z)\ww_\ell^{}(z) \hs,
\\
\label{app-eq:KKGR-RS-xoverlilne-omega-nml}
\xoverline{\omega}_{nm\ell}^{} &=\frac{1}{\,L\,} \int_0^L\!\!\td z \, e^{3A(z)} \ww'_n(z)\ww'_m(z)\ww_\ell^{}(z) \hs.
\end{align}
\eeqs
In particular, we can derive the trilinear coupling coefficients containing zero-modes, 
$\hs\al_{000}^{} \!=\! \al_{nn0}^{} \!=\! \tilde{\al}_{nn0}^{} \!=\! \uu_0^{}\hs$ and 
$\hs a_{000}^{} \!=\! a_{nn0}^{} \!=\! \tilde{a}_{nn0}^{} \!=\! \ff_0^{}\hs$, 
based on the normalization condition \eqrefe{Aeq:Normalization-GR}.\ 
We summarize these coupling coefficients as follows:
\beqs
\label{eq:alpha3-a3-eg}
\begin{align}
\al_{000}^{} &= \frac{1}{\,L\,}\!\int_0^L\!\!\hsm\td z \, e^{3A(z)}\hs\uu_0^2\hs\uu_0^{} = \uu_0^{} \,,
\\
\al_{nn0}^{} &= \frac{1}{\,L\,}\!\int_0^L\!\!\hsm\td z \, e^{3A(z)}\hs\uu_n^2\hs\uu_0^{} = \uu_0^{} \,,
\label{eq:alpha-nn0=u0}
\\
\tilde{\al}_{nn0}^{} &= \frac{1}{\,L\,}\!\int_0^L\!\!\hsm\td z\,e^{3A(z)}\hs\vv_n^2\hs\uu_0^{} 
= \uu_0^{} \,,
\\
a_{000}^{} &= \frac{1}{\,L\,}\!\int_0^L\!\!\hsm\td z \, e^{A(z)}\hs\ff_0^2\hs\ff_0^{} = \ff_0^{} \,,
\\
a_{nn0}^{} &= \frac{1}{\,L\,}\!\int_0^L\!\!\hsm\td z \, e^{A(z)}\hs\ff_n^2\hs\ff_0^{} = \ff_0^{} \,, 
\label{eq:ann0=f0}
\\
\tilde{a}_{nn0}^{} &= \frac{1}{\,L\,}\!\int_0^L\!\!\hsm\td z \, e^{A(z)}\hs\tilde{\ff}_n^2\hs\ff_0^{} 
= \ff_0^{} \,.
\end{align}
\eeqs

In addition, we further define the following quartic coupling coefficients:
\beqs
\label{app-eq:KKRS-couplings-flat}
\begin{align}
\label{app-eq:KKGR-RS-couplings-4pt}
\al_{nm\ell q}^{} &= \frac{1}{\,L\,}\!
\int_0^L\!\!\!\td z \, e^{3A(z)}\hs \uu_n^{}(z)\uu_m^{}(z)\uu_\ell^{}(z)\uu_q^{}(z)\,,
\\
\label{app-eq:KKGR-RS-couplings-4pt-bet}
\tilde{\be}_{nm\ell q}^{} &= \frac{1}{\,L\,}\!\int_0^L\!\!\!
\td z \, e^{3A(z)}\hs \ww_n^{}(z)\ww_m^{}(z)\ww_\ell^{}(z)\ww_q^{}(z)\,,
\end{align}
\eeqs
which will be used for the double-copy construction of Section\,\ref{sec:4.2} and 
for deriving the gravitational sum rules in Appendix\,\ref{app:D}.

\vs

Taking the flat space limit $k\hsm\ito\hsm 0\hs$, we see that all the coupling coefficients will be 
reduced to the simple trigonometric functions:
\begin{align}
& \ff_0^{} =\uu_0^{}=\ww_0^{}=1\,,
\nn\\
\label{app-eq:Flat-values-f0-u0}
& \ff_n^{}(z) = \uu_n^{}(z) = \ww_n^{}(z) = \sqrt{2\,}\cos\frac{\,n\pi z\,}{L}\hs,
\\
& \fft_n^{}(z) = \vv_n^{}(z) = \sqrt{2\,}\sin\frac{\,n\pi z\,}{L},
\nn 
\end{align}
where $n\in\mathbb{Z}^+$. Thus, together with the definitions 
\eqref{eq:alpha3-a3-eg}-\eqref{app-eq:KKRS-couplings-flat}, we deduce the values
of these KK coupling coefficients in the flat 5d limit, 
which are summarized in Table\,\ref{tab-app:1}.\ 
\begin{table}[t]
\centering
\renewcommand{\arraystretch}{1.5}
\begin{tabular}{l|c}
\hline\hline
\hspace*{3.5cm}
Cubic and Quartic Couplings (flat 5d) & Values
\\
\hline \hline
$ \tilde{a}_{nn2n}^{} ,\, \tilde{a}_{nm|n\pm m|}^{} ,\,\tilde{\al}_{nn2n}^{} ,\, \tilde{\rho}_{nn2n}^{},\, \tilde{\al}_{nm|n\pm m|}^{} ,\,\tilde{\rho}_{nm|n\pm m|}^{}$
& $-\frac{1}{\sqrt{2\,}\,}$
\\ \hline
$a_{nm0}^{},\, \tilde{a}_{nm0}^{},\, \al_{nm0}^{} ,\, \tilde{\al}_{nm0}^{} ,\, \tilde{\be}_{nm0}^{} ,\, \tilde{\rho}_{nm0}^{},\, \tilde{\omega}_{nm0}^{}$
& 0
\\ \hline
$a_{nm\ell q}^{} ,\,\tilde{a}_{nm\ell q}^{} ,\, \al_{nm\ell q}^{} ,\,  \tilde{\be}_{nm\ell q}^{}$ 
& $\frac{1}{\,2\,}$
\\ \hline
$a_{nn2n}^{} ,\, a_{nm|n\pm m|}^{} ,\, \al_{nn2n}^{} ,\, \tilde{\be}_{nn2n}^{} ,\, \tilde{\omega}_{nn2n}^{} ,\,
\al_{nm|n\pm m|}^{} ,\, \tilde{\be}_{nm|n\pm m|}^{} ,\, \tilde{\omega}_{nm|n\pm m|}^{}$
& $\frac{1}{\sqrt{2\,}\,}$
\\ \hline
$a_{nn0}^{} ,\, \tilde{a}_{nn0}^{} ,\, a_{nnmm}^{} ,\, \al_{nn0}^{} ,\, \tilde{\al}_{nn0}^{} ,\, \tilde{\be}_{nn0}^{} ,\, \tilde{\rho}_{nn0}^{} ,\, \tilde{\omega}_{nn0}^{} ,\, \al_{nnmm}^{} ,\, \tilde{\al}_{nnmm}^{} ,\, \tilde{\be}_{nnmm}^{}$
& 1
\\ \hline
$a_{nnnn}^{} ,\, \al_{nnnn}^{} ,\, \tilde{\al}_{nnnn}^{},\, \tilde{\be}_{nnnn}^{}$
& $\frac{\,3\,}{2}$
\\ 
\hline\hline
\end{tabular}
\vspace*{-1mm}
\caption{List of relevant cubic and quartic KK coupling coefficients of the 
flat 5d gauge and gravity theories under $S^1\!/\ZZ$ compactification.\ 
The subscripts are the relevant KK indices with $n\!\neq\! m\!\neq\!\ell\!\neq\!q$ and $(n,m,\ell,q)\in\mathbb{Z}^+$.}
\label{app-table:Flat-values}
\label{tab-app:1}
\vspace*{3mm}
\end{table}

\vs

Finally, we derive the on-shell 3-point KK gluon scattering amplitudes as follows:
\beqs 
\begin{align}
\TT[\{\ep_i^{}\}] &= g\hs f^{abc}\NN [\{\ep_j^{}\}] \hs, 
\\
\NN [\{\ep_j^{}\}] &= 
-\ii\hs 2\hs g \hs a_{nm\ell}^{} 
\big[\hs (\ep_1^{}\ep_2^{})(\ep_3^{} p_1^{}) + (\ep_2^{}\ep_3^{})(\ep_1^{}p_2^{}) + (\ep_3^{}\ep_1)(\ep_2^{}p_3^{}) \hs\big] \hs,
\end{align}
\eeqs 
which reproduces the formula \eqref{eq:AAA-3pt} in the main text.\ 

\vs 

Then, we can express the 3-point KK graviton amplitude into the following form:
\begin{align}
\M \big[h^{\si_1}_n h^{\si_2}_m	h^{\si_3}_\ell \big] & =	
\sum_{\lam^{}_j,\lam'_j}\!\!\hsm
\Big(\!\prod_j\!\hsm C_{\hsm\lam_j^{}\lam'_j}^{\hs\si_{\hsm j}^{}}\Big)
\M\big[\{e_i^{}\}^{\lam_j}\!,\{\ep_i^{}\}^{\lam'_j}\big]
\nn\\
& \equiv \frac{~\ka\hs\al_{nm\ell}^{}~}{4}\!
\sum_{\lam^{}_j,\lam'_j}\!\!\hsm
\Big(\!\prod_j\!\hsm C_{\hsm\lam_j^{}\lam'_j}^{\hs\si_{\hsm j}^{}}\Big)
\xoverline{\M}\big[\{e_i^{}\}^{\lam_j}\!,\{\ep_i^{}\}^{\lam'_j}\big]\hs, 
\end{align}
where the helicity index
$\,\si_j^{}\!=\!\{\pm 2,\pm 1,0\}
\!\equiv\!\{\pm 2,\pm 1,L\}\,$
labels the 5 helicity states of each external massive KK graviton.\
The polarization tensors (including the coefficients $C_{\hsm\lam_j^{}\lam'_j}^{\hs\si_{\hsm j}^{}}$)
of the external KK graviton states are defined as in \eqrefe{eq:plz-tensor-combination}
of the main text.\ 
Then, we explicitly compute the on-shell 3-point KK graviton scattering amplitudes as follows:
\begin{align}
\label{eq:Amp-3h-exact}
&\hspace*{-5mm} \xoverline{\M}[\{e^{}_i\},\{\ep^{}_i\}] 
= \ka\hs\al_{nm\ell}^{}\!\[ e^{}_{1\mu} \ep^{}_{1\nu} e^{}_{2\rho} \ep^{}_{2\si} e^{}_{3\al} \ep^{}_{3\be} 
\Ga^{\mn\rho\si\ab}_{nm\ell}\hsm (p^{}_1,p^{}_2,p^{}_3)\]
\nn\\
\hspace*{-2mm} 
=& 
\frac{~\ka\hs\al_{nm\ell}^{}~}{4} \hsm\Big[\hsm 
(e^{}_1\ep^{}_3 ) (e^{}_2 p^{}_3 ) (e^{}_3 p^{}_2 ) (\ep^{}_1\ep^{}_2 )\!+\! 
(e^{}_1e^{}_3 ) (e^{}_2 p^{}_3 ) (p^{}_2\ep^{}_3 ) (\ep^{}_1\ep^{}_2 )\!+\!
(e^{}_1 p^{}_3 ) (e^{}_2 p^{}_3 ) (e^{}_3\ep^{}_2 ) (\ep^{}_1\ep^{}_3 )
\nn\\
&+\!(e^{}_1\ep^{}_2 ) (e^{}_2 p^{}_3 ) (e^{}_3 p^{}_2 ) (\ep^{}_1\ep^{}_3 )\!+\!
(e^{}_1 p^{}_3 ) (e^{}_2 p^{}_3 ) (e^{}_3\ep^{}_1 ) (\ep^{}_2\ep^{}_3 )\!+\!
(e^{}_1e^{}_3 ) (e^{}_2 p^{}_3 ) (p^{}_3\ep^{}_1 ) (\ep^{}_2\ep^{}_3 )
\nn\\
&+(e^{}_1\ep^{}_2 ) (e^{}_2 p^{}_3 ) (e^{}_3\ep^{}_1 ) (p^{}_2\ep^{}_3 )\!-\!
2 (e^{}_1\ep^{}_2 ) (e^{}_2\ep^{}_1 ) (e^{}_3 p^{}_2 ) (p^{}_2\ep^{}_3 )\!-\!
2 (e^{}_1e^{}_2 ) (e^{}_3 p^{}_2 ) (p^{}_2\ep^{}_3 ) (\ep^{}_1\ep^{}_2 )
\nn\\
&+\!(e^{}_1e^{}_3 ) (e^{}_2\ep^{}_1 ) (p^{}_3\ep^{}_2 ) (p^{}_2\ep^{}_3 )\!+\!
(e^{}_1e^{}_2 ) (e^{}_3\ep^{}_1 ) (p^{}_3\ep^{}_2 ) (p^{}_2\ep^{}_3 )\!-\!
(e^{}_1 p^{}_3 ) (e^{}_2\ep^{}_3 ) (e^{}_3 p^{}_2 ) (\ep^{}_1\ep^{}_2 )
\nn\\
&-\!(e^{}_1 p^{}_3 ) (e^{}_2\ep^{}_1 ) (e^{}_3 p^{}_2 ) (\ep^{}_2\ep^{}_3 )\!-\!
(e^{}_1 p^{}_3 ) (e^{}_2\ep^{}_1 ) (e^{}_3\ep^{}_2 ) (p^{}_2\ep^{}_3 )\!-\!
(e^{}_1 p^{}_3 ) (e^{}_2e^{}_3 ) (p^{}_2\ep^{}_3 ) (\ep^{}_1\ep^{}_2 )
\nn\\
&+\!(e^{}_1\ep^{}_3 ) (e^{}_2 p^{}_3 ) (e^{}_3\ep^{}_2 ) (p^{}_3\ep^{}_1 )\!-\!
(e^{}_1\ep^{}_2 ) (e^{}_2\ep^{}_3 ) (e^{}_3 p^{}_2 ) (p^{}_3\ep^{}_1 )\!-\!
(e^{}_1e^{}_2 ) (e^{}_3 p^{}_2 ) (p^{}_3\ep^{}_1 ) (\ep^{}_2\ep^{}_3 )
\nn\\
&-\!(e^{}_1\ep^{}_2 ) (e^{}_2e^{}_3 ) (p^{}_3\ep^{}_1 ) (p^{}_2\ep^{}_3 )\!-\!
(e^{}_1e^{}_2 ) (e^{}_3\ep^{}_2 ) (p^{}_3\ep^{}_1 ) (p^{}_2\ep^{}_3 )\!-\!
2 (e^{}_1 p^{}_3 ) (e^{}_2\ep^{}_3 ) (e^{}_3\ep^{}_2 ) (p^{}_3\ep^{}_1 )
\nn\\
&-\!2 (e^{}_1 p^{}_3 ) (e^{}_2e^{}_3 ) (p^{}_3\ep^{}_1 ) (\ep^{}_2\ep^{}_3 )\!+\!
(e^{}_1 p^{}_3 ) (e^{}_2\ep^{}_3 ) (e^{}_3\ep^{}_1 ) (p^{}_3\ep^{}_2 )\!+\!
(e^{}_1 p^{}_3 ) (e^{}_2e^{}_3 ) (p^{}_3\ep^{}_2 ) (\ep^{}_1\ep^{}_3 )
\nn\\
&+(e^{}_1e^{}_2 ) (e^{}_3 p^{}_2 ) (p^{}_3\ep^{}_2 ) (\ep^{}_1\ep^{}_3 )\!+\!
(e^{}_1\ep^{}_3 ) (e^{}_2\ep^{}_1 ) (e^{}_3 p^{}_2 ) (p^{}_3\ep^{}_2 )
\!-\!2 (e^{}_1\ep^{}_3 ) (e^{}_2 p^{}_3 ) (e^{}_3\ep^{}_1 ) (p^{}_3\ep^{}_2 )
\nn\\
&-\! 2 (e^{}_1e^{}_3 ) (e^{}_2 p^{}_3 ) (p^{}_3\ep^{}_2 ) (\ep^{}_1\ep^{}_3 )\!+\!(e^{}_1\ep^{}_3 ) (e^{}_2e^{}_3 ) (p^{}_3\ep^{}_1 ) (p^{}_3\ep^{}_2 )\!+\!(e^{}_1e^{}_3 ) (e^{}_2\ep^{}_3 ) (p^{}_3\ep^{}_1 ) (p^{}_3\ep^{}_2 )\big] ,
\end{align}
where for simplicity we have introduced a shorthand notation, 
$(ab)\!\equiv\! (a \hsm\cdot\hsm b)$.\

\vspace*{1.5mm}
\section{\hspace*{-2mm}Proving 3-Point and 4-Point Warped Identities}
\label{app:D}
\vspace*{0.6mm}

In this Appendix, we will prove the relevant identities and sum rules for the cubic and quartic
coupling coefficients in the warped KK gauge theory and KK gravity theory.\

\vspace*{1.5mm}
\subsection{\hspace*{-2mm}Completeness Conditions}
\label{app:D1}
\vspace*{1mm}

By imposing the following completeness conditions,
\beqs
\label{eq-app:completeness}
\begin{align}
\label{eq-app:complete-cond}
\hspace*{-4mm}
\delta(z\!-\!z') &=\sum_{j=0}^{\infty}\! e^{A(z)} \ff_j^{}(z) \ff_j^{}(z') 
= \sum_{j=0}^{\infty}\! e^{A(z)} \tilde{\ff}_j^{}(z) \tilde{\ff}_j^{}(z')\hs,
\\[1mm]
\hspace*{-4mm}
\delta(z\!-\!z') &=\sum_{j=0}^{\infty}\! e^{3A(z)} \uu_j^{}(z) \uu_j^{}(z') 
=\sum_{j=0}^{\infty}\! e^{3A(z)} \vv_j^{}(z) \vv_j^{}(z') 
=\sum_{j=0}^{\infty}\! e^{3A(z)} \ww_j^{}(z) \ww_j^{}(z') \hs,
\end{align}
\eeqs
we can derive the sum rules between the cubic and quartic KK gauge/gravity coupling coefficients 
respectively:
\beqs
\label{app-eq:a-at-completeness}
\begin{align}
\label{app-eq:a-completeness}
&\sum_{j=0}^{\infty}a_{nmj}a_{\ell qj}^{} =\, a_{nm\ell q}^{} \hs, 
\qquad~~
\sum_{j=0}^{\infty}\al_{nmj}^{}\al_{\ell qj}^{} =\, \al_{nm\ell q}^{} \hs,
\\
\label{app-eq:at-completeness}
&\sum_{j=0}^{\infty}\tilde{a}_{nmj}^{} \tilde{a}_{\ell qj}^{} =\, \tilde{a}_{nm\ell q}^{} \hs, 
\qquad~~\,
\sum_{j=0}^{\infty}\tilde{\be}_{nmj}^{} \tilde{\be}_{\ell qj}^{} =\, \tilde{\be}_{nm\ell q}^{} \hs.
\end{align}
\eeqs
For the special cases with some or all of the KK indices being equal, we can recast 
above sum rules in the following simpler forms:
\beqs
\begin{align}
\label{eq:app-a-4pt}
\sum_{j=0}^{\infty}a_{nnj}^2 &= a_{nnnn}^{}\hs, \qquad~~
\sum_{j=0}^{\infty} a_{nnj}^{}  a_{mmj}^{}  =  \sum_{j=0}^{\infty} a_{nmj}^2 = a_{nnmm}^{} \hs,
\\
\label{eq:app-al-4pt}
\sum_{j=0}^{\infty} \al_{nnj}^2 &= \al_{nnnn}^{}\hs, \qquad~~
\sum_{j=0}^{\infty} \al_{nnj}^{}\al_{mmj}^{}  =  \sum_{j=0}^{\infty}\al_{nmj}^2 = \al_{nnmm}^{} \hs,
\\
\label{eq:app-alT-4pt}
\sum_{j=0}^{\infty}\tilde{a}_{nnj}^2 &=\tilde{a}_{nnnn}^{}\hs, \qquad~~
\sum_{j=0}^{\infty} \tilde{a}_{nnj}^{}  \tilde{a}_{mmj}^{}  =  \sum_{j=0}^{\infty} \tilde{a}_{nmj}^2 = \tilde{a}_{nnmm}^{} \hs,
\\
\label{eq:app-beT-4pt}
\sum_{j=0}^{\infty} \tilde{\be}_{nnj}^2 &=\tilde{\be}_{nnnn}^{}\hs, \qquad~~
\sum_{j=0}^{\infty} \tilde{\be}_{nnj}^{} \tilde{\be}_{mmj}^{}  =  \sum_{j=0}^{\infty}\tilde{\be}_{nmj}^2 = \tilde{\be}_{nnmm}^{} \hs. 
\end{align}
\eeqs
Many of these relations will be used in Sections\,\ref{sec:3} and \ref{sec:4} of the main text.\

\vspace*{1mm}
\subsection{\hspace*{-2mm}KK Mass and Coupling Identities form 3-Point Amplitudes}
\label{app:D2}
\vspace*{1.5mm}

\vspace*{1mm}
\subsubsection*{$\blacklozenge$~Identities Derived by Using Equations of Motion}
\vspace*{0.5mm}

By utilizing the equation of motion \eqref{Aeq:EOM-fn-Amu} and integration by parts, 
we derive the following integral relation:
\begin{align}
\label{app-eq:wavefunction-relation-gauge}
\int_0^L\!\!\td z\,e^{A(z)} M_n^2\,\ff_n^{}\hs X(z)
\,=\hs -\!\int_0^L\!\!\td z\,\pd_z(e^{A(z)}\pd_z \ff_n^{})\hs X(z)
= \int_0^L\!\!\td z\,e^{A(z)} \ff_n^{\pp} X^{\pp}(z) \hs,
\end{align}
where $X(z)$ can denote any function of $z\hs$, such as the product of certain eigenfunctions.\ 
Thus, using the above relation and the definitions given in Eqs.\eqref{eq:anml}-\eqref{eq:tanml}, 
and setting $X(z)\!=\hsm\ff_m^{} \hs \ff_{\ell}^{}$, 
we derive the following identities which are connected to each other 
by cycling the three KK indices $(n,m,\ell)\hs$:
\\[-7mm]
\begin{align}
M_n^2\, a_{nm\ell}^{} &= \Mn\Mm\, \tilde{a}_{nm\ell}^{} \hsm+\hsm \Mn\Ml\, \tilde{a}_{n\ell m}^{} \hs,
\nn\\
M_m^2\hs a_{nm\ell}^{} &= \Mn\Mm\, \tilde{a}_{nm\ell}^{} \hsm+\hsm \Mm\Ml\, \tilde{a}_{m\ell n}^{} \hs,
\label{app-eq:anml}
\\
M_\ell^2\hs a_{nm\ell}^{} &= \Mn\Ml\, \tilde{a}_{n\ell m}^{} \hsm+\hsm \Mm\Ml\, \tilde{a}_{m\ell n}^{} \hs.
\nn 
\end{align}
%
From these identities, we can further derive a relation connecting the coupling coefficient 
$a_{nm\ell}^{}$ to $\tilde{a}_{nm\ell}^{}\hs$:
\begin{equation}
\label{app-eq:a-ta-general}
\big(M_n^2\hsm +\hsm M_m^2\! -\hsm M_\ell^2\big) a_{nm\ell}^{}
\,=\,2 \hs\Mn\Mm\hs\tilde{a}_{nm\ell}^{}\,,
\end{equation}
which just gives \eqrefe{eq:a-at-3pt} in the main text.

\vs

We can derive another integral relation for the wraped KK gravity:
\begin{align}
\label{app-eq:wavefunction-relation-gravity}
\int_0^L\!\!\td z\,e^{3A(z)} \MGnn\,\uu_n^{}\hs X(z)
\,=\hs -\!\int_0^L\!\!\td z\,\pd_z(e^{3A(z)}\pd_z \uu_n^{})\hs X(z)
= \int_0^L\!\!\td z\,e^{3A(z)} \uu_n^{\pp} X^{\pp}(z) \hs.
\end{align}
We note that the gravitational integral relation \eqref{app-eq:wavefunction-relation-gravity} has the same structure as
that of Eq.\eqref{app-eq:wavefunction-relation-gauge} for the warped KK gauge theory except the replacements of 
$e^{A(z)}\ito e^{3A(z)}$ and $M_n^{}\ito \MGn\hs$.\ 
Then, we can derive the following relation between the wavefunction couplings $\al_{nm\ell}^{}$ 
and $\tilde{\al}_{nm\ell}^{}\hs$:
\begin{equation}
\label{app-eq:al-tal-general}
\big(\MGnn\hsm +\hsm \MGmm \!-\hsm \MGll\big) \al_{nm\ell}^{}
\,=\,2\hs\MGn \MGm\hs\tilde{\al}_{nm\ell}^{}\,,
\end{equation}
which further reproduces \eqrefe{eq:al-tal-3pt} in the main text.

\vs

\subsubsection*{$\blacklozenge$~Relation between Gravitational KK Couplings 
\boldsymbol{$\tilde{\be}_{nmj}^{}$} and 
\boldsymbol{$\al_{nmj}^{}$}}
\vspace*{1mm}

Next, we prove the identity \eqref{eq:al-tbe-LLT-55T} in the main text, 
which describes the relation between the wraped KK gravity couplings $\al_{nmj}^{}$ and $\tilde{\be}_{nmj}^{}\hs$.\
For this purpose, we first derive the following relation:
\begin{align}
\label{app-eq:LMnMmtbe}
& \hspace*{-3mm} L\MGn \MGm \tilde{\be}_{nm\ell}^{}
=\int_0^L \!\!\hsm\td z \, e^{3A(z)}\, (\MGn\ww_n^{}) (\MGm \ww_m^{})\uu_\ell^{}
\nn\\
& = \int_0^L\!\! \td z \hs \Big[\pd_z(e^{A(z)} \vv_n^{})\Big]\!
\Big[e^{2 A(z)}(A'\!+\pd_z^{}) \vv_m^{}\Big] \uu_\ell^{}
\nn\\
& =-\!\!\int_0^L\!\!\! \td z \Big(\hsm e^{A(z)} \vv_n^{}\hsm\Big) 
\pd_z^{}\hsm\Big[e^{2 A(z)}\hsm (A'\!+\hsm\pd_z^{}) \vv_m^{}\Big] \hsm\uu_\ell^{}
-\!\hsm\int_0^L\!\!\!\td z \Big(\hsm e^{A(z)} \vv_n^{}\hsm\Big)\!
\Big[\hsm e^{2A}(A'\!+\hsm\pd_z^{}) \vv_m^{}\Big]\hsm\uu_\ell^\pp \hspace*{6mm}
\nn\\
&=L\MGmm\tilde{\al}_{nm\ell}^{}-L\MGm\MGl\hs\tilde{\rho}_{n\ell m}^{}\hs, 
\end{align}
which leads to the identity:
\beq 
\label{app-eq:MnBe-MmAt-MLrho}
\MGn\tilde{\be}_{nm\ell}^{} = \MG_m^{}\tilde{\al}_{nm\ell}^{} - 
\MGl\hs\tilde{\rho}_{n\ell m}^{} \,. 
\eeq 
For the right-hand side of Eq.\eqref{app-eq:MnBe-MmAt-MLrho}, the second term 
contains the trilinear coupling coefficient $\tilde{\rho}_{nm\ell}^{}$ for which 
we can derive the following relation:
\begin{align}
\label{app-eq:trho-Mn-Mm-Ml}
& \hspace*{-7mm} L\MGn\tilde{\rho}_{m\ell n}^{}
=\int_0^L\!\!\!\td z \Big[\pd_z (e^{A(z)}\vv_n^{})\Big]\!
\Big(\hsm e^{A(z)}\vv_m^{}\hsm\Big)\!\Big(\hsm e^{A(z)}\vv_\ell^{}\hsm\Big)
\nn\\
& \hspace*{-4mm} =-\!\!\int_0^L\!\!\!\td z\Big(\hsm e^{A(z)}\vv_n^{}\hsm\Big)\!
\Big[\hsm\pd_z^{}\big(e^{A(z)}\vv_m^{}\big)\hsm\Big]\!\Big(\hsm e^{A(z)}\vv_\ell^{}\hsm\Big)
\!-\!\int_0^L\!\!\!\td z\Big(\! e^{A(z)}\vv_n^{}\hsm\Big)\!
\Big(\! e^{A(z)}\vv_m^{}\hsm\Big)\!\Big[\hsm\pd_z^{}\big(e^{A(z)}\vv_\ell^{}\big)\hsm\Big] 
\hspace*{5mm}
\nn\\
& \hspace*{-4mm}
=-L\hs\MGm\hs\tilde{\rho}_{n\ell m}^{}\!-\hsm L\hs\MGl\hs\tilde{\rho}_{nm\ell}^{}\,,
\end{align}
where we have used Eqs.\eqref{Aeq:uv} and \eqref{Aeq:wv}.\ 
We can reexpress Eq.\eqref{app-eq:trho-Mn-Mm-Ml} as follows:
\beq 
\MGn\tilde{\rho}_{m\ell n}^{} \!+\hsm \MGm\hs\tilde{\rho}_{n\ell m}^{}
\!+\hsm \MGl\hs\tilde{\rho}_{nm\ell}^{}
= 0 \,.
\eeq 
Using Eqs.\eqref{app-eq:MnBe-MmAt-MLrho} and \eqref{app-eq:al-tal-general} 
as well as cycling their three KK indices, we further derive the following relations:
\beqs
\label{app-eq:vvw}
\begin{align}
\label{app-eq:vvw-a}
2\hs\MGnn\MGmm\tilde{\be}_{nm\ell}^{} 
&=\MGmm\hsm\big(\MGmm\!+\!\MGnn\!-\!\MGll\big)\al_{nm\ell}^{}
\hsm -\hsm 2\hs\MGn\MGmm\MGl\hs\tilde{\rho}_{n\ell m}^{} \hs,
\\
\label{app-eq:vvw-b}
2\hs\MGnn\MGmm\tilde{\be}_{nm\ell}^{} 
&=\MGnn\hsm\big(\MGmm\!+\!\MGnn\!-\!\MGll\big)\al_{nm\ell}^{}
\hsm -\hsm 2\hs\MGnn\MGm\MGl\hs\tilde{\rho}_{m\ell n}^{} \hs,
\\
\label{app-eq:vvw-c}
2\hs\MGmm\MGll\tilde{\be}_{m\ell n}^{} 
&=\MGll\hsm\big(\MGll\!+\!\MGmm\!-\!\MGnn\big)\al_{nm\ell}
\hsm -\hsm 2\hs\MGn\MGm\MGll\tilde{\rho}_{nm\ell}^{}\hs,
\\
\label{app-eq:vvw-d}
2\hs\MGmm\MGll\tilde{\be}_{m\ell n}^{} 
&=\MGmm\hsm\big(\MGll\!+\!\MGmm\hsm\!-\!\MGnn\big)\al_{nm\ell}
\hsm -\hsm 2\hs\MGn\MGnn\MGl\hs\tilde{\rho}_{n\ell m}^{} \hs,
\\
\label{app-eq:vvw-e}
2\hs\MGll\MGnn\tilde{\be}_{\ell nm}^{} 
&=\MGll\hsm\big(\MGll\!+\!\MGnn\!-\!\MGmm\big)\al_{nm\ell}
\hsm -\hsm 2\hs\MGn\MGm \MGll\hs\tilde{\rho}_{nm\ell}^{} \hs,
\\
\label{app-eq:vvw-f}
2\hs\MGll\MGnn\tilde{\be}_{\ell nm}^{} 
& =\MGnn\hsm\big(\MGll\!+\!\MGnn-\MGmm\big)\al_{nm\ell}^{}
\hsm -\hsm 2\hs\MG_n^2\MGm\MGl\hs\tilde{\rho}_{m\ell n}^{}\hs.
\end{align}
\eeqs
With the six relations above, we compute the sum 
$\Fr{3}{2}[(a)\!+\!(b)]\!+\!\Fr{1}{2}\big[(c)\!-\!(d)\!+\!(e)\!-\!(f)\big]$ 
and impose Eq.\eqref{app-eq:trho-Mn-Mm-Ml}, with which we derive the final identity:
\begin{equation}
\label{eq:al-tbe-LLT-55T-3pt}
\Big[\big(\MGnn\!+\!\MGmm\!-\!\MGll\big)^{\!2}\!+\!2\hs\MGnn\MGmm\Big]
\al_{nm\ell}^{}\,=\,6\hs\MGnn\MGmm\hs\tilde{\be}_{nm\ell}^{}\hs.
\end{equation}
This just reproduces the identity \eqref{eq:al-tbe-LLT-55T} in the main text.\

\vspace*{1mm}
\subsubsection*{$\blacklozenge$\, Relation between Gravitational KK Couplings 
\boldsymbol{$\tilde{\rho}_{nmj}^{}$} and \boldsymbol{$\al_{nmj}^{}$}}
\vspace*{1mm}

Then, we prove the identity \eqref{eq:cond-h11L=VVphi} in the main text,
which describes the relation between the wraped KK gravity couplings $\al_{nmj}^{}$ and $\tilde{\rho}_{nmj}^{} \hs$.\ 
We can sum up the third and fifth identities in \eqrefe{app-eq:vvw} and derive the following relation:
\begin{equation}
4 \hs\MGn \MGm \MGll \tilde{\rho}_{nm\ell}^{} = 2 \hs\mathbb{M}_\ell^4 \al_{nm\ell}^{} - 2\hs\MGmm \MGll \tilde{\be}_{m\ell n}^{} -2\hs\MGnn\MGll \tilde{\be}_{n\ell m}^{} \,.
\end{equation}
With this and further using the identity \eqref{eq:al-tbe-LLT-55T-3pt}, we arrive at
\begin{equation} 
\label{eq-app:cond-h11L=VVphi}
\Big[ 2\hs\mathbb{M}_\ell^4  
\!-\hsm\MGll\big(\MGnn \!+\hsm \MGmm\big) \!-\!\big(\MGnn \!-\hsm\MGmm\big)^{\!2} \Big] \al_{nm\ell}^{} 
= 6\hs \MGn \MGm \MGll \hs\tilde{\rho}_{nm\ell}^{} \,,
\end{equation}
which reproduces the identity \eqref{eq:cond-h11L=VVphi} in the main text.\

\vspace*{1mm}
\subsection{\hspace*{-2mm}KK Mass and Coupling Identities from 4-Point Amplitudes}
\vspace*{1mm}
\label{app:D3}

Choosing $X(z)=\ff_m^{} \hs \ff_{\ell}^{} \hs \ff_q$ in \eqrefe{app-eq:wavefunction-relation-gauge}, 
we further derive the following identities 
by cycling the four KK indices $(n,m,\ell,q)\hs$:
%
\begin{align}
\label{app-eq:MnMmMlMq-anmlq}
\hspace*{-12mm}
\Mnn\,a_{nm\ell q}^{} &\hs =\hs 
\Mn\Mm\lrb{\fft_n^{}\fft_m^{}\ff_\ell^{}\hs\ff_q^{}}
\!+\!\Mn\Ml\hs\lrb{\fft_n^{}\fft_\ell^{}\hs\ff_m^{}\ff_q^{}}
\!+\!\Mn\Mq\lrb{\fft_n^{}\fft_q^{}\ff_m^{}\ff_\ell^{}},
\nn\\
\hspace*{-12mm}
\Mmm\,a_{nm\ell q}^{} &\hs =\hs 
\Mn\Mm\lrb{\fft_n^{}\fft_m^{}\ff_\ell^{}\hs\ff_q^{}}
\!+\!\Mm\Ml\lrb{\fft_m^{}\fft_\ell^{}\hs\ff_n^{}\ff_q^{}}
\!+\!\Mm\Mq\lrb{\fft_m^{}\fft_q^{}\ff_n^{}\ff_\ell^{}},
\nn\\[-3mm]
\\[-3mm]
\hspace*{-12mm}
\Mll\,a_{nm\ell q}^{} &\hs =\hs 
\Mn\Ml \lrb{\fft_n^{}\fft_\ell^{}\hs\ff_m^{}\ff_q^{}}
\!+\!\Mm\Ml\lrb{\fft_m^{}\fft_\ell^{}\hs\ff_n^{}\ff_q^{}}
\!+\!\Ml\Mq\lrb{\fft_\ell^{}\hs\fft_q^{}\ff_n^{}\ff_m^{}},
\nn\\
\hspace*{-12mm}
\Mqq\, a_{nm\ell q}^{} &\hs =\hs 
\Mn\Mq\lrb{\fft_n^{}\fft_q^{}\ff_m^{}\ff_\ell^{}}
\!+\!\Mm\Mq\lrb{\fft_m^{}\fft_q^{}\ff_n^{}\ff_\ell^{}}
\!+\!\Ml\Mq\lrb{\fft_\ell^{}\hs\fft_q^{}\ff_n^{}\ff_m^{}}.
\nn
\end{align}
%
For the sake of simplicity, we have used brackets $\lrb{\cdots}$ in the above equations
to denote integration over the relevant product of the 5d wavefunctions:
\begin{equation}
\label{app-eq:def-[[]]}
\lrb{\mathsf{X}_{n_1^{}}^{}\!\hsm\cdots\hs \mathsf{X}_{n_N^{}}^{}} =
\frac{1}{\,L\,}\!\!\int_{0}^{L}\!\!\hsm\td z\, e^{c_0^{}A(z)}\hs 
\mathsf{X}_{n_1^{}}^{}\hsm\!(z)\cdots \mathsf{X}_{n_N^{}}^{}\hsm\!(z)\hs,
\end{equation}
where $\mathsf{X}_n^{}$ represents the 5d wavefunctions and the coefficient $c_0^{}$ is chosen as
$c_0^{}\!=\!1\,(c_0^{}\!=\!3)$ for the warped KK gauge (gravity) theory.\ 

\vspace*{1mm}
\subsubsection*{$\blacklozenge$\, Summing up ${j}$-Modes with KK Mass-Squared $M_j^2$ ($\MGjj$)}
\vspace*{1.5mm}

Summing up the four identities of \eqrefe{app-eq:MnMmMlMq-anmlq} and making use of
\eqrefe{app-eq:a-ta-general}, we derive the following new sum rule:
\begin{equation}
\label{app-eq:justforfun}
\sum_{j=0}^{\infty}\! 
\Mjj (a_{nmj}^{}a_{\ell qj}^{}\!+\hsm a_{n\ell j}^{}a_{mqj}^{}\!+\hsm a_{nqj}^{}a_{m\ell j}^{})
= \big(\Mnn\!+\! \Mmm \!+\!\Mll\!+\!\Mqq\big) a_{nm\ell q}^{}\hs.
\end{equation}
For the special case of $n\!=\!m\!=\!\ell\!=\!q$, the above identity reduces to:
\beq
\label{app-eq:SumRule2-YM-4pt}
\sum_{j=0}^{\infty}\! \Mjj a_{nnj}^{2} = \frac{\,4\,}{3}M_n^2  a_{nnnn}^{}\hs,
\eeq 
which just reproduces the identity \eqref{eq:SumRule2-YM-4pt} in the main text.\
As for the warped KK gravity theory, we can derive a new sum rule in similar form:
\begin{equation}
\label{app-eq:alpha2=M2nmlq}
\sum_{j=0}^{\infty} \!\MGjj (\al_{nmj}^{}\al_{\ell qj}^{}\!+\!\al_{n\ell j}^{}\al_{mqj}^{}
\!+\!\al_{nqj}^{}\al_{m\ell j}^{})
= \big(\MGnn\!+\hsm\MGmm\!+\hsm\MGll\!+\hsm\MGqq\big)\al_{nm\ell q}^{} \hs.
\end{equation}
For the special case of $n\!=\!m\!=\!\ell\!=\!q\hs$, we simplify the above identity as follows:
\beq 
\label{app-eq:alpha2nnj=M2nnnn}
\sum_{j=0}^{\infty} \hsm \MGjj \al_{nnj}^{2} 
= \frac{\,4\,}{3}\MGnn\al_{nnnn}^{} \hs,
\eeq 
which reproduces the identity \eqref{eq:SumRule-GR-4pt-1b} in the main text.

\vs

Then, we derive two sum rules \eqref{eq:4AL-E2a-cancel} and \eqref{eq:4AL-E2b-cancel}, 
which are used for computing the inelastic scattering amplitudes of longitudinal KK gauge bosons.\
For \eqrefe{eq:4AL-E2a-cancel}, we can set 
$(\ell,q)\!=\!(n,m)$ in \eqrefe{app-eq:justforfun}, and derive the following sum rule:
\begin{equation}
\label{app-eq:nnmm-E2-cancel-a}
\sum_{j=0}^{\infty}\! \Mjj a_{nmj}^2
= \sum_{j=0}^{\infty}\!\(\hsm \Mnn \!+\! M_m^2 \!-\! \Fr{1}{2}\Mjj \)\! a_{nnj}^{} a_{mmj}^{}\,.
\end{equation}
For the wraped KK gravity theory, we can derive another sum rule in parallel: 
\begin{equation}\label{app-eq:GRET-nnmm-Mj2}
\sum_{j=0}^{\infty}\! \MGjj \al_{nmj}^2
= \sum_{j=0}^{\infty}\!\(\hsm \MGnn \!+\! \MGmm \!-\! \Fr{1}{2}\MGjj \)\! \al_{nnj}^{} \al_{mmj}^{}\,.
\end{equation}

\vspace*{0.5mm}
\subsubsection*{$\blacklozenge$~Summing up $j$-Modes with Inverse Mass-Squared $M_j^{-2}$\,($\MG_j^{-2}$)}
\vspace*{2mm}

To derive the sum rule \eqref{eq:4AL-E2b-cancel}, we take the difference 
between the first two identities in \eqrefe{app-eq:anml} and arrive at
\begin{equation}
\label{app-eq:lemma-d6}
\big(\Mnn\hsm -\!\Mmm\big) a_{nmj}^{} 
= \Mj\big(\Mn \tilde{a}_{njm}^{}\!-\!\Mm \tilde{a}_{mjn}^{}\big).
\end{equation}
Then, squaring the above formula and summing over $j\hs$, we derive the following result: 
\begin{align}
\label{app-eq:justforfun2}
&\sum_{j=1}^{\infty}\hsm\!\big(\Mnn\hsm -\!\Mmm\big)^{\!2}M_j^{-2} a_{nmj}^2
=\sum_{j=1}^{\infty}\!\big(\Mn\tilde{a}_{njm}^{}\!-\!\Mm\tilde{a}_{mjn}^{}\big)^{\!2}
\nn\\
&=\sum_{j=1}^{\infty}\!\hsm\big(\Mnn\tilde{a}_{njm}^2
\!-\!2\Mn\Mm\tilde{a}_{njm}^{}\tilde{a}_{mjn}^{}\!+\!\Mmm\tilde{a}_{mjn}^2\big)
\nn\\
&=\sum_{j=1}^{\infty}\hsm (\Mnn\tilde{a}_{nnj}^{}a_{mmj}\!-\hsm 2\Mn\Mm\tilde{a}_{nmj}^{}a_{nmj}^{}
\!+\!\Mmm\tilde{a}_{mmj}^{}a_{nnj}^{})
\nn\\
&=\sum_{j=1}^{\infty}\!\hsm\[\hsm\Fr{1}{2}(2\Mnn\!-\!\Mjj)a_{nnj}^{}a_{mmj}^{}\!-\!(\Mnn\!+\!\Mmm-\Mjj)
a_{nmj}^2\!+\!\Fr{1}{2}(2\Mmm-\Mjj)a_{mmj}^{}a_{nnj}^{}\]
\nn\\
&=\sum_{j=0}^{\infty}\!\Mjj\big(a_{nmj}^2\!-\hsm a_{nnj}^{} a_{mmj}^{}\big)\hs,
\end{align}
where for the third equality sign, we have imposed the following completeness relation:
\begin{align}
\label{app-eq:lemma-com}
\sum_{j=1}^{\infty} \!\tilde{a}_{njm}^{} \tilde{a}_{\ell jq}^{}
=\lrb{\fft_n^{}\ff_m^{}\fft_\ell^{}\ff_q^{}}
=\sum_{j=1}^{\infty}\hsm\lrb{\fft_n^{}\fft_\ell^{}\ff_j^{}}\!\lrb{\ff_m^{}\ff_q^{}\ff_j^{}}
=\sum_{j=1}^{\infty} \!\tilde{a}_{n\ell j}^{}a_{mqj}^{}\hs.
\end{align}
For the fourth equality of Eq.\eqref{app-eq:justforfun2}, we have used \eqrefe{app-eq:a-ta-general}.\
Moreover, substituting \eqrefe{app-eq:nnmm-E2-cancel-a} into \eqrefe{app-eq:justforfun2}, 
we derive the following sum rule identity: 
\begin{equation}
\label{app-eq:3.45c-f}
\sum_{j=1}^{\infty}\!\!\big(\Mnn\!-\!\Mmm\big)^{\!2}M_j^{-2} a_{nmj}^2=\sum_{j=0}^{\infty}\!\big(\Mnn\!+\!\Mmm\!-\!\Fr{3}{2}\Mjj\big)a_{nnj}^{}a_{mmj}^{} \hs,
\end{equation}
which is \eqrefe{eq:4AL-E2b-cancel} in the main text.

\vs

Next, we derive the sum rule identities in Eq.\eqref{eq:E2-cancelation-nmlq} of Sec.\,\ref{sec:3.2}.\ 
We prove the $s$-channel sum rule \eqref{eq:E2-cancelation-nmlq-s} as an example
and other sum rules in Eq.\eqref{eq:E2-cancelation-nmlq} can be readily obtained 
from Eq.\eqref{eq:E2-cancelation-nmlq-s}
by permuting the KK indices.\ 
From Eq.\eqref{app-eq:lemma-d6}, we derive the following: 
\begin{align}
\label{app-eq:justforfun3}
&\big(\Mnn\!-\!\Mmm\big)\big(\Mll\!-\!\Mqq\big)\!
\sum_{j=1}^{\infty}\!M_j^{-2}a_{nmj}^{}\hs a_{\ell qj}^{}
\nn\\
&=\sum_{j=1}^{\infty}\!\[\!\big(\Mnn\!-\!\Mmm\big)\hsm M_j^{-1} a_{nmj}^{}\hsm\]\!\!
\[\!\big(\Mll\!-\!\Mqq\big)\hsm M_j^{-1}\hsm a_{\ell qj}^{}\]
\nn\\
&=\sum_{j=1}^{\infty}\!\big(\Mn \tilde{a}_{njm}^{}\!-\!\Mm \tilde{a}_{mjn}^{}\big)\!
\big(\Ml \tilde{a}_{\ell jq}^{}\!-\!\Mq \tilde{a}_{qj\ell}^{}\big)
\nn\\
&=\sum_{j=1}^{\infty}\!\hsm\big(\Mn\Ml \tilde{a}_{njm}^{}\tilde{a}_{\ell jq}^{}\!-\!
\Mn\Mq \tilde{a}_{njm}^{}\tilde{a}_{qj\ell}^{}\!-\!\Mm\Ml \tilde{a}_{mjn}^{}\tilde{a}_{\ell jq}^{}
\!+\!\Mm\Mq \tilde{a}_{mjn}^{}\tilde{a}_{qj\ell}^{}\big)
\nn\\
&=\sum_{j=1}^{\infty}\!\!\big(\Mn\Ml \tilde{a}_{n\ell j}^{}a_{mqj}^{}\!-\!\Mn\Mq \tilde{a}_{nqj}^{}a_{m\ell j}^{}
\!-\!\Mm\Ml \tilde{a}_{m\ell j}^{}a_{nqj}^{}\!+\!\Mm\Mq \tilde{a}_{mqj}^{}a_{n\ell j}^{}\big) ,
\end{align}
where we have applied Eqs.\eqref{app-eq:anml} and \eqref{app-eq:lemma-com} 
for the second and fourth equality signs, respectively.\ 
Then, substituting \eqrefe{app-eq:a-ta-general} into \eqrefe{app-eq:justforfun3}, we finally arrive at
\begin{align}
\label{app-eq:3.57-f}
&\big(\Mnn\!-\!\Mmm\big)\!\big(\Mll\!-\!\Mqq\big)\!\sum_{j=1}^{\infty}\hsm\!M_j^{-2}a_{nmj}^{}a_{\ell qj}^{}
\nn\\
&=\sum_{j=1}^{\infty}\!\Big[\Fr{1}{2}\hsm 
\big(\Mnn\!+\!\Mll\!-\!\Mjj\big) a_{n\ell j}^{}a_{mqj}^{} 
\!-\!\Fr{1}{2}\hsm\big(\Mnn\!+\!\Mqq\!-\!\Mjj\big) a_{nqj}^{}a_{m\ell j}^{}
\nn\\
&\hspace*{10mm}
-\Fr{1}{2}\hsm\big(\Mmm\!+\!\Mll\!-\!\Mjj\big) a_{m\ell j}^{}a_{nqj}^{}
\!+\!\Fr{1}{2}\hsm\big(\Mmm\hsm\!+\hsm\!\Mqq\hsm\!-\!\hsm\Mjj\big) a_{mqj}^{}a_{n\ell j}^{} \Big]
\nn\\
&=\sum_{j=0}^{\infty}\!\Mjj\big(a_{nqj}^{}a_{m\ell j}^{}\!-\hsm a_{n\ell j}^{}a_{mqj}^{}\big).
\end{align}
This just reproduces the sum rule identity \eqref{eq:E2-cancelation-nmlq-s} given in the main text.

\vs 

\vspace*{1mm}
\subsubsection*{$\blacklozenge$~Summing up $j$-Modes with KK Masses up to $\MG_j^{\hs 6}$}
\vspace*{1mm}

Finally, we derive the sum rule \eqref{eq:SumRule-GR-4pt-2} in the main text.\ 
With \eqrefe{eq:al-tal-nnj} and \eqrefe{app-eq:alpha2nnj=M2nnnn}, we first derive the 
following identity:
\beq 
\sum_{j=0}^{\infty}\!\al_{nnj}^{}\tilde{\al}_{nnj}^{} = \frac{1}{\,3\,}\al_{nnnn}^{} \hs.
\eeq
By using \eqrefe{eq:al-tal-nnj} and \eqrefe{app-eq:alpha2nnj=M2nnnn}, we compute:
\begin{align}
\label{app-eq:rj6-alpha2}
\sum_{j=0}^{\infty}\! \rh_j^6 \al_{nnj}^2 
&= \sum_{j=0}^{\infty}\hsm\! 4\hs\rh_j^2 \big(\al_{nnj}^{} \!-\hsm \tilde{\al}_{nnj}^{}\big)^{\!2} 
=\sum_{j=0}^{\infty}\hsm\! 4\hs (4\hsm +\hsm \rh^2_j)\hs\tilde{\al}_{nnj}^2 \hs,
\end{align}
where we have used the notation $\hs \rh_j^{}\!=\!\MGj/\MGn\hs$.\ 
Then, we further compute the following:
\beqs 
\label{app-eq:rj2-Talpha-nnj}
\begin{align}
\MGjj \tilde{\al}_{nnj}^{} 
&= \MGjj \int_{0}^L \!\!\hsm\td z\hsx {e^{3A(z)}} \hs \vv_n^2 \uu_j 
\nn\\
&= \int_{0}^L \!\!\hsm\td z\hsx {e^{3A(z)}} \hs \uu_j^{} 
\Big[\hsm\!-\!12 {A''}\hs\vv_n^2 \!+\!12 {A'}\hs\uu_n^{}\vv_n^{} \!+\!2\hs\vv_n^2 
\!-\!2\hs\MGnn \uu_n^2 \Big] ,
\\
\sum_{j=0}^{\infty} \hsm\MGjj \tilde{\al}_{nnj}^2 
&= -2\hs\MGnn \al_{nnj}^{}\tilde{\al}_{nnj}^{} \hsm 
+\hsm 2\hs\MGnn \tilde{\al}_{nnj}^2 + I_1^{} +I_2^{}\hs, 
\end{align}
\eeqs 
where the $I_1^{}$ and $I_2^{}$ represent two integrals:
\beqs
\label{app-eq:I1-I2}
\begin{align}
I_1^{} &= -12\hs \MGn\!\!\int_0^L\!\!\!\td z \hs {e^{3A(z)}\hsm  A'}\hs\uu_n^{}\vv_n^3 
= -2 \hs\MGnn\!\sum_{j=0}^{\infty}\!\hsm 
\big( \tilde{\al}_{nnj}^2 \!-\!3\hs\al_{nnj}^{}\tilde{\al}_{nnj}^{}\big), 
\\
I_2^{} &= -12\! \int_0^L \!\!\!\td z \hs {e^{3A(z)}\hsm A''}\hs\vv_n^4 
= -\frac{1}{\,2\,} I_1^{} \hs. 
\end{align}
\eeqs
In the above we have denoted $A'\!\!=\!\pd_{z}^{}A\hsm(z)$ and
$A''\!\!=\!\pd_{z}^{2}A\hsm(z)$.\ 
Combining Eqs.\eqref{app-eq:rj2-Talpha-nnj} and \eqref{app-eq:I1-I2}, we further compute
the summation in Eq.\eqref{app-eq:rj6-alpha2}:
\begin{align}
\label{app-eq:3.72}
\sum_{j=0}^{\infty}\!\rh_j^6 \al_{nnj}^2 
&= \sum_{j=0}^{\infty}\!\hsm 4\big(5\hs\tilde{\al}_{nnj}^2 \!+\hsm \al_{nnj}^{}\tilde{\al}_{nnj}^{}\big) 
= \sum_{j=0}^{\infty}\!\hsm \(\!5\hs\rh_j^4 \hsm -\!\frac{\,16\,}{3}\!\)\!\al_{nnj}^2 \hs,
\end{align}
where we have made use of \eqrefe{eq:al-tal-nnj} and \eqrefe{app-eq:alpha2nnj=M2nnnn}.\ 
We can re-express Eq.\eqref{app-eq:3.72} as follows:
\begin{align}
\label{app-eq:3.72a}
\sum_{j=0}^{\infty}\!\hsm \(\!\rh_j^6\hsm -\hsm 5\hs\rh_j^4 \hsm +\hsm\frac{\,16\,}{3}\!\)\!\al_{nnj}^2 
= 0 \hs,
\end{align}
which just reproduces the identity \eqref{eq:SumRule-GR-4pt-2} in the main text.\

\section{\hspace*{-2mm}Computing Inelastic KK Graviton Amplitude of \boldmath{$(n,n)\ito (m,m)$}}  
\label{app:E}

In this Appendix, we explicitly compute the inelastic scattering amplitude of longidtudinal KK gravitons
via $h^n_L \hs h^n_L \ito h^m_L \hs h^m_L$, 
and demonstrate the nontrivial large energy cancellations of $E^{10}\!\ito E^2$
for this inelastic channel.$\!$\footnote{%
This is a highly nontrivial calculation and is harder than the previous calculation for the elastic amplitude of $h^n_L \hs h^n_L \!\ito\! h^n_L \hs h^n_L$ \cite{Chivukula:2020hvi}.}\ 
With these, we derive its leading-order amplitude 
\eqref{eq:LO-amp-nnmm} as given in the main text.\ 
Then, we prove the GRET \eqref{eq:GRET-nnmm} for this inelastic scattering process.

\subsection{\hspace*{-2mm}Proving Energy Cancellations for Inelastic Channel \boldmath{$(n,n)\ito (m,m)$}}
\label{app:E1}
\vspace*{1mm}

We make high energy expansion for the inelastic scatering amplitude of longitudinal KK gravitons 
in terms of the scattering energy $s_0^{}$, as defined in Eq.\eqref{eq:s0-t0-u0}:
\beq
\M[h^n_L \hs h^n_L \ito h^m_L \hs h^m_L] \,=\, 
\sum_{k=0}^5\! \kappa^2\hs s_0^k\hs \xoverline{\M}_k^{} +O(s_0^{-1}) \hs.
\eeq 

We first compute the leading individual contributions of $O(s_0^5)$ as follows:
\begin{align}
\xoverline{\M}_5^{} =\frac{s_\theta^2 }{\,576\hs\MG_n^4\MG_m^4\,} 
\bigg(\!\al_{nnmm}^{} \!-\! \sum_{j=1}^{\infty}\!\al_{nnj}^{}\al_{mmj}^{}\!\bigg)
\!+\!\frac{ (3\!+\!\ctt) s_\theta^2}{\,2304\hs\MG_n^4 \MG_m^4\,}
\!\bigg(\!\al_{nnmm}^{} \!-\! \sum_{j=1}^{\infty}\al_{nmj}^2\!\bigg) .
\end{align}
We find that 
the amplitude $\xoverline{\M}_5^{}$ vanishes by applying the completeness condition \eqref{eq:app-al-4pt}.\ 

\vs 

Then, we compute the next sub-amplitude at $O(s_0^4)\hs$:
\begin{align}
\hspace{-8mm} 
\xoverline{\M}_4^{}\!=\frac{\,-45\!-\!20\hs\ctt \!+\! c_{4\theta}^{}\,}{18432\hs\MG_n^4\hs\MG_m^4}\!
\sum_{j=1}^{\infty} \! 
\[ \MGjj (\al_{nnj}^{}\al_{mmj}^{} \!-\! \al_{nmj}^2) \!+\! (\MGnn \hsm\!-\hsm \MGmm)^2 
\frac{\,\al_{nmj}^2\,}{\MG_j^2}\!\] \hsm\!. 
\end{align}
In order to prove that this sub-amplitude vanishes,    
we carry out a derivation similar to that of Eq.\eqref{app-eq:justforfun2},  
where we replace the wraped KK gauge coupling $a_{nmj}^{}$  by the KK gravity coupling $\al_{nmj}^{}$
in each step.\ Thus, we derive the following sum rule identity:
\begin{align}
(\MGnn \!-\hsm \MGmm)^2 \sum_{j=1}^{\infty}\hsm\!\frac{\,\al_{nmj}^2\,}{\MGjj} 
= \sum_{j=1}^{\infty}\!\MGjj ( \al_{nmj}^2 \!-\! \al_{nnj}^{}\al_{mmj}^{} )\hs, 
\end{align}
which ensures the sub-amplitude $\xoverline{\M}_4^{}$ to vanish.\ 

\vs

Next, we compute the scattering amplitde at $O(s_0^3)$ and derive the following:
{\small 
\begin{align}
\hspace*{-0.5cm}
\xoverline{\M}_3^{}
= &\frac{\,47 \!+\! 10\rh^2 \!+\hsm 47\rh^4 \!+\! (9\!+\hsm 6\rh^2 \!+\hsm 9\rh^4)\ctt\,}{6912 \hs \MG_n^4 \hs \rh^4 (1\!+\hsm \rh^2)}\! 
\sum_{j=1}^{\infty}\!\rh_j^2 \al_{nnj}^{}\al_{mmj}^{} +\!\frac{\,9\tilde{\rho}_{nn0}^{}\tilde{\rho}_{mm0}^{} \!-\! \al_{nn0}^{}\al_{mm0}^{}\,}{216 \MG_n^4 \rh^2}
\nn \\
& +\!\frac{\,22 \!-\hsm\! 124\rh^2 \!+\! 22\rh^4 \!-\hsm 6(1 \!+\hsm 6\rh^2 \!+\hsm \rh^4)\ctt\,}
{6912 \hs \mathbb{M}_{n}^4 \hs \rh^4(1 \!+\hsm \rh^2)}\!
\sum_{j=1}^{\infty}\!\rh_j^2\al_{nmj}^2 
\!-\!\frac{\,(1\!-\hsm \rh^2)^2 (1 \hsm\!+\! 23\ctt)\,}{2304\MG_n^4 \hs \rh^4}\al_{nm0}^2
\nn \\
& -\frac{5 \!+\hsm 3\ctt}{\,1728\MG_n^4 \hs \rh^4\,}\!
\bigg(\!9\rh^2\tilde{\rho}_{nm0}^2 \!+\!\frac{1}{4}(1 \!-\hsm \rh^2)^4 \!\sum_{j=1}^{\infty}\rh_j^{-4} \al_{nmj}^2\!\bigg)\!
+\!\frac{1 \!+\hsm 3\ctt}{\,3456 \MG_n^4 \hs \rh^4\,}\!\! \sum_{j=1}^{\infty}\rh_j^4\al_{nnj}^{}\al_{mmj}^{}
\nn \\
& +\frac{-17 \!+\hsm 9\ctt}{\,6912\hs\MG_n^4 \hs\rh^4\,}
\!\sum_{j=1}^{\infty}\!\rh_j^4\al_{nmj}^2 \hs. 
\end{align}
}
We can prove that the above amplitude $\xoverline{\M}_3^{}$ vanishes by using the following relations:
{\small 
\beqs
\label{app-eq:GRET-nnmm-Mj4}
\begin{align}
\label{app-eq:GRET-nnmm-Mj4-a}
& \hspace*{-0.5cm}
\rh^2(9\tilde{\rho}_{nn0}^{}\tilde{\rho}_{mm0}^{} \!-\hsm \al_{nn0}^{}\al_{mm0}^{})
=\frac{5}{4}\!\sum_{j=1}^{\infty}\rh_j^4\al_{nnj}^{}\al_{mmj}^{} \!+\!8\hs\rh^2\al_{nnmm}^{} 
\!-\!\frac{7}{2} (1 \!+\hsm \rh^2) \hsm\!\sum_{j=1}^{\infty}\!\rh_j^2\al_{nnj}^{}\al_{mmj}^{} \hs ,
\\[0mm]
\label{app-eq:GRET-nnmm-Mj4-b}
&\hspace*{-0.5cm}
9\hs\rh^2\tilde{\rho}_{nm0}^2 + \frac{1}{4} (1\!-\hsm \rh^2)^4 \!\sum_{j=1}^{\infty} \!\rh_j^{-4} \al_{nmj}^2
= \frac{5}{4}\!\sum_{j=1}^{\infty}\!\rh_j^4 \al_{nmj}^2 \!+\! 
\frac{1}{4}\Big[\hsm (1\!+\hsm\rh^2)^2 \!-\! 4(1\!-\hsm\rh^2)^2\Big] \al_{nm0}^2 
\nn \\
&\hspace*{-2mm} 
+\!\Big[2(1\!+\hsm\rh^2)^2 \!+\!(1\!-\hsm\rh^2)^2\Big] \al_{nnmm}^{}
\!-\!\frac{7}{2} (1\!+\hsm\rh^2)\! \sum_{j=1}^{\infty} \rh_j^2\al_{nmj}^2 
\!-\!\frac{1}{2}(1\!-\hsm\rh^2)^2 (1\!+\hsm\rh^2) 
\!\sum_{j=1}^{\infty} \!\rh_j^{-2} \al_{nmj}^2 \hs ,
\\
\label{app-eq:GRET-nnmm-Mj4-c}
&\hspace{-0.5cm} 
\sum_{j=1}^{\infty}\!\rh_j^4 (\al_{nnj}^{}\al_{mmj}^{} \!-\! \al_{nmj}^2)
=(1\!-\hsm\rh^2)^2\al_{nnmm}^{} \!+\!2(1\!+\hsm\rh^2) \!
\sum_{j=1}^{\infty} \!\rh_j^2 (\al_{nnj}^{}\al_{mmj}^{}\!-\!\al_{nmj}^2) \hs ,
\end{align}
\eeqs
}
which will be proved in Appendix\,\ref{app:E3}.\ 
Then, we calculate the $O(s_0^2)$ term as follows:
{\small 
\begin{align}
\xoverline{\M}_2^{} 
=& -\!\frac{~23\!+\hsm20\rh^2\!+\hsm23\rh^4 \!+\! 10(1\!+\hsm\rh^2\!+\hsm\rh^4)\ctt\,}
{864\hs\MGnn\hs\rh^4} \!\sum_{j=1}^{\infty}\rh_j^2\al_{nnj}^{}\al_{mmj}^{}
\nn\\
& -\!\frac{1\!+\!\rh^2}{\,54\MGnn \hs \rh^2\,}(9\tilde{\rho}_{nn0}^{}\tilde{\rho}_{mm0}^{} \!-\hsm \al_{nn0}^{}\al_{mm0}^{})
\!+\!\frac{\,3\!+\hsm 314\rh^2\!+\hsm3\rh^4 \!+\!(17\!+\hsm 94\hs\rh^2\!+\hsm 17\rh^4)\ctt\,}{3456 \MGnn \hs \rh^4}
\!\sum_{j=1}^{\infty}\!\rh_j^2\al_{nmj}^2
\nn\\
& +\!\frac{(1\!-\hsm\rh^2)^2 (1\!+\hsm\rh^2)}{3456\hs\MGnn\hs \rh^4}(87\!+\hsm 97\ctt) \al_{nm0}^2 
\hsm +\!\frac{\,(1\!+\hsm\rh^2)(9\!-\hsm\ctt)\,}{864\MGnn \hs \rh^4} \!\!\left[\hsm 9\hs\rh^2\tilde{\rho}_{nm0}^2
\!+\!\frac{1}{4} (1\!-\!\rh^2)^4\!\sum_{j=1}^{\infty} \!\rh_j^{-4} \al_{nmj}^2\!\right]
\nn\\
& +\!\frac{\,(1\!+\hsm\rh^2)(1\!+\hsm3 \ctt)\,}{864\hs \MGnn \hs \rh^4} \!
\sum_{j=1}^{\infty}\!\rh_j^4\al_{nnj}^{}\al_{mmj}^{}
\!-\!\frac{\,(1\!+\hsm\rh^2)(39\!+\!17 \ctt)\,}{3456 \hs\MGnn\hs \rh^4} \!
\sum_{j=1}^{\infty}\!\rh_j^4 \al_{nmj}^2
\nn\\
&+\! \frac{\,(1\!+\!3\ctt)\,}{\,3456\hs \MGnn\hs \rh^4\,}\!\sum_{j=1}^{\infty}\!\rh_j^6\al_{nnj}^{}\al_{mmj}^{} 
\!+\!\frac{(27\!+\!\ctt)}{\,3456\hs\MGnn\hs\rh^4\,}\!\!
\sum_{j=1}^{\infty}\!\rh_j^6\al_{nmj}^2 \hs.
\label{eq:M2-nnmm}
\end{align}
}
We find that the above amplitude \eqref{eq:M2-nnmm} vanishes  
by applying the following identities:
\beqs
\label{app-eq:GRET-nnmm-Mj6}
\begin{align}
\label{app-eq:GRET-nnmm-Mj6-a}
\hspace*{-10mm}
\sum_{j=1}^{\infty}\!\rh_j^6\al_{nnj}^{}\al_{mmj}^{}
& =\frac{5}{2} (1\!+\!\rh^2) \!\sum_{j=1}^{\infty}\!\rh_j^4 \al_{nnj}^{}\al_{mmj}^{}
\!+\!4\rh^2 (1\!+\!\rh^2) \al_{nnmm}^{}
\nn\\
&\hspace*{4mm} 
-6\hs\rh^2 \!\sum_{j=1}^{\infty} \!\rh_j^2 \al_{nnj}^{}\al_{mmj}^{}
\!-\!4\hs\rh^2 \!\sum_{j=1}^{\infty} \!\rh_j^2 \al_{nmj}^2 \hs,
\\
\hspace*{-10mm}
\label{app-eq:GRET-nnmm-Mj6-b}
\sum_{j=1}^{\infty}\!\rh_j^6 \al_{nmj}^2
&=\frac{5}{2} (1\!+\!\rh^2)\!\sum_{j=1}^{\infty}\!\rh_j^4 \al_{nmj}^2
\!+\!\left[\frac{1}{2} (1\!+\!\rh^2)^3 \!+\! 2\hs\rh^2 (1\!+\!\rh^2)\hsm\right] \hsm\!\al_{nnmm}^{}
\nn\\
&\hspace*{4mm} 
-(1\!+\!\rh^4)\!\sum_{j=1}^{\infty}\!\rh_j^2\al_{nnj}^{}\al_{mmj}^{}
\!-\!2(1\!+\!\rh^2)^2 \!\sum_{j=1}^{\infty}\!\rh_j^2\al_{nmj}^2 \hs ,
\end{align}
\eeqs
which will be proved in Appendix\,\ref{app:E3}.\ 

Finally, under high energy expansion the leading-order inelastic scattering amplitude is of $O(s_0^1)$ and 
has been presented in Eq.\eqref{eq:LO-amp-nnmm} of the main text:
\begin{equation}
\ka^2 s_0^{}\hs \xoverline{\M}_1^{} = \M_0^{}[h^n_L\hs h^n_L \ito h^m_L\hs h^m_L] \,.
\end{equation}

\subsection{\hspace*{-2mm}Proving GRET for Inelastic Scattering \boldmath{$(n,n)\ito (m,m)$}}
\vspace*{1.5mm}
\label{app:E2}

As discussed in the mian text, we first connect the KK Goldstone coupling $\tilde{\be}_{nmj}^{}$ 
to the KK graviton coupling $\al_{nmj}^{}$ by using Eq.\eqref{eq:al-tbe-LLT-55T}.\
Then, we derive the following identities:
{\small 
\beqs
\begin{align}
&\hspace*{-10mm}
36\hs\rh^4 \!\sum_{j=0}^{\infty}\!\tilde{\be}_{nnj}^{}\tilde{\be}_{mmj}^{}
=\sum_{j=1}\!\rh_j^8 \al_{nnj}^{}\al_{mmj}^{} 
\!-\!4(1\!+\hsm\rh^2) \!\sum_{j=1}^{\infty}\!\rh_j^6 \al_{nnj}^{}\al_{mmj}^{} 
\!+\!36\hs\rh^4 \!\sum_{j=0}^{\infty}\!\al_{nnj}^{}\al_{mmj}^{} 
\nn\\
& +\!\Big[6(1\!+\hsm\rh^4) \!+\hsm 16\hs\rh^2\Big]\!\sum_{j=1}^{\infty}\!\rh_j^4 \al_{nnj}^{}\al_{mmj}^{}
\!-\!24\hs\rh^2(1\!+\hsm\rh^2) \!\sum_{j=1}^{\infty} \!\rh_j^2 \al_{nnj}^{}\al_{mmj}^{} \hs,
\\
&\hspace*{-10mm}36\hs\rh^4 \!\sum_{j=0}^{\infty}\!\tilde{\be}_{nmj}^2
=\sum_{j=1} \!\rh_j^8 \al_{nmj}^2 \!-\!4(1\!+\hsm\rh^2) \!\sum_{j=1}^{\infty} \!\rh_j^6 \al_{nmj}^2
\!+\!\Big[6(1\!+\hsm\rh^4)\!+\! 16\hs\rh^2\Big] 
\!\sum_{j=1}^{\infty} \!\rh_j^4 \al_{nmj}^2
\nn \\
& -\!4(1\!+\hsm\rh^2)(1\!+\hsm 4\rh^2\!+\hsm\rh^4) \!\sum_{j=1}^{\infty} \!\rh_j^2\al_{nmj}^2\!+\!(1\!+\hsm4\rh^2\!+\hsm\rh^4)^2 \!\sum_{j=0}^{\infty}\!\al_{nmj}^2 \hs.
\end{align}
\eeqs
}
Using Eq.\eqref{app-eq:GRET-nnmm-Mj2} and Eq.\eqref{app-eq:GRET-nnmm-Mj6}, we derive the following relations:
{\small 
\beqs
\label{app-eq:GRET-nnmm-2beta}
\begin{align}
\label{app-eq:GRET-nnmm-2beta-a}
&\hspace*{-4mm} 
\sum_{j=1}^{\infty}\!\rh_j^8 \al_{nnj}^{}\al_{mmj}^{}
= 36\hs\rh^4 \!\sum_{j=0}^{\infty}\!\tilde{\be}_{nnj}^{}\tilde{\be}_{mmj}^{}
\!+\!4(1\!+\hsm\rh^2\!+\hsm\rh^4) \sum_{j=1}^{\infty} \!\rh_j^4\al_{nnj}^{}\al_{mmj}^{} 
\nn\\
& +\!8\hs\rh^2(1\!+\hsm\rh^2) \!\sum_{j=1}^{\infty} \!\rh_j^2 \al_{nnj}^{}\al_{mmj}^{}
\!-\!\frac{18\hs\rh^2}{\,1\!+\hsm\rh^2\,}\!\sum_{j=1}^{\infty}\!
\rh_j^2(\al_{nnj}^{}\al_{mmj}^{} \!+\! 2\al_{nmj}^2)\hs,
\\
\label{app-eq:GRET-nnmm-2beta-b}
&\hspace{-4mm} 
\sum_{j=1}^{\infty} \!\rh_j^8 \al_{nmj}^2
= 36\hs\rh^4 \!\sum_{j=0}^{\infty}\!\tilde{\be}_{nmj}^2 \!+\! 
4(1\!+\hsm\rh^2\!+\hsm\rh^4) \!\sum_{j=1}^{\infty}\!\rh_j^4 \al_{nmj}^2
\!-\!3(1\!+\!\rh^2)(1\!-\hsm\rh^2)^2 \!\sum_{j=1}^{\infty}\!\rh_j^2 \al_{nnj}^{}\al_{mmj}^{}
\nn\\
& -\!2(1\!+\hsm\rh^2)(1\!-\hsm6\rh^2\!+\hsm\rh^4) \sum_{j=1}^{\infty} \rh_j^2 \al_{nmj}^2
-\frac{\,(1\!+\hsm4\rh^2\!+\hsm \rh^4)^2\,}{2(1\!+\hsm\rh^2)}\! 
\sum_{j=1}^{\infty} \!\rh_j^2 (\al_{nnj}^{}\al_{mmj}^{}\!+\!2\al_{nmj}^2)\hs.
\end{align}
\eeqs
}
\hspace*{-2mm}
Then, we substitute Eq.\eqref{app-eq:GRET-nnmm-Mj4-c} into the above formulas 
to eliminate $\rh_j^4$ terms 
and re-express Eq.\eqref{app-eq:GRET-nnmm-2beta-a} as follows:
{\small 
\begin{align}
\hspace*{-5mm}
\sum_{j=1}^{\infty}\!\rh_j^8\al_{nnj}^{}\al_{mmj}^{}
&= 36\hs\rh^4 \!\sum_{j=0}^{\infty}\!
\tilde{\be}_{nnj}^{}\tilde{\be}_{mmj}^{}
\!+\!4(1\!+\hsm\rh^2\!+\hsm\rh^4) \!\sum_{j=1}^{\infty}\! 
\rh_j^4 \al_{nmj}^2 
\!+\!8\hs\rh^2 (1\!+\hsm\rh^2)\!\sum_{j=1}^{\infty}\!\rh_j^2\al_{nnj}^{}\al_{mmj}^{}
\nn\\
&\hspace{0.5cm} 
+\!4(1\!+\hsm\rh^2\!+\hsm\rh^4) \!\!\[2(1\!+\hsm\rh^2) \!+\!\frac{(1\!-\hsm\rh^2)^2}{\,2(1\!+\hsm\rh^2)\,}\]\! 
\sum_{j=1}^{\infty} \!\rh_j^2 
\(\al_{nnj}^{}\al_{mmj}^{} \!-\! \al_{nmj}^2\)
\nn\\
&\hspace{0.5cm} 
-\frac{18\hs\rh^2}{\,1\!+\hsm\rh^2\,} \!\sum_{j=1}^{\infty} \!\rh_j^2 \big(\al_{nnj}^{}\al_{mmj}^{} \!+\!2\al_{nmj}^2\big)\hs. 
\end{align}
}
\hspace*{-3.5mm}
With these, we can further simplify the leading-order amplitude $\xoverline{\M}_1^{}$
for the inelastic scattering channel $h^n_Lh^n_L\!\ito\! h^m_Lh^m_L$
(as given by Eq.\eqref{eq:LO-amp-nnmm} in the main text):
\begin{align}
\xoverline{\M}_1^{} = \frac{\,1\!+\hsm3\ctt\,}{96} \sum_{j=0}^{\infty}\!\tilde{\be}_{nnj}^{}\tilde{\be}_{mmj}^{}
-\frac{\,37\!+\hsm 11\ctt\,}{48 s_{\theta}^2} \!\sum_{j=0}^{\infty}\!\tilde{\be}_{nmj}^2
=-\frac{\,(7\!+\hsm\ctt)^2\,}{64 s_{\theta}^2}\hs\tilde{\be}_{nnmm}^{} \hs.
\end{align}
This agrees to the leading-order inelastic scattering amplitude 
\eqref{eq:LO-Amp-4phi-nnmm} of the KK gravitational Goldstone bosons.\ 
Hence, we have proved the GRET for the inelastic scattering channel
$h^n_Lh^n_L\hsm\ito\hsm h^m_Lh^m_L$.

\vspace*{1mm}
\subsection{\hspace*{-2mm}Proving Sum Rules for Inelastic Scattering \boldmath{$(n,n)\ito (m,m)$}}
\vspace*{1mm}
\label{app:E3}

In this sub-Appendix, we present a proof of the identities 
\eqref{app-eq:GRET-nnmm-Mj4} and \eqref{app-eq:GRET-nnmm-Mj6}
which are used in Appendix\,\ref{app:E1}.

\vs 

\subsubsection*{$\blacklozenge$~Derivation of the Inelastic Sum Rule \eqref{app-eq:GRET-nnmm-Mj4}}
\vspace*{1mm}

We first compute the following sums:
\beqs 
\begin{align}
\label{app-eq:derive-Mj4-1}
& \sum_{j=1}^{\infty}\!\mathbb{M}_j^4\al_{nnj}^{}\al_{mmj}^{}
= \sum_{j=1}^{\infty}\!2\hs\MGn (\al_{nnj}^{} \!-\!\tilde{\al}_{nnj}^{})  2\hs\MGm(\al_{mmj}^{}\!-\!\tilde{\al}_{mmj}^{}) 
\nn\\
& =\sum_{j=1}^{\infty}\! 
4\MGnn\MGmm (\al_{nnj}^{}\al_{mmj}^{} \!-\hsm\al_{nnj}^{}\tilde{\al}_{mmj}^{}
\!-\hsm\tilde{\al}_{nnj}^{}\al_{mmj}^{} \!+\hsm 
\tilde{\al}_{nnj}^{}\tilde{\al}_{mmj}^{})
\nn \\
& =-4\MGnn\MGmm \al_{nnmm}^{} 
+2(\MGnn\!+\!\MGmm)\!\hsm\sum_{j=1}^{\infty}\!\MGjj \al_{nnj}^{}\al_{mmj}^{}
\!+\!4\MGnn\MGmm \lrb{\vv_n^2 \vv_m^2} \hs ,\hspace*{8mm}
\\
\label{app-eq:derive-Mj4-2}
&\sum_{j=1}^{\infty}\!\mathbb{M}_j^4\al_{nmj}^2
=\sum_{j=1}^{\infty}\!
\Big[\hsm (\MGnn\!+\!\MGmm)^2 \al_{nmj}^{} 
\!-\! 2\hs\MGnn\MGmm\tilde{\al}_{nmj}^{}\Big]^{\!2}
\nn\\
&=-\big(\MGnn\!+\!\MGmm\big)^2\al_{nnmm}^{}
\!+\!2\big(\MGnn\!+\!\MGmm\big)\!\hsm 
\sum_{j=1}^{\infty}\!\MGjj\al_{nmj}^2 \!+\! 
4\hs\MGnn\MGmm \lrb{\vv_n^2 \vv_m^2} \hs. 
\end{align}
\eeqs 
Then, we take the difference between the above equations and derive the following:
\begin{equation}
\hspace*{-3mm}
\sum_{j=1}^{\infty}\!\MG_j^4 \hsm
\big( \al_{nnj}^{}\al_{mmj}^{}\!-\hsm\al_{nmj}^2 \big)
\hsm =\hsm (\MGnn\!-\hsm\MGmm)^{\hsm 2}\hsm\al_{nnmm}^{} 
\!+\hsm 2(\MGnn\!+\hsm\MGmm) \!\!\sum_{j=1}^{\infty}\!\MGjj 
(\al_{nnj}^{}\al_{mmj}^{} \!-\hsm \al_{nmj}^2),~~
\end{equation}
which just reproduces the identity \eqref{app-eq:GRET-nnmm-Mj4-c}.

\vs

Next, we compute the following products of KK masses and couplings:
\begin{align}\label{app-eq:derive-Mj4-3}
& \MGnn\MGmm \lrb{\vv_n^2 \vv_m^2} 
=\MGnn\MGmm \!\sum_{j=0}^{\infty} \!\tilde{\rho}_{nnj}^{}\tilde{\rho}_{mmj}^{}  
\nn \\
& =\MGnn\MGmm\tilde{\rho}_{nn0}^{}\tilde{\rho}_{mm0}^{}+\sum_{j=1}^{\infty} \!\frac{1}{\,9\,}{(\MGnn\!-\!\MGjj)
{(\MGmm\!-\!\MGjj)\al_{nnj}^{}}\al_{mmj}^{}}
\nn\\
& =\MGnn\MGmm\tilde{\rho}_{nn0}^{}\tilde{\rho}_{mm0}^{} +\!\frac{1}{\,9\,} \MGnn\MGmm (\al_{nnmm}^{}\!-\!\al_{nn0}^{}\al_{mm0}^{})
\nn\\
& \hspace*{4mm}-\!\frac{1}{\,9\,}(\MGnn\!+\!\MGmm) 
\hsm\!\sum_{j=1}^{\infty}\!\MGjj\al_{nnj}^{}\al_{mmj}^{}\!+\!\frac{1}{\,9\,}\!\sum_{j=1}^{\infty}\!\mathbb{M}_j^4\al_{nnj}^{}\al_{mmj}^{} \hs.
\end{align}
Substituting Eq.\eqref{app-eq:derive-Mj4-3} into Eq.\eqref{app-eq:derive-Mj4-1},
we can readily deduce Eq.\eqref{app-eq:GRET-nnmm-Mj4-a}.\ 
Then, we further rewrite Eq.\eqref{app-eq:derive-Mj4-3} as follows:
\begin{align}
\label{app-eq:derive-Mj4-4}
& \MGnn\MGmm \lrb{\vv_n^2 \vv_m^2} 
= \MGnn\MGmm\!\sum_{j=0}^{\infty}\!\tilde{\rho}_{nmj}^2
\nn \\
& = \MGnn\MGmm\tilde{\rho}_{nm0}^2
\!+\hsm\frac{1}{\,9\,}\!\sum_{j=1}^{\infty}\!\mathbb{M}_j^4\al_{nmj}^2
\!+\hsm\frac{1}{\,36\,}{\big(\MGnn\!-\!\MGmm\big)^{\!4}}
\hsm\sum_{j=1}^{\infty}\!\frac{\,\al_{nmj}^2\,}{\mathbb{M}_j^4}
\nn \\
&\hspace*{4mm}
+\!\frac{1}{\,36\,}\big[(\MGnn\hsm\!+\!\MGmm)^2
\!-\!4(\MGnn\!\!-\hsm\MGmm)^2\big]
(\al_{nnmm}^{}\hsm\!-\!\al_{nm0}^2)
\!-\!\frac{1}{\hs 9\hs}(\MGnn\!+\!\MGmm)\hsm\!
\sum_{j=1}^{\infty}\!\MGjj\al_{nmj}^2
\nn\\
&\hspace*{4mm}
+\frac{1}{\,18\,}
{(\MGnn\!-\!\MGmm)^2(\MGnn\!+\!\MGmm)}
\!\sum_{j=1}^{\infty}\!\frac{\,\al_{nmj}^2\,}{\MGjj} \hs.
\end{align}
Thus, substituting Eq.\eqref{app-eq:derive-Mj4-4} into Eq.\eqref{app-eq:derive-Mj4-2},
we can derive Eq.\eqref{app-eq:GRET-nnmm-Mj4-b}.

\vs 
\subsubsection*{$\blacklozenge$~Derivation of Inelastic Sum Rule 
\boldmath{\eqref{app-eq:GRET-nnmm-Mj6}} }
\vs 

Next, we compute the following sums:
\beqs 
\begin{align}
\hspace*{-9mm}
\sum_{j=1}^{\infty}\mathbb{M}_j^6\al_{nnj}^{}\al_{mmj}^{}
&= 4\hs\MGnn\MGmm
\sum_{j=0}^{\infty}\MGjj 
(\al_{nnj}^{}\!-\!\tilde{\al}_{nnj}^{}) 
(\al_{mmj}^{}\!-\!\tilde{\al}_{mmj}^{}) 
\nn \\
\hspace*{-9mm}
&= 4\hs\MGnn\MGmm\!\sum_{j=0}^{\infty}\!
\MGjj\al_{nnj}^{}\al_{mmj}^{}
\!-\!8\hs\MGnn\MGmm\hsm
\Big(\MGnn \lrb{\uu_n^2 \vv_m^2} 
\!+\hsm\MGmm \lrb{\vv_n^2 \uu_m^2}\hsm\Big)
\nn \\
& \hspace*{4mm}
+\!8\MGnn\MGmm(\MGnn\!+\!\MGmm) 
\lrb{\vv_n^2 \vv_m^2}
\!+\!4\hs\MGnn\MGmm\!\sum_{j=0}^{\infty}\!
\MGjj\tilde{\al}_{nnj}^{}\tilde{\al}_{mmj}^{} \hs, 
\label{eq:18a}
\\
\sum_{j=1}^{\infty}\!\mathbb{M}_j^6\al_{nmj}^2
&= \sum_{j=0}^{\infty}\!
\MGjj\Big[(\MGnn\!+\!\MGmm)^2\al_{nmj}
\!-\!2\MGnn\MGmm\tilde{\al}_{nmj}\Big]^{\!2}
\nn\\
&= \big(\MGnn\!+\!\MGmm)^2\!
\sum_{j=0}^{\infty}\!\MGjj\al_{nmj}^2
\!-\!4\hs\MGn\MGm
\big(\MGnn\!+\!\MGmm\big)^2 \lrb{\uu_n^{}\vv_n^{}\uu_m^{}\vv_m^{}}
\nn \\
&\hspace*{4mm}
+\!8\hs\MGnn\MGmm\big(\MGnn\!+\!\MGmm\big) 
\lrb{\vv_n^2 \uu_n^2}
\!+\!4\hs\MGnn\MGmm\!\sum_{j=0}^{\infty}\!
\MGjj \tilde{\al}_{nmj}^2 \hs .
\label{eq:18b}
\end{align}
\eeqs 
We note that Eq.\eqref{app-eq:MnMmMlMq-anmlq} can be directly extended to the warped KK gravity theory
with the same structure except the replacements $\Mn\!\ito\hsm \MGn$ and
$(\ff_n^{},\hs \fft_n^{})\hsm\ito\hsm (\uu_n^{},\hs \vv_n^{})$.\ 
Using this extended equation, we can solve 
$\lrb{\uu_n^{} \uu_m^{} \vv_{\ell}^{} \vv_q^{}}$ terms.\ 
Thus, we need to compute the sum $\dis\sum_{j=0}^{\infty}\!\MGjj\tilde{\al}_{nnj}^{}\tilde{\al}_{mmj}^{}$
in Eq.\eqref{eq:18a} and
the sum $\dis\sum_{j=0}^{\infty}\!\MGjj\tilde{\al}_{nmj}^2$ in \eqref{eq:18b}.\ 
Using Eq.\eqref{app-eq:al-tal-general}, we derive the following relation:
\begin{align}
\hspace*{-3mm}
&\sum_{j=0}^{\infty}\!\MGjj\tilde{\al}_{nnj}^{}\tilde{\al}_{mmj}^{} 
=\sum_{j=0}^{\infty}
\lrb{\hsm v_n^2(3A'\!+\!\pd_z^{})v_j^{}\hsm}\!
\lrb{\hsm v_m^2(3A'\!+\!\pd_z^{})v_j^{}\hsm}
\nn\\
\hspace*{-3mm}
&= 4\lrb{v_n^{}(\MGn u_n^{}\!-\!3A'v_n^{})v_m^{}(\MGm u_m^{}\!-\!3A'v_m)}
\\
\hspace*{-3mm} 
&= 4\hs\MGn\MGm \lrb{\uu_n \vv_n \uu_m \vv_m} 
\!-\! 12\hs\MGn \lrb{A' \uu_n \vv_n \vv_m^2}
\!+\!12\hs\MGm \lrb{\hsm A'\vv_n^2 \uu_m \vv_m} 
\!+\! 36\hs\lrb{\hsm A'' \vv_n^2 \vv_m^2} \hs.
\hspace*{8mm} 
\nn 
\end{align}
Then, we can derive the following identities:
\beqs 
\begin{align}
\lrb{\hsm A'' \vv_n^2 \vv_m^2} 
&=\frac{1}{\hs 4\hs}\hsm\Big(\MGn\lrb{A' \uu_n \vv_n \vv_m^2} \!+\hsm\MGm\lrb{A'\vv_n^2 \uu_m \vv_m}\!\Big) \hs,
\\
6\lrb{A'\uu_n \vv_n \vv_m^2} &=\,
\MGn\lrb{\uu_n^2 \vv_m^2} \!+\!2\hs\MGm
\lrb{\uu_n^{}\vv_n^{}\uu_m^{}\vv_m^{}} \!-\!\MGn \lrb{\vv_n^2 \vv_m^2} \hs,
\\
6\lrb{A'\vv_n^2\uu_m^{}\vv_m^{}\hsm} &=\, 
\MGm\lrb{\uu_n^2 \vv_m^2} \!+\!2\hs\MGn
\lrb{\uu_n \vv_n \uu_m \vv_m} \!-\!\MGm \lrb{\vv_n^2 \vv_m^2} \hs. 
\end{align}
\eeqs 
With these identities, we can prove Eq.\eqref{app-eq:GRET-nnmm-Mj6}.

\vs 

In passing, for the current archive version of this paper [arXiv:2406.12713.v3], we have corrected
some minor typos in its journal version\,\cite{Hang:2024uny}.\  
For example, we have corrected a typo in Eq.(3.29b) of Ref.\,\cite{Hang:2024uny},
as shown in the present Eq.\eqref{eq:Amp3phi-E0}.\  
For Appendix\,\ref{app:A}, in the second line just below Eq.(A.5) of Ref.\,\cite{Hang:2024uny}, 
the limit $\theta\ito\infty$ should be replaced by $k\ito\infty$
as in the present text below Eq.\eqref{eq:KCond-M123}.\ 
Finally, in Appendix\,\ref{app:C} we have removed two typos in 
Eqs.(C.5c) and (C.5e) of Ref.\,\cite{Hang:2024uny}, and the corrected formulas are given
in the current Eqs.\eqref{app-eq:FR-pph} and \eqref{app-eq:FR-ppp}.\ 





\end{document}